\documentclass[preprint,number,sort&compress,longtitle]{elsarticle}   % version for ArXiV
\pdfoutput=1 % only if pdf/png/jpg images are used -- try this for ArXiV
\journal{Nuclear Instruments and Methods in Physics Research A}

\usepackage[switch]{lineno}

\usepackage[english]{babel}
\usepackage{times}
\usepackage{graphicx}
\usepackage{fontenc}
\usepackage{mathptmx}
\usepackage{subfigure}
\usepackage{amsmath,amssymb}
\usepackage{upgreek}
\usepackage{xspace}
\usepackage{color}
\definecolor{darkred}{rgb}{0.5,0,0}
\definecolor{darkblue}{rgb}{0,0,0.5}
\definecolor{firebrick}{rgb}{0.75,0.125,0.125}
\definecolor{darkgreen}{rgb}{0,0.5,0}
\usepackage[pdftex,colorlinks=true]{hyperref}
\usepackage{units}
\usepackage{booktabs}
\usepackage{epstopdf}
\usepackage{amsmath}

\graphicspath{{plots/}}
\makeatletter
\def\elsartstyle{%
   \def\normalsize{\@setfontsize\normalsize\@xiipt{14.5}}
   \def\small{\@setfontsize\small\@xipt{13.6}}
    \let\footnotesize=\small
    \def\large{\@setfontsize\large\@xivpt{18}}
    \def\Large{\@setfontsize\Large\@xviipt{22}}
    \skip\@mpfootins = 18\p@ \@plus 2\p@
    \normalsize
}
\@ifundefined{square}{}{\let\Box\square}
\makeatother

\def\Offline{\mbox{$\overline{\textrm%
{Off}}$\hspace{.05em}\protect\raisebox{.4ex}%
{$\protect\underline{\textrm{line}}$}}\xspace}

\makeatletter
\def\blfootnote{\xdef\@thefnmark{}\@footnotetext}
\makeatother

\begin{document}

%%%%%%%%%
\begin{frontmatter}
\title{The Pierre Auger Cosmic Ray Observatory\tnoteref{mytitlenote}}
\tnotetext[mytitlenote]{Published in Nuclear Instruments and Methods in Physics Research A \\ as \href{http://dx.doi.org/10.1016/j.nima.2015.06.058}{DOI:10.1016/j.nima.2015.06.058}}

%\date{\today}

\author[auger]{\bf The Pierre Auger Collaboration: \rm}
\ead{auger\_spokespersons@fnal.gov}
\ead[url]{http://www.auger.org}

\author[43]{A.~Aab}
\author[65]{P.~Abreu}
\author[54]{M.~Aglietta}
\author[82]{E.J.~Ahn}
\author[29]{I.~Al Samarai}
\author[30]{J.N.~Albert}
\author[17]{I.F.M.~Albuquerque}
\author[1]{I.~Allekotte}
\author[85]{J.~Allen}
% ...

\cortext[cor1]{Corresponding author}
\fntext[mort]{Deceased.}
\fntext[fna]{Now at Konan University}
\fntext[fnb]{Also at the Universidad Autonoma de Chiapas; on leave of absence from Cinvestav}
\fntext[fnc]{Now at Unidad Profesional Interdisciplinaria de Ingenier\'{\i}a y Tecnolog\'{\i}as
Avanzadas del IPN, M\'{e}xico, D.F., M\'{e}xico}
\fntext[fnd]{Now at Universidad Aut\'{o}noma de Chiapas, Tuxtla Guti\'{e}rrez, Chiapas, M\'{e}xico}
\fntext[fne]{Also at Vrije Universiteit Brussels, Belgium}

\address[auger]{Pierre Auger Collaboration, Av.\ San Mart\'{\i}n Norte 306, 5613 Malarg\"ue, Mendoza, Argentina}
\address[1]{Centro At\'{o}mico Bariloche and Instituto Balseiro (CNEA-UNCuyo-CONICET), San
Carlos de Bariloche,
Argentina}
% ...
\address[17]{Universidade de S\~{a}o Paulo, Instituto de F\'{\i}sica, S\~{a}o Paulo, SP,
Brazil}
% ...
\address[29]{Institut de Physique Nucl\'{e}aire d'Orsay (IPNO), Universit\'{e} Paris 11, CNRS-IN2P3,
France}
\address[30]{Laboratoire de l'Acc\'{e}l\'{e}rateur Lin\'{e}aire (LAL), Universit\'{e} Paris 11, CNRS-IN2P3,
France}
% ...
\address[43]{Universit\"{a}t Siegen, Siegen,
Germany}
% ...
\address[54]{Osservatorio Astrofisico di Torino  (INAF), Universit\`{a} di Torino and Sezione INFN,
Torino,
Italy}
% ...
\address[65]{Laborat\'{o}rio de Instrumenta\c{c}\~{a}o e F\'{\i}sica Experimental de Part\'{\i}culas - LIP and
Instituto Superior T\'{e}cnico - IST, Universidade de Lisboa - UL,
Portugal}
% ...
\address[82]{Fermilab, Batavia, IL,
USA}
% ...
\address[85]{New York University, New York, NY,
USA}
% ...

%\author{The Pierre Auger Collaboration \\
%Pierre Auger Collaboration, Av.\ San Mart\'{\i}n Norte 306, 5613 Malarg\"ue, Mendoza, Argentina \\
%E-mail: auger\_spokespersons@fnal.gov \\
%URL: http://www.auger.org}

\begin{abstract}
The Pierre Auger Observatory, located on a vast, high plain in western Argentina, is the world's largest 
cosmic ray observatory.  The objectives of the Observatory are to probe the origin and characteristics 
of cosmic rays above $10^{17}$ eV and to study the interactions of these, the most energetic particles observed in nature.  
The Auger design features an array of 1660 water Cherenkov particle detector stations spread over 3000\,km$^2$ overlooked by
24 air fluorescence telescopes.  
In addition, three high elevation fluorescence telescopes overlook a 23.5\,km$^2$, 
61-detector infilled array with 750~m spacing.
The Observatory has been in successful operation since completion 
in 2008 and has recorded data from an exposure %of 30\,000\,km$^2$\,sr\,yr.  
exceeding 40,000\,km$^2$\,sr\,yr.
This paper describes the design and performance of the detectors, related subsystems and infrastructure that make up the  Observatory.
\end{abstract}


\end{frontmatter}

\setcounter{footnote}{0}
%\modulolinenumbers[5]
%\setlength\linenumbersep{0.8ex}
%\linenumbers

\tableofcontents			% do we want one?

\section{Introduction}

%\red{(Paul M)}

The origin of high energy cosmic rays has been a mystery since the discovery of extensive air showers 
in the late 1930s  
\cite{Rossi:1934, Schmeiser:1938, Kolhorster:1938, Auger:1939a, Auger:1939sh}.
%The origin of high energy cosmic rays has been a mystery since 1938 when 
%Pierre Auger and colleagues first recorded extensive air showers
%initiated by particles with energies estimated to be greater than
%$10^{15}$\,eV \cite{Auger:1939sh}.  
In 1962, John Linsley recorded a
cosmic ray event with an energy of $10^{20}$\,eV \cite{Linsley:1963km}.
Subsequent work found more events near and above $10^{20}$\,eV
\cite{Afanasiev:1993, Lawrence:1991cc, Takeda:2002at, Bird:1994uy}.
%\cite{Nagano:2004am}.  
The Pierre Auger Observatory was proposed to
discover and understand the source or sources of cosmic rays with
the highest energies.  

A unique partnership of 18 countries, the Pierre Auger Collaboration came together to pursue this science.  
Construction of the Pierre Auger Observatory was started in 2002 and completed in 2008.  
The purpose of this article is to review the design and performance of the detector systems 
and associated infrastructure that constitute the Observatory.

To achieve the scientific goals, the Collaboration designed the Pierre Auger
Observatory for a high statistics study of cosmic rays at
the highest energies.  Measured properties of the air showers
determine the energy and arrival direction of each cosmic ray.  These properties
also provide a statistical determination of the distribution of
primary masses (cosmic ray composition).  The Pierre Auger Observatory in the 
Province of Mendoza, Argentina, has been taking data since 2004, adding detectors 
as they became active until completion in 2008.

\begin{figure}[t]
\centering
\includegraphics[width=0.48\textwidth]{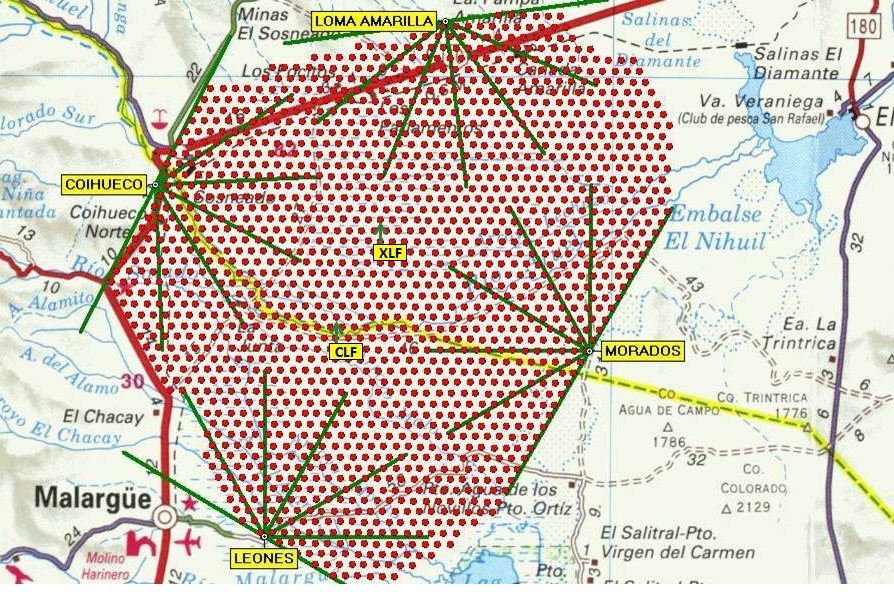}
\caption{The Auger Observatory.  Each dot corresponds to one of the 1660 surface detector stations.  
The four fluorescence detector enclosures are shown, each with the $30^\circ$ field of view of its six telescopes.
Also shown are the two laser facilities, CLF and XLF, near the Observatory center.}
\label{southern_site}
\end{figure}

\begin{figure}[t]
\centering
\centerline{\includegraphics[width=0.48\textwidth]{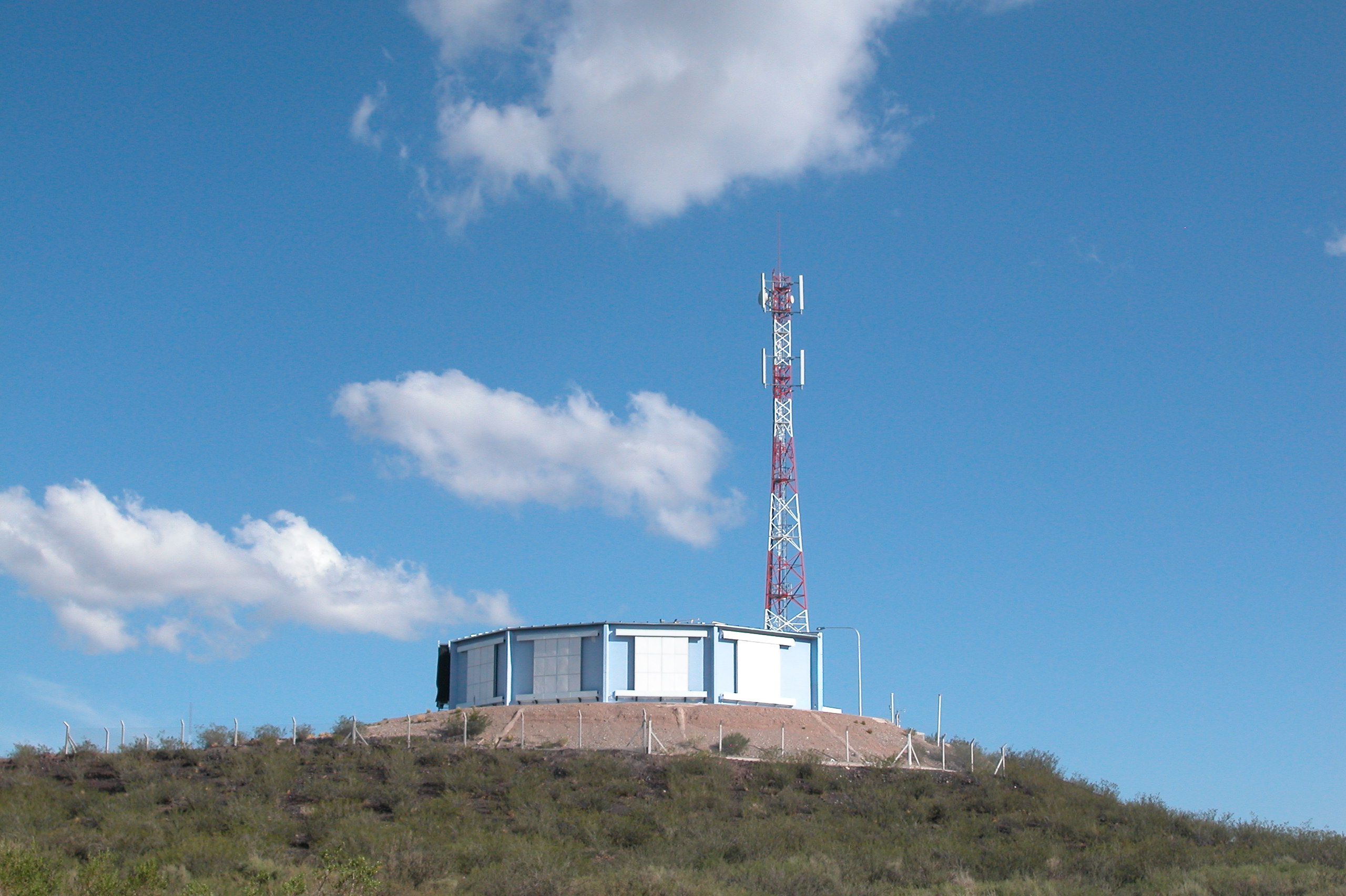}}
\vspace*{1mm}
%\centerline{\includegraphics[width=0.48\textwidth]{SD_station}}   % for NIM
\centerline{\includegraphics[width=0.48\textwidth]{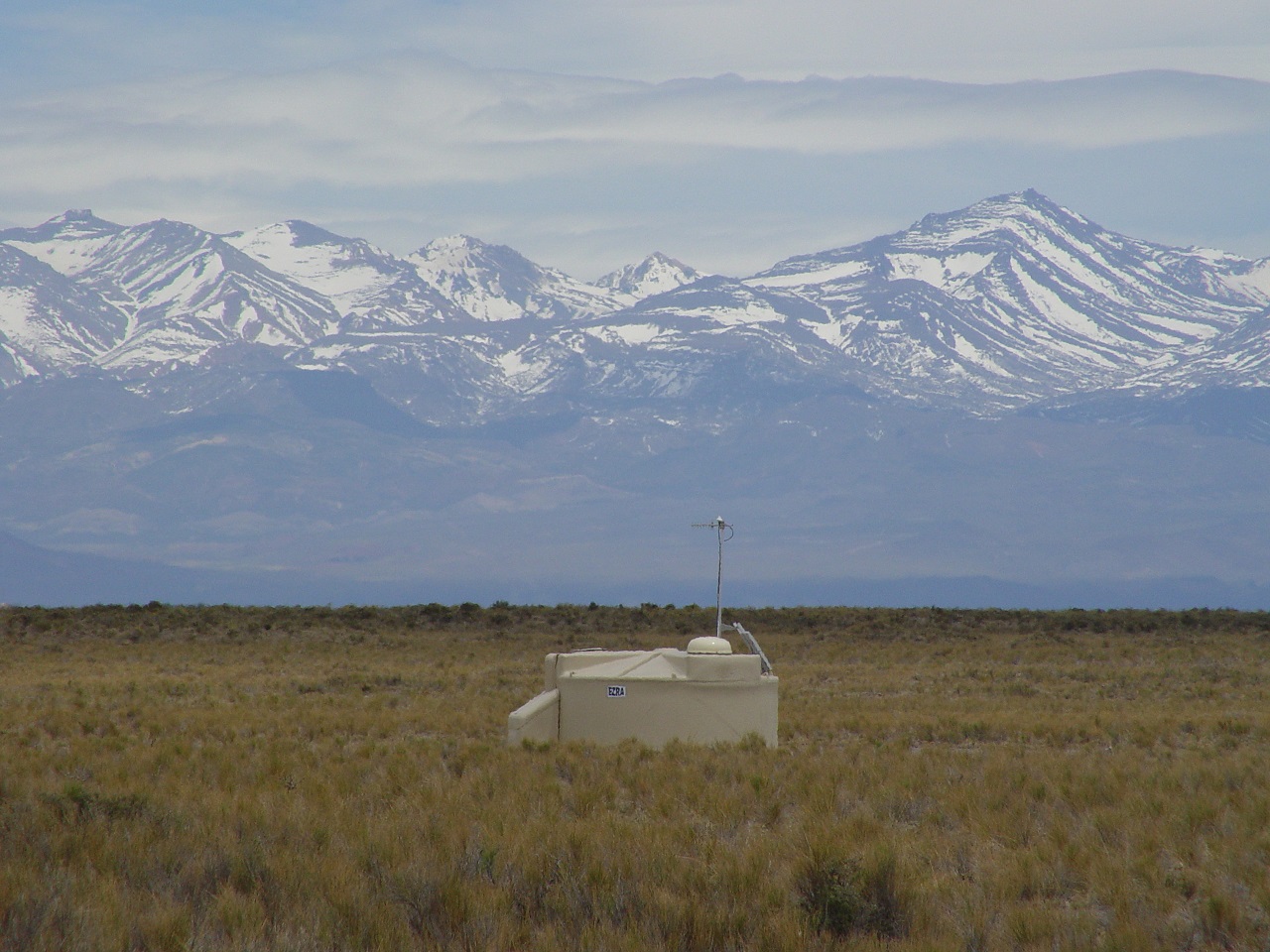}}   % for ArXiV
\caption{The fluorescence detector enclosure Los Leones (top); a surface detector station (bottom).}
\label{fig:FD_SD_stations}
\end{figure}

The Observatory is a hybrid detector, a combination of a large surface detector (SD) and a 
fluorescence detector (FD). The SD  is composed of a baseline array, comprising 1660 water Cherenkov stations placed in a triangular grid with nearest
neighbors separated by 1500 m, and a smaller array (stations separated by 750\,m). 
The surface array is spread over an area of $\sim 3000$~km$^2$, and is depicted in 
Figure~\ref{southern_site}. Figure~\ref{fig:FD_SD_stations} shows examples of FD (left) and SD (right) detector 
elements.
This area is generally flat, with detectors located at altitudes between 1340~m and 1610~m. 
The mean altitude is $\sim 1400$~m, corresponding to an atmospheric overburden of $\sim 875$~g~cm$^{-2}$.
The array is located between latitudes $35.0^\circ$ and $35.3^\circ$~S and 
between longitudes $69.0^\circ$ and $69.4^\circ$~W. 

As an aid to the reader we provide here a list of acronyms used throughout this 
paper\footnote{AERA,	Auger Engineering Radio Array;
AMIGA, Auger Muon and Infilled Ground Array;
APF, Aerosol phase function;
ARQ, Automatic repeat request;
BLS, Balloon Launching Station;
CDAS, Central Data Acquisition System;
CIC, Constant Intensity Cut method;
CLF, Central Laser Facility;
EAS, Extensive air shower;
ELVES: (defined in Section~\ref{sec:Elves});
FADC, Flash Analog-to-Digital Convertor;
FD, Fluorescence Detector;
FRAM, Photometric Robotic Atmospheric Monitor;
GDAS, Global Data Assimilation System;
HAM, Horizontal Attenuation Monitor;
HEAT, High-Elevation Auger Telescopes;
LDF, Lateral distribution function;
MD, Muon Detector;
PLD, Programmable logic device;
PMT, Photomultiplier tube;
SCS, Slow Control System;
SD, Surface Detector;
SDP, Shower-detector plane;
TDMA, Time Division Multiple Access;
ToT, Time over threshold;
VEM, Vertical Equivalent Muon;
XLF, eXtreme Laser Facility
}.

%\noindent[FIGURE] Caption: The Auger Observatory.  Each dot
%corresponds to one of the 1600 surface detector stations.  The four
%fluorescence detector stations are shown along each with the
%$30^\circ$ fields of view of its six telescopes.

%\noindent [FIGURE] Caption: A fluorescence detector station.  A
%surface detector station.

\subsection{Highlights of science results}
The Pierre Auger Observatory has made key measurements of the highest energy cosmic rays.
Cosmic ray showers with zenith angle $<$ 60$^{\circ}$ are defined as vertical showers, while those with
60$^{\circ} $ $ <$  zenith angle $< $ 80$^{\circ} $ are defined as horizontal showers. 
The  energy spectrum is measured with unprecedented precision  using four datasets: hybrid (events measured simultaneously by the SD and FD), SD 750\,m array, SD baseline vertical,  and SD baseline horizontal   \cite{Schulz-ICRC:2013,Abraham:2008ru}.
Thanks to the high statistics of the SD data, a first harmonic analysis was performed in different energy ranges starting from $2.5\times 10^{17}$ eV in a search for dipolar modulations in right ascension \cite{Abreu:2011ve,Sidelnik-ICRC:2013}. The upper limits in the dipole amplitude impose stringent constraints in astrophysical models \cite{Almeida-ICRC:2013,Abreu:2011ve, Abreu:2012ybu}. 

The Auger data provide evidence for a weak  correlation between arrival directions of cosmic rays above 55 EeV and the positions of AGNs with $z < 0.018$ in the VCV catalogue \cite{Abraham:2007si, Abraham:2007bb}.  
 The Collaboration also has performed the measurement of the proton-air cross-section at 57 TeV \cite{Collaboration:2012wt} that favors a moderately slow rise of the cross-section towards higher energies, 
and inferred the proton-proton cross-section, whose value is within one sigma of the best extrapolation from the recent LHC data points  \cite{Csorgo:2012dm}.  The composition measurements could be interpreted as an evolution from light to heavier nuclei if current hadronic interaction models describe well  the air shower physics \cite{Abraham:2010yv, Abreu:2013env,
Aab:2014pza, Aab:2014dua}.

Upper limits have been obtained on the photon flux integrated above an energy threshold which impose stringent limits for top-down models \cite{Settimo:2013, Aab:2014bha}. Also, competitive neutrino limits were published \cite{Pieroni-ICRC:2013,
Abreu:2012zz, Abreu:2013zbq}, as well as searches for Galactic neutron signals \cite{Auger:2012yc, Aab:2014caa}.

\subsection{Observatory design}
  
Design targets for the surface detector array included 100\% duty cycle, 
a well-defined aperture independent of energy above $10^{18.5}$\,eV, measurement of 
the time structure of the signals of the shower particles, sensitivity to showers 
arriving at large zenith angles, self-contained detector stations 
and {\it in situ} calibration of detector stations by cosmic ray muons.  
The fluorescence detector design required that every event above $10^{19}$\,eV arriving within the FD 
on-time should be recorded by at 
least one fluorescence telescope camera, direct measurement of the longitudinal 
development profile and timing synchronization for simultaneous measurement of showers with the surface 
detector array \cite{design-report}. 

Each water Cherenkov surface detector is self-powered and communicates with the 
central data acquisition system using wireless technology. Air
fluorescence telescopes record air sho\-wer development in the
atmosphere above the surface array on dark moonless nights.  
There are four air fluorescence sites on the perimeter of the array, each with six
telescopes.

An essential feature of this Auger hybrid design is the capability of
observing air showers simultaneously by two different but
complementary techniques.  The SD operates continuously, measuring the
particle densities as the shower strikes the ground just beyond its
maximum development.  On dark nights, the FD records the development of the air shower
via the amount of nitrogen fluorescence produced along its path.  Since the intensity of
fluorescence light is proportional to the energy dissipated
by the shower, integrating the intensity of light produced along the shower axis yields a
nearly calorimetric measurement of the cosmic ray energy. 
Using the observation of hybrid events, this energy
calibration can then be transferred to the surface array with its
100\% duty factor and large event gathering power.  Moreover,
independent measurements by the surface array and the fluorescence
detectors alone have limitations that can be overcome by comparing
their measurements in the set of showers measured by both.  The hybrid
dataset  provides a thorough
understanding of the capabilities and the systematic uncertainties of
both components.

The FD always operates  in
conjunction with the SD.  Its primary purpose is to
measure the longitudinal profile of showers recorded by the SD
whenever it is dark and clear enough to make reliable measurements of
atmospheric fluorescence from air showers.  The integral of the
longitudinal profile is used to determine the shower energy, and the
rate of shower development is indicative of the primary particle's
mass.  The hybrid detector has better angular resolution than the
surface array by itself.
       
Subsequent to the completion of construction, two significant enhancements 
have been incorporated into the baseline detectors that significantly extend the 
Observatory's science capability.  The HEAT (High Elevation Auger Telescopes) 
fluorescence detectors together with a 750\,m array, 
part of AMIGA (Auger Muon and Infilled Ground Array) 
extend the sensitivity down to $10^{17}$\,eV, in keeping with the hybrid detection 
strategy of the original Observatory. 
       
\subsection{Background}

The Pierre Auger Observatory was conceived during the International Cosmic Ray
Conference in Dublin in 1991 by Jim Cronin of the University of
Chicago and Alan Watson of the University of Leeds.  It had become
clear to them that only the construction of a very large air shower
array would yield the
statistical power and complete sky coverage necessary to address the
question of the origin of the highest energy cosmic rays.  

A six-month design workshop was held in 1995 that produced a Design
Report \cite{design-report} with a
discussion of the science, a conceptual design and cost estimate.  The
design report became the basis for funding proposals by the
collaborating countries.  Subsequent to the workshop a team of
scientists evaluated numerous prospective sites in both hemispheres.
Preferred sites were selected in the southern and northern hemispheres
by the collaboration in 1995 and 1996, respectively.  At the direction of the 
funding agencies, the project was to begin by building the Observatory in the southern hemisphere. 

After a period of research and development, the Engineering Array,
consisting of 32 prototype surface array detectors and two prototype
fluorescence telescopes, was built to validate the design
\cite{Abraham:2004dt}. 
At the end of 2001, before the end of the scheduled
two years, the Engineering Array was able to record and reconstruct
air shower events  simultaneously
by the surface array and the fluorescence detectors.  The Engineering
Array demonstrated the validity of the design and the performance of
all of the detector systems, communications equipment and data
acquisition as well as the deployment methodology.  The detectors
performed better than the design requirements, substantially
increasing the physics reach of the Observatory.

Installation of production detectors began in 2002.  While the
Engineering Array was assembled and deployed almost completely by
Auger collaborators, production deployment was accomplished by
trained Observatory staff.  Scientists monitored the quality of
the work and carried out the commissioning of completed detectors. The
Observatory started collecting data in January 2004 with 154 active detector stations.  
The first physics results were presented during the 2005 summer conference
season.  

Many important results have now been published by the Auger
Collaboration that have had a major impact on the field of cosmic ray physics.  
As of this writing, 60 full author list papers have been published or accepted,
with another 2 submitted and about 7 more in preparation. The Auger Collaboration 
is also training a cadre of future scientists, with 238 students 
granted PhDs based on their work on Auger. Another 161 PhD students
are in the pipeline.
Publications and other technical reports are available online at
%http://www.auger.org/technical\_info/.
\url{http://www.auger.org/technical_info/}.

\section{Hybrid design}

%\red{(Bruce D.)}

As indicated above, a key feature of the Pierre Auger Observatory is its hybrid design, in
which ultrahigh energy cosmic rays are detected simultaneously by
a surface array and by fluorescence telescopes.  The two techniques are used to observe
air showers in complementary ways, providing important cross-checks
and measurement redundancy \cite{Sommers:1995dm, Dawson:1996ci}. 

The surface detector array views a %two-dimensional 
slice of an air shower at
ground level, with robust and sensitive water Cherenkov stations
which respond to both the electromagnetic and muonic components of the
shower.  Well-established methods exist for determining arrival
directions and for estimating primary energy (see Section~\ref{sec:SDreco}).  The
SD operates 24 hours per day and thus provides uniform coverage in right ascension with
a huge 3000\,km$^2$ collecting area.  The instantaneous aperture 
of the array is easily calculable, especially for energies
above $3\times10^{18}$\,eV, where a shower falling on any part of the array 
is detected with 100\% efficiency independently of the mass of the primary 
particle that initiated the shower.  The aperture
is found simply by counting the number of hexagons of active
surface stations at any time, and multiplying by the aperture, $A\Omega$,
of a hexagonal cell, 4.59\,km$^2$\,sr (for shower zenith angles
${<}60^\circ$) \cite{Abraham:2010zz}.  The SD has the important
property that the quality of the measurements improves with the shower
energy.

The fluorescence detector is used to image the longitudinal
development of the shower cascade in the atmosphere.  The fluorescence
light is emitted isotropically in the ultraviolet part of the spectrum
and is produced predominantly by the electromagnetic component of the
shower.  Observation periods are limited to dark nights of good
weather, representing a duty cycle which has increased from 12\%
during early years \cite{Abreu:2010aa} up to $\sim$15\% at the present
time (see Section~\ref{sec:performance}).
This disadvantage is balanced by the considerable gain of being
able to view the development of the shower profile.  Firstly, since
fluorescence light production is proportional to the collisional
energy deposit in the atmosphere, the technique provides a
near-calorimetric method for determining the primary cosmic ray
energy.  Secondly, the depth at which a shower reaches maximum size,
$X_\text{max}$, is observable: this is the most direct of all
accessible mass composition indicators.

The aim is to use the FD and the SD to measure the same properties of primary
cosmic rays (energy, mass composition, direction) but to do so
using different techniques with {\em very different systematic uncertainties}.
Thus, part of the function of the fluorescence detector is to enable 
cross-checks to be made and to train the surface array, providing confidence in the
SD measurements made during the majority of the time when no fluorescence detector is operating.
However, the fluorescence detector is much more than a calibration
tool.  The data set collected during hybrid observations is of high
quality, being especially useful for those studies that require more
precise shower directions than are available from the surface array alone
and for studies where longitudinal profile measurements are vital.

An example of the synergy between the two techniques is illustrated by
measurements of the cosmic ray energy spectrum for
showers arriving with zenith angles less than $60^\circ$
\cite{Abraham:2010mj}.  The exposure achievable with the surface 
array is much larger than is possible with hybrid measurements so 
that in principle a greater number of events can be used in the 
determination of the spectrum.  However, with the surface array 
alone, there is a serious problem in that the relationship between 
the primary energy and the SD observable chosen to mirror it, namely the 
signal measured in the water Cherenkov detectors at 1000\,m from 
the shower axis, $S(1000)$, can only be found using cascade simulations.  
This method is inherently unreliable as the necessary hadronic physics 
is unknown at the energies of interest and it is therefore not even 
practical to assign a reliable systematic uncertainty.  Using the 
hybrid system, it has proved possible to develop an alternative method 
for estimating the primary energy that is essentially free from simulations.

The first step is to quantify the dependence of $S(1000)$ with zenith angle.  
This is done using the ``constant intensity'' method \cite{Hersil:1961zz},
where the attenuation of the typical
air shower with increasing atmospheric depth is mapped out using SD
data alone.  The conversion to primary energy is then achieved using a
subset of SD events also observed with the FD.  The only
simulation input to the determination of primary energy with the FD is in
estimating the small fraction (${\sim}10$\%) that goes into neutrinos
and high energy muons that continue into the ground.  Atmospheric variability (mostly
changing aerosol properties) complicates the analysis, due to
essential corrections for atmospheric attenuation of the
fluorescence light and because of the allowance that must be made for
scattered Cherenkov light in the FD signal.  The event reconstructions
utilize extensive atmospheric monitoring that is performed at the
Observatory site whenever the FD operates.  The final step in measuring
the energy spectrum is a precise determination of the exposure
(km$^2$\,sr\,yr) for the observations.  As already mentioned, the
instantaneous aperture of the SD array is straightforward to
calculate, even during the period of construction when it was
continually growing.  This example of an analysis procedure
illustrates the complementary strengths of the SD and FD, and how a
robust result can be achieved by drawing on them.

A key to the success of the hybrid technique is that it allows a
precise determination of the position of a shower axis in space with
an accuracy better than could be achieved independently with either
the surface array detectors or a single fluorescence telescope
\cite{Dawson:1996ci}.  The first step in geometrical reconstruction
makes use of the known orientations of the pixels of the fluorescence
detector and of the light intensities registered at the pixels.  This
enables the shower-detector plane (SDP), the plane in space that
contains the shower axis and the FD site, to be determined.  Timing
information is then used to find the orientation of the shower axis
within the SDP (see Section~\ref{sec:HybridReconstruction}).  With a
FD alone, the accuracy of determining the shower geometry may be poor
if the angular length of the observed shower track is short, say less than
$15^\circ$.  In this case, the apparent angular speed of the shower
image in the telescope is approximately constant, leading to a
degeneracy in the geometry solution.  This degeneracy is broken if
the angular track length is long enough for a \emph{change} in the
angular speed to be detected, or more robustly, with a measurement of
the arrival time of the shower at any point on the ground: thus a
timing measurement at a single station of the SD suffices.  Using this
hybrid reconstruction method, a directional resolution of $0.5^\circ$
is routinely achieved \cite{Bonifazi:2009ma}.  Since only the timing
information from a single SD station is needed, the hybrid geometry
constitutes an independent and sensitive cross-check on the
directional and core location assignments made with the SD.  The
precise geometry of a hybrid event is also the first step towards a
high quality measurement of the longitudinal profile of a shower which,
in turn, yields the energy of the primary particle and the depth of
maximum $X_\text{max}$.

Many experimental challenges exist in fully realizing the promise of
the hybrid technique to provide high quality measurements of shower
observables.  The Collaboration employs a series of cross-checks and
measurement redundancies to understand the systematic uncertainties in
each measurement.  These cross-checks include comparisons between SD
and FD measured parameters.  There are also important redundancies in
various calibration measurements and redundancies in measurements of
those atmospheric properties that are critical for accurate FD event
reconstruction.  Of particular importance is the aerosol content of
the atmosphere.  The concentration of aerosols is variable over
timescales of hours and it can vary over the area of the Observatory.
The concentration affects the transmission of fluorescence light from the shower
to the FD telescopes, and the scattering of Cherenkov light into the
fields of view of the telescopes.  While the choice of the number and location
of fluorescence sites around the surface array was driven by the
desire to minimize the effects of atmospheric uncertainties on
reconstruction, a great deal of effort is still required in
atmospheric monitoring (see Section~\ref{sec:atmosMonit} for details).  The aerosol concentration 
and distribution are monitored on timescales of 15 minutes using the two laser 
facilities (see Figure~\ref{southern_site}) within the detection volume 
and by lidar systems at each FD site.  In
addition the directional properties of aerosol scattering are measured
at two sites, and the wavelength dependence of the scattering is
obtained (Section~\ref{sec:atmosMonit}).

While the threshold energy for a fully-efficient trigger
for showers with zenith angles smaller than
$60^\circ$ is $3\times10^{18}$\,eV, hybrid observations
require only one SD station to have triggered and thus the threshold for hybrid events is
significantly lower.  For example, a hybrid measurement of the
``elongation rate'', the energy dependence of $X_\text{max}$, has been
published for energies above $10^{18}$\,eV \cite{Abraham:2010yv}, with
work underway to push the threshold energy lower with the HEAT
fluorescence telescopes (Section~\ref{sec:HEAT}).  Hybrid showers are
collected with full efficiency over the entire SD array at energies
above $10^{19}$\,eV.  Around 6000 quality hybrid
events are recorded per year above $10^{18}$\,eV, with 300 per year above $10^{19}$\,eV.
Quality cuts normally applied to these events include the requirement that the
depth of shower maximum is in the field of view of one of the FD telescopes.

One final experimental cross-check is worth mentioning.  Though the FD was not designed as a
``stereo'' instrument, a significant number of showers
at the higher energies are observed by more than one FD site.  At
$10^{19}$\,eV over 60\% of FD showers are viewed in stereo, increasing to
90\% at $3\times10^{19}$\,eV.  Stereo observations provide two or more
independent hybrid reconstructions of the shower geometry, and of
profile parameters such as energy and $X_\text{max}$.  This allows
cross-checks of atmospheric corrections, and of simulations of the
detector resolution.  This has confirmed the statistical resolutions 
for a single site to be ${\sim}10\%$ and 20\,g/cm$^2$ for energy and $X_\text{max}$
respectively, at around $10^{19}$\,eV \cite{Dawson:2007di}.

\section{The surface detector}

%\red{(Peter M)}

\subsection{Overview}

A surface detector station consists of a 3.6\,m diameter water
tank containing a sealed liner with a reflective inner surface. The
liner contains 12,000 liters of ultra-pure water. Three 9-inch diameter 
Photonis XP1805/D1 photomultiplier tubes (PMTs) are
symmetrically distributed on the surface of the liner
at a distance of 1.20\,m from the tank center axis
and look downward through windows of clear polyethylene
into the water.  They record Cherenkov light produced by the
passage of relativistic charged particles through the water. The
tank height of 1.2\,m makes it also sensitive to high energy
photons, which convert to electron-positron pairs in the water
volume.

\begin{figure}[t]
\centering
\includegraphics[width=0.48\textwidth]{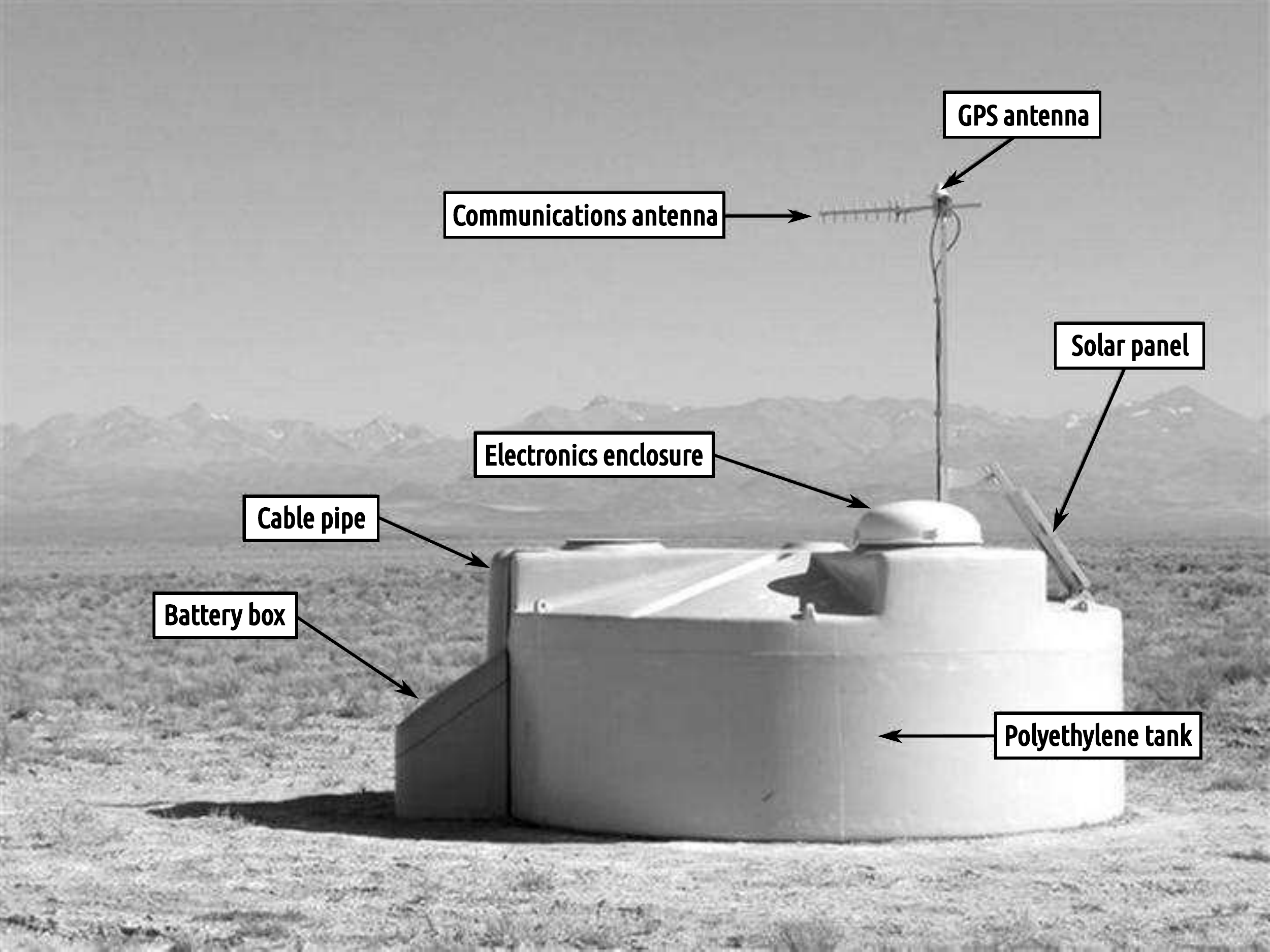}
\caption{A schematic view of a surface detector station in the field,
showing its main components.}
\label{fig-sd}
\end{figure}

Each surface detector station is self-contained. A solar power system
provides an average of 10\,W for the PMTs and electronics package
consisting of a processor, GPS receiver, radio transceiver and power
controller. The components of a surface detector station are shown
in Figure~\ref{fig-sd}.  Ref.~\cite{Allekotte:2007sf} describes
the surface detector in detail. 
%The electronics is described in \cite{sde-paper}.

Figure~\ref{southern_site} shows the layout of the surface array and the
FD buildings at its periphery.

\subsection{The SD station}

%The tanks are made of carefully selected, custom compounded high density polyethylene by the rotomolding
%process. The exterior is colored beige to blend with the landscape.
%The resins are compounded with other additives to enhance ultraviolet
%protection. The interior two-thirds of the wall thickness has added carbon black to guarantee
%light tightness. Tanks have an average wall thickness of 1.3\,cm
%and a nominal weight of 530\,kg. The tanks do not exceed 1.6\,m in height
%so they can be shipped over the roads within transportation regulations.
%
%The technique of rotational molding, or rotomolding, was chosen for its
%low cost and tank uniformity. Polyethylene meets the requirement of robustness
%against the environmental elements.

The tanks are made of polyethylene using the rotational molding, or rotomolding, process.  This process, 
in simplified form, consists of putting a set amount of polyethylene resin inside a mold, 
then rotating the mold and heating it until the resin has melted and uniformly coated the interior walls of the mold. 
The result is a low cost, tough, and uniform tank with robustness against the environmental elements. 
The carefully selected, custom compounded polyethylene resins contained additives to enhance ultraviolet protection. 
The interior two-thirds of the wall thickness was compounded with 1\% carbon black to guarantee light-tightness. 
The outer one-third was colored beige to blend with the landscape. The tanks have an average wall thickness of 1.3\,cm 
and a nominal weight of 530\,kg. The tanks do not exceed 1.6\,m in height so that they can be shipped over the 
roads within transportation regulations.

Three hatches, located above the PMTs,  provide access to the
interior  of the tank for water filling.
They also provide access for installation and
servicing of the interior parts. 
The hatches are covered with light-
and water-tight polyethylene hatchcovers. 
%The hatches are elevated to prevent rainwater from 
%accumulating around the hatchcover and leaking into the tank in case of gasket leakage.
For reasons of cost, durability, and ease of installation, the gaskets sealing the hatchcovers to the tanks 
may not be perfectly leak-tight, so the hatches are elevated to prevent accumulated water from collecting at the gasket.
One hatchcover is larger than the other two and
accommodates the electronics on its top surface. The electronics is
protected by an aluminium dome that keeps out rain and dust. The tanks also possess molded-in lugs,
six for lifting and four additional lugs to support the solar panel and
antenna mast assembly.

Electrical power is provided by two 55\,Wp (watt-peak)
solar panels which feed two 12\,V, 105\,Ah, lead-acid, low
maintenance batteries wired in series to produce a 24 V system. 
Power is expected to be available over 99\% of the time.
Batteries are charged through a commercial
charge controller, which is epoxy encapsulated and has robust surge protection. 
The electronics assembly at each SD station possesses a Tank Power Control Board (TPCB) which 
monitors the power system operation. The TPCB also has a control function which allows the 
remote operator to set into hibernation any (or all) of the SD stations if the charge of the batteries 
falls below a critical level. There is enough reserve in the solar power system
that this feature has not yet been employed.

The batteries are accommodated in a rotationally molded polyethylene battery box. 
Since battery lifetime is reduced with increased temperature, the battery box is protected from direct sunlight by 
installing it on the shaded side of the tank. It is also insulated with poly\-styrene foam plates 
to minimize high temperature excursions during the day.
The box is anchored by a plate which extends under the tank.
Power cables run through the
tank interior top from feed\-throughs in the large hatch to the far side
of the tank, where they exit the tank to run to the battery box.
The cables are protected from the point where they exit the tank to the entry of the battery box
by a polyethylene pipe.
%[Need to explain the feed-through for the cables?]

The solar panels are mounted on aluminium brackets, which also support
a mast.  Antennas for radio communication and GPS reception are mounted at the 
top of this mast. The mast-and-bracket system is designed to withstand 160\,km/h winds.

The tank liners are right circular cylinders made of a flexible
plastic material conforming approximately to the inside surface of
the tanks up to a height of 1.2 m. They enclose the water volume, provide a light-tight
environment and diffusively reflect the Cherenkov light produced in
the water volume. The liners are produced from a laminate composed
of an opaque three-layer coextruded low density polyethylene (LDPE)
film bonded to a 5.6\,mils (0.14\,mm) thick layer of DuPont\texttrademark 
Tyvek\textsuperscript{\textregistered} 1025-BL\footnote{E.I.\ du Pont
de Nemours and Co., Wilmington, Delaware, U.S.A., 
%http://www.dupont.com} by
\url{www.dupont.com}} by
a layer of titanium dioxide pigmented LDPE of 1.1\,mils (28\,$\upmu$m) thickness.
%(see Figure~\ref{fig:laminate}). 
The three-layer coextruded film consists
of a 4.5\,mils (0.11\,mm) thick carbon black loaded LDPE formulated to be opaque
to single photons, sandwiched between layers of clear LDPE to
prevent any carbon black from migrating into the water volume.
Custom designing the laminate materials has resulted in a durable, flexible liner.

\begin{figure}[t]
\centering
\includegraphics[width=0.4\textwidth]{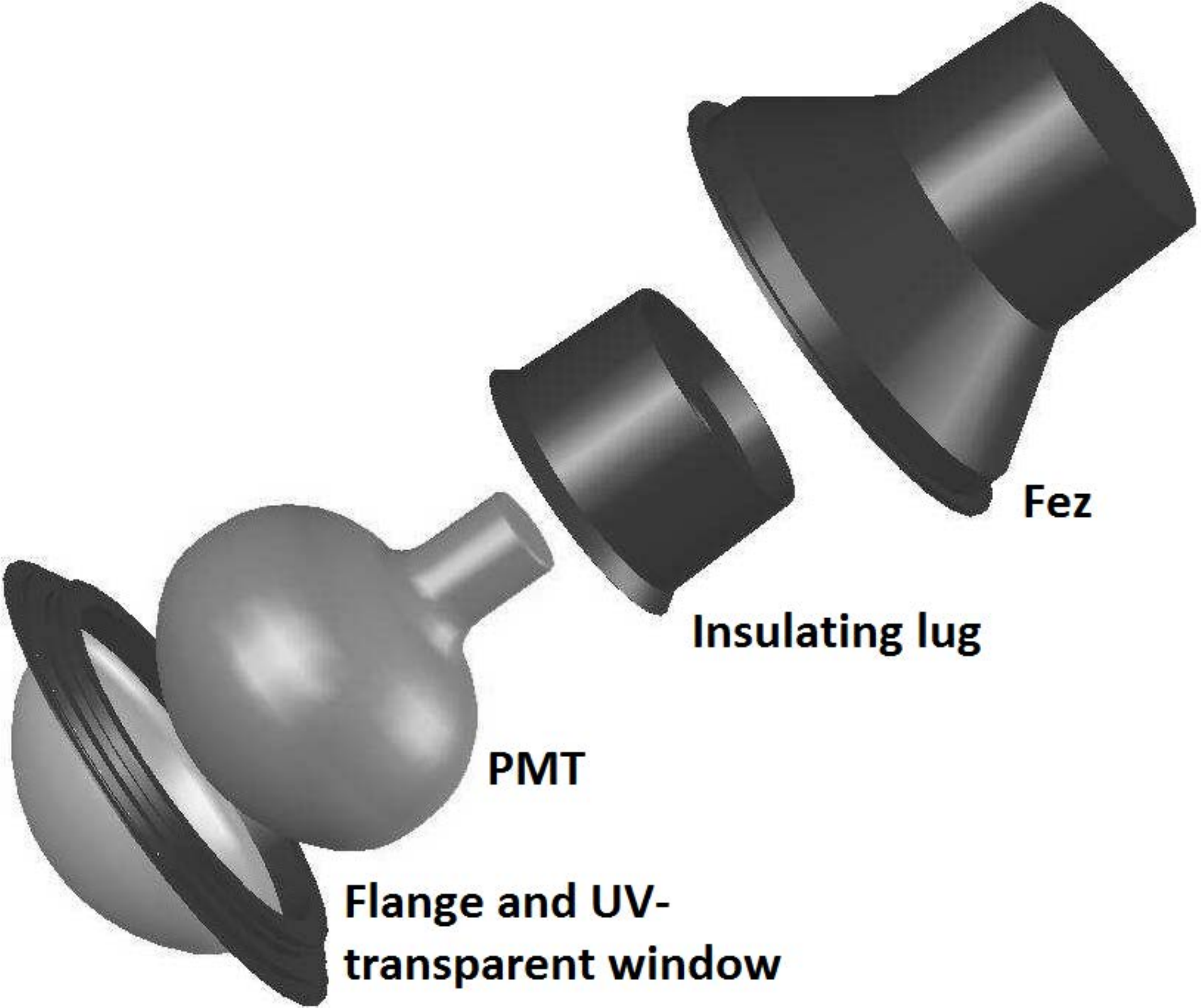}
\caption{Mechanical housing for the SD PMT. Top to bottom: outer plastic housing (fez); 
insulating lug; PMT; flange; UV-transparent window.}
\label{fig:PMT_encl}
\end{figure}

The liner has three windows through which the PMTs look into the water
volume from above. These windows are made of UV-transparent linear
low density polyethylene.  Each PMT is optically coupled to a
window with optical GE silicone RTV-6196 and shielded above by a light-tight
plastic cover, designated as the ``fez". Figure~\ref{fig:PMT_encl} shows the PMT enclosure.
The fez has four ports, including a light-tight air vent for pressure relief.
The other ports are for cable feedthroughs. Two foam insulation rings that
fit inside the fez serve to prevent ice buildup near the PMT.

Once deployed in their correct positions in the field, the tanks are
filled with ultra-pure water produced at a water plant owned and operated by the
Auger Collaboration. Water quality (resistivity) exceeds 15\,M$\Omega$\,cm at the output of the
water plant, and the water is transported in clean specialized transport tanks. 
%Because of its high purity, the
%water is expected to maintain its clarity without significant degradation for
%the lifetime of the Observatory, estimated at 20 years.
Tests have indicated that ultra-pure water does not support bacterial growth which could 
lead to reduced water clarity. Because of its high purity, the water is expected to maintain its 
clarity without significant degradation for the lifetime of the Observatory, estimated at 20 years. 
Occasional testing of the water in a sampling of detectors has shown no significant bacterial growth.

\subsection{The SD electronics}

%\red{(Jim Beatty)}

To collect the Cherenkov light produced in the water volume of the
detectors by the air showers, three PMTs view the water volume from
above. The PMTs have a 9\,inch diameter
photocathode and eight dynodes, with the chemical composition of the
dynode surfaces optimized by the manufacturer to maximize
linearity. Due to their proximity to water they are operated with a
positive anode voltage, the photocathode being grounded. The high
voltage is provided locally from a module integrated in the PMT base,
and is proportional to a DC control voltage provided by the slow
control system. The PMTs are operated at a nominal gain of $2\times
10^{5}$, and are specified for operation at gains up to $10^6$.  The
PMTs are required to be linear within 5\% up to 50\,mA anode
current. To minimize the effect of the geomagnetic field on the PMTs,
their orientation is aligned with respect to the azimuth of the
Earth's magnetic field at deployment.  The base, including the high
voltage supply, is attached to the tube by soldering to flying leads
and is potted in GE silicone RTV-6136 to protect it from the high
humidity present in the tank.

Each PMT has two outputs. An AC coupled anode signal is provided.  In
addition, the signal at the last dynode is amplified and inverted by
the PMT base electronics to provide a signal with 32 times the charge
gain of the anode.  No shaping of the signal is performed on the PMT
base.

Six identical channels of electronics are provided to digitize the
anode and amplified dynode signals from each of the PMTs.  Each
channel consists of a 5-pole Bessel filter with a $-3$\,dB cutoff at
20\,MHz and a voltage gain of $-0.5$.  This filter is implemented
using a pair of Analog Devices AD8012 current feedback op-amps.  The
filtered analog signals are fed to Analog Devices AD9203 10\,bit
40\,MHz semi-flash ADCs.  The ADC negative inputs are biased to
$-50$\,mV to bring the input pedestal on scale and allow for amplifier
section offsets.  The choice of filter cutoff results in 5\% aliasing
noise while preserving the time structure of the signals.  The use of
two 10\,bit ADCs with a gain difference of 32 extends the dynamic
range of the system to 15\,bits with a 3\% precision at the end of the
overlap region.  The maximum signal recorded before saturation
corresponds to about 650 times the peak current from a vertical muon
traversing the tank, which corresponds to the signal from a 100 EeV
cosmic ray at about 500 meters from the shower core.

An LED flasher is mounted in a test port of the water tank liner.  The
LED flasher incorporates two LEDs which can be pulsed independently or
simultaneously and with variable amplitude.  This allows testing of
the linearity of the photomultipliers to be conducted remotely.

Each SD station contains a GPS receiver with its corresponding antenna
mounted at the top of the communications mast for event timing and
communications synchronization. The receiver is a Motorola (OEM)
Oncore UT+. This receiver outputs a timed one-pulse-per-second (1\,PPS).
The GPS 1\,PPS signal is offset from the true GPS second by up to
50\,ns, and a correction for this offset is provided periodically by
the receiver.  Event timing is determined using a custom ASIC which
references the timing of shower triggers to the GPS 1\,PPS clock.  The
ASIC implements a 27\,bit clock operating at 100\,MHz.  This clock is
latched on the GPS 1\,PPS signal at the time of each shower
trigger.  A counter operating at the 40\,MHz ADC clock is also latched
on the GPS 1\,PPS clock.  These data, together with the timing
corrections provided by the GPS receiver, are used to calibrate the
frequencies of the 40\,MHz and 100\,MHz clocks and to synchronize the
ADC data to GPS time within 10\,ns RMS.

The digital data from the ADCs are clocked into a programmable logic
device (PLD).  In the first half of the deployment, we employed two
ALTERA ACEX PLDs (model EP1\-K100QI208-2) with 16K x 36 bits
additional external static RAM.  In later stations, an Altera Cyclone
FGPA replaced the two ACEX devices and external memory.  The PLD
implements firmware that monitors the ADC outputs for interesting
trigger patterns, stores the data in a buffer memory, and informs the
station microcontroller when a trigger occurs.  There are two local
trigger levels (T1 and T2) and a global third level trigger, T3.
Details of the local triggers are described in Section~\ref{SDLocalTrigger}.

The front end is interfaced to a unified board which implements the
station controller, event timing, and slow control functions,
together with a serial interface to the communications system.
The slow control system consists of DACs and ADCs used to measure
temperatures, voltages, and currents relevant to assessment of the
operation of the station.

The station controller consists of an IBM PowerPC 403 GCX-80MHz, with
a 32\,MB DRAM bank to store data and executable code, and a 2\,MB Flash
EPROM for the bootstrap and storing of the OS9 operating system.  The
data acquisition system implemented on the station controller transmits
the time stamps of the ${\sim}20$ T2 events collected each second to CDAS
(Central Data Acquisition System; see Section~\ref{sec:CDAS}).
CDAS returns T3 requests to the station within ${\sim}8$ seconds of the event
(including communications delays due to retransmission).  The station
controller then selects the T1 and T2 data corresponding to the T3
requests and builds it into an event for transmission to
CDAS. Calibration data are included in each transmitted event.

\subsection{Calibration of the surface detector\label{subsec:sde_calib}}

%\red{(Xavier)}

\begin{figure}[t]
\centering
\includegraphics[width=0.48\textwidth]{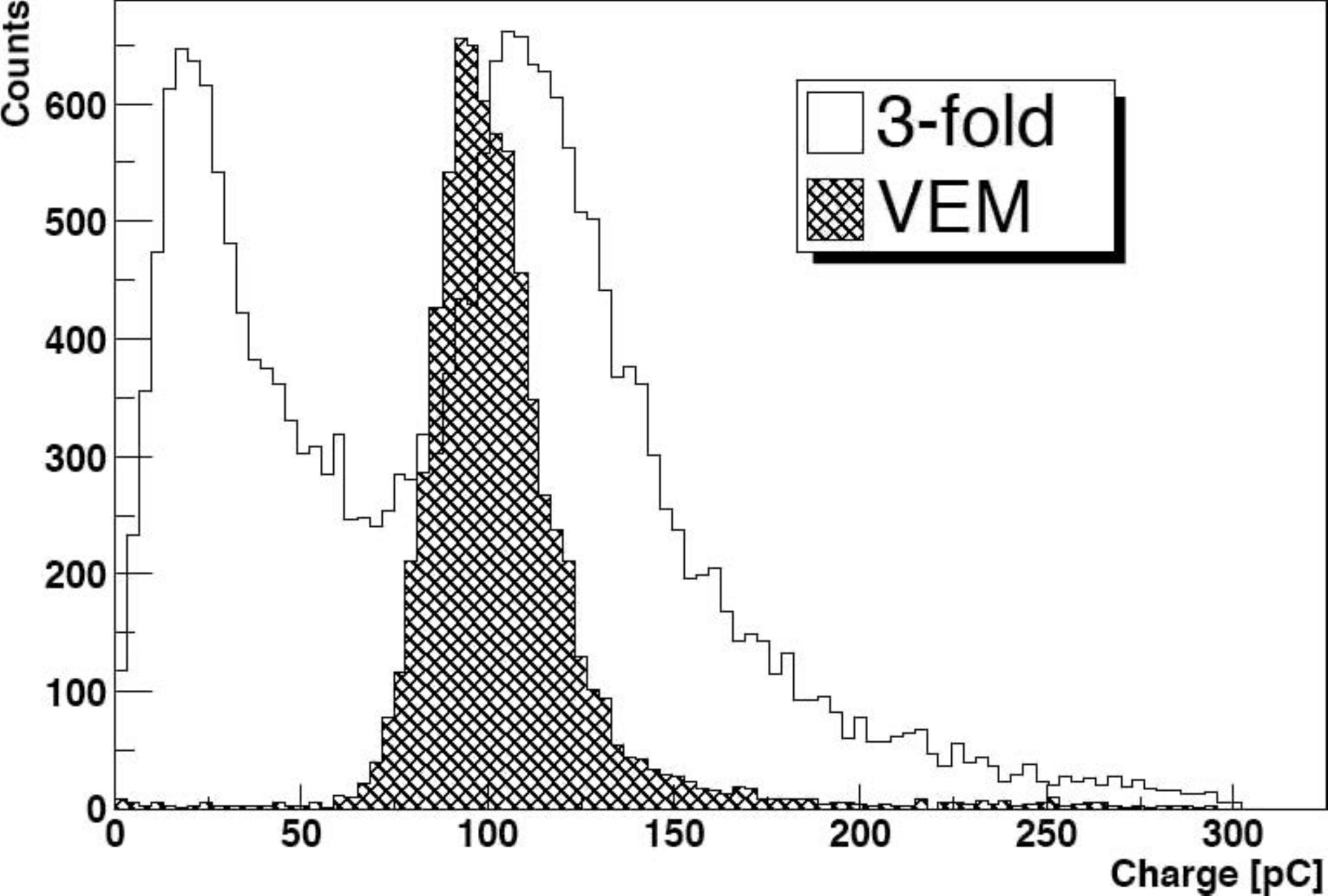}
\caption{Charge spectrum obtained when a surface detector is triggered by a threefold coincidence among its photomultipliers 
(open histogram). The hatched histogram shows the spectrum when triggered on central vertically aligned plastic scintillators. The bin containing the peak of the scintillator triggered spectrum is defined as a 
vertical equivalent muon.  The leftmost peak in the open histogram is due to low energy and corner-clipping muons convolved with the threefold low threshold coincidence.}
\label{vemhisto}
\end{figure}

%Calibration of the surface detector. The VEM.
The Cherenkov light recorded by a surface detector is measured in 
units of the signal produced by a muon traversing the tank on a vertical trajectory (Figure~\ref{vemhisto}).
This unit is termed the Vertical Equivalent Muon (VEM). The goal of
the surface detector calibration is to measure the value of 1\,VEM
in hardware units (i.e., in integrated FADC channels). During shower
reconstruction, the signal recorded by the tanks is converted into units of
VEM, and the total shower energy and arrival direction are fitted using
a lateral distribution function and energy conversion based upon
hybrid analysis using the FD. The conversion to units of VEM
is done both to provide a common reference level between tanks
and to calibrate against the detector simulations.

The total bandwidth available from each SD station to the CDAS is ${\sim}1200$\,bit/s,
which requires that the calibration be done by the local electronics.
Also, the remoteness of the detectors implies that the calibration procedure
be robust, allowing for failures of individual PMTs.

We define $Q_\text{VEM}^\text{peak}$ (denoted simply by $Q_\text{VEM}$  hereafter) as the bin containing the peak in the charge histogram
of an individual PMT response, and 
$I_\text{VEM}^\text{peak}$ (denoted by $I_\text{VEM}$  hereafter) as the bin containing the peak in the pulse height histogram. These 
quantities are used in the three main steps in the calibration procedure:
\begin{enumerate}
\item Set up the end-to-end gains of each of the three PMTs to have
$I_\text{VEM}$ at 50 channels. The choice of 50\,ch/$I_\text{VEM}$
results in a mean gain of ${\sim}3.4\times10^5$ for a mean
$n_\text{pe}$/VEM of ${\sim}94$\,photo\-electrons.
\item To compensate for drifts, adjust the electronics level trigger by
continually performing a local calibration to determine $I_\text{VEM}$
in channels.
\item Determine the value of $Q_\text{VEM}$ to high accuracy
using char\-ge histograms, and use the known conversion from $Q_\text{VEM}$
to 1.0\,VEM to obtain a conversion from the integrated signal of the PMT
to VEM units.
\end{enumerate} 

The high voltages, and thus the gains of each of the three PMTs, are tuned to match
a reference event rate. This tuning implies that the PMTs in the SD stations will not have equivalent
gains, even for PMTs in the same tank. If, for example, a particular SD station 
has more detected photons per vertical muon than the average station, then the PMTs in this station 
will be operated at a lower gain than average to compensate. Conversely,
if a PMT has a worse optical coupling than others in the same tank,
resulting in fewer observed photons per vertical muon, the PMT
will be run at a higher gain.

In addition to the primary conversion from integrated channels
to VEM units, the calibration must also be able to convert the
raw FADC traces into integrated channels. The primary parameters needed
for this are the baselines of all six FADC inputs, and the gain ratio
between the dynode and anode.
The dynode/anode, or $D/A$, ratio is determined by averaging large pulses
and performing a linear time-shifted fit to obtain both $D/A$
and the phase delay between the dynode and anode. This method 
determines $D/A$ to 2\%.

The calibration parameters are determined every 60\,s. The most recently
determined parameters are returned
to CDAS with each event and stored along with the event data.
Each event therefore contains information about the state of
each SD station in the minute preceding the trigger,
allowing for an accurate calibration of the data. 
Ref.~\cite{Bertou:2005ze} describes in detail the 
calibration method of the surface detector.

\subsection{The SD local triggers}
\label{SDLocalTrigger}

Several independent local trigger functions are implemented in the front-end electronics: the scaler trigger, the calibration trigger, and the main shower trigger.

The scaler trigger records pulses with a very low threshold for auxiliary physics purposes such as space weather. 
The calibration trigger collects low threshold pulses using a small number of bins (20), which is one bin 
above 0.1 $I_\text{VEM}$, thus providing high rate cosmic ray data. 
Data from the three
high gain channels are stored from three samples before the trigger to 20
samples after the trigger.
These data are used to build calibration histograms such as the one shown in Figure~\ref{vemhisto}, and are also used to convert offline the six FADC traces into VEM units.  
It was lowered during the deployment period for investigating signals from the tails of showers 
and to measure muon decay in the SD water volume~\cite{Allison:2005ge}. 

The main trigger is the shower trigger that results in the recording of 768 samples (19.2\,$\upmu$s) of the six FADCs. It has two levels of selection. The first level, called T1, has 2 independent modes. The first one is a simple threshold trigger (TH) requiring the coincidence of all three PMTs being above 1.75\,$I_\text{VEM}$.  This trigger is used to select large signals that are not necessarily spread in time. It is particularly effective for the detection of very inclined showers that have penetrated through a large atmospheric depth and are consequently dominantly muonic. The threshold has been adjusted to reduce the rate of atmospheric muon triggers from about 3\,kHz to 100\,Hz. The second T1 mode is a time-over-threshold trigger (ToT) requiring that at least 13 bins within a 3\,$\upmu$s window (120 samples) exceed a threshold of 0.2\,$I_\text{VEM}$ in coincidence for two out of the three PMTs. The ToT trigger selects sequences of small signals spread in time, and is thus efficient for the detection of vertical events, and more specifically for stations near the core of low-energy showers, or stations far from the core of high-energy showers. The rate of the ToT trigger depends on the shape of the muon pulse in the tank and averages 1.2\,Hz with a rather large spread (about 1\,Hz rms).
The second trigger level, called T2, is applied to decrease the global rate of the T1 trigger down to about 23\,Hz. 
While all T1-ToT triggers are promoted T2-ToT, only T1-TH triggers passing a single threshold of 3.2\,$I_\text{VEM}$ in coincidence for the three PMTs will pass this second level and become T2-TH.  All T2s send their timestamp to CDAS for the global trigger (T3) determination.  More details on the triggers can be found in Ref.~\cite{Abraham:2010zz} .

In June 2013, the Observatory installed across the entire array
two additional SD T1 triggers. These triggers build upon the ToT
trigger in two ways, applying more sophisticated analysis to the
FADC traces. The time-over-threshold-decon\-vol\-ved (ToTd) trigger
deconvolves the exponential tail of the diffusely reflected
Cherenkov light pulses before applying the ToT condition. This has the
effect of reducing the influence of muons in the trigger, since the
typical signal from a muon, with fast rise time and
${\sim}60$\,ns decay constant, is compressed into one or two
time bins. The multiplicity-of-positive-steps trigger (MoPS), on the
other hand, counts the number of positive-going signal steps in two of three
PMTs within a $3\,\mu\mathrm{s}$ sliding window. The steps are
required to be above a small FADC value ($\approx5\times$ RMS noise)
and below a moderate value ($\approx\frac{1}{2}$ vertical muon
step). This reduces the influence of muons in the trigger. Both the
ToTd and MoPS triggers also require the integrated signal to be above
$\approx0.5$ VEM. Because these triggers minimize the influence of
single muons, they reduce the energy threshold of the array, while
keeping random triggers at an acceptable level. Thus they improve the
energy reach of the SD, as well as improve the trigger efficiency for
photon and neutrino showers.

%The purpose of the muon triggers is to accumulate data for
%self-calibration of the SD stations using the incident cosmic ray flux
%at ground level.  They are based on a single bin trigger with a threshold
%of ${\sim}0.1\,I_\text{VEM}$.  When the trigger occurs, data from the three
%high gain channels are stored from 3 samples before the trigger to 20
%samples after the trigger. The threshold is then temporarily lowered
%to allow for investigating signals from the tails of showers and to
%detect muon decays in the tank.  Time stamps are recorded for each
%block of data.

\subsection{Operation and maintenance}

Currently more than 1660 surface detector stations are operational. Concerning
the water Cherenkov detectors themselves, only very few failures have been detected.  Only a few liners
were observed to leak shortly after installation.  In this case, which
constitutes the worst failure mode, the tank is emptied and brought back to the
Assembly Building for replacement of the interior components.

The electronics of the surface detector operates using solar power. A tank power control
board incorporates protection circuits, signal conditioning for the
monitoring of the solar power system, and a circuit allowing for
orderly shutdown and wakeup of the station in the event of an
extended cloudy period during winter when there could be inadequate solar
power available to operate the station continuously.
The solar power system has not yet experienced a dark period long enough to require shutting
down the array for battery recharging.
The most probable battery lifetime is 4.5--6 years~\cite{Sato-ICRC:2011}, and batteries are changed during
regular maintenance trips. 

The PMTs and electronic boards are the most critical elements of the SD stations. 
They are subject to very severe environmental conditions:
temperature variations, humidity, salinity and dust.  The failure rate of the
PMTs is about 20 per year (about 0.5\%). Some high voltage (HV) module and base problems
have been detected as well as some problems due to bad connections. All other
failures except those concerning the  PMTs (such as a broken photocathode)  can
be repaired in the field.  It is currently estimated that the number of spare PMTs
is sufficient for about 10 to 15 more years of operation.  The failure rate of
electronic boards is about 1\% per year. Some of the problems are repaired
simply by reflashing the software.  Most of the electronic problems can also
be repaired on site. All the spare parts are stored in Malarg\"ue.

The operation of the array is monitored online and alarms are set on various
parameters \cite{Rautenberg:2011zz}. The maintenance goal is to have no more that 20 detector stations
out of operation at any time. Currently the achieved number is less that 10
detector stations out of operation. It is currently estimated that the
long-term maintenance (including the battery change) requires about 3 field
trips per week. This maintenance rate is within the original expectations.  The
maintenance is organized by the Science Operation Coordinator and performed by
local technicians. The surface detector does not require a permanent presence
of physicists from other Auger institutions on site.

\section{The fluorescence detector}

%\red{(Radomir S.)}

\subsection{Overview}

\begin{figure}[t]
\centering
\includegraphics[width=0.48\textwidth]{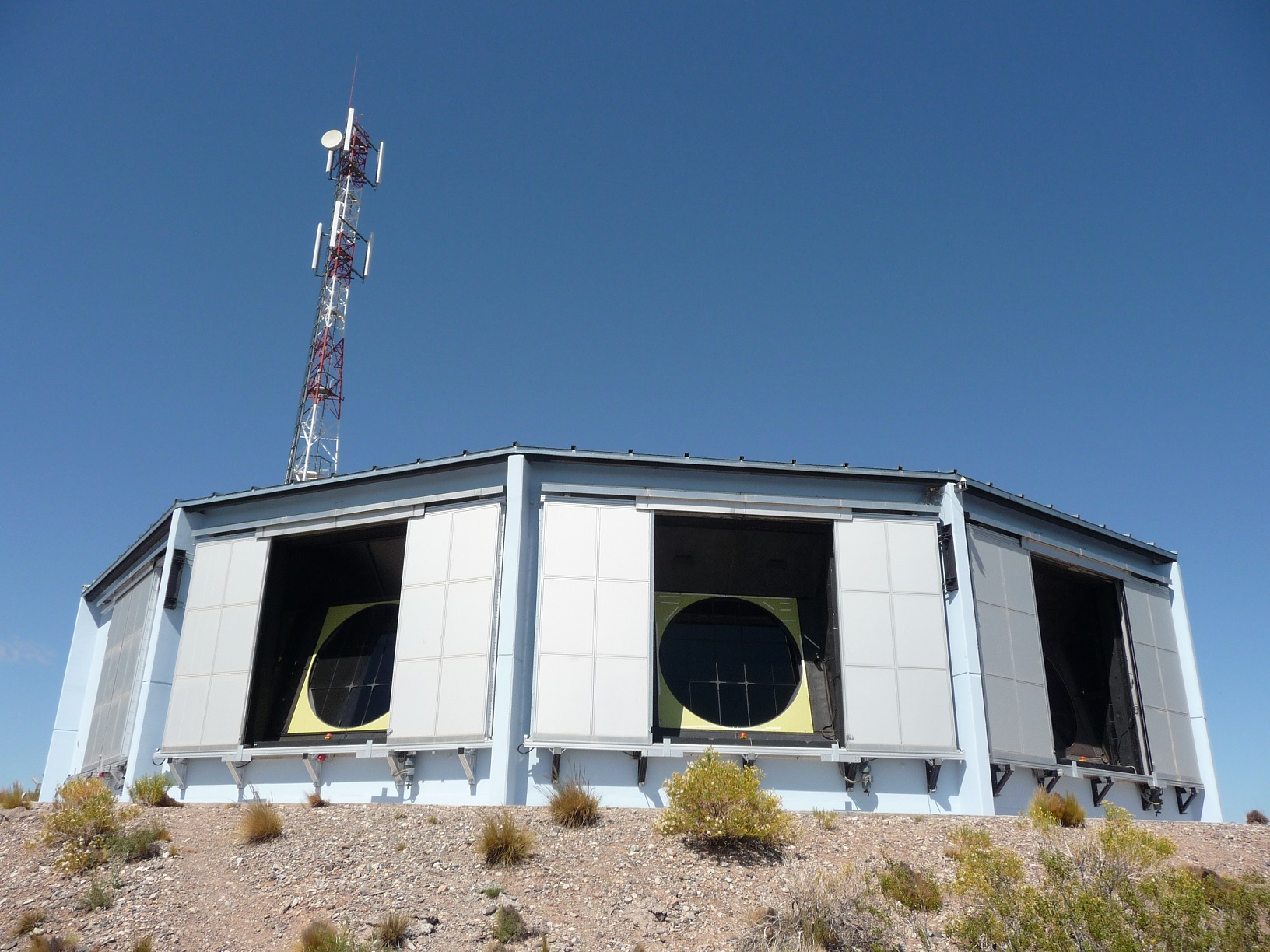}  % for ArXiV
\caption{FD building at Los Leones during the day. Behind the building
is a communication tower. This photo was taken during daytime when
shutters were opened because of maintenance.}
\label{fig:FD-losleones-building}
\end{figure}

The $24$ telescopes of the FD 
overlook the SD array from four sites -- Los Leones, Los Morados,
Loma Amarilla and Coihueco \cite{Abraham:2009pm}. Six independent telescopes are located
at each FD site in a clean climate controlled building~\cite{Abraham:2004dt},
an example of which is seen in Figure~\ref{fig:FD-losleones-building}.
A single telescope has a field of view of
$30^\circ\times30^\circ$ in azimuth and elevation, with a minimum
elevation of $1.5^\circ$ above the horizon. The telescopes face towards
the interior of the array so that the combination of the six telescopes
provides $180^\circ$ coverage in azimuth. 

\subsection{FD telescopes}
\label{sec:FDtelescopes}

\begin{figure}[t]
\centering
\includegraphics[width=0.4\textwidth]{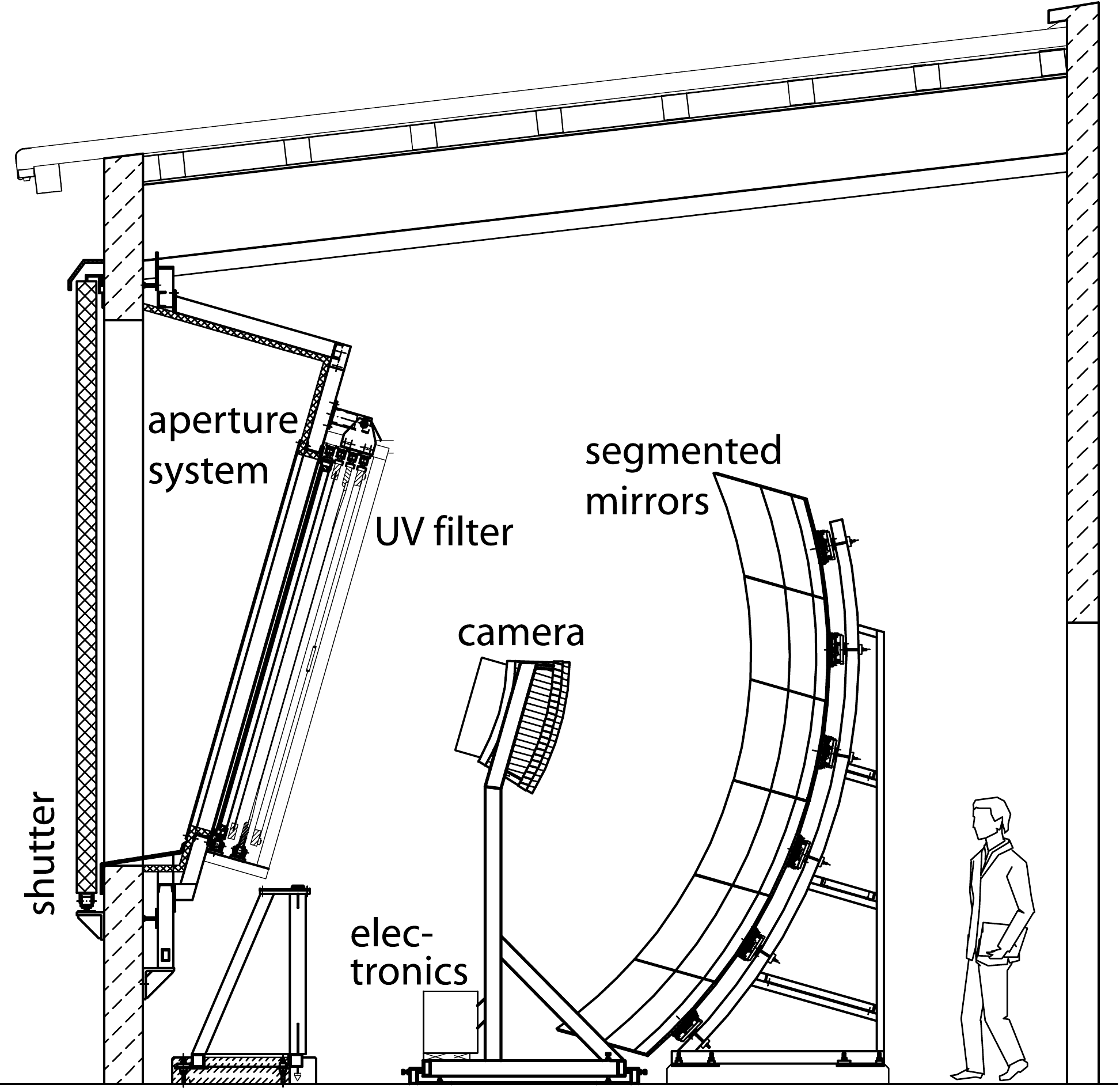}
\caption{Schematic view of a fluorescence telescope with a description of its main components.}
\label{fig:Telescope}
\end{figure}

\begin{figure}[t]
\centering
\includegraphics[width=0.48\textwidth]{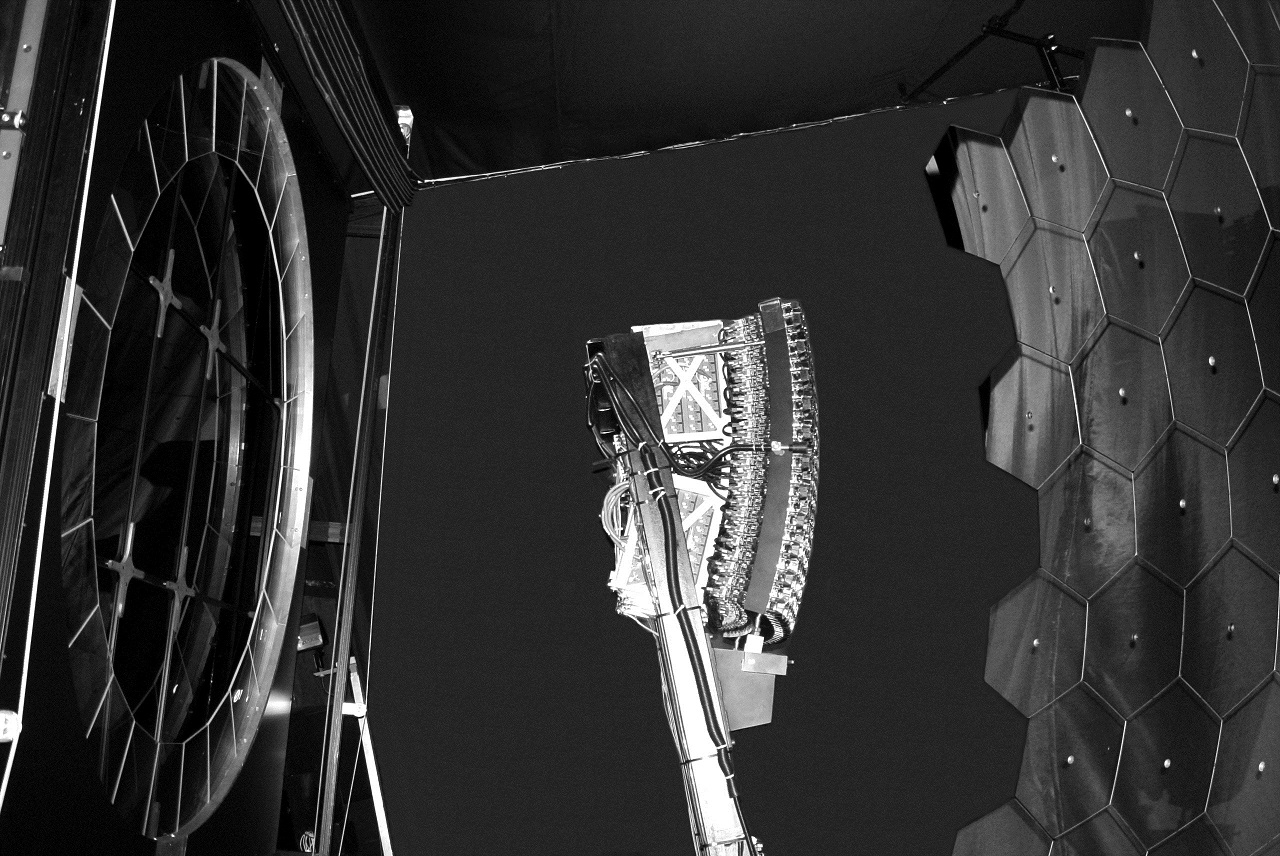}   % for ArXiV
\caption{Photo of a fluorescence telescope at Coihueco.}
\label{fig:Fluorescence-telescope}
\end{figure}

The details of the fluorescence detector telescope are shown in
Figure~\ref{fig:Telescope} and an actual view of an installed telescope
in Figure~\ref{fig:Fluorescence-telescope}. The telescope design is based
on Schmidt optics because it reduces the coma aberration of large optical
systems.
Nitrogen fluorescence light, emitted isotropically
by an air shower, enters through
a circular diaphragm of $1.1$\,m radius covered with a Schott
MUG-6 filter glass window. The filter transmission
is above $50$\% ($80$\%) between $310$ and $390$\,nm
($330$ and $380$\,nm) in the UV range.
The filter reduces the background light flux and thus improves
the signal-to-noise ratio of the measured air shower signal. It also serves 
as a window over the aperture which keeps the space containing
the telescopes and electronics clean and climate controlled.
The shutters seen in Figure~\ref{fig:Telescope} are closed during daylight and also 
close automatically at night when
the wind becomes too high or rain is detected. 
In addition, a fail-safe curtain is mounted behind the diaphragm to prevent
daylight from illuminating a camera in case of a malfunction of the shutter
or a failure of the Slow Control System.

A simplified annular lens, which corrects spherical aberration
and eliminates coma aberration,
is mounted in the outer part of the
aperture. The segmented corrector ring has inner and outer radii
of $850$ and $1100$\,mm, respectively. Six corrector rings were
made from Schott BK7 glass and Borofloat was used for the rest.
More details about the corrector ring can be found
in~\cite{Abraham:2004dt,deOliveira:2004dh}.

The light is focused by a spherical mirror of ${\sim}3400$\,mm
radius of curvature onto a spherical focal surface with radius of curvature
%$1743$\,mm.
${\sim}1700$\,mm.
Due to its large area (${\sim}13$\,m$^2$), the primary mirror 
is segmented to reduce the cost and weight of the
optical system. Two alternative segmentation configurations
are used: one is a tessellation of $36$ rectangular anodized
aluminium mirrors of three different sizes; the other is a
structure of $60$ hexagonal glass mirrors (of four shapes and
sizes) with vacuum deposited reflective coatings~\cite{Abraham:2004dt}.
The average reflectivity of cleaned mirror segments at a wavelength
$\lambda=370$\,nm is more than $90$\%.

The camera body is machined from a single aluminium block of
$60$\,mm thickness, with an outer radius of curvature of
$1701$\,mm and an inner curvature radius of $1641$\,mm. The
hexagonal photomultiplier tubes, model XP3062 manufactured by
Photonis, are positioned inside $40$\,mm diameter holes drilled
through the camera block at the locations of the pixel centers. The
pixels are arranged in a matrix of $22$ rows by $20$ columns.

The PMT boundaries are approximate hexagons with a side to side
distance of $45.6$\,mm. The PMTs are separated by simplified
Winston cones secured to the camera body which collect the light
to the active cathode of the photomultiplier tube. The light
collectors serve to prevent photons from landing in the dead
spaces between the PMT cathodes. The upper edge of the light
collectors lie on the focal surface of $1743$\,mm radius. The
pixel field of view defined by the upper edges corresponds
to an angular size of $1.5^\circ$.

All support structures and cables are distributed so as to
minimize any obscuration in the light path.
The contribution of reflection and scattering inside the optical
system of the telescope has been measured in situ and with
an airborne remotely controlled platform carrying an isotropic
and stabilized UV light source~\cite{Baeuml-ICRC:2013}.
The measured point spread function of the light distribution
in pixels has been implemented in the software used in the air
shower reconstruction.

Cleaning and maintenance work has been required during years
of detector operation. The cleaning of the UV filter from outside
has been performed several times because of deposited dust layers.
Currently, the cleaning of all UV filters from outside is done
three times per year. The equipment inside the building (i.e. the inner
side of the filter, the corrector ring, the dust curtain) is cleaned
less frequently, because it is not exposed to the outside environment. Dry and
wet methods have been adopted for cleaning the mirror segments and
they both improve the reflectivity of mirrors. For telescopes, where
the first cleaning took place six years after their installation, the
reflectivity increased by $\leq1$\% in the case of mirror segments in
the upper rows and $\sim5$\% for mirror segments in the bottom rows,
where the segments are turned slightly upward (see, e.g.,
Figure~\ref{fig:Fluorescence-telescope}).
The reflectivity of a few selected mirror segments is measured once
or twice each year and it changes less than $1\%$ per year.

Alignment of individual mirror segments was cross-checked
with a laser on site. Moreover, additional methods using
data measured by telescopes were used, such as star tracking,
Central Laser Facility (CLF) and eXtreme Laser Facility (XLF)
shots (Section~\ref{sec:CLFXLF}), or a comparison of FD and SD geometry
reconstruction. Only in two cases were a realignment of a telescope
and a readjustment of camera position needed.

\subsection{FD electronics}

%\red{(M.\ Kleifges)}

The FD electronics must provide a large dynamic range and
strong background rejection, while accepting any physically
plausible air shower. Moreover, the electronics is responsible
for anti-alias filtering, digitizing, and storing signals
from the PMTs.

The XP3062 photomultiplier tube is an 8-stage unit with a bialkaline
photocathode with quantum efficiency of about $25$\% in the
wavelength range $350$ to $400$\,nm.
The PMT high voltage is provided by a HV divider
chain which forms a single physical unit together with the signal
driver circuitry. This head electronics unit is soldered to the 
flying leads of the PMT \cite{Becker:2007zza}.

The nominal gain for standard operation of the FD is set to
$5\times10^4$. Stabilization of the HV potential for large
pulses, and in the presence of the low but not negligible light
intensity of the dark sky background, is realized by employing an active
network that uses bipolar transistors in the last three stages
of the PMT. The active divider ensures that the gain shift due
to the divider chain is less than $1$\% for anode currents up to
about $10$\,mA. The normal dark sky background on moonless
nights induces an anode current of about $0.8\,\upmu$A on each PMT.

The head electronics for each PMT is connected to a distribution
board located just behind the camera body. Each board serves 44
PMTs, providing high and low voltage and receiving the output
signals. The signal is then shaped and digitized in the front-end
electronics (FE) unit, where  threshold and geometry
triggers are also generated. Analog boards in the FE unit are designed
to handle the large dynamic range required for air fluorescence
measurements; this means a range of $15$\,bits and $100$\,ns timing.
%or $50$\ ns in new electronic used in HEAT.

As the PMT data are processed, they are passed through a flexible
three-stage trigger system implemented in firmware and software.
The trigger rate of each pixel in a camera (first level trigger)
is kept around $100$\,Hz by adjusting the pixel threshold level.
The algorithm of the second level trigger searches for track
segments at least five pixels in length within a camera. The typical
trigger rate per camera fluctuates between $0.1$ and $10$\,Hz. The third level
trigger is a software algorithm designed to clean the air shower
data stream of noise events that survive the low-level hardware
triggers. It is optimized for the fast rejection of triggers caused
by lightning, triggers caused by cosmic ray muon impacts on the camera and
randomly triggered pixels.

A rugged commercial computer (MPC) is associated with each telescope 
and serves to readout
the event data from the front-end electronics through a FireWire interface.
The MPCs at each FD site are connected through a 100~Mbit Ethernet LAN
switch to the site's central readout computer, called an ``EyePC".
This PC provides a connection between the communications network and 
the MPCs. The MPCs are diskless, thus they boot their Linux
system through the network and store their data on the EyePC’s hard drive.

The events surviving all trigger levels are sent through the MPC to the EyePC,
which builds an event from the coincident data in all telescopes at a given site
and generates a hybrid trigger (FD-T3) for the surface array. The event
rate is about $0.012$\,Hz per site for the $24$ baseline telescopes
(see Sec.~\ref{sec:performance}).
%The HEAT telescopes have higher rate, particularly because of the
%Cherenkov light from low energetic showers. Therefore the T4 trigger
%has been implemented to reduce the readout of the SD array for these
%low energetic showers.

\subsection{FD test and calibration}
\label{sec:FDbeams}

\subsubsection{Laser test beams}

Throughout each night of FD operation, thousands of collimated UV laser pulses
are directed into the atmosphere from two facilities located near the center of
the SD (see Figure~\ref{southern_site} and \ref{fig:atmosOverview} in Section~\ref{sec:atmosMonit}). 
Light scattered out of the laser pulses generates tracks in the same FD
telescopes that also record  the tracks generated by air showers.  In
contrast to high energy air showers, the direction, rate, and energy of the
laser pulses can be preprogrammed as desired.  Laser pulses can be fired at
specific directions relative to the ground, for example vertically, or in
specific directions relative to the sky, for example aimed at the galactic
center, Cen A or other potential sources of cosmic rays.  An optical fiber at
each laser directs a small amount of light into an adjacent SD 
station to provide hybrid laser events. 

Laser data recorded by the FD telescopes are used to measure FD performance, measure SD-FD
time offsets, check FD pointing, and make the hourly measurements of aerosol
optical depth vertical profiles for the atmospheric database. 
In addition, a low power roving system is available for use on a
campaign basis.

\subsubsection{CLF and XLF}
\label{sec:CLFXLF}
Laser test pulses are provided by the CLF \cite{Arqueros:2005yn} and XLF,  each of which features 
a Q-switched frequency tripled YAG laser. The spectral purity of the 355\,nm  light pulses delivered to the sky is better than 99.9\%. This wavelength falls near the middle of the nitrogen UV fluorescence spectrum, between the two major N$_2$ fluorescence bands of 337\,nm and 357\,nm. The maximum energy per pulse is nominally 7\,mJ.  Pulses of this energy produce tracks in the FD that have an approximate optical equivalence of 100\,EeV air showers. 
The CLF has been in operation since 2003.  The XLF was installed in 2008 and includes an automated calibration system that measures the beam energy and polarization.  A major upgrade to the CLF in 2013 added a beam calibration system and a backscatter Raman lidar receiver. %\cite{CLFUPGRADE}.

\subsubsection{FD detector calibration}
\label{FdCalibration}

%\red{(Jeff B.)}

% relative
% absolute

The precise reconstruction of air shower longitudinal profiles requires the
conversion of an ADC count to a light flux for each pixel that receives a
portion of the signal from a shower. To this end, the absolute calibration of
the detector response is essential. A calibrated large diameter, drum shaped
light source provides an absolute, end-to-end calibration for each pixel of the
fluorescence telescopes, with independent verification for some pixels by
atmospheric Rayleigh scattering from vertical laser pulses.

For these absolute methods, the flux of photons on the telescope aperture is
independently measured.  The effects of diaphragm area projection, optical
filter transmittance, mirror reflectivity, pixel light collection efficiency
and area, cathode quantum efficiency, PMT gain, preamp and amplifier gains,
and digital conversion are all included in the end-to-end calibration procedure.

\begin{figure}[t]
\centering
\includegraphics[width=0.6\textwidth]{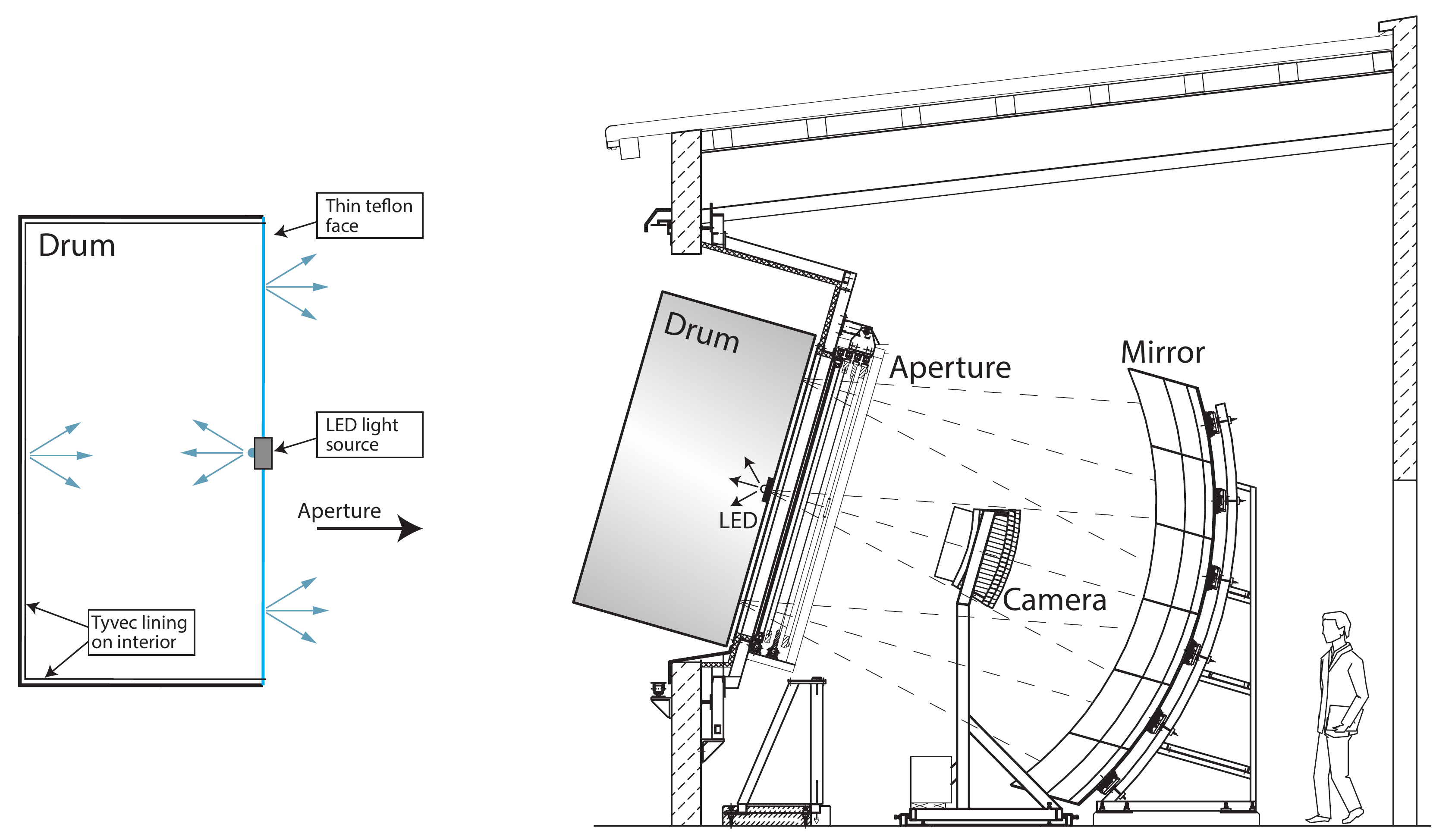}
\caption{Drum calibration source for the FD telescopes. Left: Detail of the drum (side view). 
Right: Sketch of the drum mounted on the telescope window.}
\label{fig:drum}
\end{figure}

The drum light source consists of a pulsed UV LED, emitting in a narrow band
around 365\,nm, mounted in a cylindrical shell of Teflon\textsuperscript{\textregistered}, 
illuminating the
interior of the 2.5\,m diameter cylindrical drum, 1.4\,m deep. The sides and
back surfaces of the drum are lined with Tyvek, 
while the front face is made of
a thin sheet of Teflon, which transmits light diffusively. The drum is
positioned at the entrance of the telescope under calibration, filling the
aperture.  Emission from the front face is Lambertian (within 3\%), and
provides uniform illumination to each pixel over the full acceptance of the
telescope. A schematic of the drum calibration source is shown in 
Figure~\ref{fig:drum}.

The drum light source intensity is calibrated~\cite{Brack:2013bta} to a
precision of better than 4\% in a dark room, using a NIST calibrated photodiode
as a reference.  Absolute calibration constants are obtained from
the ratio of the known pulsed flux of photons emitted by the drum and the
corresponding ADC pulse integrals of the camera pixels.

Periodically, a Rayleigh calibration system~\cite{Abraham:2009pm} is used as an
independent check on the drum light source calibration.  The 355\,nm roving laser is
positioned a few kilometers from the fluorescence telescope to be calibrated.
The laser is directed vertically. The laser beam is depolarized and the
pulse-to-pulse intensity monitored to a precision of 5\%. The scattered light,
mainly from Rayleigh scattering by the molecular atmosphere, is then used to
calibrate the fluorescence telescope.

From the end-to-end calibration, the appropriate constants are found to be
approximately 4.5\,photons/ADC count for each pixel.  To derive a flux of
photons for observed physics events, the integrated ADC number is multiplied by
this constant and divided by the area of the aperture.  The flux in
photons per m$^2$ perpendicular to the arrival direction is thus obtained.

An optical system for relative calibration~\cite{Abraham:2009pm} is used to monitor the
long-term time variations in the calibration of telescopes.  In each building,
three light sources coupled to optical fibers distribute light signals to three
destinations on each telescope.  Signals from a pulsed LED light source are
brought to a Teflon diffuser at the center of the mirror with the light
directed towards the camera.  Fibers from a second xenon flash lamp light
source end in 1\,mm thick Teflon diffusers at the center of two sides of the
camera, with the light directed at the mirror.  The signals from the third
source, also a Xenon flash lamp, are sent to ports on the sides of the
entrance aperture where the light is directed toward reflective Tyvek targets
mounted on the telescope doors, from which it is reflected back into the
telescopes. Drifts of the temporal performance of pixels, mirror and aperture
components can be identified by comparing measurements from the three light
sources. The sources are also equipped with neutral density filters to permit
linearity measurements, or with interference filters to monitor stability at
wavelengths in the range of 330 to 410\,nm.

\begin{figure}[t]
\centering
\includegraphics[width=0.6\linewidth]{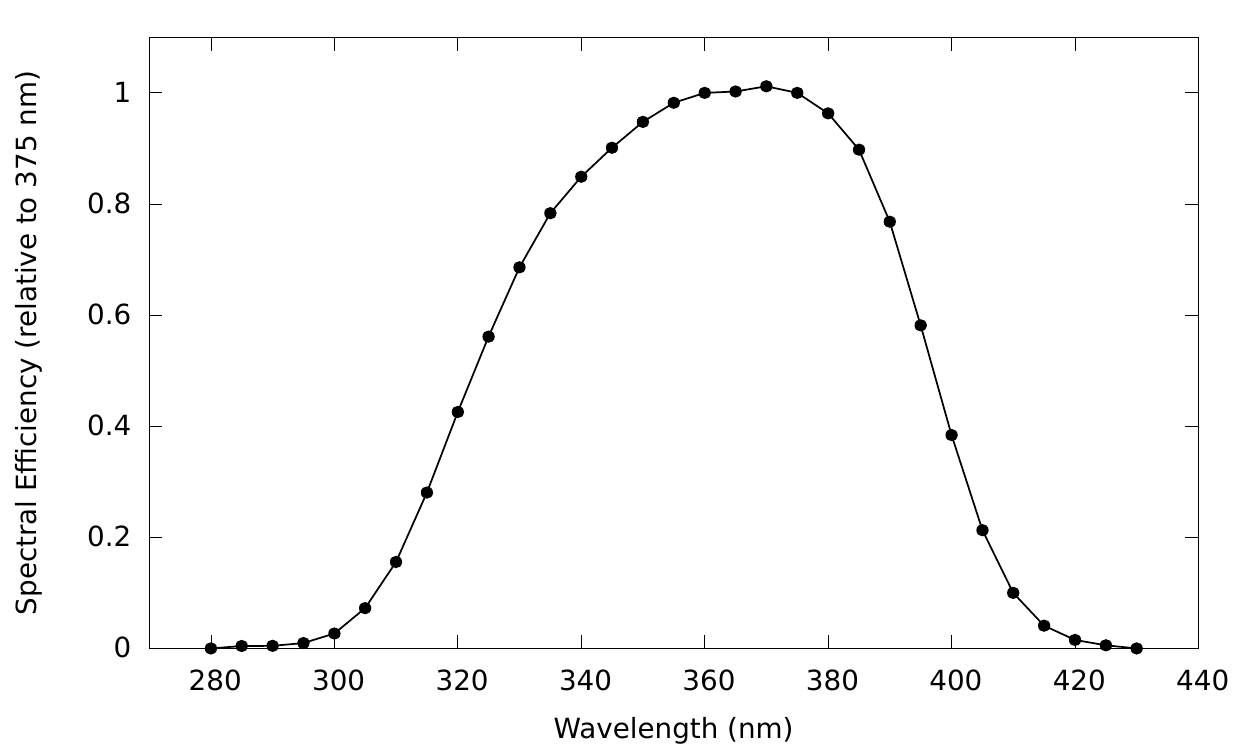}
\caption{
The relative efficiency between 280\,nm and 430\,nm measured for the telescope~3
at Coihueco. The curve is taken relative to the efficiency of the telescope at 375\,nm.
}
\label{Co3-efficiency}
\end{figure}

The relative spectral efficiencies, or multi-wavelength calibrations,
of FD telescopes were measured using a monochromator-based drum light source
with a xenon flasher. The measurement was done in steps of 5\,nm from 270\,nm
to 430\,nm. As described in Sec.~\ref{sec:FDtelescopes} there are two types
of mirrors and two different glass materials used for the corrector rings in
the FD telescopes. In total eight telescopes were measured to have a complete
coverage of the different components and a redundant measure of each combination.
The uncertainty of these measurements is $\sim3$\%.
An example of measured relative efficiency of an FD telescope is shown in
Fig.~\ref{Co3-efficiency}.

\subsection{FD operation}
\label{sec:FDperformance}

%The FD has been operated since 2005 with $12$ telescopes at two
%sites. The other sites were finished during the following
%years. 
Beginning in 2005, the FD initially operated with 12 telescopes
at two sites, Los Leones and Coihueco (see Figure~\ref{southern_site}).
In the following years, two additional sites, Los Morados and Loma Amarilla, with six telescopes
at each site, were brought into operation. 
All FD telescopes are operated remotely from the central
campus by shift personnel. Their responsibilities include
preparation of the FD for a run, making relative calibrations,
starting and stopping runs and online checking of the quality
of measured data \cite{Rautenberg:2011zz}.

The Slow Control System (SCS) assures a secure remote operation
of the FD system. The SCS works autonomously and continuously
monitors detector and weather conditions. Commands from
the remote operator are accepted only if they do not violate
safety rules that depend on the actual experimental conditions:
high voltage, wind speed, rain, light levels inside/outside
the buildings, etc. In case of external problems, such as
power failures or communication breakdowns, the SCS performs
an orderly shutdown, and also a subsequent start up of the
fluorescence detector system if the conditions have changed.
If parts of the SCS itself fail, the system automatically
reverts to a secure mode so that all critical
system states (open shutters, high voltage on, etc.) are actively maintained.

The observation of air showers via fluorescence light
is possible only at night. Moreover, night sky brightness
should be low and thus nights without a significant amount of
direct or scattered moonlight are required. We also require that
the sun be lower than $18^\circ$ below the horizon, the moon remain
below the horizon for longer than three hours, and that the illuminated fraction
of the moon be less than $70$\% in the middle of the night. The
mean length of the dark observation period is then $17$ nights each month.

The on-time of the FD telescopes is currently ${\sim}15$\%. The value
varies slightly between telescopes depending on the telescope
pointing and various hardware or software factors. The main remaining
source of downtime is weather. The telescopes are not
operated when the weather conditions become dangerous
(high wind speed, rain, snow, etc.) and when the observed
sky brightness (caused mainly by scattered moonlight) is
too high.

\section{Data communications system}

%\red{(Corbin)}
%
% Corbin Covault, submitted Nov 4, 2012
%
% Modified November 2013, mostly to include performance details
%
% Mofified May 2014 in response to reviewer comments
%
% corbin.covault@cwru.edu
%
% Graphics list:
%
% comms_overall_schematic.jpg  -- REPLACED!
% comms_overall_schematic_revised.pdf  -- NEW VERSION
% comms_backbone.jpg
% comms_tower_photo.jpg
% comms_su_photo.jpg
% comms_bsu_photo.jpg  -- DELETED May 2014
% comms_tdma_slots.jpg
% comms_radio_signal_strength.jpg
% comms_arq_plot_mar_2011.jpg -- CHANGED to ARQtotals_7s.pdf
%

%
%
%  Not yet done:  References to prior publication on the Comms system
%
% Revised 6/10/14 by C. Covault

\begin{figure*}[t]
\centering
\includegraphics[width=0.75\textwidth]{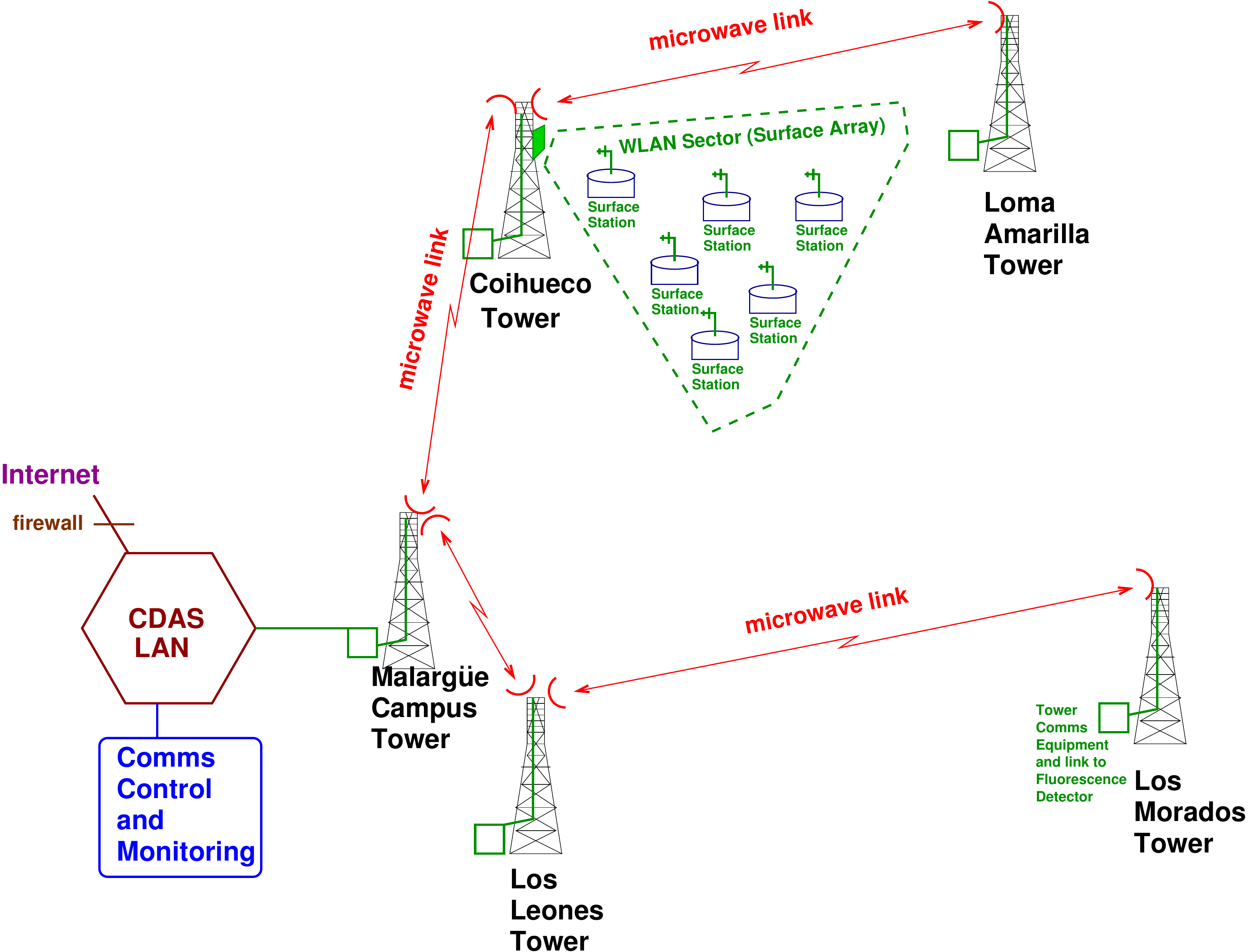}
\caption{Conceptual schematic of the overall radio telecommunications
  system for the Pierre Auger Observatory.}
\label{comms_schematic}
\end{figure*}

The detector systems of the Observatory are deployed at widely
dispersed positions over a very large area.  To send commands
and receive data from the four individual FD
sites and from 1660 SD stations in the field,
a bidirectional radio frequency telecommunication
network has been designed and deployed. The reliability of the network is
critical to the function of the Observatory, particularly in the
context of controlling the experiment, identifying event triggers, and
collecting data recorded at each detector for each air 
shower event.

For Auger, a custom designed system based on a two-layer hierarchy has
been implemented. Individual surface detector stations are connected
by a custom WLAN which is sectorized and
supported by four concentration nodes.  The WLAN is serviced by a high capacity 
microwave backbone network which also supports communications
between the four fluorescence detector sites and the main campus data
acquisition and control center.  Figure~\ref{comms_schematic} shows a
conceptual schematic of the overall layout of the data communication
system  for the Observatory.  Table~\ref{table:comms_performance} lists
the main performance characteristics. 

\begin{table*}[t]
\caption{Performance summary for the radio data communications
system for the Pierre Auger Observatory.}
\centering
\begin{tabular}{ll}
\toprule
\textbf{Microwave backbone network}
\\
\midrule
Links & 4
\\
Frequency & 7\,GHz
\\ 
Data rate & 24\,Mbps
\\
\toprule
\textbf{Wireless LAN}
\\
\midrule
Nodes & 1660
\\
Frequency & 902 to 928\,MHz ISM band
\\
Protocol & TDMA, custom
\\
Subscriber Unit over-air rate & 200\,kbps
\\
Effective payload rate & 1200\,bps uplink
\\
Typical daily data packet loss rate & less than 0.002\%
\\
\bottomrule
\end{tabular}

\label{table:comms_performance}
\end{table*}

\subsection{The microwave backbone network}

\begin{figure}[ht!]
\centering
\includegraphics[width=0.3\textwidth]{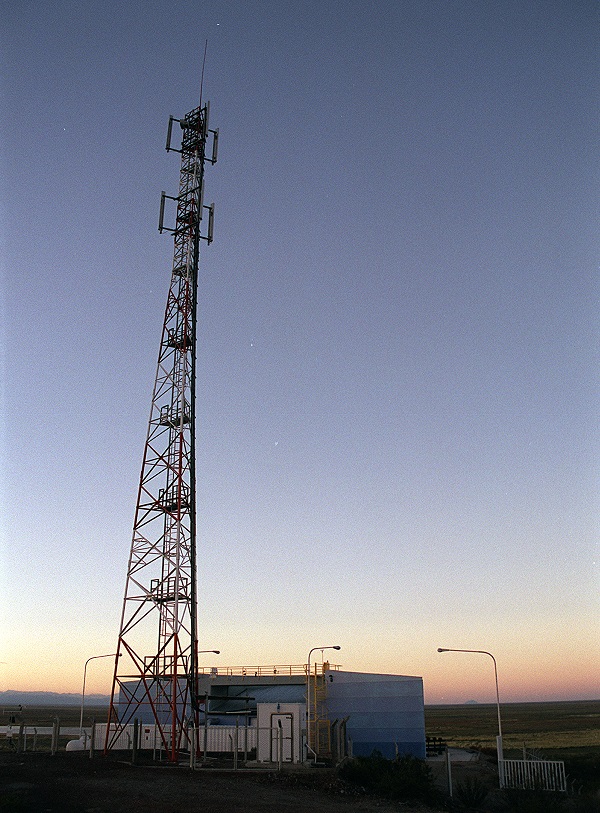}    % for ArXiV
\caption{One of the five communications towers: the one shown is deployed at
the Los Leones site; see Figure~\ref{comms_schematic}.}
\label{fig:tower}
\end{figure}

The top layer of the Auger data communications system is a 34\,Mbps
backbone network made from commercial, point-to-point, dish mounted
equipment operating in the 7\,GHz band.  Receivers and transmitters are
mounted on five communications towers located at the perimeter of
the array as depicted in
Figure~\ref{fig:tower}.  The microwave backbone provides high speed network
communications to nodes at all four FD sites and with the main campus.

\begin{figure}[t]
\centering
\includegraphics[width=0.48\textwidth]{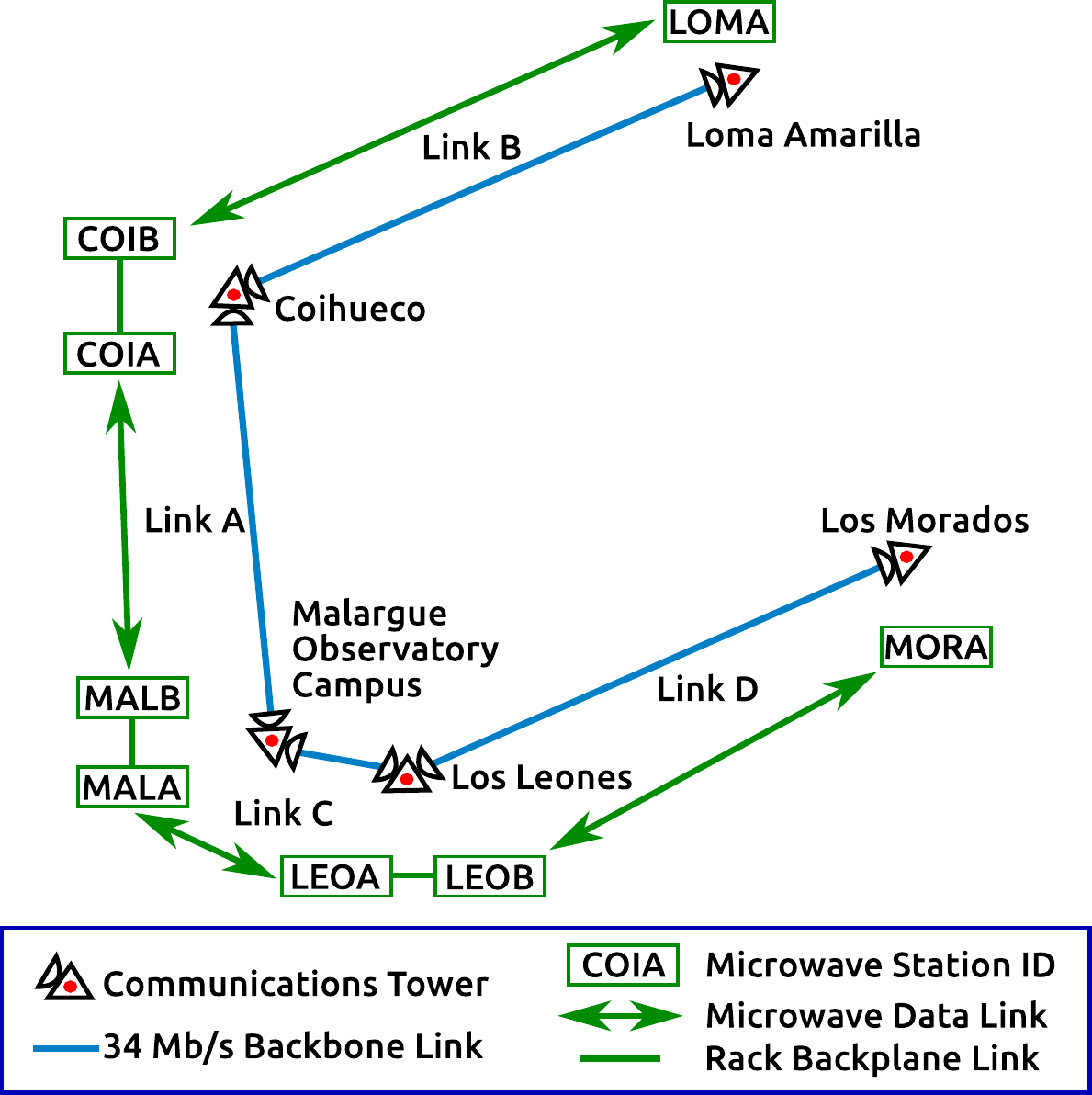}
\caption{Configuration of the high capacity microwave backbone
  network that connects the four FD sites with the main campus control
  and data acquisition center in Malarg\"ue.}
\label{fig:comms_backbone}
\end{figure}

The microwave backbone, depicted schematically in Figure~\ref{fig:comms_backbone}, 
consists of a set of paired links providing
sufficient capacity to stream data to and from each of the FD sites
as well as for collecting data from the individual surface stations.

\subsection{Wireless LAN}

The bottom layer of the Auger communication system consists of an extended
WLAN comprising custom designed units
operating in the 902 to 928\,MHz ISM band.  A point-to-point
bidirectional communications link is established between each surface
detector station and one of four communication concentrator nodes
mounted on the four towers located at each of the fluorescence detector
sites.  Communication is achieved in a manner similar to a cellular
telephone system by dividing the array into 28~sectors, each of which
contains up to 68 stations.

Communications operations at each surface station are governed by a
custom-built programmable Subscriber Unit used to mediate the
transmission and reception of digital data between the electronics board 
of a surface detector and the concentrator node.  An analogous
custom-built unit, called a Base Station Unit, mediates data
transfer between each concentrator node and the backbone network
connection at each tower.

\subsection{Time division multiple access}

\begin{figure}[t]
\centering
\includegraphics[width=0.48\textwidth]{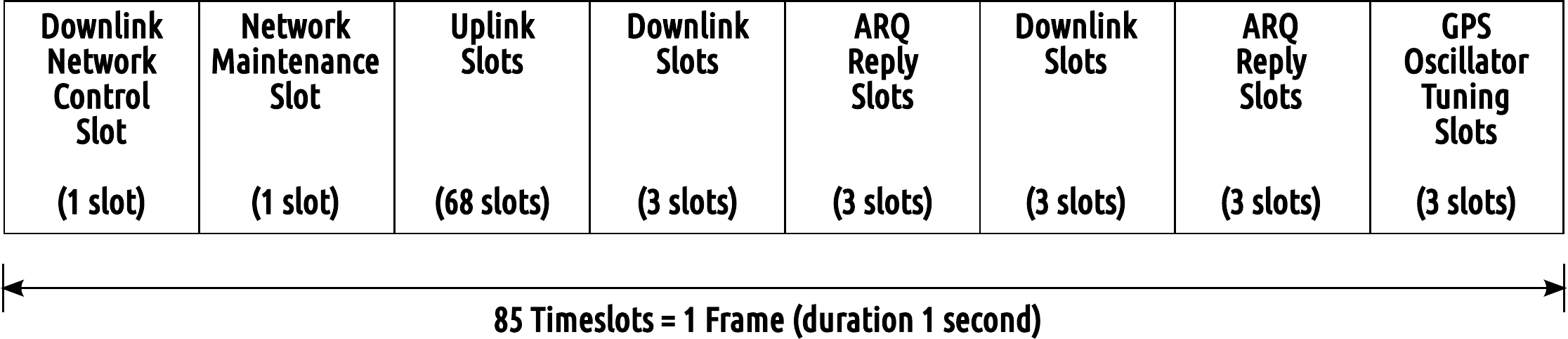}
\caption{A single GPS synchronized one-second TDMA frame is broken in time slots as shown.}
\label{comms_tdma}
\end{figure}

Transmissions to and from the stations are synchronized by GPS timing
so that each station is assigned a particular time slot during which
it is available to send and receive data.  This Time Division Multiple
Access (TDMA) scheme provides a contention free communication
environment within the array. A one-second data frame includes
68~uplink slots for collecting data from the array and 6 downlink
slots for sending trigger requests and other commands to the
stations. An additional 11 slots are reserved for network management,
monitoring, and packet error control. The assignment of individual time slots
within the one-second TDMA frame is shown in Figure~\ref{comms_tdma}.  
This makes available an effective bandwidth of at least
1200\,bps uplink for each surface station and a 2400\,bps for broadcast
downlink.

\subsection{Error handling}

A central requirement of the Auger WLAN system for collecting data
from the surface detector stations is very high reliability.  In the
wake of a typical trigger, digitized data from PMT traces and other
detector information must be relayed promptly to the CDAS so that the event can be built and
recorded. Data from a single event trigger will typically be broken
into several dozens packets transmitted by each station on request, a
process that can continue for as long as two minutes.  If even a
single data packet is missing or corrupt, the entire trace from the
station is lost.  A custom packetization protocol that includes Cyclic
Redundancy Checking to detect transmission errors is used at
every level.  An advanced retransmit-on-error scheme, commonly called
an Automatic Repeat Request (ARQ), is also employed.  The ARQ scheme
is especially designed to prevent data loss in the case of variable
signal fluctuations, external sources of interference or any other
episodic environmental influences.  If a packet is flagged as missing
or corrupt at the monitoring concentrator of the central network, a
request to retransmit the packet is automatically initiated and
collected via the subsequent data frame reserved TDMA time slots.  The
ARQ request will be sent once per frame and will be repeated so that
at least six attempts are made to retrieve each missing or corrupt
packet.
%This scheme results in a nearly error-free data stream from the array.

%
%
%  Corbin Covault: 26 May 2014: Removed following figure per reviewer comment.
%
%\begin{figure}[t]
%\centering
%\includegraphics[width=0.4\textwidth]{comms_su_photo}
%\includegraphics[width=0.4\textwidth]{comms_bsu_photo}
%\caption{Top: Photographs of the custom-built radio Subscriber Unit
%  which controls wireless communications at each surface detector
%  station.  Bottom: Photo of the Base Station Unit (BUS) that controls
%  wireless communications for each sector, located at the concentrator
%  nodes on each of four communications towers.}
%\label{comms_su}
%\end{figure}

%!TEX root =  AugerSouth.tex
\section{Central data acquisition system}
\label{sec:CDAS}

%\red{(Antoine;
%11Dec12 - Ruben S: text added)}

\subsection{Overview}

The CDAS has been running since March 2001.  The system was designed to 
assemble the triggers from the surface array detectors, to allow control of 
these detectors and to organize the storage of data.  It is constructed using 
a combination of commercial hardware and custom made, high level, software 
components. The system is designed to run continuously, with minimum 
intervention, with the full 1660 detector array, and can manage many more. 
Data from the FD are recorded separately at the FD locations and  transferred 
daily to the Computer Center at Malarg\"{u}e, although hybrid coincidences 
are identified online within the SD data stream.

The primary role for the CDAS is to combine local trigger information from the SD stations in order to identify 
potential physical events generating an SD higher level trigger (T3). These triggers combined with the T3 from FD 
sites (FD T3) are used to generate a request for the relevant data from SD stations for these events. The CDAS is 
then used to combine and store these data to form a shower event. The CDAS also contains configuration and control 
mechanisms, the means to monitor system performance, and the tools to access and download SD monitoring, calibration, 
control and configuration data. 

Except for triggering information (see Section~\ref{event-triggering}), the CD\-AS and the FD data acquisition 
systems are completely independent. The merging of  FD and SD data is made offline during the daytime following an FD 
run.  Data are synchronized on the central storage hardware after each night of observation.  The newly acquired data 
within the central storage are mirrored at the primary data mirror located at the Lyon HEP Computer Center (France) 
every 3 hours; later these data can be transferred to secondary mirror sites.  The data may then be 
transferred from a convenient mirror site to over 50 
participating institutions.

The communication between applications within the CDAS is controlled using a central message routine manager 
called the `Information Kernel'. This manager allows formatted messages to be broadcast by producer applications 
(applications that need to advertise their status), and for consumer applications (applications that need to know 
about the status of others) to receive them on demand.  All data, with one exception, are exchanged between the CDAS 
applications in human readable formatted ASCII and go through the `Information Kernel' manager.  The exception is the 
large binary block of raw data coming from the SD stations.  Data exchanged in raw format are calibration blocks and 
FADC traces (these comprise the event data), data from local triggers as well as the monitoring data.

Since the beginning of 2013, the CDAS runs on a virtual machine using resources within a private Cloud installation. 
The Cloud is composed of 6 servers, summing up 42 CPUs and 112\,GB of RAM. CPUs are 2\,GHz or faster. This scheme 
allows the live migration and automatic failover of virtual machines, to minimize the impact of critical failures. 
The disk storage system is comprised of redundant disk arrays installed in each server plus some standalone devices, 
making a total storage space near 8.5\,TB, using a shared, replicated and distributed scheme. A Network Time 
Protocol (NTP) GPS clock is used to synchronize the system times.  We have adopted the GNU/Linux Debian latest stable distribution as the operating system, currently v6.0r4.  Only a very small fraction of the CPU power is used by the CDAS 
application processes.  Most of the software was developed in C or C++. 
The whole system is installed in a Computer Center at the Observatory campus, with controlled temperature and 
redundant uninterruptible power supply.

\subsection{SD data collection}

The data flow over the radio network, from individual SD stations to the central campus, is controlled by a dedicated 
application called the `Post Master'. The Post Master is the end point of the communication backbone at the 
Observatory Campus, and is designed to dispatch information extracted from the different data streams of a local 
station to the other applications of the CDAS.  As its name suggests, the Post Master application is used to read the 
data type contained in a radio frame and to forward it to the proper application within the CDAS so that specific 
data can be handled.  When the data received from individual SD stations are split into several radio frames, they 
are reassembled and forwarded to clients by the Post Master after all the frames have been received. 

The Post Master is used also to route data between the applications of the CDAS and the SD.  Commands and 
configuration parameters can be transmitted, along with event requests, such as the level 3 trigger identified by the 
`Central Trigger' processor. Software downloads over the communications link are also possible, thus enabling 
upgrades of the local DAQ software at the stations without the need to travel many kilometers to each one.

Data received from each SD station belong to different data streams:

\begin{enumerate}
 \item {\em Local triggers, T2}: the highest priority stream, with a list of time stamps and the type of trigger (threshold or time over threshold), is forwarded to the `Central Trigger' application.
 \item {\em Shower data with its calibration data}:  data in this high priority stream are sent only when a request is received from the CDAS at an SD station.  Shower events are split into small pieces and sent together with the T2 packets so that the available bandwidth is fully used. These data are forwarded to the `Event Builder' application. 
 \item {\em Control}: this is a medium priority stream that describes the state of the detector. It is forwarded to the central `Information Kernel' of the CDAS.
 \item {\em Calibration and monitoring information}: this is a low priority data stream.  It is forwarded to the `Monitoring Recorder' application.  
\end{enumerate}

\subsection{The event triggering system}
\label{event-triggering}

The triggering system of the Observatory fulfills two conditions. First, it detects showers with high efficiency 
across the SD array, that is more than 99\%  for vertical showers with energy above  $3 \times 10^{18}$\,eV.
Second, it allows and identifies cross-triggers (hybrid events) between the FD and SD systems. Triggers from the FD 
use  separate algorithms but are forwarded to the SD system to construct the hybrid data set (see below).

The local DAQ system of each SD station is designed to generate low level triggers (T2) as described in 
Section~\ref{SDLocalTrigger}.  The time stamps of these triggers are sent every second to the CDAS. The T2 requirements are such 
that the average rate is always around 20 to 25\,Hz so that at least 50\% of the bandwidth is free for data 
transmission.  This limitation does not reduce the global trigger efficiency (see below). At the CDAS, the T2s 
received from all stations are stored in a data block stamped according to the second to which they correspond. This 
information also allows us to acquire the status of all stations of the array at each second. Once a block has 
existed for a time greater than the maximum transit time across the radio network (five seconds), it is transmitted 
to the `Central Trigger' processor and discarded.

The Central Trigger, or third level trigger {\bf T3}, initiates the central data
acquisition from the array in the three following conditions: 

\begin{enumerate}

\item  A {\em main trigger} condition, corresponding to shower candidates, is based on both the time and the spatial clustering of the local triggers received from stations, and is described in detail below. 
\item A {\em random trigger} is generated every $N$ minutes by selecting one of the T2s in an arbitrary manner, and promoting it to a T3.  Currently, $N=30$ but values of 3 and 15 have also been used.  The purpose of this trigger is to randomly monitor  the FADC traces that satisfy the local trigger conditions and thus to verify the efficiency of the global trigger processor. 
\item A {\em 2-fold coincidence} within 1\,$\upmu$s of one of the two doublet stations, two couples of neighboring stations (10 m distant).  These occur at 0.8\,Hz and are scaled to 0.0017\,Hz for transmission. 
\end{enumerate}

To apply the  {\em main trigger } condition, the system defines concentric hexagons centered in each station. 
The `Central Trigger' processor is used to identify groups of stations that are clustered in time and space as SD events.  
First, time clusters are sought by centering a window of $\pm25\,\upmu$s on each T2. Clusters, with multiplicity of 
three or more, are then examined for spatial coincidences.

\begin{figure}[t]
\centering
\includegraphics[width=0.8\textwidth]{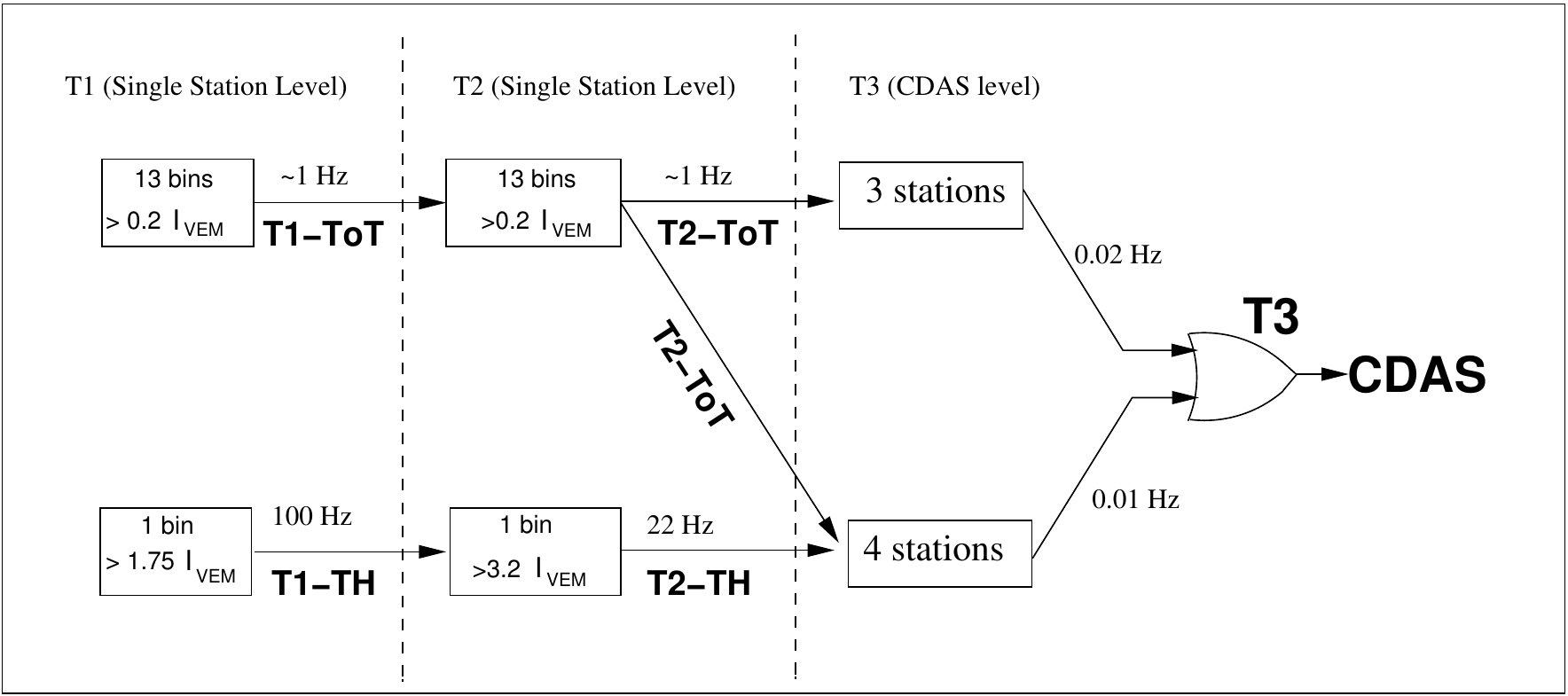}
\caption{Schematics of the hierarchy of the trigger system of the Auger surface detector.}
\label{fig:trigger}
\end{figure}

\begin{figure}[hbt]
\centering
\includegraphics[width=0.35\textwidth]{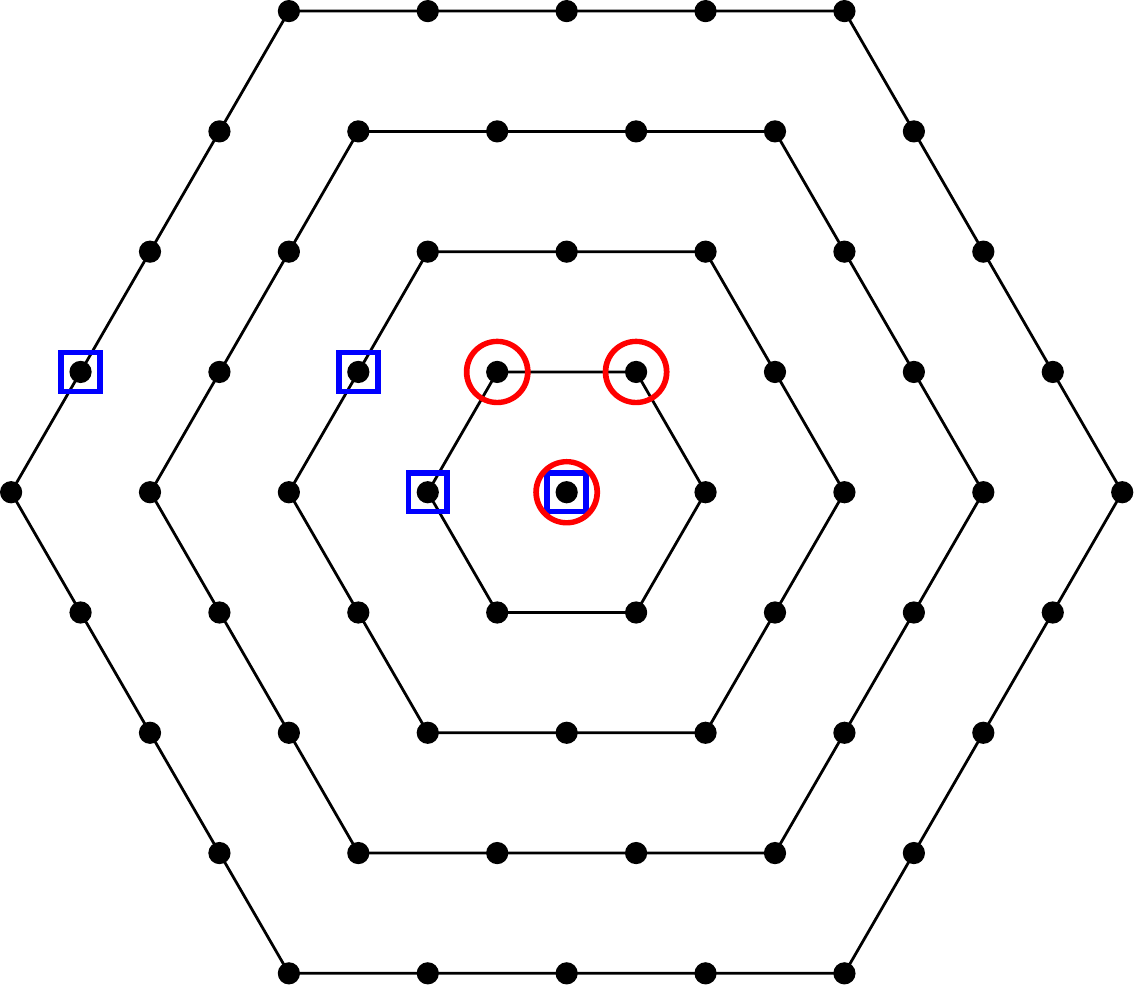}
\caption{Four hexagons, containing stations, are illustrated around a central surface station, for a portion of an ideal array. For a 3-fold coincidence, a T3 is issued if  the 3 T2s are ToT, and if one of them is found in the first hexagon of the central station, and the other one no further than the second hexagon. A 4-fold coincidence applies to any kind of T2s and the additional station may be as distant as in the 4th hexagon. Two examples of the topology of triggers are shown: a 4-fold coincidence in which the triggered stations are identified by blue squares, and a 3-fold coincidence identified by red circles.}
\label{fig:crowns}
\end{figure}

The  {\em main trigger} condition is satisfied in two modes depending on the local shower trigger conditions.  
A block diagram illustrating the logic chain and approximate rates of these trigger modes is shown in
Figure~\ref{fig:trigger}. The first mode requires the coincidence of at least three 
detectors that have passed the T2-ToT trigger condition (described in Section~\ref{SDLocalTrigger}) and that meet the 
requirement of a minimum of compactness, namely,  one of
the fired detectors must have one of the other fired detectors in the first hexagon of neighbors while another one 
is no further than the second hexagon. 
The second mode is more permissive. It requires a four-fold coincidence
of any type of T2 (T2-ToT or T2-TH) with moderate compactness. Namely, among the four fired detectors, one station may be as far away as  
the fourth hexagon, if a station is within the first hexagon and another station is no further than the second hexagon.
 Figure~\ref{fig:crowns} illustrates the geometric requirements of the two trigger modes.

Once the spatial 
coincidence is verified, final timing criteria are imposed: each T2 must be within $(6+5n)\,\upmu$s of the central 
station, where $n$ represents the hexagon number.

Once a trigger has been identified, a T3 message requesting that all FADC trace information recorded within 
30\,$\upmu$s of the central T2 is built by the `Central Trigger' processor and forwarded to the `Event Builder' and to the 
stations by the `Post Master'. The trigger message is also stored locally by the `Information Kernel'. To select 
which stations are asked for their traces the system takes each station within six hexagons of each of the stations 
whose T2 participated in the T3 construction. Additionally, the `Central Trigger' process also stores the number of 
T2s for each station recorded for monitoring purposes.

With the arrangements described above, the total T3 trigger rate of the Observatory is presently of the order of  
0.1\,Hz and about 3 million SD events are recorded yearly.

The DAQ system of the fluorescence detector is completely independent of the CDAS. Local triggers are generated at 
each FD site and those identified as T3 FD event triggers are logged by a local processor if a shower track can be 
found. T3 FD event triggers are transmitted online (within one second) from the local FD site to the CDAS system at 
the central site. The trigger information sent describes the geometry of the shower candidate. This includes the 
estimated time of arrival of the light front of the shower at the camera as well as the geometry of the SDP (see 
Section~\ref{sec:SDP}). From this information, the time of impact of the shower at a ground position in the region of 
the SD stations is computed and a corresponding SD event T3 is constructed. All FADC traces recorded within 
20\,$\upmu$s of the computed time are assembled as a normal `SD only' event, but with the addition of the identification of the 
corresponding FD T3 trigger.  

After each night of operation, details of events recorded at the FD telescopes for each T3 FD event triggers are 
transferred to the CDAS.  Data from these triggers are then merged with the data collected by the SD DAQ and form the 
hybrid data set. A hybrid event is therefore an `FD only' event together with a special SD event that contains all 
the information from the surface stations that were in space and time coincidence with the FD event. 

Cross calibration of the SD and FD clocks is achieved by firing a laser into the sky and, at the same time, injecting 
a portion of the laser signal into one of the SD stations via an optical fiber. The time of laser pulse emission and 
the local time stamp recorded in the station are then compared. The former is reconstructed from the laser track 
recorded in the telescopes and the latter from the local trigger generated by the light going through the tank.

\subsection{Operation and control}

The CDAS has been designed to provide the means to monitor both its own operation as well as the slow control system 
at SD stations and various environmental parameters. The operation of the CDAS is monitored using a low level 
application that routinely checks that all software components are running correctly. This `watch dog' system is used 
to reinitialize and relaunch any application that may have failed. Over the Observatory's life, the CDAS has been 
operational more than 99.9\% of the time. Most of the downtime is due to system tests, upgrades and debugging with a 
minimal impact of critical failures.

The manager for the `Information Kernel' system, used to route messages, can also serve as an offline monitoring 
tool. Its architecture is based upon one central daemon, the message dispatcher, to which all messages are sent 
automatically and transparently. Any application wishing to distribute its status information, or wishing to know 
about the state of other applications in the system, connects to the `Information Kernel' and issues a monitoring 
request defining the class of message it wants to hear.

The `message listener' applications range from monitoring applications to system oriented ones, such as the message 
logger that is used to make all messages persistent. Applications have been developed that allow these messages to be 
browsed. Thus, the behavior of the system can be monitored both online and offline. The storage capacity is 
sufficient to keep the complete history of all messages exchanged in the system (around 3\,GB per year of 
uncompressed ASCII files) for several years.

The `Monitoring Recorder' application of the CDAS is the core of all subsequent monitoring applications that will be 
described in next section. 

\begin{figure}[t]
\centering
\includegraphics[width=0.48\textwidth]{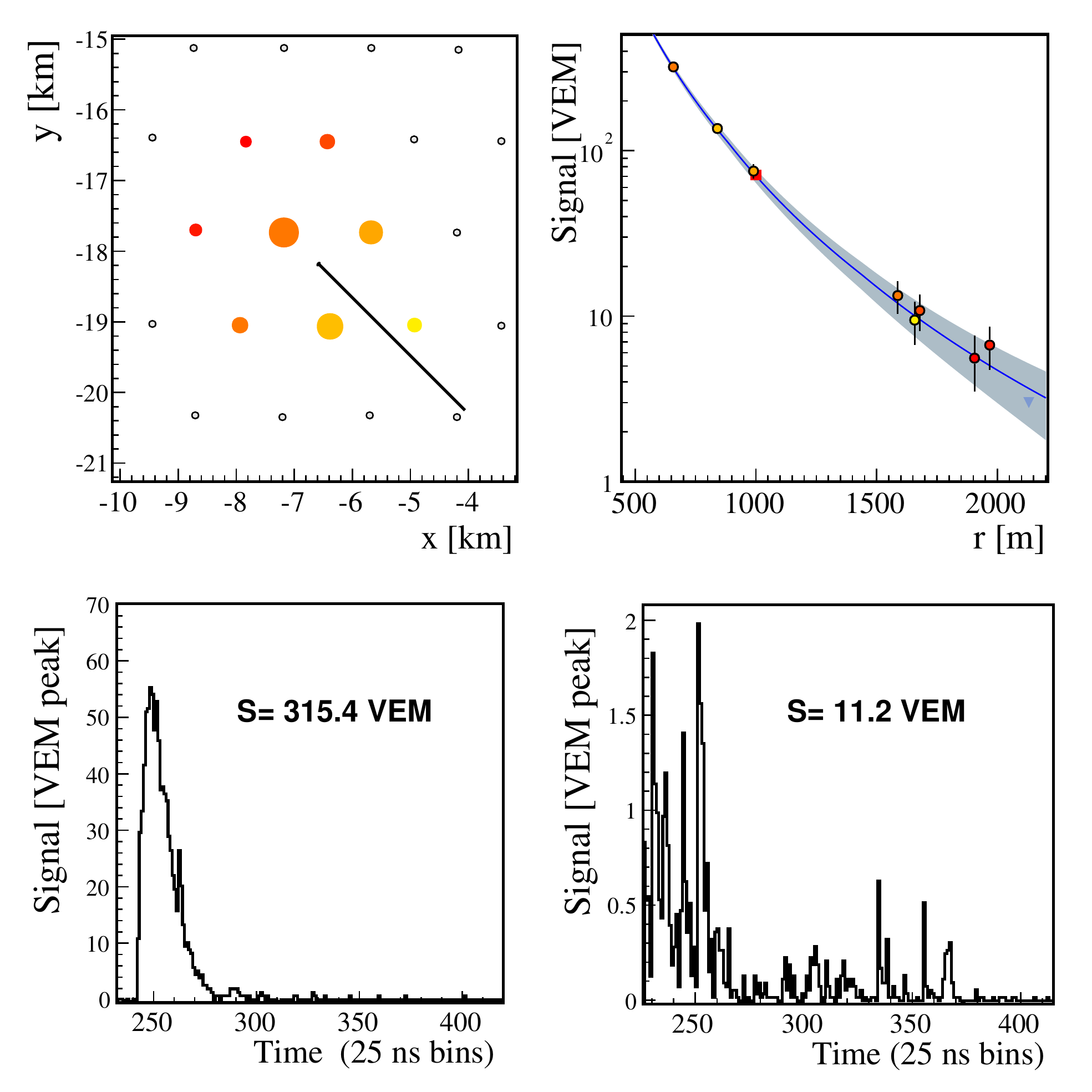}
\caption{Event 13357690: a typical vertical event of about $3\times10^{19}$\,eV . {\em Top left}: The array seen from above with the 8 triggered stations. {\em Top right}: The fit to the lateral distribution function (LDF) for this shower of zenith angle $28^\circ$. {\em Bottom}: The FADC traces from 2 detectors at distances of 650 and 1780 meters from the shower core. The signal sizes are in units of VEM.}
\label{fig:Evt_13357690}
\end{figure}

An event display program (part of which can be seen in Figure~\ref{fig:Evt_13357690})
% and \ref{fig:Evt_850018}) 
 has also been developed.  This allows the selection, viewing and reconstruction of SD events that are stored on disk. This program, and the input/output and reconstruction libraries that compose it, have also been used extensively for preliminary data analysis.

%\begin{figure}[t]
%\centering
%\includegraphics[width=0.48\textwidth]{CDAS_EVT850018}
%\caption{Event 850018: A high multiplicity inclined event .  {\em Left}:  Display of the array on the shower plane with the contour plot
%of the fitted distribution superimposed, indicating the signal measured in the 37 triggered stations
%and the position of the reconstructed core. The reconstructed zenith angle is 71$^{\circ}$ and the best fit
%value of N19 is 9.2 which corresponds to an energy of 54.6EeV after the calibration procedure. The
%color code indicates the start time from early (blue) to late (red) stations. {\em Right}:  signal sizes in the
%triggered stations as a function of the distance to the shower core in the shower plane. Filled and
%open symbols indicate measured and expected signals, respectively.
%}
%\label{fig:Evt_8500018}
%\end{figure}

\section{Monitoring}
\label{sec:monitoring}

%\red{(Julian, Corinne)}

\subsection{Overview}

For the optimal scientific output of the Observatory the status of the detector array as well as its measured data 
have to be monitored.
In normal operating conditions, shift personnel or ``shifters" monitor
the performance of the Auger detector systems.
The Auger Monitoring tool %~\cite{AugerMON} 
has been developed
to support the shifter in judging and supervising 
the status of the detector components, the electronics, communications, and the data acquisition. It is also useful to study 
``offline" the behavior of the different components of our detectors and to define quality cuts on our data when necessary. 

The two hybrid detector systems, SD and FD, are operated differently and therefore the monitoring of their status has different requirements. 
%SD
The stations of the SD array operate continuously in semi-automated mode. 
Data acquisition must be monitored and failures of 
%stations or of their communication 
any station component (power system, PMT, communication device)
must be detected.
%FD
The data taking of the FD 
can only take place under specific environmental conditions 
and is organized in shifts. 
The sensitive cameras can only be operated on dark nights without strong winds or rain.
This makes the operation a busy task for the shifters who have to judge the operating mode 
on the basis of the information given by the monitoring system.

The technical description of the implementation is given 
separately for the services of the central server and the subcomponents.
The subcomponents show the different ways 
the data flow to the central database are organized.

\subsection{Server techniques}
\label{server}

The basis of the monitoring system is a database. 
We have chosen the widespread and publicly available MySQL database system.
It includes all necessary features, e.g., replication and stored functions/procedures.
The front end is based on a web server running Apache. The web site uses mainly PHP, CSS and JavaScript.
An interface has been developed for generating visualizations. Currently the usage of Gnuplot is implemented via an internal system call. 
The second option used for generation of visualizations is the JPGraph package 
which is implemented with direct PHP calls on an object oriented basis with the interface 
defined in the inherited classes.

Alarms, occurrences of states that require immediate action,
are first entered into a specified table of the database.
The web front end checks this table for new entries and displays them on the web page.
The shifter is trained to observe and acknowledge the alarm;
when the problem is solved, he/she can declare the corresponding alarm as resolved.
%and, after solving the problem, can declare this alarm as resolved.

In addition to  the detector performance and data quality monitoring, 
the functionality of the monitoring server itself has to be guaranteed.
Therefore, the central services of the computer have to be monitored 
to assure that an alarm will definitely be noticed by the shifter.

\subsection{Surface detector}
\label{sd}

%\begin{figure}[t]
%\centering
%\includegraphics[width=0.48\textwidth]{T2array}
%\caption{Graphical representation 
%of the T2 trigger information for the array of the SD tanks.}
%\label{sd_t2}
%\end{figure}
%
%\begin{figure}[t]
%\centering
%\includegraphics[width=0.48\textwidth]{CDAS_Fig_monitoring}
%\caption{Monitoring plot, as a function of time, of the VEM ADC value (top) and the board temperature (bottom) of PMT2 in one of the SD stations.  Day/night variations, although at only the few percent level, are clearly visible, as is the correlation between the two parameters.}
%\label{fig:monitoring}
%\end{figure}
%

With large ambient temperature variations, high salinity, dus\-ty air, high humidity inside the tank and remoteness of access, monitoring the performance and reliability of the SD array is a challenge.
The temperature ranges  from $-15^\circ$C in winter up to 60$^\circ$C inside
the electronics box in summer, with typical daily cycling of about 20$^\circ$C.
Inside the water tank the relative humidity can reach about 90\%. In addition, thunderstorms are rather strong, with lightning that
can damage electronic components. 

To monitor the whole array accurately, various sensors are installed in every station.
Temperature is measured on each PMT base, on the electronics board, and on
each battery. PMT voltage and current are also monitored, as well as solar panel voltages, individual battery voltage and charge current.
The calibration described in Section~\ref{subsec:sde_calib} is performed online every minute. 
A number of quantities are computed to check the behavior of each water Cherenkov detector: 
baseline values, single muon peak signal, single muon average charge, dynode/anode ratio and PMT stability.
The monitoring and calibration data are sent to the CDAS every six minutes. 
Dedicated software constantly parses  the information sent to the CDAS, independently of the acquisition processes, and exports the data to the MySQL server.

\begin{figure}[htbp]
\centering
\includegraphics[width=0.48\textwidth]{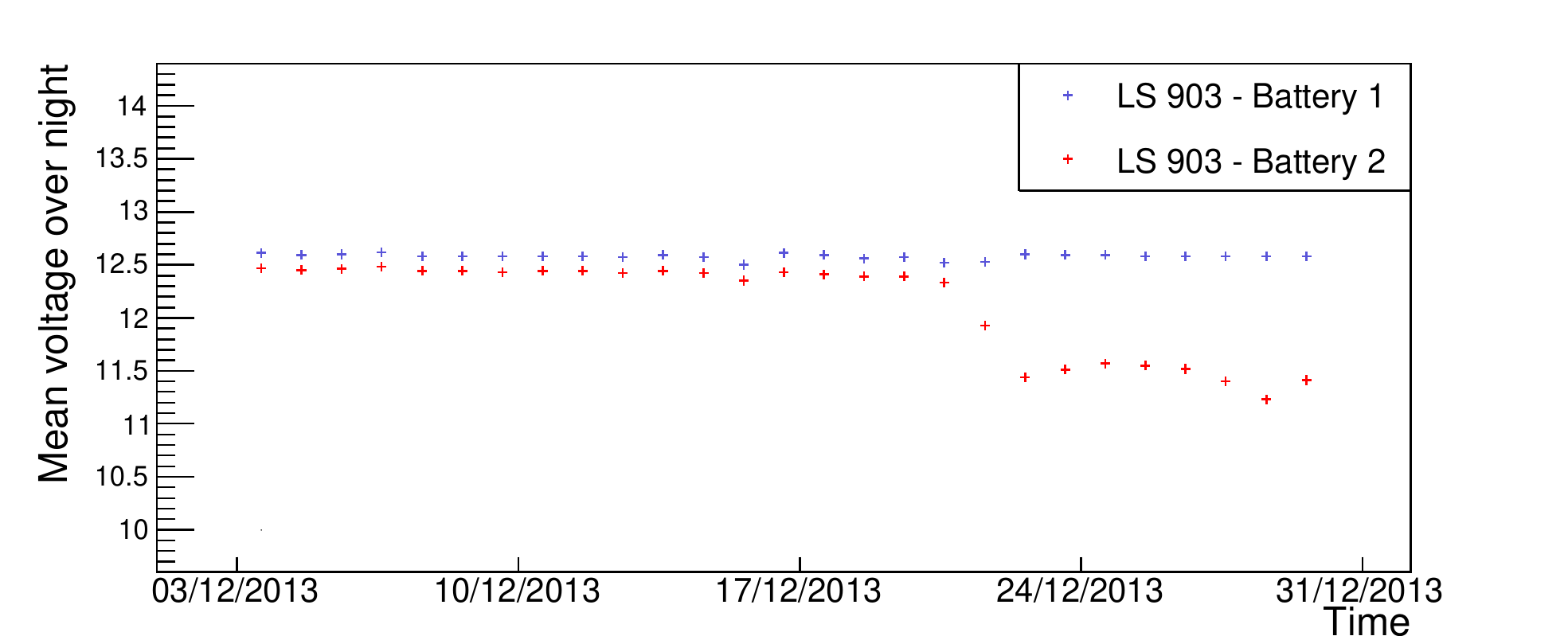}
\caption{Evolution with time of the battery average nightly voltage of station 903 in December 2013.}
\label{mon_bat}
\end{figure}

As an example, we show in  Figure~\ref{mon_bat} the monitoring of the daily average of the voltages of the two 
batteries of a tank. While the value for Battery 1 is stable above 12 V, it can be seen that since 22 December 2013, 
the value of Battery 2 has been decaying and is always below 12~V. When the voltage drops below 12~V for the first 
time, an alarm is triggered for shifters, so that they can make further checks on the history of the battery to 
understand the origin of the decay. Once the value becomes lower than 11.5 V, a more severe alarm is triggered to 
indicate that the battery should  be repaired or replaced. 

\begin{figure}[htbp]
\centering
\includegraphics[width=0.48\textwidth]{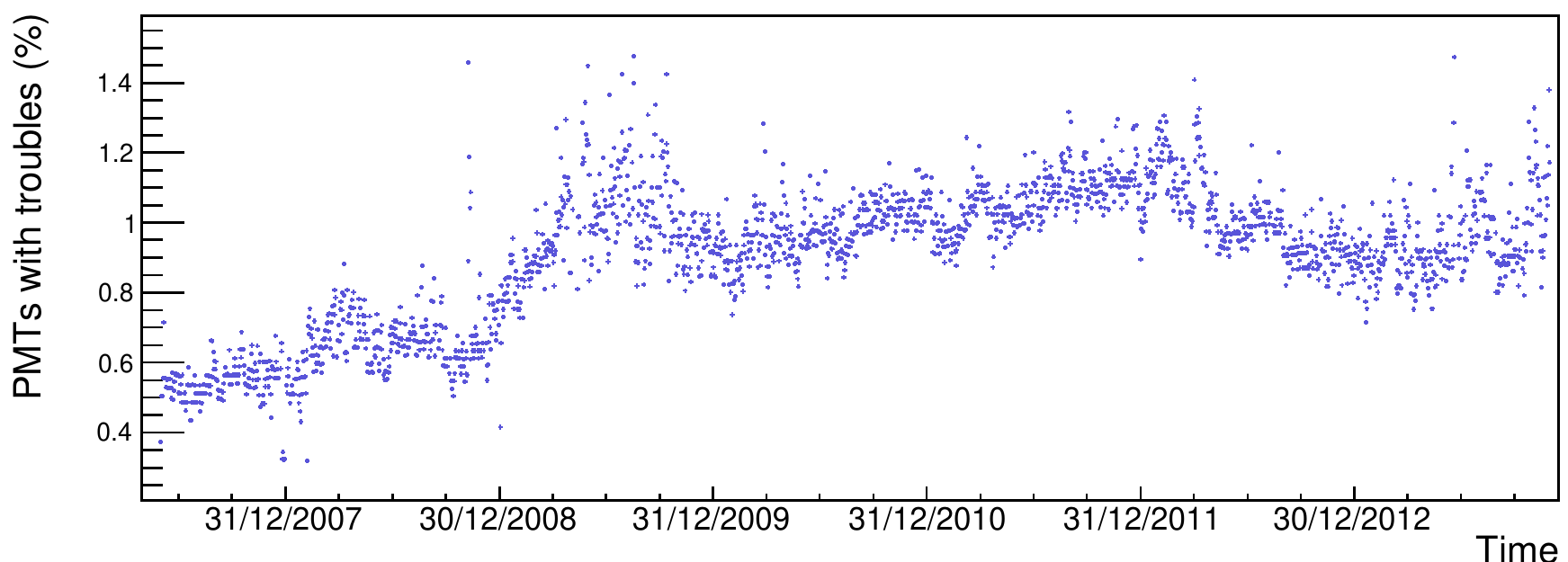}
\caption{Percentage of PMTs which do not satisfy the quality criteria among the functioning ones, as a function of time.}
\label{mon_pm}
\end{figure}

A second example shows how  the monitoring system is used to clean the data. 
Figure~\ref{mon_pm} displays the evolution of the fraction of PMTs that are rejected by the data analysis due to troubles detected by studying the monitoring data.
This number itself is controlled within the monitoring system. The number of low-quality PMTs increased after 2008 because the full array of nearly 5,000 PMTs was now deployed, and the Observatory staff needed to carefully balance maintenance priorities.  Since 2008 the rate of low-quality PMTs has been rather stable around 1\%, except for a specific period in 2009 when communication problems did not allow reliable monitoring of the array.  We chose to be conservative and disregard  doubtful data during this period (see Section~\ref{sec:performance}). 

\begin{figure}[htbp]
\centering
\includegraphics[width=0.4\textwidth]{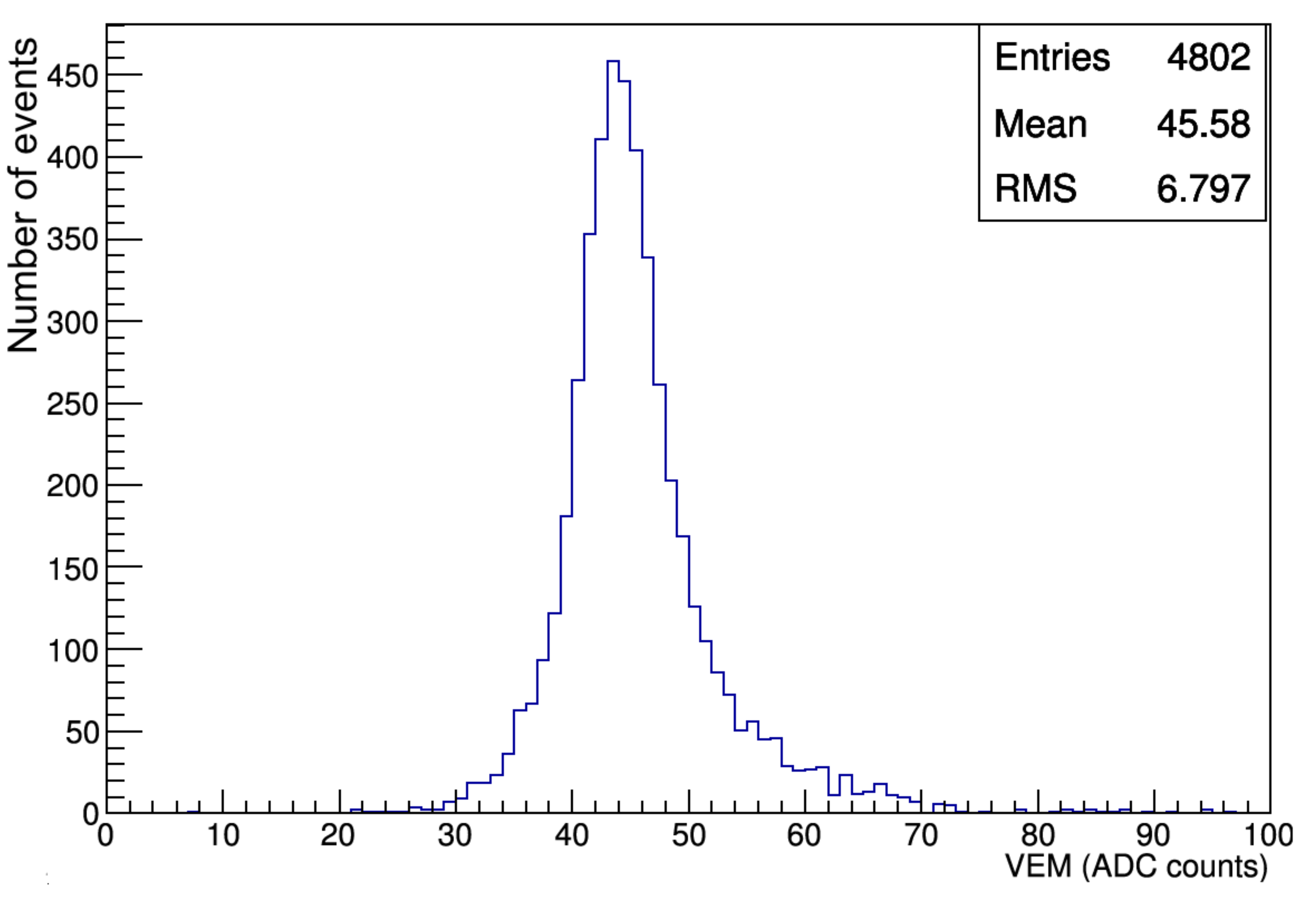}
\caption{VEM measured for 4802 PMTs.}
\label{fig:VEM}
\end{figure}
The monitoring tool also allows  a general control of the behavior of the array. Figure \ref{fig:VEM} shows the muon peak current ($I_\text{VEM}$)  values for 4802 PMTs.  The mean value of the muon peak ($I_\text{VEM}$) is at
channel 45.6 with an RMS of 6.8 showing good uniformity of the detector
response.  The typical day/night variations are of the order of two ADC channels. This is mainly due to the sensitivity of the PMTs to temperature.  The muon calibration is made  online every minute,  allowing the correction for temperature effects.

\begin{figure}[htbp]
\centering
\includegraphics[width=0.8\textwidth]{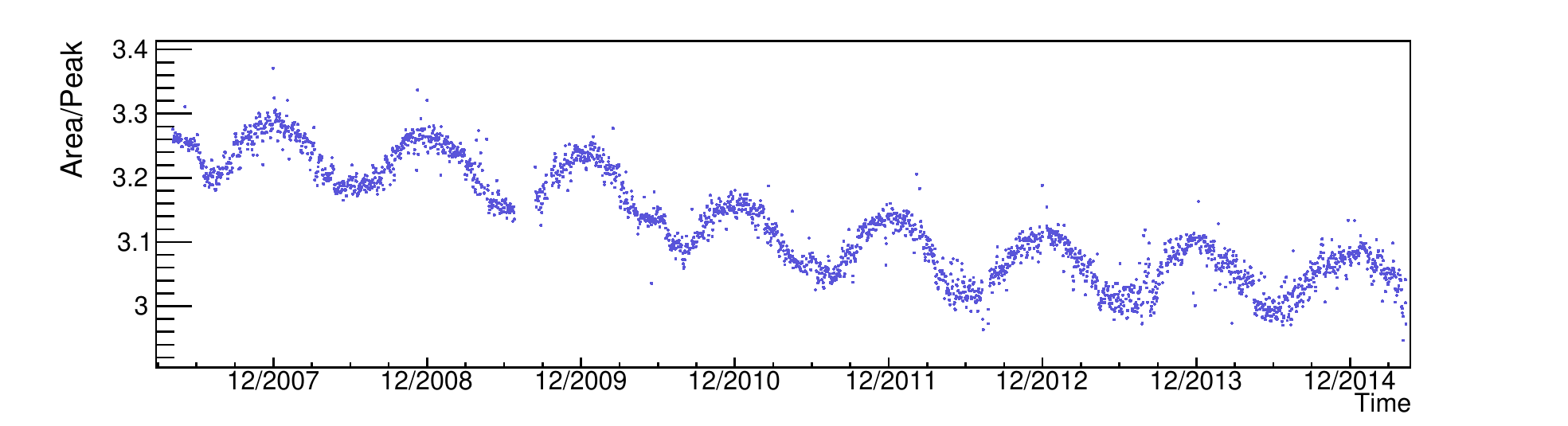}
\caption{A/P as a function of time for station 116. The dots are the average of the A/P value over one day.}
\label{fig:ATOP}
\end{figure}

The ratio Area/Peak (A/P), i.e., the ratio between the integrated charge and maximum of the atmospheric muon signals recorded with the calibration trigger, is also a monitored quantity directly available from the local station software. It is related to the water transparency and the reflectivity of Cherenkov light on the inner liner of the SD station. These properties
control the absorption length of the light and thus the signal decay constant. Figure~\ref{fig:ATOP} shows the decay of the A/P ratio of a typical station in the first seven years following deployment, coupled to a seasonal modulation. 
After 10 years of deployment, the A/P tends to be stable.  This behavior is described in detail in Ref.~\cite{Sato-ICRC:2011}. 

The two contributions of the T2 shower triggers (see Section~\ref{SDLocalTrigger}) are also monitored. While the 
T2-TH mode has a mean value of 22\,Hz with a low dispersion of less than 2\%, the T2-ToT mode contributes only 1\,Hz but 
with a much larger spread. Indeed, the ToT mode is directly related to the A/P, since it is by construction sensitive 
to the signal shape and thus to the characteristics of the detector. Therefore the T2-ToT rate also decreases with 
time. It has been observed that, even if the rates of the different stations show a large initial spread across the array, most 
of them stabilize after a few years to about 1\,Hz. Temperature variations also slightly affect the ToT 
trigger. Fortunately, these variations do not affect the uniformity and the stability of the data, since the event 
rate above the threshold energy of the experiment has not been affected, as will be shown in 
Section~\ref{sec:performance}. 

%Trigger rates  are   uniform over all detector  stations,  implying good calibration and 
%baseline determination. 
%The two contributions to the T2 (see Section~\ref{SDLocalTrigger}) can be studied with the monitoring.
%While the single bin threshold trigger has a mean value of 22\,Hz with a low dispersion of less than 2\%, the time-over-threshold trigger contributes only 1\,Hz but with a much larger  spread. 
%Indeed  this trigger  is sensitive to the pulse shape and thus is more sensitive to the characteristics of the detector. It is 
%observed that the newly installed detectors often have ToT values which are higher and then stabilize after a few months 
%to about 1\,Hz. Both  the trigger rates and the muon response studies show that the detectors have, after a few months' stabilization, good uniformity.  Temperature variations also slightly affect the ToT trigger. 

The monitoring tool also includes performance metrics  to control the overall performance of the Surface Detector array. 
One of the metrics is the number of stations sending a signal divided by the number of deployed tanks as a function of time,
indicating the efficiency of data collection with the SD array, which is typically better than 98\%.

\subsection{Fluorescence detector}
\label{fd}

\begin{figure}[htbp]
\centering
\includegraphics[width=0.53\textwidth]{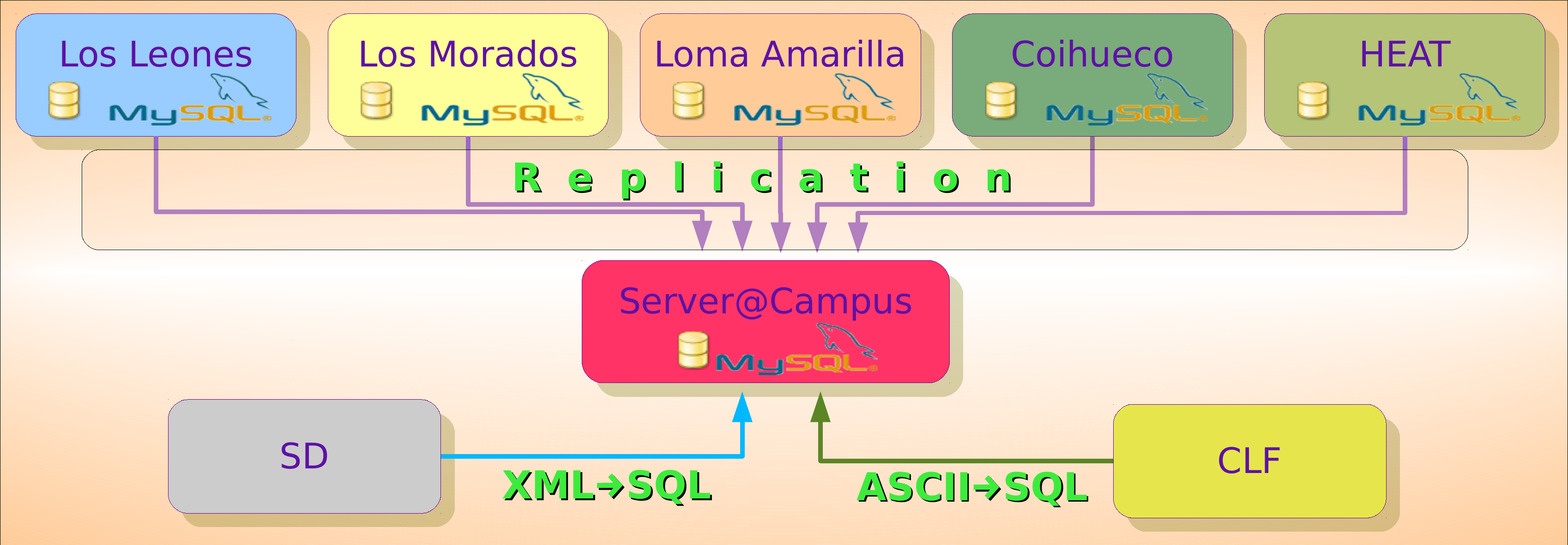}
\caption{Organization of the monitoring system databases: 
   The single databases at each FD site are replicated to the 
   database server at the central campus, while other sources like 
   SD fill directly into the database.}
\label{db-setup}
\end{figure}

The data acquisition for the FD telescopes is organized by site
to insure against disruption of data collection due to possible communication losses 
between the CDAS and the remote detectors.
The data transport for FD monitoring is organized via a database internal replication mechanism.
This mechanism recognizes communication problems and recovers  
submitted database changes when the connection is reestablished.
This guarantees completeness of the dataset on the central server,
even if the information is not immediately available online during network failures.
Figure~\ref{db-setup} shows a schematic layout of the databases.

\begin{figure*}[htbp]
\centering
\includegraphics[width=0.98\textwidth]{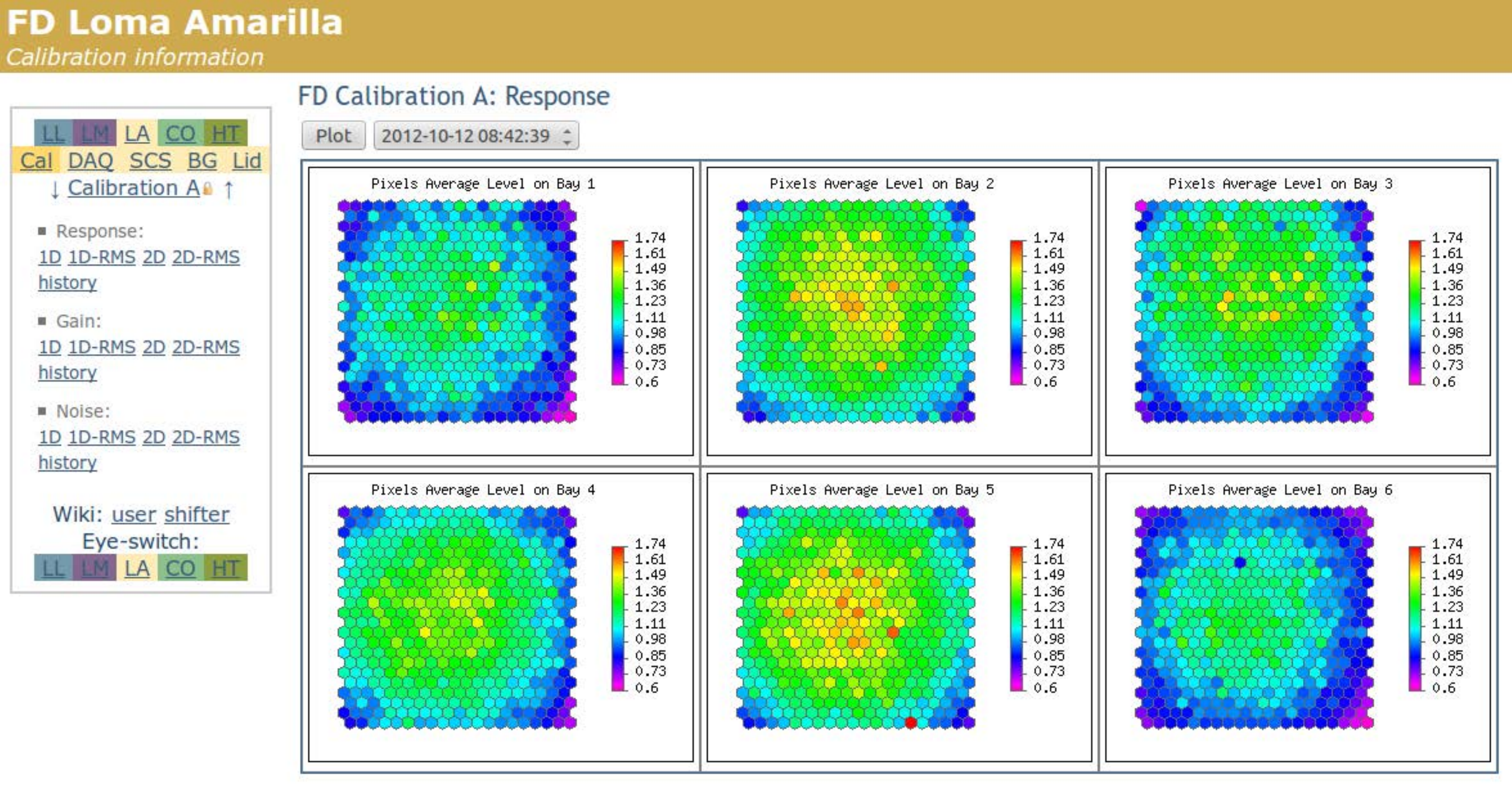}
\caption{Screenshot of the web interface
showing a selection of FD calibration data
for the six cameras in an FD site 
in a specialized view
representing the PMT geometry.}
\label{calib2D}
\end{figure*}

The information collected for the supervision of the FD operation is organized into five
sections. 
The calibration section contains the information from the different levels of calibration 
as described in Section~\ref{FdCalibration}, with an example representation given in Figure~\ref{calib2D}.
The background data section contains the information obtained from each 30 second readout of the full camera,  
which is valuable as an unbiased observation of background.
%     It can be used to verify the uptime as typical non-measuring states 
%     such as closed shutter lead to well identifiable pattern of low variances.
The section on DAQ and trigger shows the frequency of fired triggers that 
indicates the status of the telescopes at an advanced stage.
Information from the Slow Control System such as rain, wind, and outside and inside temperatures is displayed in the fourth section.
The lidars~\cite{BenZvi:2006xb} monitor the atmosphere close to the telescopes.
Their information helps judging the atmospheric conditions 
at the site, which is vital for the operation of the telescopes.

The data collected in the database can be used to derive higher level quantities such as 
the on-time of the FD telescopes.
This quantity is of major importance since it is a necessary ingredient of flux measurements.
The dead time of each telescope is also recorded in the database. 
Together with the run information and other corrections retrieved from the database, 
the total on-time for each telescope can be determined individually.
The on-time is calculated only for time intervals of ten minutes, 
balancing the statistical precision of the calculated on-time due to statistics
with the information frequency.
A program to execute the calculation runs on the database server and is used to  fill 
the appropriate tables in the database continuously. 
The web interface displays the stored quantities.
%An example of one night of data taking is given in figure~\ref{fUptime}.
The on-time is available in near real time for the shifter as a diagnostic 
and figure of merit.

\subsection{Communications}

\begin{figure}[htbp]
\centering
\includegraphics[width=0.48\textwidth]{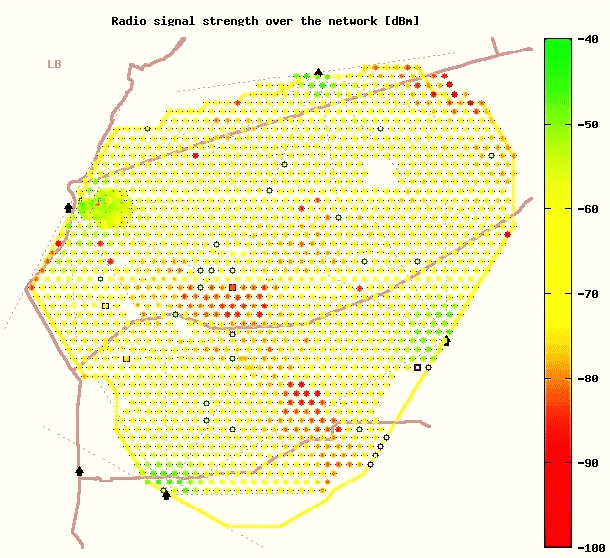}
\caption{Example of near real time performance monitoring of
  the Auger data communications system showing a map of the radio signal strengths (in dB) for each detector in the array.}
\label{comms_radio_signal}
\end{figure}

\begin{figure}[htbp]
\centering
\includegraphics[width=0.7\textwidth]{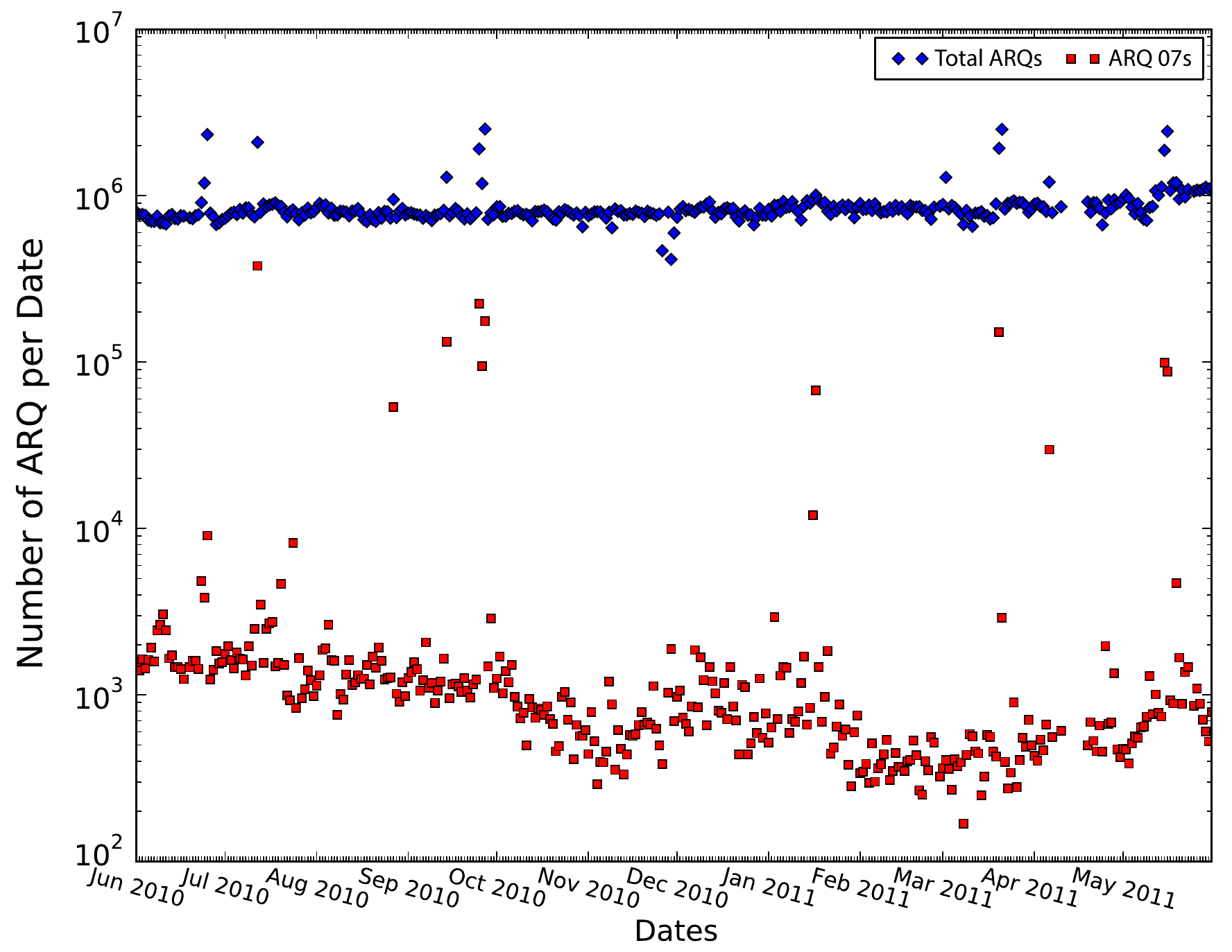}   %new version of plot to address reviewer comment
\caption{Example of benchmark long term performance monitoring of
  the Auger data communications system showing the daily rate of
  ARQs over a period of 12 months from June
  2010 to May 2011.  Upper points indicate the rate of any ARQ; almost
  all of these result in successful retrieval of a lost or corrupt
  data packet.  Lower points (ARQ\,7) correspond to packets which were
  lost after at least six attempts to retransmit.}
\label{comms_arq_rates}
\end{figure}

All aspects of the Auger data communications system control,
operations, and performance housekeeping information are coordinated
and reported via a central data concentrator node called NetMon, which
also serves as the relay for all data transferred between the
detectors and the array control center.  NetMon enables regular
monitoring of the system performance, including critical details
on the status of data links, the status of sectors and packet error
rates.  All of this information is integrated into the main monitoring
system so that the on site operator can be made aware of any
difficulties that may arise in near real time.  For example,
Figure~\ref{comms_radio_signal} shows a monitoring event display of the
radio signal strength (in dB) of the uplink receiver as reported to
the operator for each surface detector station in the array. 
Long term performance benchmarks for the data communications system are also
recorded and monitored.  Figure~\ref{comms_arq_rates}
shows the daily number of ARQs
over the course of four months of data collected during 2010 and
2011. Here, the ARQ rate serves as a global diagnostic for the overall
health of the system indicating the extent to which packets are
missing or corrupt.  Shown are the rates for single ARQ generating
packet errors and also the rates for ``ARQ\,7'' errors corresponding to
those packets for which at least six retransmission attempts were made
before abandoning the packet; the ARQ\,7 rate
thus represents the rate at which data are irretrievably lost.   
We had for a period in 2009 a large rate of error rate leading to event loss.
It has been fixed and since then, the typical loss rate has been approximately 1000 to 2000 packets out of
over 140 million packets per day, corresponding to a data loss rate of
less than 0.002\%.  The data from a typical event involving
several stations corresponds to approximately 1000 packets. 

Therefore, the overall event loss rate due to communication failures is less than 2\%.

\section{Data processing and \Offline framework}
\label{sec:offline}

%\red{(Tom P)}

The \Offline software of the Pierre Auger Observatory
provides both an implementation of simulation and reconstruction algorithms,
discussed later, and an infrastructure to support the development of such
algorithms leading ultimately to a complete simulation, reconstruction
and analysis pipeline. Indeed, when the \Offline code was originally devised,
the only existing systems were the SD and FD. It has since been extended to handle the radio
and AMIGA extensions without requiring dramatic framework changes.
The software has been designed to accommodate
contributions from a large number of
physicists developing C++ applications
over the long lifetime of the experiment.  The essential
features include a ``plug-in'' mechanism for physics
algorithms together with machinery which assists users in
retrieving event data and detector conditions from various
data sources, as well as a reasonably straightforward
way of configuring the abundance of different applications and logging all configuration
data. A detailed description of the \Offline software design,
including some example applications, is available in~\cite{Argiro:2007qg}.

\begin{figure}[t]
\centering
\includegraphics[width=0.4\textwidth]{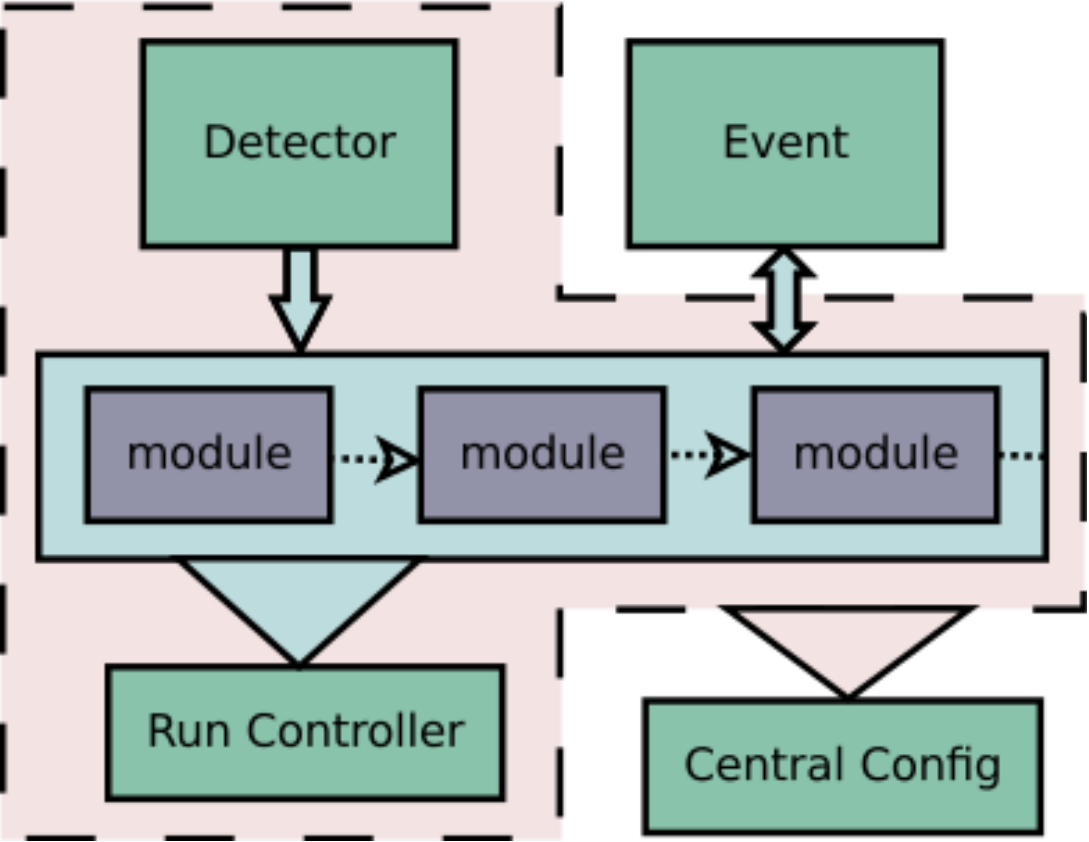}
\caption{General structure of the \Offline framework.
Simulation and reconstruction tasks are encased
in modules.  Each module is able to read information
from the detector description and/or the event,
process the information, and write the results back
into the event under command of a {\em Run Controller}.
A {\em Central Config} object is responsible for handing
modules and framework components their configuration
data and for tracking provenance.}
\label{f:general}
\end{figure}

The overall organization of the \Offline framework
is depicted in Figure~\ref{f:general}.
A collection of processing {\em modules} is
assembled and sequenced through instructions
contained in an XML file~\cite{xml} or in a Python~\cite{python} script.
An \emph{event} data model allows
modules to relay data to one another,
accumulates all simulation and reconstruction
information, and converts between various formats
used to store data on file.
Finally, a \emph{detector description}
provides a gateway to data on the detector conditions, including calibration constants and
atmospheric properties as a function of time.

\subsection{Physics modules}

Most simulation and reconstruction tasks can be
factorized into sequences of processing steps which
can simply be pipeli\-ned. Physicists prepare processing algorithms
in modules, which they register with the
\Offline framework via a one line macro.  This modular design allows
collaborators to exchange code, compare algorithms and
build up a variety of applications by combining modules in various
sequences. Run time control over module sequences
is implemented with a {\em Run Controller}, which invokes the various
processing steps within the modules according to a set of
user provided instructions. We devised
an XML-based language as one option for
specifying sequencing instructions; this
approach has proved sufficiently flexible for
the majority of our applications, and it is quite simple
to use.  An example of the structure of a sequence file
is shown in Figure~\ref{f:xml}.

\begin{figure}[t]
\centering
\begin{small}
\begin{verbatim}
<sequenceFile>
  <loop numTimes="unbounded">
    <module> SimulatedShowerReader </module>
    <loop numTimes="10">
      <module> EventGenerator   </module>
      <module> TankSimulator    </module>
      <module> TriggerSimulator </module>
      <module> EventExporter    </module>
    </loop>
  </loop>
</sequenceFile>
\end{verbatim}
\end{small}
\caption{A simplified example in which an XML file
sets a sequence of modules to conduct a simulation
of the surface array.  {\tt <loop>} and {\tt <module>}
tags are interpreted by the run controller, which
invokes the modules in the proper sequence.  In this
example, simulated showers are read from a file, and
each shower is thrown onto the array in 10
random positions by an {\tt EventGenerator}.  Subsequent
modules simulate the response of the surface detectors
and trigger, and export the simulated data to file.
Note that XML naturally accommodates common sequencing
requirements such as nested loops.}
\label{f:xml}
\end{figure}

\subsection{Data access}

The \Offline framework includes two parallel hierarchies for accessing
data: the detector description for retrieving data on conditions, including
detector geometry, calibration constants, and atmospheric conditions, and
an event data model for reading and writing information
that changes for each event.

\begin{figure}[t]
\centering
\includegraphics[width=0.48\textwidth]{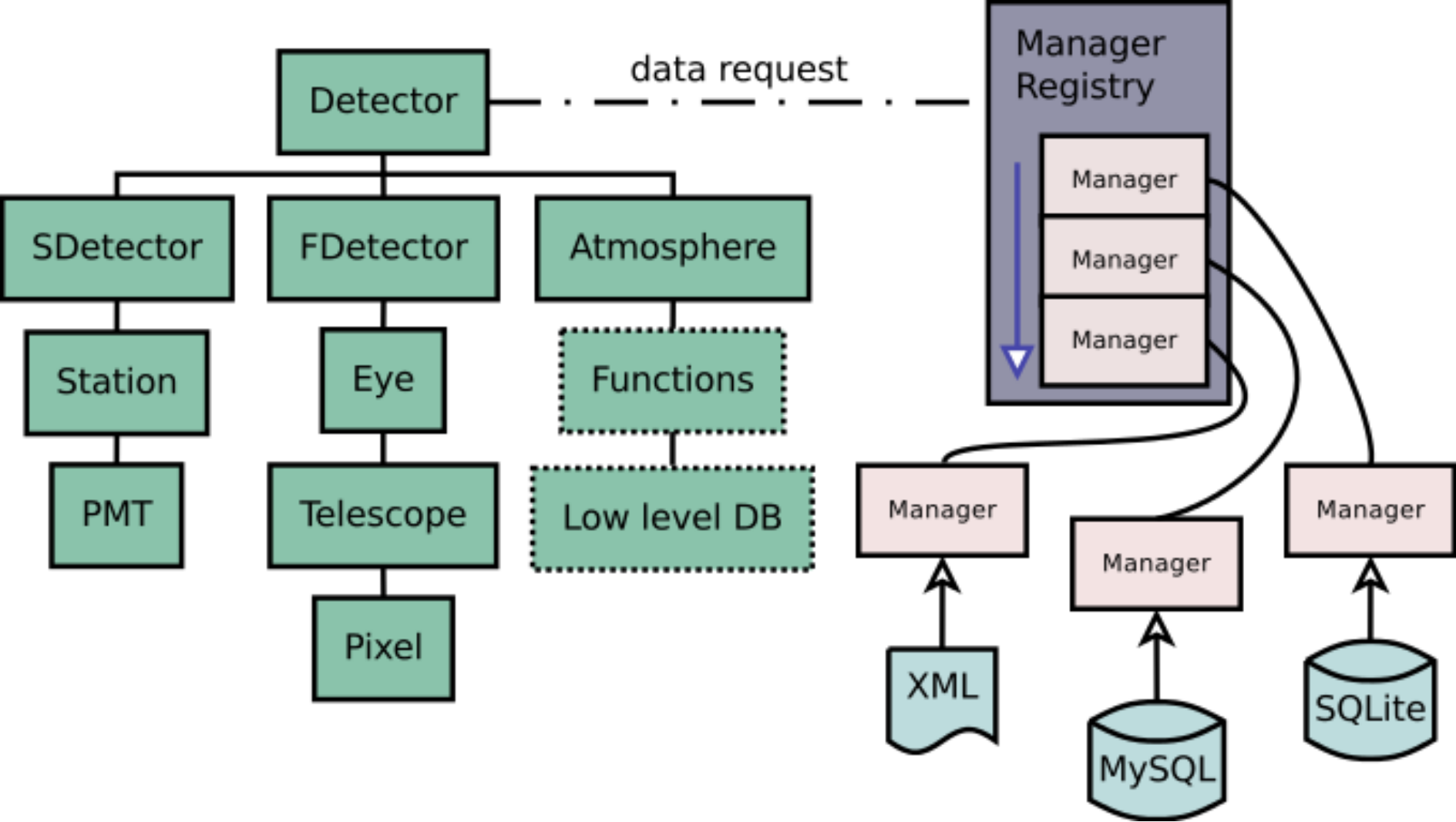}
\caption{
Machinery of the detector description.  The user interface
(left) comprises a hierarchy of objects describing the
various components of the Observatory. These objects relay
requests for data to a registry of managers
(right) which handle multiple data sources and formats.
}
\label{f:detector}
\end{figure}

The {\em detector description} provides a unified interface
from which module authors can retrieve conditions data.
Data requests are passed by this interface to a back end
comprising a registry of so-called {\em managers}, each of which is capable of
extracting a particular sort of information from
a given data source.
%Generally we choose to store
%static detector information in XML files, and
%time-varying monitoring and calibration data in MySQL~\cite{mysql}
%databases.  However, as the project evolves it sometimes
%happens that access to detector data in some other format is required,
%perhaps as a stop-gap measure.  This manager mechanism allows
%one to provide simple interfaces quickly in such cases, keeping
%the complexity of accessing multiple formats hidden from the user.
%The structure of the detector description
%machinery is illustrated in Figure~\ref{f:detector}.
The manager mechanism allows for a simple interface to a
potentially complex collection of different data sources and formats.
The general structure of the detector description
machinery is illustrated in Figure~\ref{f:detector}.

%The detector description also contains
%set of plug-in functions, called {\em models} which can be used for
%additional processing of data retrieved through the detector interface.
%These are used primarily to interpret atmospheric
%monitoring data and, like modules, are meant to be prepared
%by the user community, rather than (just) framework developers.

%As an example of use, analysis code can invoke a function
%to evaluate attenuation of light due to aerosols
%between two points in the atmosphere.  This
%request is passed to a model, which
%interrogates the detector interface to find the
%atmospheric conditions at the specified time, and
%computes the attenuation.  Models can also be
%placed under command of a {\em super-model} which
%can attempt various methods of computing the desired
%result, depending on what data are available for the
%specified time.

%The choice of which model(s) to use
%for a particular application is specified in a
%configuration file.

The event data model contains the raw, calibrated,
reconstructed and Monte Carlo information and serves as the
backbone for communication between modules.
%The overall structure comprises encapsulated
%classes organized following the hierarchy normally
%associated with the observatory instruments, with
%further subdivisions for accessing such information as  Monte Carlo truth,
%reconstructed quantities, calibration information and raw data.
%User modules access the event through
%a reference to the top of the hierarchy which is
%passed to the module interface by the run cotronller.
The event is instrumented with a protocol allowing
modules to discover its constituents at any point in processing, and
thereby determine whether the input data required to carry
out the desired processing are available.

The transient (in memory) and persistent (on disk) event models are decoupled
in order to avoid locking to a single provi\-der solution for
serialization, the process by which C++ objects are
converted to a form that can be written to disk.
When a request is made to write
event contents to file, the data are transferred from the
transient event through
a \emph{file interface} to the persistent event, which is instrumented
with serialization machinery.
We currently use the input/output portion of the ROOT~\cite{root} toolkit
to implement serialization.
Various file formats are interpreted using the
file interface, including numerous raw event and
monitoring formats as well as the different formats employed by the
AIRES~\cite{Sciutto:1999jh}, CORSIKA~\cite{Heck:1998vt}, CONEX~\cite{Bergmann:2006yz}
and SENECA~\cite{Drescher:2002cr} air shower simulation packages.
%
%\begin{figure}[t]
%\centering
%\includegraphics[width=0.5\textwidth]{eventBackend}
%\caption{Event input/output.  The section labeled ``Event Interface''
%portrays a portion of the transient event.
%Data are transferred between this transient event and
%persistent objects through a common file interface.
%Different file
%implementations are able to read and/or write
%in different formats, including those used by the
%data acquisition systems (\texttt{DAS} formats),
%formats used by other simulation
%packages, as well as a ``native'' format (\texttt{ROOTEventFile})
%which accommodates  all raw data, reconstructed quantities, and Monte Carlo truth.
%}
%\label{f:eventBackend}
%\end{figure}

\subsection{Configuration}
\label{sec:configuration}

The \Offline framework includes a system to organize and track
data used to configure the software for different applications
as well as parameters used in the physics modules.
A \emph{Central Config} configuration tool (Figure~\ref{f:general})
points modules and framework components to the location of
their configuration data, and creates Xerces-based~\cite{xerces}
XML parsers to assist in reading information
from these locations.  We have wrapped Xerces
with our own interface which provides ease of use at the cost
of somewhat reduced flexibility, and which also
adds functionality such as automatic units conversion and casting
to various types, including commonly used containers.
%The locations of configuration data are declared
%in a so-called {\em bootstrap} file, and may comprise local filenames,
%URIs~\cite{URI} or XPath~\cite{xpath} addresses.

The {\em Central Config} keeps track of all configuration data
accessed during a run and stores them in an XML log file, which
can subsequently be used to
reproduce a run with an identical configuration. This allows
collaborators to easily exchange and use configuration data for result comparisons.
The logging mechanism is also used to record the versions
of modules and external libraries which are used for each run.

Syntax and content checking of the configuration files is afforded through
W3C XML Schema \cite{xml-schema} standard
validation.  Schema validation is used
not only for internal configuration prepared by framework developers,
but also to check the contents of physics module configuration files.
This approach reduces the amount of code users and developers must prepare and
supports very robust checking.
%
%The standard schema types are complemented by a
%collection of types commonly used in our applications,
%allowing for quite detailed checking with minimal investment in schema preparation.
%Even casual users have demonstrated a willingness to invest the (small) time required to learn
%enough XML schema to check simple configuration files for modules.

%The configuration machinery can also verify configuration file
%contents against a set of default files by employing MD5 digests~\cite{md5}.
%The default configuration files are prepared by the framework developers and the analysis
%teams, and reference digests are computed from these files.
%At run time, the digest for each configuration file is recomputed
%and compared to the reference value.
%This provides a means for those managing
%production campaigns to quickly verify that configurations in use are
%the ones which have been recommended for the task at hand.

\subsection{Utilities, testing and quality control, and build system}

The \Offline framework is complemented by a collection of utilities, including
an XML parser, an error logger and various mathematics and physics services.  We
have also developed a novel geometry package which allows the manipulation of
abstract geometrical objects independent of coordinate system choice. This is
particularly helpful for our applications since the Observatory comprises many
instruments spread over a large area and oriented in different directions, and
hence possesses no naturally preferred coordinate system. Furthermore, the
geometry package supports conversions to and from geodetic coordinates based on
a reference ellipsoid.

As in many large scale software development efforts, each low level component of
the framework is verified with a small test program, known as a {\em unit test}.
We have adopted the CppUnit~\cite{cppunit} testing framework as an aid in
implementing these tests.  In addition to unit tests, a set of higher level
acceptance tests has been developed which is used to verify that complete
applications continue to function as expected, within some tolerance, during
ongoing development.  We employ a BuildBot system~\cite{buildbot} to
automatically compile the \Offline software, run the unit and acceptance tests,
and inform developers of any problems each time the code is modified.

The \Offline build system is based on the CMake~\cite{cmake} cross-platform
build tool, which has proven sufficient for this project. In order to
ease installation of \Offline and its various external dependencies,
we have prepared a tool known as
APE (Auger Package and Environment)~\cite{ape}. APE is a python-based
dependency resolution engine, which downloads the external packages
required for a particular application, builds them in whatever native
build system applies for each package, and sets up the user's environment
accordingly. APE is freely available, and has been adopted by other
experiments, including HAWC, NA61/SHINE and JEM-EUSO.

\subsection{\Offline summary}
At the time of writing, the \Offline software comprises over 350,000 lines
of code, corresponding to some 95 person years of effort, according to the
Constructive Cost Model~\cite{boehm}.  The framework has been enhanced for simulation
and reconstruction of the Observatory extensions discussed in
Sections~\ref{sec:HEAT} and~\ref{sec:amiga} and for the
radio research program described in Section~\ref{sec:developments_radio}, for which
substantial additions to the \Offline functionality were developed~\cite{Abreu:2011fb}.
The code is available under an open source BSD license upon request.
Other experiments have adopted portions of the \Offline code, including
Tunka-Rex~\cite{Kostunin:2013iaa}, 
HAWC~\cite{Abeysekara:2013tka}, JEM-EUSO~\cite{Ebisuzaki:2014wka}, CODALEMA~\cite{Ardouin:2006gj}
and NA61\-/SHINE~\cite{Wyszynski:2012fa, Sipos:2012hs}.

\section{Atmospheric monitoring}
\label{sec:atmosMonit}

%\red{(L.\ Wiencke, B.\ Keilhauer)}

The Observatory makes use of the atmosphere as a giant calorimeter. This motivated 
the selection of a site with generally good viewing conditions and 
the implementation of an extensive program to monitor the troposphere 
above the site.  A detailed knowledge of the atmosphere 
is required for the accurate reconstruction of air showers observed 
by the FD  \cite{Abraham:2010pf,Abreu:2012zg,Keilhauer:2012yp,Abreu:2012oza} and for the 
accurate estimation of the exposure of the detectors ~\cite{Abreu:2010aa}.
  
The atmospheric state variables, including temperature, pressure and
humidity, are needed to assess the longitudinal 
development of extensive air showers \cite{Keilhauer:2004jr,Abreu:2012zg}
as well as the amount of the isotropically emitted fluorescence light
induced by the air showers 
\cite{Arqueros:2008cx,Keilhauer:2008sy,Monasor:2010fn,Keilhauer:2012hu}. The SD observations 
are altered by different atmospheric conditions \cite{Abraham:2009bc}. Varying air 
densities close to the ground modify the Moli\`ere radius 
affecting the lateral distribution of the electromagnetic component of the 
extensive air shower (EAS). 
Varying air pressure affects the trigger probability and the rate of events 
detected above a fixed energy. Furthermore, the atmospheric state 
variables are used to determine the Rayleigh (pure molecular) scattering
of the fluorescence and Cherenkov light. Installations for recording
local conditions of the state variables are described in Section~\ref{sec:statevariables}.

\begin{figure}[htbp]
\centering
\includegraphics[width=0.48\textwidth]{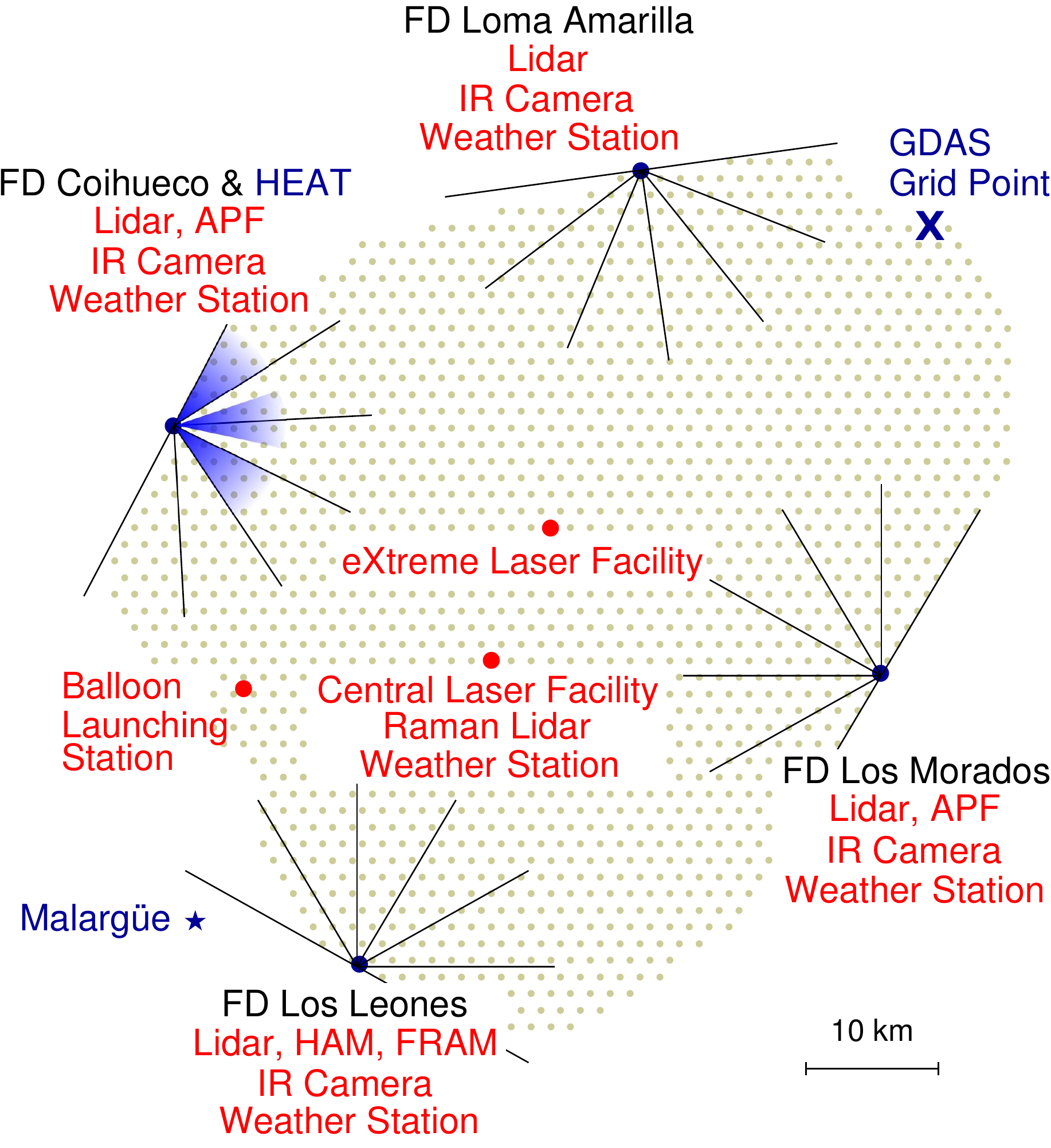}
\caption{Schematic overview of the atmospheric monitoring devices installed at the 
    Pierre Auger Observatory. At each FD site, there is
    a lidar station, a ground-based weather station, and an infrared camera for cloud
    cover detection. In addition, there are devices for measuring the Aerosol Phase 
    Function (APF) at FD Coihueco and Los Morados, a Horizontal Attenuation Monitor     (HAM) at FD Los Leones, and a ph(F)otometric Robotic Atmospheric Monitor (FRAM)  also at Los Leones.  A steerable backscatter elastic lidar system is installed at each of the 4 FD sites  to
    measure aerosols and the positions of clouds near each site.  At central positions 
    within the surface detector array, two laser facilities are
    installed (CLF and XLF). These instruments, together with the FD, are used to measure
    $\tau_{\rm aer}(h)$ in the line of sight of each FD telescope 4 times per hour. 
    In April of 2013 the CLF was upgraded with a Raman lidar receiver. 
    Near the western boundary of the array, the Balloon Launching Station (BLS) was 
    assembled together with a weather station as a base unit for an electric field meter.
    From this launch station, the weather balloons were typically carried
		across the entire array by westerly winds.
\label{fig:atmosOverview}}
\end{figure}

Aerosols and clouds represent the most dynamic monitoring and calibration 
challenges at the Observatory. The optical transmission properties 
of the atmosphere,
including the vertical aerosol optical depth profile $\tau_{\rm aer}(h)$, 
have to be measured across the Observatory during FD data taking. 
In the air shower reconstruction, the atmospheric transmission between the 
FD and an air shower must be taken into account to properly reconstruct the light
generated along the shower axis  from the light recorded at the 
telescope(s) \cite{Abraham:2010pf,Abreu:2012oza}. Moreover, Cherenkov light
induced by the air showers is also detected with the FD and needs to be
reconstructed as a function of atmospheric conditions at the time of the
event.
Installations dedicated for determining the optical scattering
and absorption behavior of the atmosphere in the field of view are described in 
Section~\ref{sec:scattering} and those for identifying and determining clouds
and the general extinction above the Observatory in Section~\ref{sec:clouds}. 

An extensive system of atmospheric monitoring devices has been installed
(Figure~\ref{fig:atmosOverview}). The types of measurements
possible with these instruments are listed in Table~\ref{fig:Atmos_Table}.

\begin{table}[htbp]
\caption{Atmospheric measurements performed and the instruments that are used.}
\begin{minipage}{\columnwidth}
\centering
{\footnotesize
\begin{tabular}{llll}
\toprule
\textbf{Category} & \textbf{Variable} & \textbf{Frequency} & \textbf{Instrument(s)}
\\
\midrule
State & At ground:& 5\,min & Weather Stations
\\
 & Pressure, Temp., & & \\
 & Wind, Humidity & & \\
 & Profile: Pressure, & 3 hours & GDAS\footnote{atmospheric model developed at the
National Centers for Environmental Prediction, operated by NOAA; provided via READY - Real time Environmental
Applications and Display sYstem.} \\
 & Temp., Humidity & & 
\\
\midrule
Aerosols & Vert.\ Optical Depth ($z$) & hourly & CLF, XLF + FD
\\
 & Phase Function & hourly & 2 APF units
\\
 & {\AA}ngstr\"om Coefficent & hourly & FRAM (HAM)
\\
\midrule
Clouds & Presence in FD pixels & 15\,min & 4 Cloud Cameras
\\
 & Behind FD sites & 15\,min & 4 lidar stations
\\
 & Along select tracks & avg.\ 1/night & FRAM, lidar 
\\
 & Above CLF/XLF & hourly & CLF, XLF + FD
\\
\bottomrule
\end{tabular}
}
\label{fig:Atmos_Table}
\end{minipage}
\end{table}

\subsection{Installations for atmospheric state variables 
\label{sec:statevariables}}

\subsubsection{Ground-based weather stations}

The Auger Collaboration operates several weather stations, 
as indicated in Figure~\ref{fig:atmosOverview}.
Some of these stations are used for operational
control of the nearby installations. The data from the 
weather stations at each FD site and at the CLF
additionally serve as atmospheric ground information
in several parts of the air shower reconstruction. Typically,
those data are transferred via the central campus in Malarg\"ue,
processed and stored in our databases for atmospheric monitoring
information (cf.\ Section~\ref{sec:atmosDB}) within a couple of days.

The weather stations are commercial 
products\footnote{Campbell Scientific, 
\url{http://www.campbellsci.com}} 
%http://www.campbellsci.com} 
equipped with temperature, pressure, humidity, and wind speed 
sensors recording data every 5 minutes. The stations at 
FD buildings Los Leones and Coihueco and
at the laser facilities are additionally equipped with a
sensor for wind direction. Formerly at the Balloon Launching Station (BLS) site and now at the
AERA site (cf.\ Section~\ref{sec:developments_radio}), the weather 
station serves as a base unit for an
electric field meter. The values of the electric field are 
recorded every second for lightning and thunderstorm detection
which is particularly important for the radio detection technique.

\subsubsection{Balloon Launching Station}

For a proper reconstruction of the fluorescence telescope signals,
not only are ground-based atmospheric data  needed, but also atmospheric profiles
of the state variables temperature, pressure, and humidity up to about
20 to 25\,km~a.s.l.\ \cite{Keilhauer:2004jr,Arqueros:2008cx,Keilhauer:2008sy,Keilhauer:2009ax,
Abreu:2012zg}. From these directly measured values, the derived quantities
air density and atmospheric depth are calculated. The program of 
launching meteorological radiosondes
attached to helium filled weather balloons was started at the Observatory
site in August 2002. After 331 successfully measured profiles, the 
routine operation was terminated in December 2010 \cite{Keilhauer:2012yp}
and then replaced by the meanwhile validated GDAS data. 
During the first years, 
campaigns of about three weeks with an average of nine launches per campaign
were done roughly three times a year. The starts of the soundings were usually placed
at some FD buildings, mostly at Los Leones and Coihueco. 
In 2005, a dedicated BLS, cf.\ 
Figure~\ref{fig:atmosOverview}, was installed at a suitable position
to optimally cover the large area above the surface
detector array and in the field of view of the FD telescopes by the weather
balloons. From this fully equipped station, more regular launches
could be managed, in particular during the night. 
Between July 2005 and March 2009,
roughly one launch was performed about every five
days. Between 2009 and 2011, the program was part of the rapid
atmospheric monitoring system of the Pierre Auger Observatory (see
Section~\ref{sec:atmos-adds}). A radiosonde launch was triggered
shortly after the detection of particularly interesting air showers
such as very energetic events. Since 2011, the BLS is used for 
dedicated measurement campaigns.

The radiosondes and the receiver station are standard meteorological
products\footnote{\url{http://www.graw.de}}
%products\footnote{http://www.graw.de}
providing data on the temperature, pressure, humidity, and GPS position
including altitude.
Typically, a set of measurement values is recorded every 2 to 8 seconds. 
The upper limit
of the profile is given by the height of the burst of the weather 
balloon, typically at about 23\,km, with a few balloons reaching a
maximum altitude of 27\,km.

\begin{figure}
  \begin{minipage}[t]{.32\textwidth}
    \centering
    \includegraphics*[width=.99\linewidth,clip]{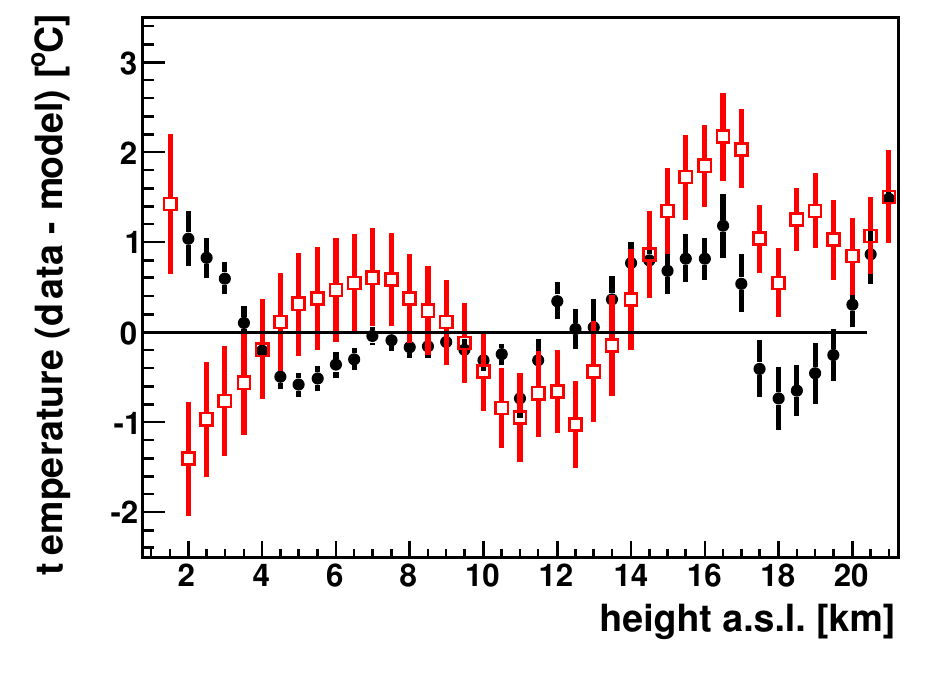}
  \end{minipage}
  \hfill
  \begin{minipage}[t]{.32\textwidth}
    \centering
    \includegraphics*[width=.99\linewidth,clip]{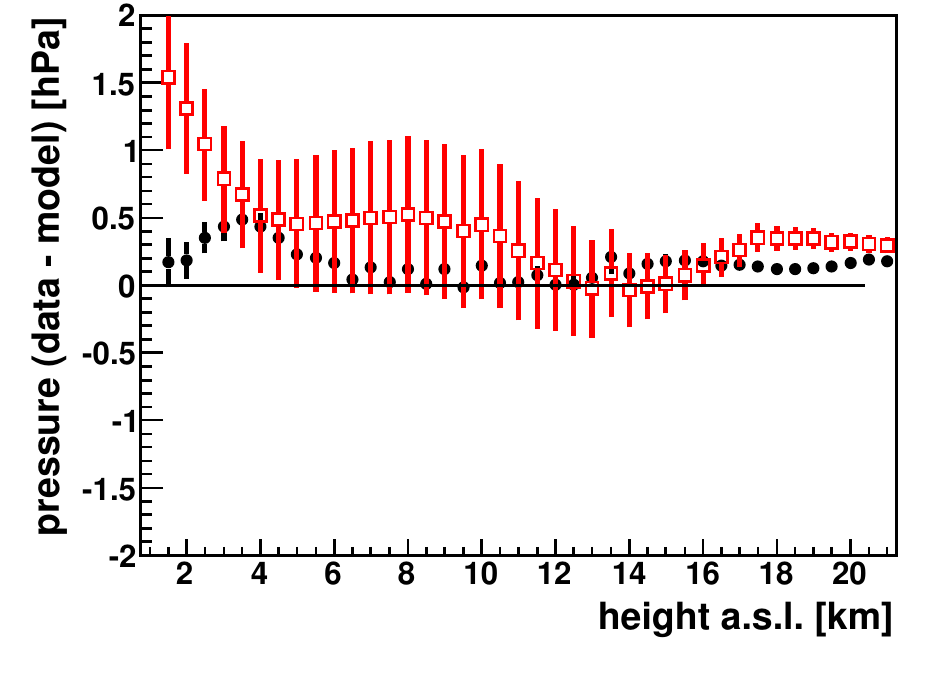}
  \end{minipage}
  \begin{minipage}[t]{.32\textwidth}
    \centering
    \includegraphics*[width=.99\linewidth,clip]{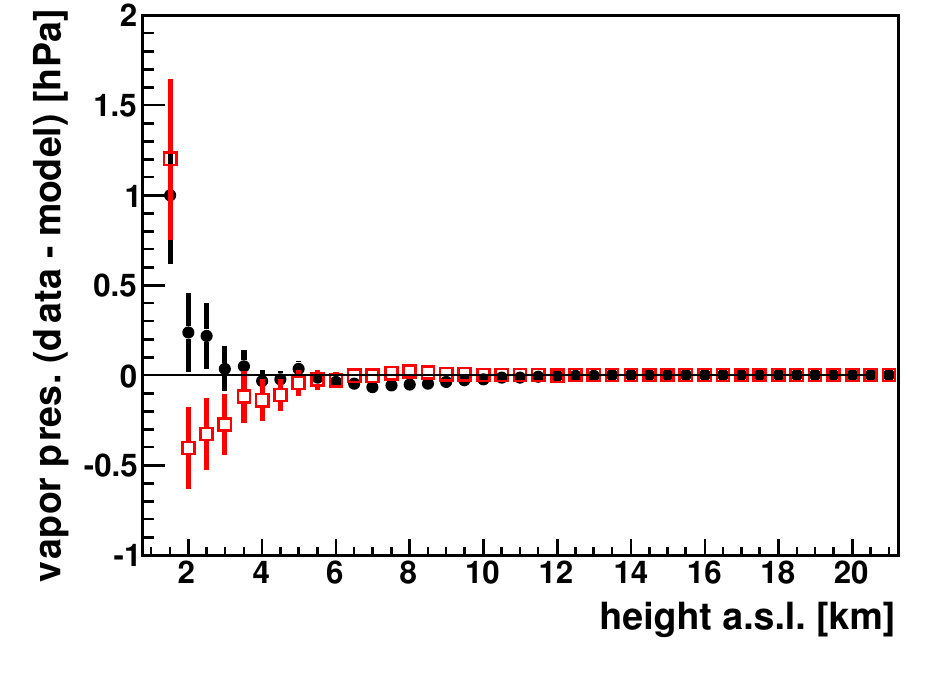}
  \end{minipage}
  \hfill
 \vspace{-12pt}
  \caption{\label{fig:SondeVsGDAS_nMMM_2009}
    Difference of locally measured radiosonde data to according GDAS profiles 
    versus height for all radio soundings performed in 2009 and 2010 (black dots).
    In addition, the difference of the same radiosonde data to according
    monthly mean profiles (red squares) are plotted. 
    The monthly mean profiles used for this 
   comparison are those developed end 2008, representing a totally independent
   dataset of local records than the plotted radiosonde data~\cite{Abreu:2012zg}. 
  }
\end{figure}

Based on the locally measured atmospheric profiles, monthly models of
atmospheric conditions at the Pampa Amarilla were derived in
December 2008 \cite{Abreu:2012oza,Keilhauer:2012yp}. The monthly models
are also compiled for application in air shower simulations
like CORSIKA \cite{Heck:1998vt}. Finally, these measurements were
used to validate the utility of data from the Global Data 
Assimilation System (GDAS) for the purpose of air shower
reconstruction at the Pierre Auger Observatory \cite{Abreu:2012zg}.
GDAS is the result of atmospheric computer analyses and forcasts which
are run serveral times per day and are based on meteorological 
measurements from all around the world. The data are available in
3-hourly, global, 1$^\circ$ latitude-longitude (360$^\circ$ by 180$^\circ$)
datasets. The position of the chosen GDAS grid point is marked in 
Fig.~\ref{fig:atmosOverview}. Each GDAS dataset consists of surface data
together with data for 23 constant pressure levels reaching up to about 
26\,km. For the purpose of the Auger Observatory, maily the information
for temperature, pressure and humidity are used. These GDAS data
have been compared with according local radio soundings and the
records from the ground-based weather stations. The agreement of the
locally recorded data with the GDAS data for the given grid point is
well enough for the application in air shower physics,
%%%----------------add a comment and a figure for this
c.f.\ Fig.~\ref{fig:SondeVsGDAS_nMMM_2009}
%%%--------end of additional comment and figure
%%%%%%%%%Lawrence, Paul: If you think the figures are too small,
%%%% we could remove the vapor pressure plot and enlarge the width of the 
%%%% first two 'minipage' to 0.49\textwidth.
. The variation for
temperature is below $\pm 1$~K for altitudes between ground and about 
20~km a.s.l., for pressure, the variation is below 0.5~hPa at worst, but
for most of the altitude range well below 0.3~hPa. Even the vapor
pressure is well below 0.3~hPa except for the data point closest to 
ground where the difference goes up to 1.0~hPa. After validating the
utility of the GDAS data for the site of the Pierre Auger Observatory
\cite{Abreu:2012zg}, advanced monthly models were compiled from GDAS
data for this location for air shower simulations and 
reconstructions~\cite{Keilhauer:2012yp,corsika_usersguide}.

\subsection{Installations for atmospheric transmission{\label{sec:scattering}}}

The transition of fluorescence light, incduced by extensive air showers, in the atmosphere 
is reduced by absorption and scattering of the UV photons. For a correct 
reconstruction of the energy deposit of an extensive air shower in the atmosphere, the
attenuation of the fluorescence photons has to be know for the time of the 
air shower.

The attenuation of light in the wavelengths range of interest here, is dominated
by scattering rather than by absorption. The scattering of photons in air
can be described analytically from molecular scattering theory. 
Once the vertical profiles
of atmospheric temperature, pressure, and humidity are known, the molecular transmission
factor $T_{mol}(\lambda,s)$ is a function of the total wavelength-dependent Rayleigh
scattering cross section along the line of sight $s$. The scattering of photons by 
aerosols can be described by
Mie scattering theory, but for real conditions with strongly varying shapes and amounts
of aerosols, local measurements are needed. The knowledge of the aerosol transmission
factor $T_{aer}(\lambda,s)$ depends on frequent field measurements of the vertical
aerosol optical depth $\tau_{aer}(h)$, the integral of the aerosol extinction
$\alpha_{aer}(z)$ from the ground to a point at altitude $h$ observed at an given 
elevation angle.

\subsubsection{Aerosol optical depth profiles and clouds: CLF and XLF}

\begin{figure}
  \begin{minipage}[t]{.44\textwidth}
    \centering
    \includegraphics*[width=.99\linewidth,clip]{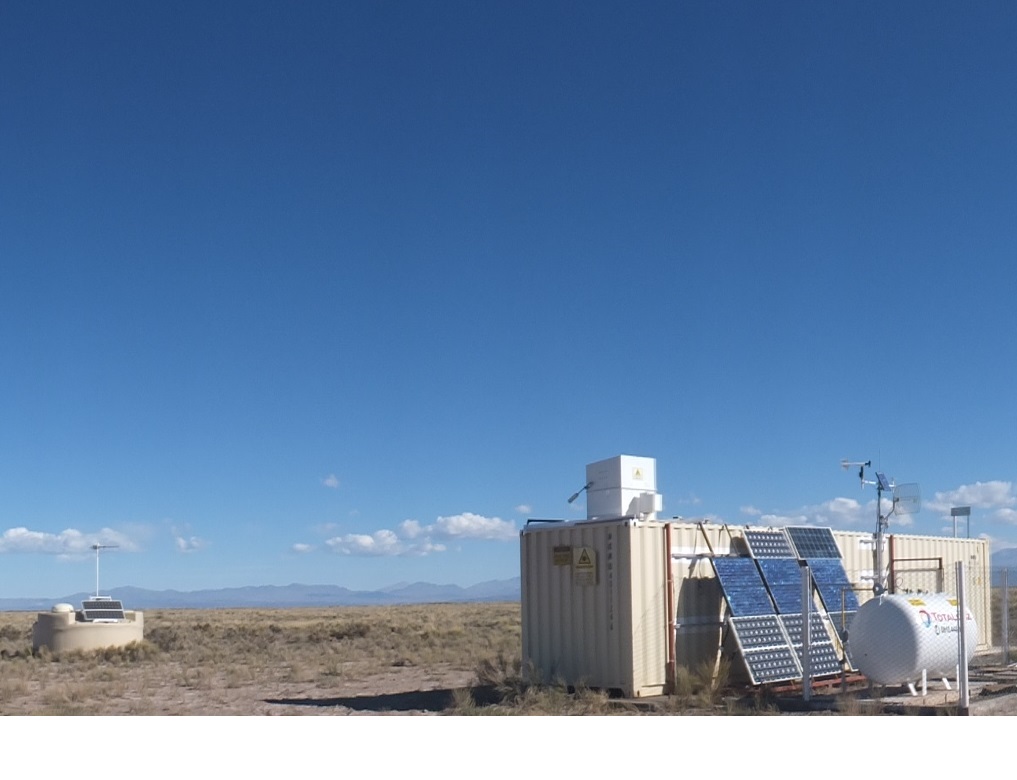}
  \end{minipage}
  \hfill
  \begin{minipage}[t]{.55\textwidth}
    \centering
    \includegraphics*[width=.99\linewidth,clip]{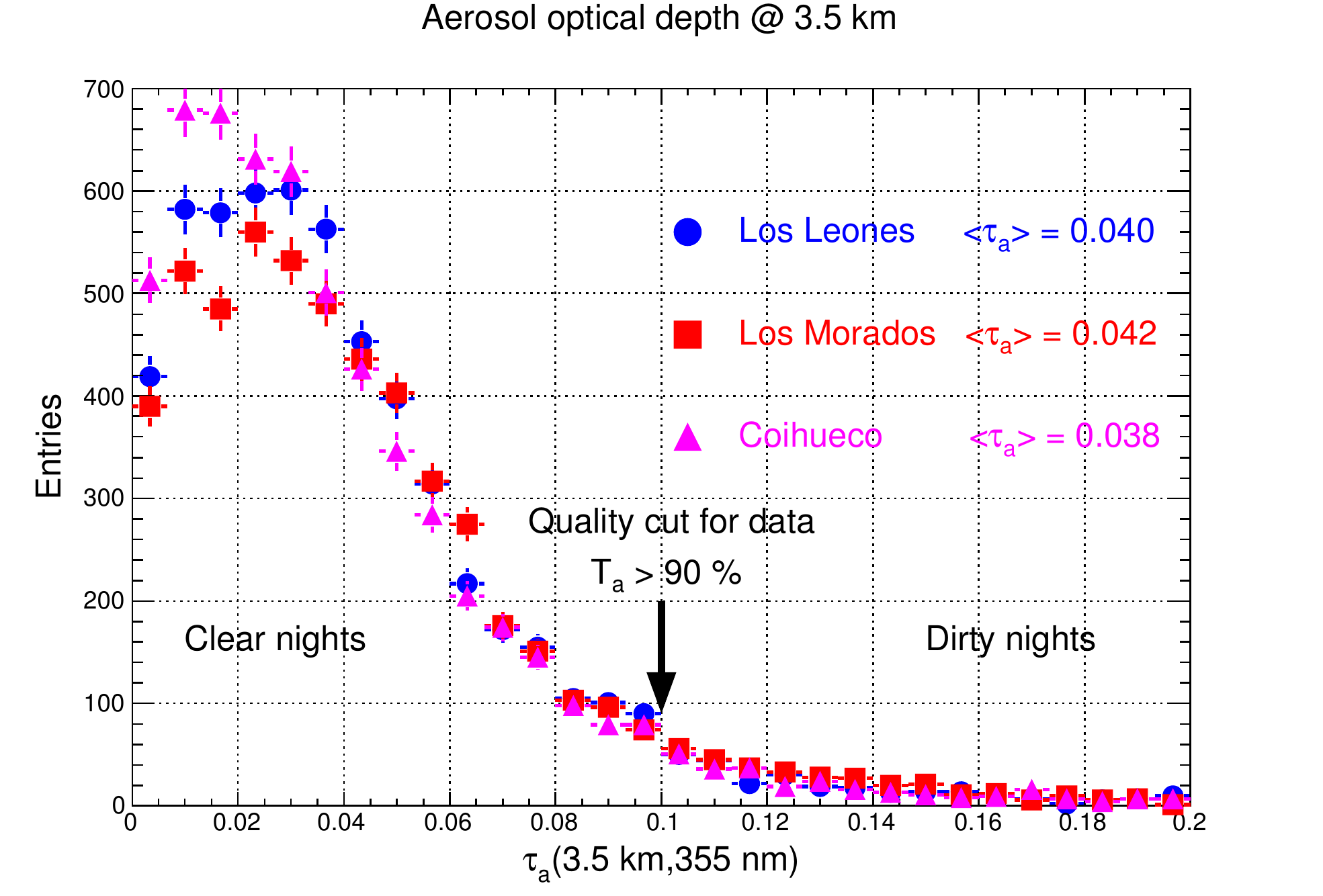}
  \end{minipage}
  \hfill
 \vspace{-20pt}
\centering
\caption{Left: The CLF.  Right: Vertical aerosol optical depth at 3.5~km above the observatory obtained from FD measurements of CLF vertical laser shots recorded between January 2004  and December 2010. Data from 3 fluorescence stations are shown and the quality cut between clear and dirty nights is indicated \cite{Abreu:2011pg}.
\label{fig:vaod}}
\end{figure}

Laser tracks from the CLF and XLF (Section~\ref{sec:CLFXLF}) are recorded by the 4 FD sites.
They are used to obtain hourly measurements of the aerosol optical depth profiles 
\cite{Abreu:2013qtw} that are used in the reconstruction of each FD air shower event. 
%The vertical aerosol optical depth, defined as the 
%integral of the aerosol extinction from the ground to a point at a given 
%altitude observed at a given elevation angle, is required to determine the 
%aerosol transmission factor. 
Sets of 50 vertical shots are measured every 15 minutes by the FD telescopes 
throughout each night.  The polarization of the CLF beam is randomized so that
the amount of light scattered out of the beam is azimuthally symmetric about the beam axis.
The CLF data sample began in 2004 and that for the XLF in 2009.
These samples contain more than 1.5 million laser shots corresponding to 
more than 4 million tracks in the FD telescopes. 
To obtain the aerosol optical depth profile for each FD site, two techniques are used:
the 50-shot averages are compared to averages collected under clear conditions 
(Data Normalized Analysis \cite{Abreu:2013qtw}), and to simulations generated 
with different aerosol attenuation conditions (Laser Simulation Analysis \cite{Abreu:2013qtw}).
Data from the aerosol phase function
monitors (next paragraph) can be used to cross check the clear reference periods. 
The technique is independent of the absolute photometric calibrations of the lasers and of the FD,
and provides essential information in the reconstruction of hybrid and FD data.
In figure~\ref{fig:vaod}, data from the vertical aerosol optical depth at 3.5~km above
the fluorescence telescopes are shown.

The analysis of the light profile of laser shots, as detected by the FD, also provides 
measurements of the heights of clouds directly above the CLF or XLF~\cite{Abreu:2013qtw} . 

\subsubsection{Aerosol Phase Function monitors}

The atmospheric scattering of both
fluorescence and Cherenkov light from extensive air showers
occurs over a range of scattering angles. The scattering angular distribution 
(phase function) 
can be estimated analytically for the atmospheric molecular component.  
For the aerosol component this function depends on
the size and shape of the aerosols~\cite{Abraham:2010pf}.
The scattering function is characterized in-situ to implement a suitable parametrization
of the scattering behavior by the air shower reconstruction.

The Aerosol Phase Function monitors  use a
collimated Xe\-non flash lamp to direct light between 350 and 390\,nm
horizontally across the FD field of view at Coihueco and Los Morados.
 The FD measures the light as a function of scattering angle (30$^\circ$ to 150$^\circ$).
An analysis including data over several years revealed that a parameterization as
\begin{equation}
P_a(\theta) = \frac{1-g^2}{4\pi}\cdot \biggl( \frac{1}{(1+g^2-2g \cos \theta)^{3/2}} + f
\frac{3 \cos^2 \theta - 1}{2(1+g^2)^{3/2}} \biggr)
\end{equation}
 describes the aerosol scattering at the 
Observatory site reasonably well. The first term is a Henyey-Greenstein 
function~\cite{Henyey-Greenstein}, corresponding to forward scattering, and the second term
accounts for the peak at large $\theta$ typically found in the angular distribution
of aerosol-scattered light~\cite{BenZvi:2007px}. The quantity $g$ describes the asymmetry of scattering, and 
$f$ determines the relative strength of the forward and backward scattering peaks. 
An average value of the
phase function asymmetry parameter $g$ of $0.56\pm0.10$ is used in 
the Auger air shower analysis for nights with Mie scattering. For clear nights
without any aerosols, $g$ is set to zero. To also allow  for very small aerosol 
content during almost clear nights, causing only small asymmetries in the phase
function, an uncertainty of 0.2 is estimated and attached to the value of $g$
equals zero. 

\subsubsection{Horizontal Attenuation Monitor}

%The wavelength dependence of the aerosol
%attenuation is typically a parametrization by a falling power law in which
%the exponential factor, $\gamma$. Larger the value of $\gamma$ correspond
%to smaller particle size with a clear air limit of $\gamma \approx 4$.
The wavelength dependence of the aerosol attenuation is modeled
by a falling power law with an exponential parameter $\gamma$. The value of $\gamma$
varies inversely with the typical size of the aerosol particles. In the limit of clean air, $\gamma \approx 4$. 
At the Observatory, $\gamma$ is obtained by the Horizontal
Attenuation Monitor (HAM).
The HAM consists of a high intensity discharge lamp installed close to the 
FD site at Coihueco.  Light from this lamp is measured by a filtered 
CCD camera at the  Los Leones FD site, about 45\,km away \cite{Abraham:2010pf}.
Total horizontal atmospheric attenuation is measured over this path
at  five wavelengths
between 350 and 550\,nm. The data indicate that the atmosphere of the 
Observatory is quite desert-like with weak wavelength dependence.
An average $\gamma$ of about 0.7 with an RMS of 0.5 is used
as a parameter in the air shower reconstruction.

\subsection{Installations for clouds and extinction{\label{sec:clouds}}}

\subsubsection{Cloud identification}
The presence of clouds can alter the observed optical signatures of an EAS and reduce the aperture of the FD.
Clouds can attenuate or block light from an air shower, producing a dip in the 
longitudinal profile observed by the FD.  Conversely, if a shower passes through 
a cloud layer, the cloud can
enhance the scattering of the intense Cherenkov light beam, producing a 
bump~\cite{Chirinos-ICRC:2013}.

The Observatory uses measurements from infrared cameras and
lidar systems and FD measurements of CLF and XLF tracks to detect clouds.
A cloud is warmer than the
surrounding atmosphere and produces an infrared signal that depends
on the cloud temperature and emissivity (or optical depth). An infrared 
camera (Raytheon ControlIR 2000B) mounted on a pan-and-tilt scanning 
platform, operates at each FD building. The
cameras are sensitive in the 7 to 13\,$\mu$m band, appropriate for the peak
of the blackbody radiation from thick clouds, but unfortunately also
sensitive to water emission bands even in clear sky.  Every 5 minutes
each camera scans the field of view (FOV) of the telescopes, and every
15 minutes the entire hemisphere is imaged.  The raw FOV images are
converted into a binary image (cloudy/clear) after a series of image
processing steps designed to remove camera artefacts and account for
the expected elevation angle dependence of the clear sky intensity
\cite{Winnick:2010}. These data are then mapped onto FD pixel 
directions to indicate the presence or absence of clouds in each FD pixel.
At the time of writing, the Raytheon IR cameras are being replaced by 
Xenics Gobi 384 radiometric
IR cameras  to improve image processing and reduce cable maintenance.

Data from the
Geostationary Operational Environmental Sa\-tellites (GOES) are also being
analyzed~\cite{Abreu:2013qfa}. GOES instruments provide radiance 
data in one visible and four infrared bands from which brightness 
temperatures are derived. The GOES-12 imager instrument
 covers the area of the Pierre Auger
Observatory every 30 minutes. Cloud
probability maps with a grid of 2.4~km by 5.5~km pixel size are
derived for the area of the Observatory. Cross checks of
cloud identification as derived from GOES measurements for the 
pixels viewing the CLF and as derived  from FD measurements of CLF
vertical tracks show a reliable correlation.

\subsubsection{Clouds and aerosols: FD lidar stations}

Four elastic lidars, installed next to each FD station, are used 
to measure cloud cover, cloud height and 
aerosols~\cite{Rizi:2012dn,BenZvi:2006xb}. Each lidar has a Nd:YLF
laser that produces 0.1~mJ pulses at a wavelength of 351~nm.
Three 80~cm mirrors and a 20~cm mirror collect the backscattered light
which is measured with Hamamatsu R7400U photomultipliers. A UG-1 optical
filter reduces background light. The lasers are operated remotely at a
repetition rate of 333~Hz. Thousands of pulses are averaged by analog
and photon counting readout systems. The two traces from these parallel
readout paths are then combined to cover from 200~m up to 25-30~km.

The lidars are steerable and perform discrete and continuous scanning patterns
automatically during the FD operation. To avoid interference with FD data
collection, most of the scanning shots are aimed outside the FD field of view.
Two exceptions are horizontal shots fired in the direction of the CLF to
measure ground level aerosol horizontal attenuation length and
shoot-the-shower scans to probe the detector shower plane shortly after
especially interesting cosmic ray candidates have been observed.

Long term measurements from these lidars find a mean aero\-sol extinction
length at ground level of 0.028\,km, which indicates that the atmosphere at
the site is quite clear. Furthermore, it can be derived that about 62\% of
the FD data taking time is quite clear with a mean cloud cover
of less than 20\%.
 
\subsubsection{FRAM}

The ph(F)otometric Robotic Atmospheric Monitor \\
(FRAM) is an optical
telescope (0.3\,m diameter mirror) that measures starlight to 
determine the wavelength dependence of 
Rayleigh and Mie scattering. It is also used to make automatic 
observations of light curves of optical
transients associated with gamma ray bursts \cite{Jelinek:2006gf}.

FRAM was installed at the site of the Los Leones fluorescence detector
in 2005. Since the end of 2009, it has been part of the rapid atmospheric monitoring 
program, cf.\ Section~\ref{sec:atmos-adds}.
Because the FRAM is a passive instrument, it can operate in the field 
of view of the FD.  A
photometric observation of selected standard (i.e., non-variable) stars
has been supplemented with a photometric analysis of CCD images since 
2011 \cite{BenZvi:2007uj,Abreu:2012oza}. 
A wide field camera is used to measure the atmospheric extinction along the
shower detector plane. Its field of view is 240' (4$^\circ$) in azimuth
(aligned with right ascension) and 160' (2.67$^\circ$) in elevation (aligned
with declination). This instrument is a Finger Lake Instrumentation 
(FLI) MaxCam CM8 with Carl Zeiss Sonnar 200\,mm $f/2.8$ telephoto lens. 
A second, narrow field camera is used to calibrate the images of the 
wide field camera. Before June 2010, this camera was also a FLI instrument, 
but was exchanged for a Moravian Instruments CCD camera G2.

\subsection{Rapid monitoring}
\label{sec:atmos-adds}

In 2009, the atmospheric monitoring program was upgraded \cite{Abreu:2012oza} to probe the
shower FD detector plane with the FRAM and lidar systems a few minutes after the FD detected any 
extremely high energy EAS or an EAS with an unusual longitudinal profile. All
atmospheric subsystems involved use individual and modifiable trigger settings to identify these
kind of FD detected air showers.  The motivation was 
to check for clouds or aerosol layers that might distort the observed profile. Between 
March 2009 and December 2010, a weather balloon was also released from the BLS site to 
measure the pressure, temperature and humidity profiles within a few hours of the 
EAS detection above the Auger array with an energy above 10$^{19}$~eV. During this period,
100 FD events were triggered for a weather balloon launch. Some of the triggers were
received while a radiosonde was already in flight, due to the tendency of high-quality,
high-energy observations to cluster during very clear, cloudless nights. Thus, 62 triggers
were covered by 52 flights. The remaining triggers were lost due to technical issues such as
hardware failures at the BLS, problems with the transmission of the text message to the technician who
needed to launch the weather balloon, or other failures in the radiosonde flights.

\subsection{Atmospheric databases}
\label{sec:atmosDB}
The atmospheric monitoring data are organized into MySQL databases 
that are accessed by the Auger offline analysis package 
for air shower reconstruction, condition assessment, and aperture estimates.

The cloud information includes the IR cloud camera distributions mapped onto the FD pixels
and  the hourly measurement of the fraction of the sky covered by 
cloud measured by the lidars. An hourly cloud coverage below 20\% is required for hybrid 
events to be used in the analysis of the mass composition and energy 
spectrum of the cosmic rays ~\cite{Abraham:2010yv,Abraham:2010mj}.

The aerosol optical database contains, for each FD site, 
hour\-ly $\tau_{\rm aer}(h)$ profiles in 200 m steps and
derived from CLF or XLF laser shots.  FD data used for cosmic ray publications is required
to be collected during hours with $\tau_{\rm aer}$(3\,km a.g.l.)$ < 0.1$.  FD data recorded 
during hours without $\tau_{\rm aer}(h)$ profiles due to extremely  poor viewing conditions 
or technical problems are not used for publications.  This database spans nearly 10 years.

The molecular database contains the atmospheric state variables 
measured on the ground by the weather stations and the vertical profiles
derived from BLS weather balloons and more recently 
from the Global Data Assimilation System, provided from 
NOAA's\footnote{National Oceanic and Atmospheric Administration.} National
Centers for Environmental Prediction (NCEP)~\cite{Abreu:2012zg}. These
atmospheric data are available for a position close  to the
Auger array in three hour intervals. 
Data updates can be obtained
once per week and are filled automatically into the Auger
atmospheric monitoring database. 
Such atmospheric 
state variables are applied as standard in the air shower
reconstruction, mainly during the description of the fluorescence
light emission and the Rayleigh scattering of that light on its path
between emission point at the air shower and the fluorescence telescopes,
since mid 2011.

\subsection{Interdisciplinary atmospheric measurements}
\label{sec:atmos-science}

Through its secondary role as an Earth observatory, the interdisciplinary science program of the 
Observatory is quite extensive \cite {Bueno-Wiencke:2012}. For example, the FD has turned out to be the 
world's best detector for measuring atmospheric transient luminous events known as ELVES
(see Section~\ref{sec:Elves}) \cite{Mussa:2012dq}. 
%Elves are created above some thunderstorms and form part of the planet's electrical system.
%Their detailed measurement probes a little understood region of the ionosphere.

To characterize local aerosol particles, an Andersen-Graseby 240 dichotomous sampler 
was installed at the Coihueco FD building for 6 months. The particles collected were studied to 
determine their sizes and shapes.  Low aerosol concentrations were 
observed during the winter with an increasing concentration in spring. An elemental 
composition analysis was also performed \cite{Micheletti:2012fs}.

The \emph{HYbrid Single Particle Lagrangian Integrated Trajectory} HySPLIT model,
(developed from NOAA), that estimates the trajectories of air mass displacements was
used to study the source of aerosols at the Observatory.  A possible correlation was 
observed between clear conditions at the Observatory, 
determined from measured $\tau_{\rm aer}(h)$ profiles,
and air mass sources at the Pacific ocean \cite{Micheletti-ICRC:2013}.

\section{Hybrid event reconstruction}
\label{sec:HybridReconstruction}

%\red{(Michael Unger \& Ralf Ulrich)}

The hybrid reconstruction is based on fluorescence detector data with
additional timing information from the surface detector. In the
following, the individual reconstruction steps will be described.

\subsection{Pulse reconstruction at the FD}

At the beginning of the reconstruction, the baseline is subtracted
from the ADC trace of each pixel and the background noise is estimated
from the variance of the ADC signals at early time bins that are free
from any shower signal. Each ADC count is then converted to photons at the
aperture using the calibration constants obtained from the drum and
relative calibrations.

Subsequently, each triggered FD pixel is searched for a sho\-wer signal
by scanning the signal trace for pulse start and stop times that
maximize the signal to noise ratio. Only pulses with a signal to noise
ratio $\ge5$ are considered in the geometrical reconstruction.

The pulse time of the $i$th pixel is given by the centroid time
(``signal weighted time'') of all trace bins belonging to the pulse
\begin{equation}
  \label{eq:PulseTime}
  t_i = \frac{\sum\tau_k^i \; s_k^i}{q_i}\,,
\end{equation}
where the sum runs over the time bins defined by the aforementioned
signal to noise maximization. $\tau_k$ and $s_k$ are the time and
charge of the $k$th ADC bin, respectively, and the pixels' integrated signal is given
by
\begin{equation}
  \label{eq:PulseSignal}
  q_i = \sum s_k^i.
\end{equation}
The uncertainties of $q_i$ and $t_i$ are obtained by propagating the
noise variance and Poissonian photoelectron fluctuations into
equations~(\ref{eq:PulseTime}) and (\ref{eq:PulseSignal}).

\subsection{Shower detector plane}
\label{sec:SDP}

The \emph{shower detector plane} is the plane containing the
shower axis and the triggered fluorescence telescope.  It can be
reconstructed from the data of a telescope by minimizing
\begin{equation}
  S =  \frac{1}{\sum_i q_i}\sum_i q_i \; \left(\frac{\frac{\pi}{2}-\arccos(\vec{p}_i \cdot \vec{n}^\text{SDP}_\perp)
  }{ \sigma_\text{SDP}} \right)^2
\end{equation}
over all pulses $i$, with the two free parameters $\theta_\text{SDP}$
and $\phi_\text{SDP}$ to define the vector $\vec{n}^\text{SDP}_\perp$
normal to the plane in spherical coordinates and the pixel pointing
direction $\vec{p}_i$.  The pointing uncertainty for the SDP fit,
$\sigma_\text{SDP}$, was determined to be 0.35$^\circ$ by studying SDP
fits of CLF laser shots with a known geometry. The normalization of the fit
function $S$ to $\sigma_\text{SDP}$ allows one to interpret it as a $\chi^2$
function and to derive the uncertainties of the SDP parameters from
the $S+1$ contours.

\subsection{Hybrid time fit}
\label{sec:TimeFit}

\begin{figure}[t]
\centering
\includegraphics[width=0.98\linewidth]{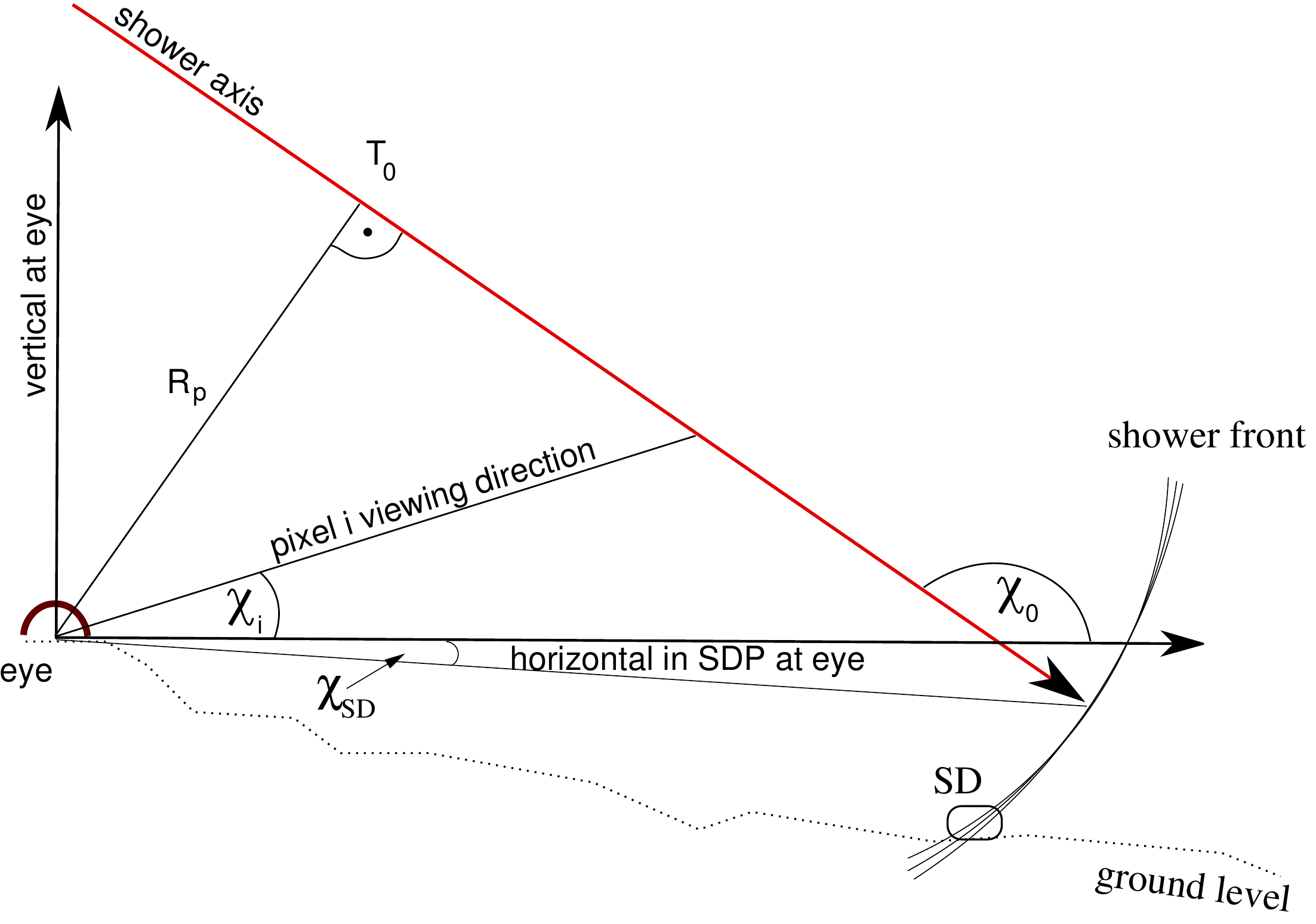}
\caption{Geometry of an air shower within the shower detector plane.}
\label{f:TimeFit}
\end{figure}

From the perspective of a telescope, the projection of a shower on the
camera evolves along the SDP.  Each pulse pixel can be associated
with an angle $\chi_i$ along the SDP with respect to the horizontal
axis at the telescope (see Figure~\ref{f:TimeFit}).  The angular
movement of the shower within the SDP in this representation is~\cite{Porter:1970et}
\begin{equation}
  t(\chi_i) = T_0 + \frac{R_\text{p}}{c}\;\tan\left(\frac{\chi_0-\chi_i}{2}\right).
\end{equation}
To determine the three free parameters $T_0$, $R_\text{p}$ and
$\chi_0$ the minimum of the function
\begin{equation}
  \chi^2 = \sum_i \frac{(t_i-t(\chi_i))^2}{\sigma(t_i)^2} + \frac{(t_\text{SD} - t(\chi_\text{SD}))^2}{\sigma(t_\text{SD})^2}
\end{equation}
has to be found~\cite{Sommers:1995dm, Dawson:1996ci}. The sum runs
over all pulse pixels $i$ with the centroid pulse time $t_i$ and the
associated uncertainty $\sigma(t_i)$, adding the additional SD station
time $t_\text{SD}$ with the uncertainty $\sigma(t_\text{SD})$.  The
shower front containing the surface detector station meets the (trial)
shower axis at a point that would be seen at angle $\chi_SD$, and
$t(\chi_SD)$ is the expected time when the shower center would pass
that point.

\begin{figure}[t]
\centering
\includegraphics[width=0.9\linewidth]{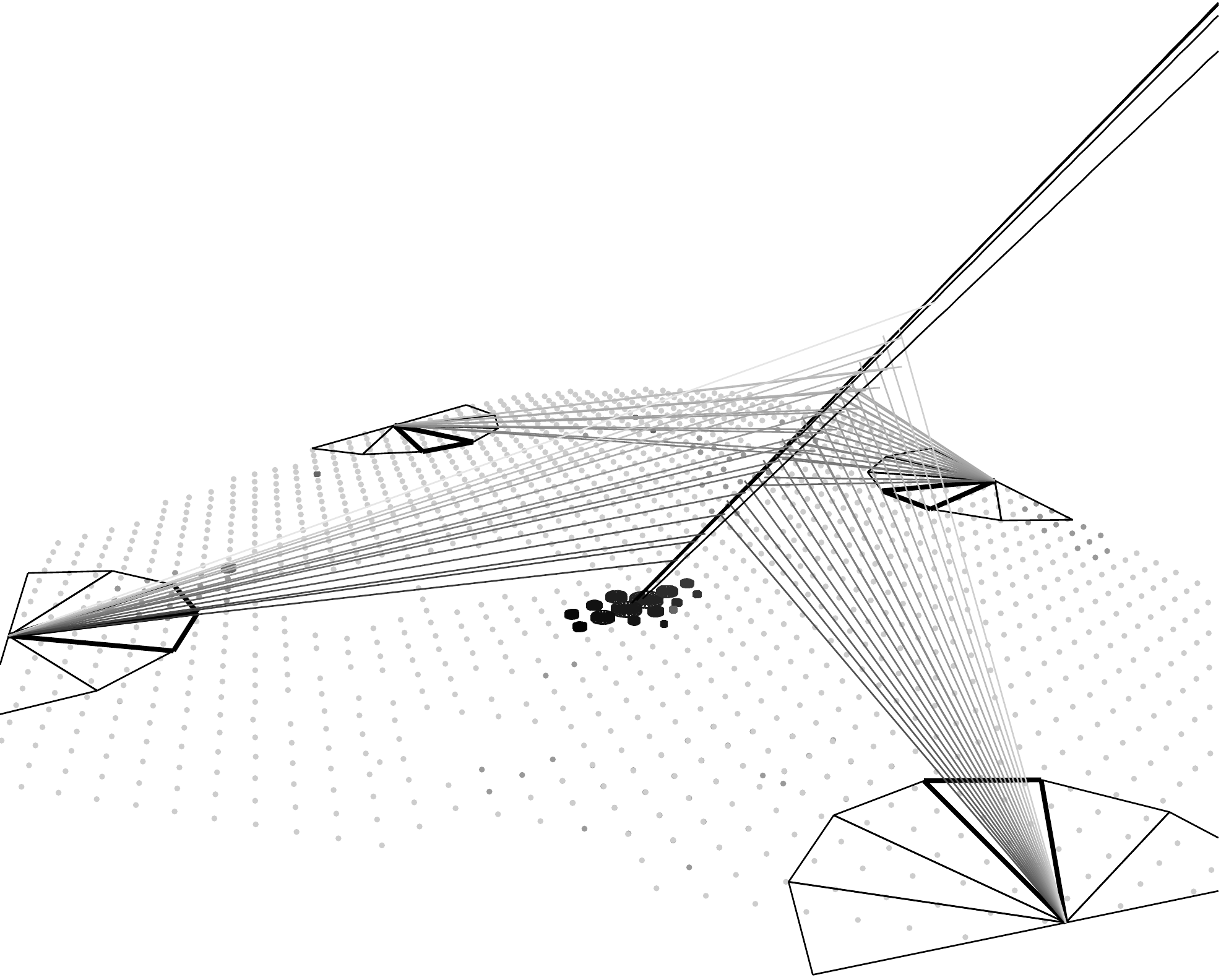}
\caption{Geometry reconstruction of an event observed by four
               telescopes and the surface detector.}
\label{fig:Quadruple}
\end{figure}

An example of an event that has been observed by four telescopes
is shown  in Figure~\ref{fig:Quadruple}. The individual four reconstructions
of the geometry using the hybrid approach are indicated by black lines.

\subsection{Light collection}

The total light at the aperture as a function of time is obtained by
adding the signals $s$ of the camera pixels $j$ at each time bin $i$.
For this purpose, only the pixels with pointing directions that are within
a certain angular distance $\zeta$ to the vector from the telescope
to the shower center at time $i$ are included.
$\zeta$ is chosen such that the signal to noise ratio is maximized. The
light flux arriving at the detector in time bin $i$ is
\begin{equation}
  \label{eqn:signalZeta}
  F_i = \frac{1}{A_\text{dia}} \sum_{j=1}^{N_\text{pix}} s_{ij},
\end{equation}
where the sum runs over all pixels $N_\text{pix}$ within
$\zeta$ at time bin $i$ and $A_\text{dia}$ is the area of
the diaphragm opening.

\subsection{Profile reconstruction}
\label{sec::profileReconstruction}

Once the geometry of the shower is known, the light collected at the
aperture as a function of time can be converted to the energy
deposited by the shower as a function of slant depth. For this
purpose, the light attenuation from the shower to the detector needs
to be accounted for and all contributing light sources need to be
disentangled: fluorescence light~\cite{Arqueros:2008cx}, direct and
scattered Cherenkov light~\cite{Giller:2004,Nerling:2005fj} as well as
multiply scattered light~\cite{Roberts:2005xv, Pekala:2009fe,
  Giller:2012tt}.

The proportionality between the fluorescence intensity and the energy
deposit is given by the fluorescence yield. A good knowledge
of its absolute value as
well as its dependence on wavelength, temperature, pressure and
humidity is essential to reconstruct the longitudinal profile.
We use the most precise of the measurements available to date
(cf.~\cite{Rosado:2014bya}) as provided by the Airfly
Collaboration~\cite{Ave:2008zza, Ave:2012ifa}.

The Cherenkov and fluorescence light produced by an air shower are
connected to the energy deposit by a linear set of equations and
therefore the shower profile is obtained by an analytic linear least squares
minimization~\cite{Unger:2008uq}.  Due to the lateral extent of air
showers, a small fraction of shower light is not contained within the
optimal light collection area. To correct this, the universal lateral
fluorescence~\cite{Gora:2005sq} and Cherenkov
light~\cite{Dawson-Giller:2007} distributions must be taken into
account.  The full longitudinal energy deposit profile and its maximum
$(\frac{{\rm d}E}{{\rm d}X})_\mathrm{max}$ at depth $X=X_\mathrm{max}$
are estimated by
fitting a Gaisser-Hillas function~\cite{Gaisser-Hillas:1977},
\begin{equation}
%\begin{multline}
    f_\mathrm{GH}(X)=\\\left(\frac{{\rm d}E}{{\rm d}X}\right)_\mathrm{max}
      \left(\frac{X-X_0}{X_\mathrm{max}-X_0}\right)^\frac{X_\mathrm{max}-X_0}{\lambda}
      e^{\frac{X_\mathrm{max}-X}{\lambda}}\,,
 \label{eq:GH}
%\end{multline}
 \end{equation}
to the photoelectrons detected in the PMTs of the FD cameras. For
this purpose, a log-likelihood fit is used in which the number of
photoelectrons detected by the PMTs of the FD cameras is compared to
the expectation from equation~(\ref{eq:GH}) after folding it with the light
yields, atmospheric transmission, lateral distributions and detector
response.  The two shape parameters $X_0$ and $\lambda$ are
constrained to their average values to allow for a gradual transition
from a two- to a four-parameter fit depending on the observed track
length and number of detected photons of the respective event
(cf.~\cite{Unger:2008uq}).

\begin{figure}[t]
\subfigure[Light at aperture.]
 {\label{fig:lightAtAperture}
\includegraphics[width=0.48\textwidth]{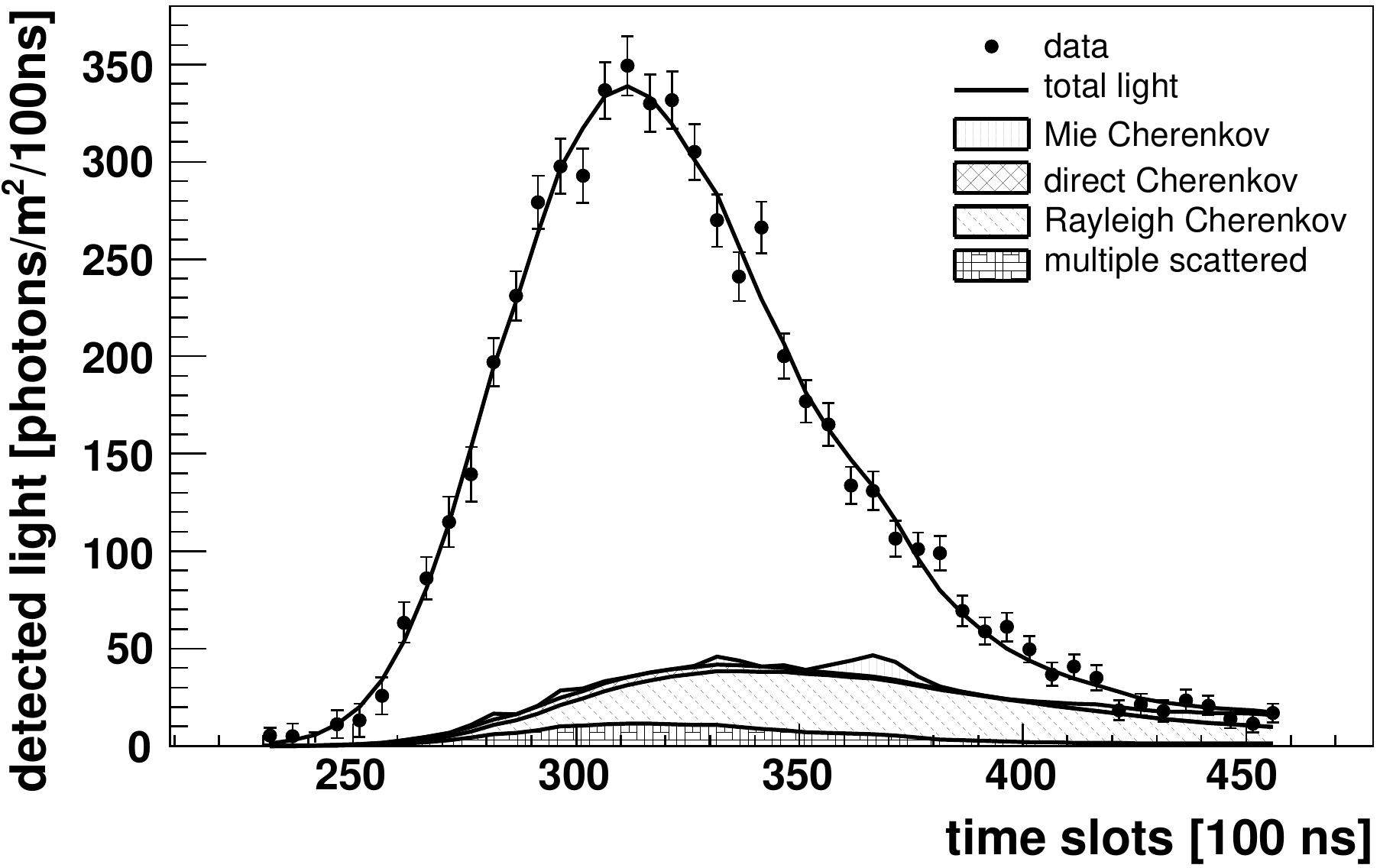}}
\subfigure[Energy deposit.]
 {\label{fig:dEdXProfile}
 \includegraphics[width=0.48\textwidth]{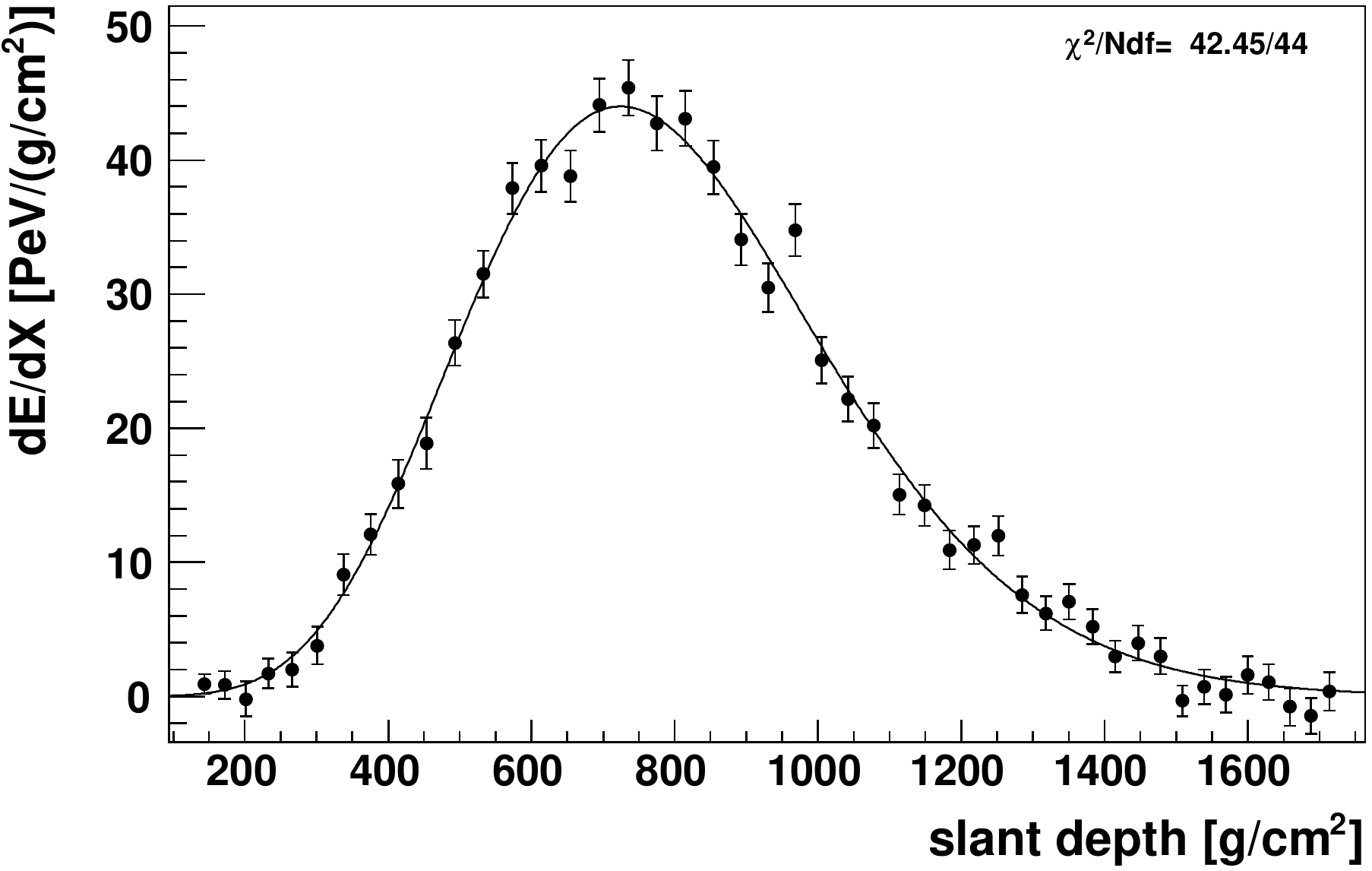}}
  \caption{Example of a reconstructed shower profile.}
        \label{fig:fdreco}
\end{figure}

Finally, the calorimetric energy of the shower is obtained by
integrating equation~(\ref{eq:GH}) and the total energy is estimated by
correcting for the `invisible energy' carried away by neutrinos and
high energy muons \cite{Tueros-ICRC:2013}. An example of the measured light at aperture and the reconstructed
light contributions, and energy deposit profile is shown in
Figs.~\ref{fig:lightAtAperture} and~\ref{fig:dEdXProfile}.

\section{SD event reconstruction}
\label{sec:SDreco}

%\red{(Ioana \& Darko)}

The reconstruction of the energy and the arrival direction of the
cosmic rays producing air showers that have triggered the surface
detector array is based on the sizes and times of signals registered
from individual SD stations. At the highest energies, above
\unit[10]{EeV}, the footprint of the air shower on the ground extends
over more than \unit[25]{km$^2$}. By sampling both the arrival times
and the deposited signal in the detector array, the shower geometry,
i.e., the shower core, the arrival direction of the incident cosmic
ray, and the shower size can be determined.

\subsection{Event selection}

To ensure good data quality for physics analysis there are two
additional off-line triggers.  The physics trigger, T4, is needed to
select real showers from the set of stored T3 data (see
Section~\ref{event-triggering}) that also contain background signals
from low energy air showers.  This trigger is mainly based on a
coincidence between adjacent detector stations within the propagation
time of the shower front.  In selected events, random stations are
identified by their time incompatibility with the estimated shower
front. The time cuts were determined such that 99\,\% of the stations
containing a physical signal from the shower are kept. An algorithm
for the signal search in the time traces is used to reject signals
produced by random muons by searching for time-compatible peaks.

To guarantee the selection of well-contained events, a fiducial cut
(called the 6T5 trigger) is applied so that only events in which the
station with the highest signal is surrounded by all 6 operating
neighbors (i.e.,\ a working hexagon) are accepted. This condition
assures an accurate reconstruction of the impact point on the ground,
and at the same time allowing for a simple geometrical calculation of
the aperture/exposure \cite{Abraham:2010zz}, important for, e.g., the
spectrum analysis \cite{Abraham:2010mj}. For arrival-direction studies
a less strict cut can be used (5T5 or even 4T5).

%Atmospheric muons and low energy air-showers (with energies smaller than
%$10^{14}$\,eV) can generate a background signal in the time window
%around the arrival of particles from the large air-shower. Accordingly stations are
%selected by their time compatibility with the estimated
%shower front. The time cuts were determined such that 99\% of the
%stations containing a physical signal from the shower are kept. An
%algorithm for the signal search in the time traces is used to reject
%accidental signals by searching for time-compatible peaks.

\begin{figure}[t]
\centering
\includegraphics[width=0.48\textwidth]{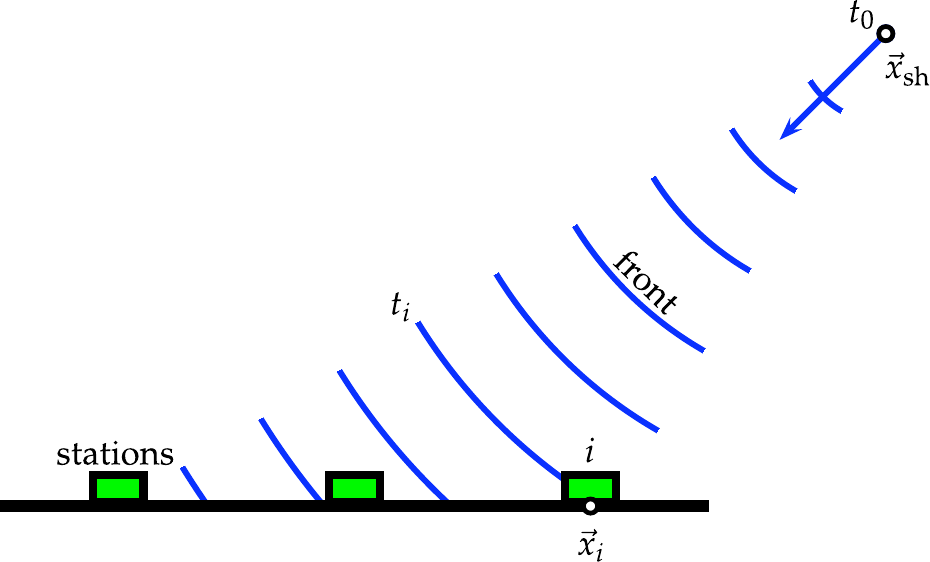}
\caption{Reconstruction of shower geometry: schematic representation
  of the evolution of the shower front.}
\label{fig:curvature}
\end{figure}

\begin{figure}[t]
\centering
\includegraphics[width=0.45\textwidth]{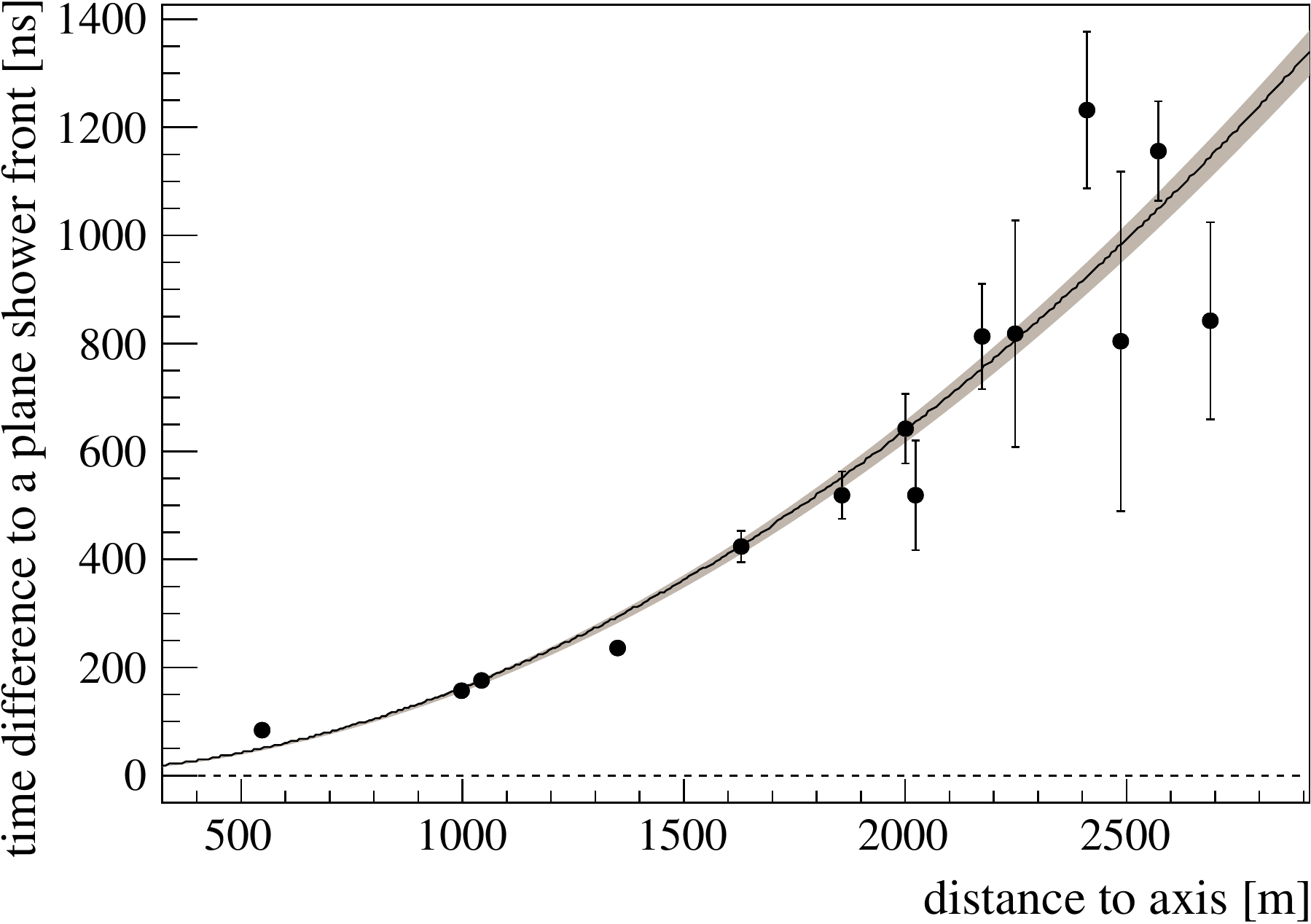}
\caption{Reconstruction of shower geometry: dependence of signal start
  times (relative to the timing of a plane shower front) on
  perpendicular distance to the shower axis. The shaded line is the
  resulting fit of the evolution model and its uncertainty.}
\label{fig:timeFit}
\end{figure}

\subsection{Shower geometry}

A rough approximation for the arrival direction of the shower is
obtained by fitting the start times of the signals, $t_i$, in
individual SD stations to a plane front. For events with enough
triggered stations, these times are described by a more detailed
concentric-spherical model, see Figure~\ref{fig:curvature}, which
approximates the evolution of the shower front with a speed-of-light
inflating sphere,
\begin{equation}
c(t_i - t_0) = |\vec{x}_\text{sh} - \vec{x}_i|,
\end{equation}
where $\vec{x}_i$ are positions of the stations on the ground and
where $\vec{x}_\text{sh}$ and $t_0$ are a \emph{virtual} origin and a
start-time of the shower development (see
Figure~\ref{fig:timeFit}). From this 4-parameter fit the radius of
curvature of the inflating sphere is determined from the time at which
the core of the shower is inferred to hit the ground.

\subsection{Lateral distribution function}

The impact points of the air showers on the ground,
$\vec{x}_\text{gr}$, are obtained from fits of the signals in SD
stations. This fit of the lateral distribution function (LDF) is based
on a maximum likelihood method which also takes into account the
probabilities for the stations that did not trigger and the stations
close to the shower axis with saturated signal traces. The saturation
is caused by the overflow of the FADC read-out electronics with finite
dynamic range and a modification of the signal due to the transition
of the PMTs from a linear to a non-linear behavior. In the majority of
cases the missing part of the signals are recovered using the
procedure described in~\cite{Veberic-ICRC:2013}.

\begin{figure}[t]
\centering
\includegraphics[width=0.4\textwidth]{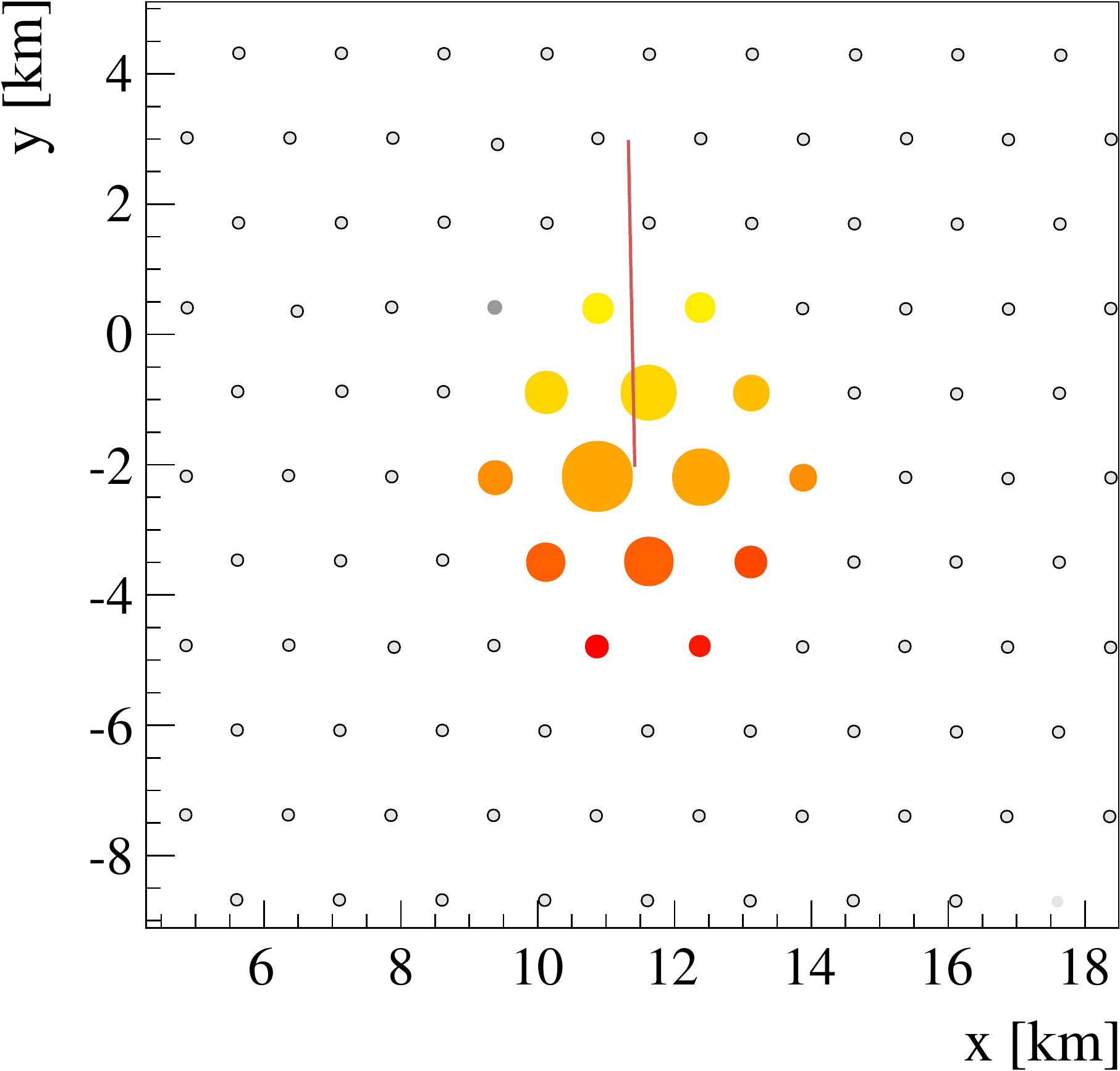}
\caption{Example of signal sizes an extensive air shower induces in
  the stations of the surface detector array. Note that the spacing of
  the regular grid is 1.5\,km. Colors represent the arrival time of
  the shower front from early (yellow) to late (red) and the size of the
  markers is proportional to the logarithm of the signal. The line represents
	the shower arrival direction.}
\label{fig:footprint}
\end{figure}

\begin{figure}[t]
\centering
\includegraphics[width=0.48\textwidth]{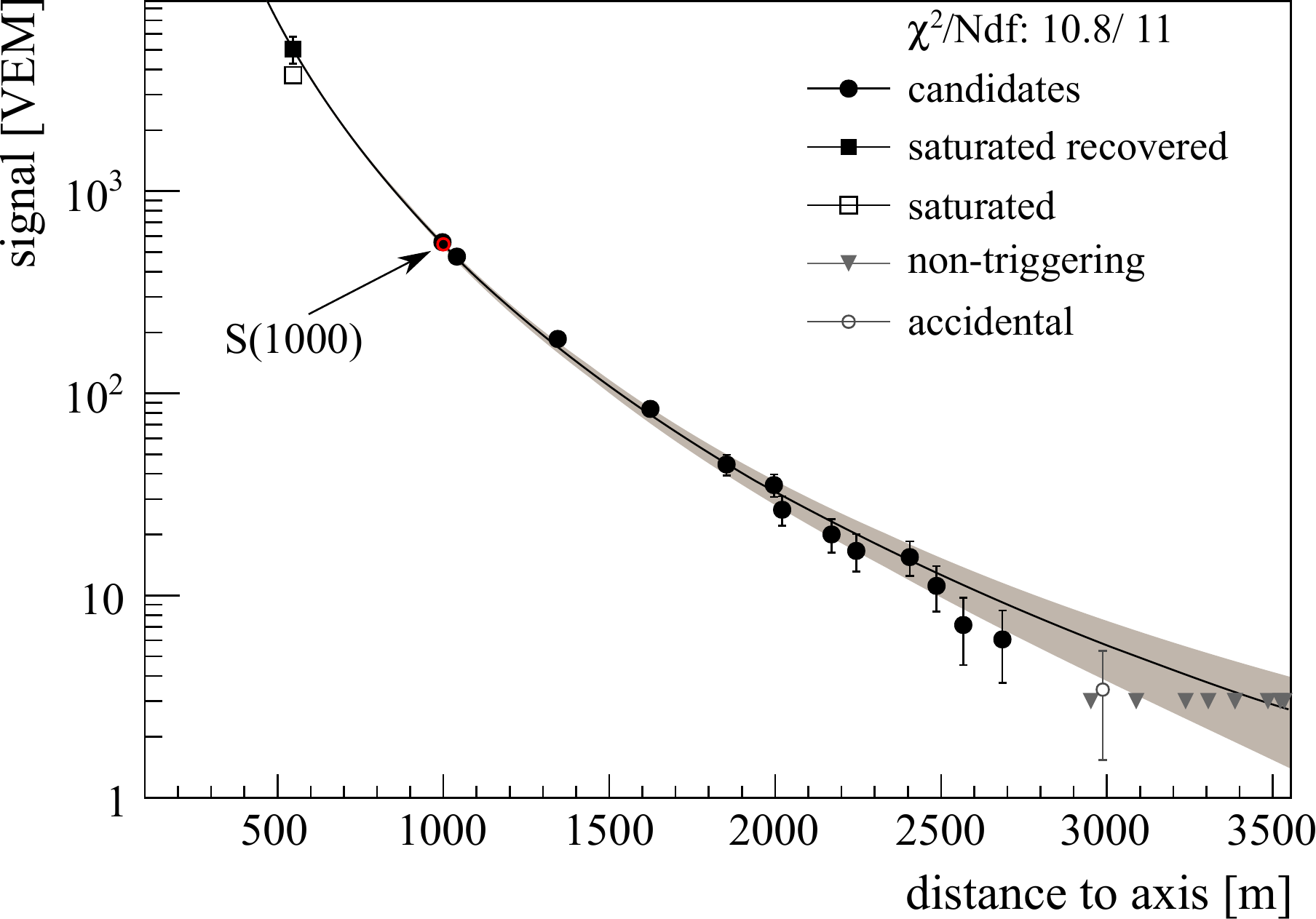}
\caption{Dependence of the signal size on distance from the shower core.}
\label{fig:ldf}
\end{figure}

An example of the footprint on the array of an event produced by a
cosmic ray with an energy of ($104 \pm 11$) EeV and a zenith angle of
($25.1\pm 0.1$) degrees is shown in Figure~\ref{fig:footprint}. The
lateral distribution of the signals is depicted in
Figure~\ref{fig:ldf}. The function employed to describe the lateral
distribution of the signals on the ground is a modified
Nishimura-Kamata-Greisen function
\cite{Kamata:1958,Greisen:1956},
\begin{equation}
S(r) = S(r_\text{opt})
  \left(\frac{r}{r_\text{opt}}\right)^\beta
  \left(\frac{r+r_1}{r_\text{opt}+r_1}\right)^{\beta+\gamma}
\end{equation}
where $r_\text{opt}$ is the optimum distance, $r_1=700$\,m and
$S(r_\text{opt})$ is an estimator of the shower size used in an energy
assignment. For the SD array with station spacing of 1.5\,km the
optimum distance \cite{Newton:2007} is $r_\text{opt}=1000$\,m and the
shower size is thus $S(1000)$.  The parameter $\beta$ depends on the
zenith angle and shower size. Events up to zenith angle $60^\circ$ are observed at an
earlier shower age than more inclined ones, thus having a steeper LDF
due to the different contributions from the muonic and the
electromagnetic components at the ground.  For events with only 3
stations, the reconstruction of the air showers can be obtained only by
fixing the two parameters, $\beta$ and $\gamma$ to a parametrization
obtained using events with a number of stations larger than 4.

The reconstruction accuracy of $S(1000)$, $\sigma_{S(1000)}$, is
composed of three contributions: a statistical uncertainty due to the
finite number of particles producing a signal in a given SD station and
the limited dynamic range of the signal detection; a systematic
uncertainty due to assumptions on the shape of the lateral
distribution function; and an uncertainty due to shower-to-shower
fluctuations \cite{Ave:2007wf}. The last term contributes a factor of
about 10\,\%, while the contribution of the first two terms depends on
energy and varies from 20\,\% (at low energies) to 6\,\% (at the highest
energies).

\subsection{Shower arrival direction}

Shower axis $\hat{a}$ is obtained from the virtual shower origin (of the
geometrical reconstruction) and the shower impact point on the ground (from
the LDF reconstruction),
\begin{equation}
\hat{a} =
  \frac{\vec{x}_\text{sh}-\vec{x}_\text{gr}}
       {|\vec{x}_\text{sh}-\vec{x}_\text{gr}|}.
\end{equation}

\begin{figure}[t]
\centering
\includegraphics[width=0.4\textwidth]{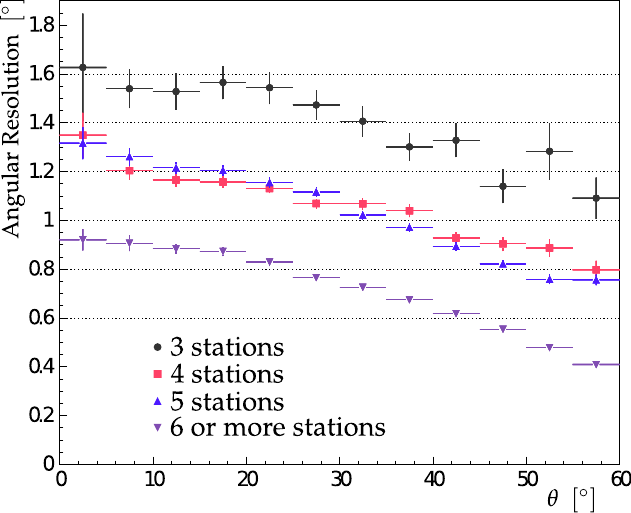}
\caption{Angular resolution as a function of the zenith angle $\theta$ for
events with an energy above 3\,EeV, and for various station multiplicities.
\cite{Bonifazi:2009ma}.}
\label{fig:angular_resolution}
\end{figure}

To estimate an angular resolution of the whole reconstruction procedure a
single station time variance is modeled \cite{Bonifazi:2007ck} to take into account the size of
the total signal and the time evolution of the signal trace. As shown in
Figure~\ref{fig:angular_resolution}, the angular resolution
achieved for events with more than three stations is better than $1.6^\circ$,
and better than $0.9^\circ$ for events with more than six stations
\cite{Bonifazi:2009ma}.

\subsection{Energy calibration}

\begin{figure}[t]
\centering
\includegraphics[width=0.4\textwidth]{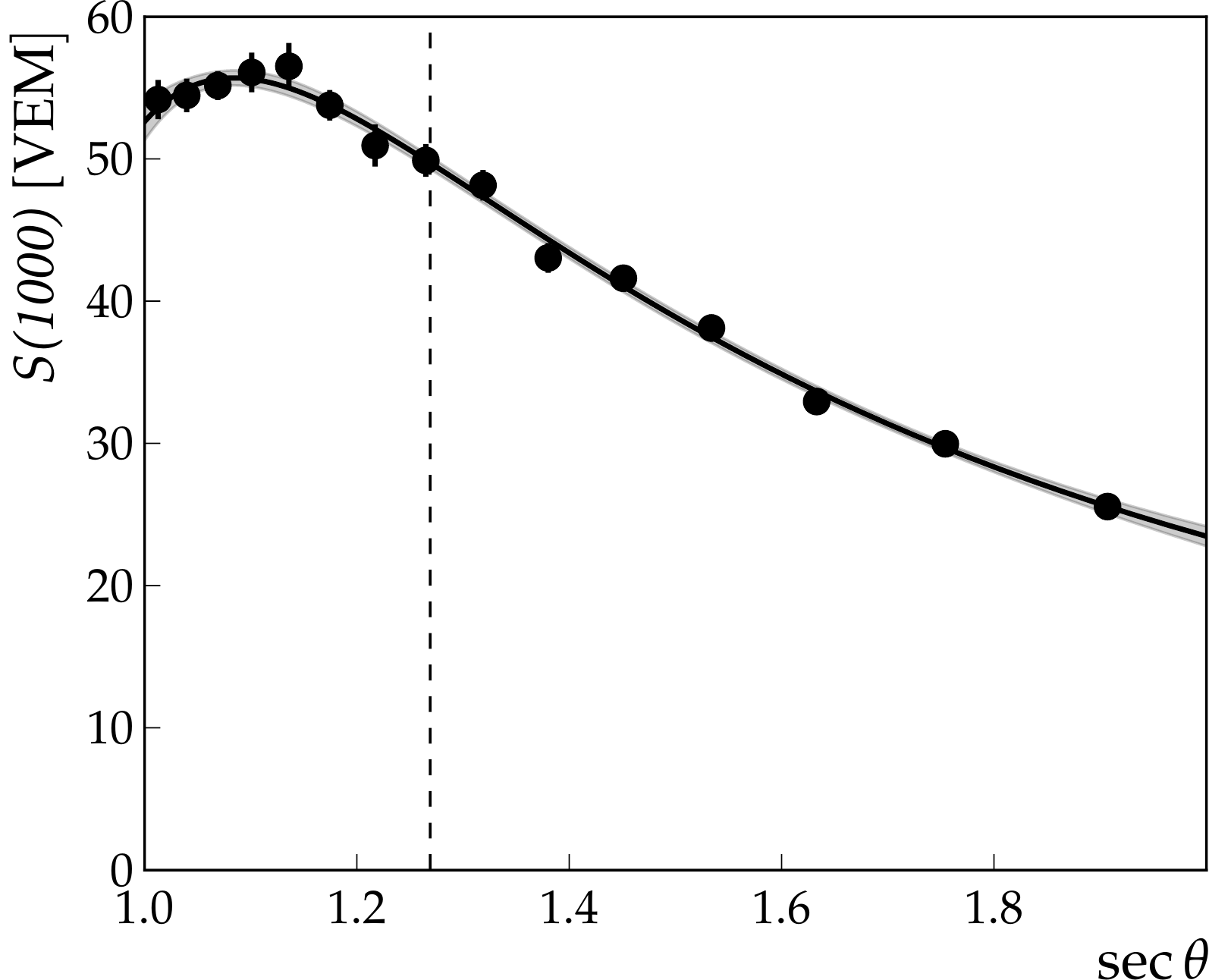}
\caption{Attenuation curve described by a third degree polynomial in
  $x=\cos^2\theta-\cos^2\bar\theta$ where $\bar\theta=38^\circ$
  (denoted by the dashed vertical line). In this example the
  polynomial coefficients are deduced from $S(1000)$ dependence at
  $S_{38}\approx 50$\,VEM which corresponds to an energy of about 10.5\,EeV.}
\label{fig:cic}
\end{figure}

For a given energy, the value of $S(1000)$ decreases with the zenith angle
$\theta$ due to the attenuation of the shower particles and geometrical
effects. Assuming an isotropic flux of primary cosmic rays at the top of the
atmosphere, we extract the shape of the attenuation curve (see
Figure~\ref{fig:cic}) from the data using the Constant Intensity Cut (CIC)
method \cite{Hersil:1961zz}. The attenuation curve $f_\text{CIC}(\theta)$ has
been fitted with a third degree polynomial in
$x=\cos^2\theta-\cos^2\bar\theta$, i.e.,\
$f_\text{CIC}(\theta)=1+a\,x+b\,x^2+c\,x^3$, where $a=0.980\pm0.004$,
%what is the third decimal on a? we should give it because the error is 0.004
$b=-1.68\pm0.01$, and $c=-1.30\pm0.45$~\cite{Schulz-ICRC:2013}.

The median angle, $\bar\theta=38^\circ$, is taken as a reference point to
convert $S(1000)$ to \\
$S_{38}\equiv S(1000)/f_\text{CIC}(\theta)$.  $S_{38}$ may
be regarded as the signal a particular shower with size $S(1000)$ would have
produced had it arrived at $\theta=38^\circ$.

To estimate the energy of the primary particle producing the
air-showers recorded with the SD, the advantage comes from the hybrid
detection: the air-showers that have triggered independently the FD
and SD are used for the cross-calibration. High-quality hybrid events,
as defined below, with reconstructed zenith angles less than
$60^\circ$ are used to relate the shower size from SD to the
almost-calorimetric measurement of the shower energy from FD,
$E_\text{FD}$.

These hybrid events must be such that the reconstruction of an energy
estimator can be derived independently from both the SD and FD parts
of the event \cite{Pesce-ICRC:2011,Tueros-ICRC:2013}.

Only a subsample of events that passes strict quality and field of view
cuts is used. For the FD part of the event, we require an accurate
fit of the longitudinal profile to the Gaisser-Hillas function.
Furthermore, the depth of the shower maximum, $X_\text{max}$, must be
contained within the telescope field-of-view and measured with an
accuracy better than 40\,g/cm$^2$.

The uncertainty on the reconstructed $E_\text{FD}$ is required to be
less than 18\,\%.  The final criteria for defining the calibration data
sample include a selection of clear atmosphere conditions based on the
measurements of the vertical aerosol optical depth, and on the cloud
fraction measured by the lidar systems of the Observatory.  To avoid
any potential bias of the event selection on the mass of the primary
particle, a fiducial cut on the slant depth range observed by the
telescopes is also added, ensuring that the field of view is large
enough to observe all plausible values of $X_\text{max}$
for the geometry of each individual shower
\cite{Pesce-ICRC:2011}.

\begin{figure}[t]
\centering
\includegraphics[width=0.4\textwidth]{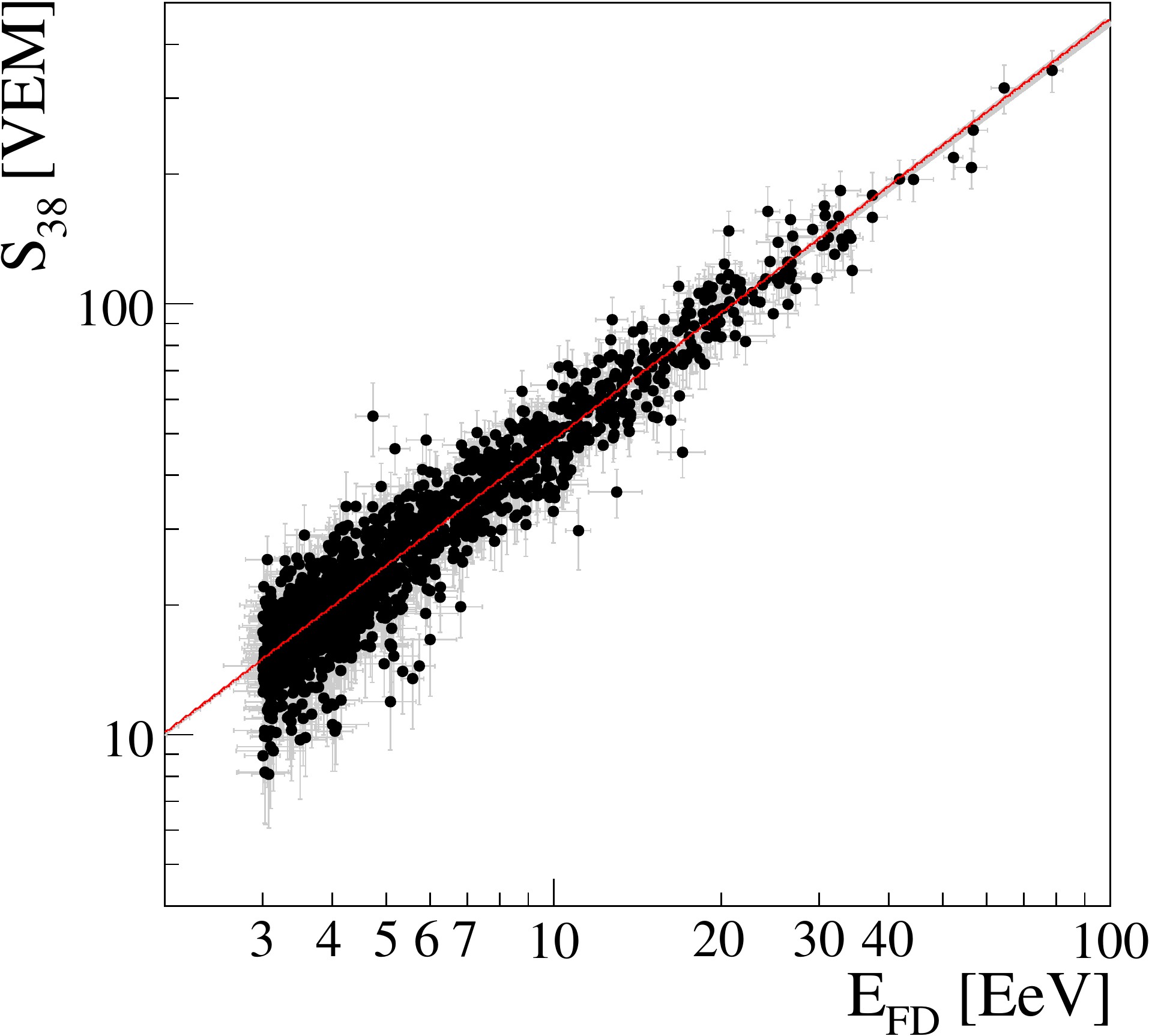}
\caption{Correlation between $S_{38}$ and $E_{FD}$
\cite{Schulz-ICRC:2013,Pesce-ICRC:2011}.}
\label{fig:energy_calibration}
\end{figure}

The final step in the calibration analysis leads to a relation between
$S_{38}$ and $E_\text{FD}$.  The 1475 high quality hybrid events
recorded between Jan 2004 and Dec 2012 which have an energy above the
SD full efficiency trigger threshold \cite{Abraham:2010zz} are used
in the calibration.  The correlation between the two variables is
obtained from a maximum likelihood method
\cite{Pesce-ICRC:2011,Dembinski-ICRC:2011} which takes into account
the evolution of uncertainties with energy, as well as event
migrations due to the finite energy resolution of the SD. The relation
between $S_{38}$ and $E_\text{FD}$ is well described by a single
power-law function,
\begin{equation}
E_\text{FD} = A\,(S_{38}/\text{VEM})^B
\end{equation}
where the resulting parameters from the data fit are
$A=(1.90\pm0.05)\times10^{17}$\,eV and
$B=1.025\pm0.007$~\cite{Schulz-ICRC:2013,Verzi-ICRC:2013}.  As can be seen in
Figure~\ref{fig:energy_calibration}, the most energetic event used in this
analysis has an energy of 79\,EeV.

The resolution of the final SD energy estimator,
\begin{equation}
E_\text{SD} =
  A(S(1000)/f_\text{CIC}(\theta)/\text{VEM})^B,
\end{equation}
can be inferred from the distribution of the ratio
$E_\text{SD}/E_\text{FD}$.  Using the FD energy resolution of 7.6\,\%,
the resulting SD energy resolution with its statistical uncertainty is
$\sigma_{E_\text{SD}}/E_\text{SD}=(16\pm1)$\% at the lower energy
edge in Figure~\ref{fig:energy_calibration} and $(12\pm1)$\% at the
highest energies. Due to the large number of events accumulated until
December 2012, the systematic uncertainty on the SD energy due to the
calibration is better than 2\,\% over the whole energy range. 
%The systematic
%uncertainties are dominated by the FD energy scale uncertainty of
%14\%~\cite{Verzi-ICRC:2013}.
%The main contributions to this uncertainty are related to
%the knowledge of the fluorescence yield (3.6\%), the atmospheric
%conditions (3.4\%- 6.2\%), the absolute calibration of the telescopes
%(9.9\%), the shower profile reconstruction(6.5\%-5.6\%) and the
%invisible energy (3\%-1.5\%).
The systematic uncertainties in the energy scale, shown in Table~\ref{table:systematics},
are dominated by the absolute FD calibration~\cite{Verzi-ICRC:2013}.
Further consistency checks are performed by joint calibration campaigns
with the Telescope Array~\cite{Tokuno:2012mi, Array:2013dra}.

\begin{table*}[th]
\caption{Systematic uncertainties in the energy scale.}
\label{table:systematics}
\begin{center}
\begin{tabular}{ll}
\toprule
Absolute fluorescence yield    & 3.4\% \\
Fluorescence spectrum and quenching parameters		& 1.1\,\% \\
{\em Subtotal, Fluorescence yield}									& {\bf 3.6\,\%} \\
\midrule
Aerosol optical depth						& 3--6\,\% \\
Aerosol phase function					& 1\,\% \\
Wavelength dependence of aerosol scattering			& 0.5\,\% \\
Atmospheric density profile											& 1\,\% \\
{\em Subtotal, Atmosphere}												& {\bf 3.4--6.2\,\%} \\
\midrule
Absolute FD calibration				& 9\,\% \\
Nightly relative calibration		& 2\,\% \\
Optical efficiency							& 3.5\,\% \\
{\em Subtotal, FD calibration}		& {\bf 9.9\,\%} \\
\midrule
Folding with point spread function		& 5\,\% \\
Multiple scattering model							& 1\,\% \\
Simulation bias												& 2\,\% \\
Constraints in the Gaisser-Hillas fit		& 3.5--1\,\% \\
{\em Subtotal, FD profile reconstruction} & {\bf 6.5--5.6\,\%} \\
\midrule
Invisible energy & {\bf 3--1.5\,\%} \\
\midrule
Statistical error of SD calibration fit		& {\bf 0.7--1.8\,\% } \\
\midrule
Stability of the energy scale			& {\bf 5\,\%} \\
\midrule
{\bf Total} & {\bf 14\,\%} \\
\midrule
\bottomrule
\end{tabular}
\end{center}
\end{table*}

The dataset recorded extends up to larger angles of $90^\circ$. For
the inclined events, with zenith angles larger than $60^\circ$  we
employ a different reconstruction method. More details on the
reconstruction of inclined events can be found
in~\cite{Dembinski-ICRC:2011,Valino-ICRC:2013,Aab:2014gua}. The energy range of
full efficiency of the surface detector has been extended down to
$3{\times}10^{17}$\,eV using the events recorded by the 750\,m array
(see Section~\ref{sec:Enhancements}). The reconstruction of this subsample of
events is described in~\cite{Maris-ICRC:2011,Ravignani-ICRC:2013, Aab:2014gua}.

\section{Performance characteristics of the Observatory}

%\red{(Tiina S \& Lorenzo P)}

\label{sec:performance}

\subsection{Key performance parameters}

In Table~\ref{tab:key_perf} are summarized some of the important parameters that characterize the performance of the Observatory. These parameters include the event rate  of  the detectors and the resolutions  of the different reconstructed observables.

\begin{table*}
\caption{Key performance parameters for the Auger Observatory.}
\label{tab:key_perf}
\begin{center}
\begin{tabular}{ll}
\toprule
\multicolumn{2}{c}{\bf SD}
\\
\midrule
SD Annual Exposure               & ${\sim}5500$\,km$^2$\,sr\,yr
\\
\midrule
T3 rate                   & 0.1\,Hz
\\
T5 events/yr, $E>3$\,EeV & ${\sim}14,500$
\\
T5 events/yr, $E>10$\,EeV & ${\sim}1500$
\\
Reconstruction accuracy ($S_{1000}$) & 22\,\% (low $E$) to 12\,\% (high $E$)
\\
Angular resolution        & $1.6^{\circ}$ (3 stations)
\\
                          & $0.9^{\circ}$ (${>}5$ stations)
\\
Energy resolution         & 16\,\% (low $E$) to 12\,\% (high $E$)
\\
\midrule
\multicolumn{2}{c}{\bf FD}
\\
\midrule
On-time                    & ${\sim}15$\,\%
\\
Rate per building               & 0.012\,Hz
\\
Rate per HEAT  & 0.026\,Hz
\\
\midrule
\multicolumn{2}{c}{\bf Hybrid}
\\
\midrule
Core resolution           & 50\,m
\\
Angular resolution        & $0.6^\circ$
\\
Energy resolution (FD)    & 8\,\%
\\
$X_\text{max}$ resolution      & ${<}20$\,g/cm$^2$
\\
\bottomrule
\end{tabular}
\end{center}
\end{table*}

\subsection{Surface detector performance}

Stable data taking with the surface detector array started in January 2004 and the
Observatory has been running in its full configuration since 2008. As described in Section~\ref{sec:monitoring}, various
parameters are continuously monitored to optimize the performance of the detectors and ensure reliable data.

\begin{figure}[t]
\centering
\includegraphics[width=0.9\columnwidth]{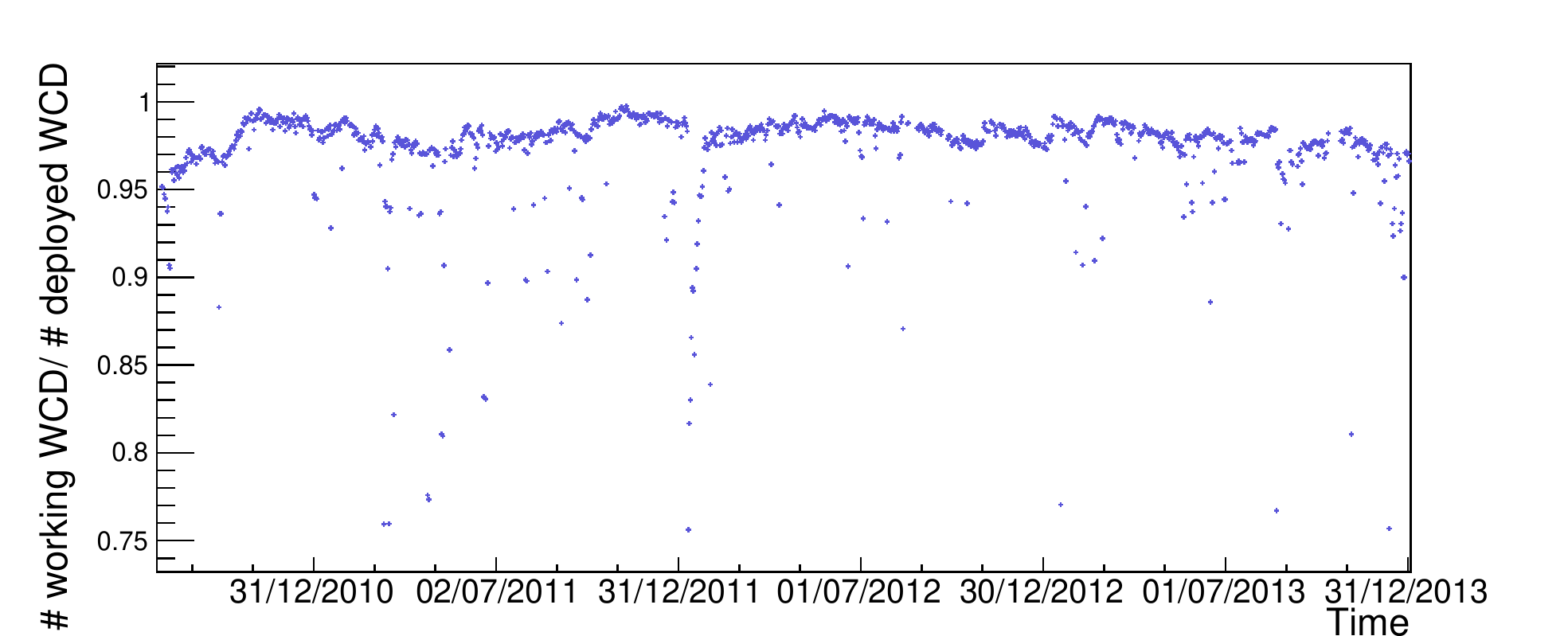}
\caption{Number of active SD stations normalized to the nominal number of SD stations in the array, as a function of time. 
(Note: WCD = water Cherenkov detector.)}
\label{fig:ratiotank}
\end{figure}

The monitoring tool includes so-called performance metrics  to monitor the overall performance of the surface detector array.
Relevant data useful for long term studies and for quality checks are stored in the Auger Monitoring database on a one-day basis.
For example, mean values  over one day of
the number of active SD detectors and the number of active hexagons
as well as the nominal value (expected value if all the detectors deployed were active) are available.
As an example, Figure~\ref{fig:ratiotank} shows the number of active SD stations normalized to the
nominal number of stations in the array  for the last 4 years. This plot is a convolution of the status of the active stations and of the efficiency of the CDAS, which since the beginning is better than 99.5\,\%.

\begin{figure}[t]
\centering
\includegraphics[width=0.9\columnwidth]{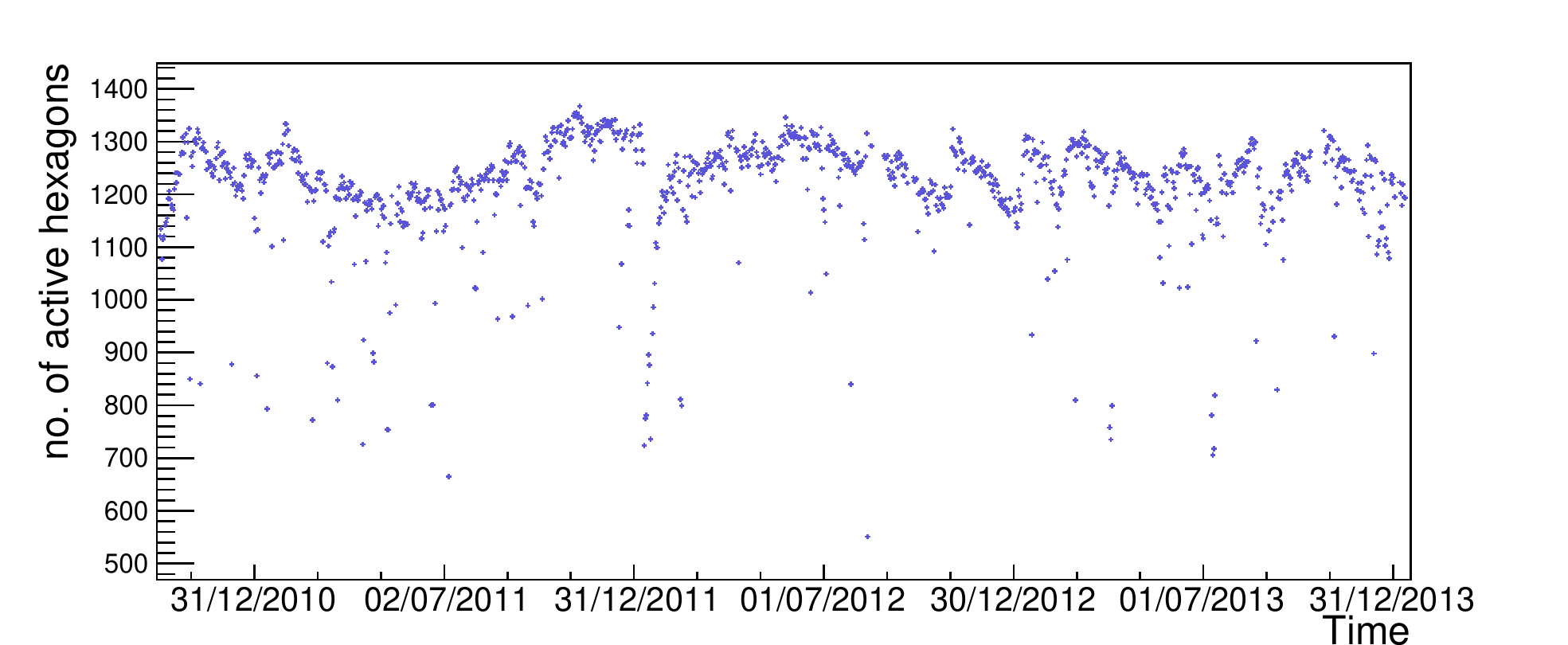}
\caption{Number of active hexagons as a function of time.  }
\label{fig:hexa}
\end{figure}

Figure~\ref{fig:hexa}  shows the number of active hexagons for the same period. This variable is a key parameter since it is the basis of the exposure evaluation. Indeed, the offline T5 fiducial  trigger, described in Section~\ref{sec:SDreco} selects only events for which the hottest station is surrounded by an active hexagon. Thus, above $3\times10^{18}$\,eV, when the full efficiency of detection of the array is reached (at least three triggered tanks), the exposure is simply proportional to the integrated number of active hexagons during the period. The number of active hexagons fluctuates because of intermittent
outages in electronics, communications, weather, and other factors~\cite{Abraham:2010zz}. Larger numbers of hexagons can be affected when the problems occur at the WLAN sector level. These outages can usually be resolved quickly.

\begin{figure}[t]
\centering
\includegraphics[width=0.99\columnwidth]{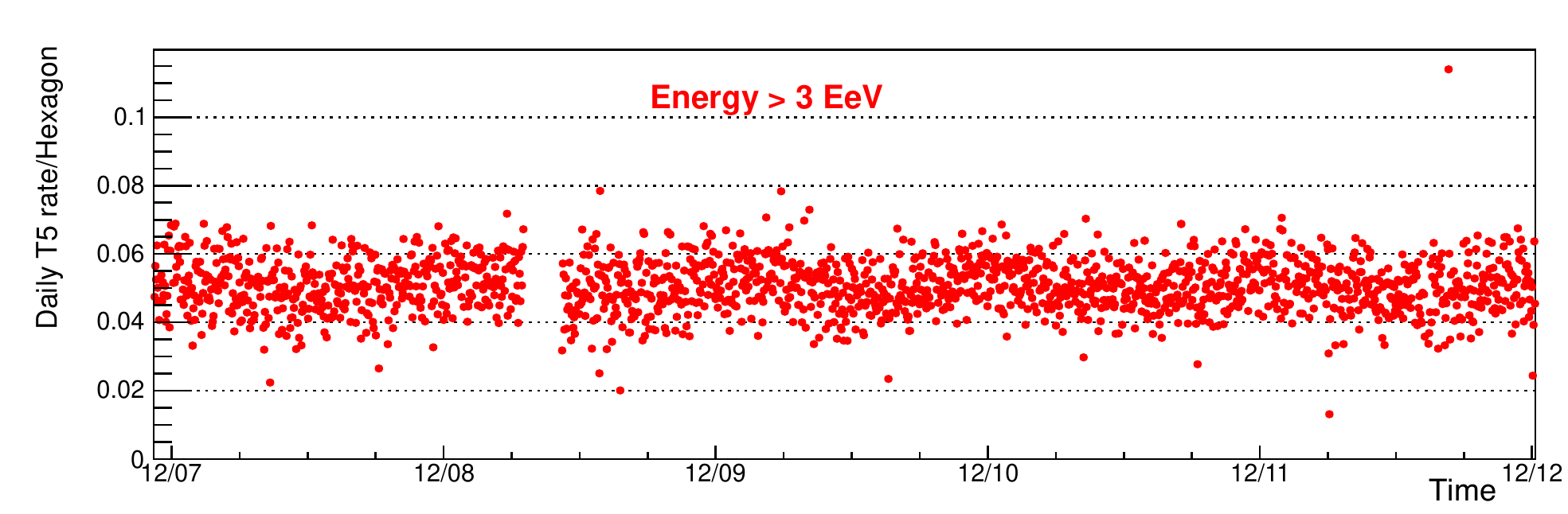}
\caption{Evolution of  the daily T5 rate normalized to the number of hexagons for the period 2008--2012. The data gap in 2009 corresponds to a period during which the communications system experienced technical problems.  }
\label{fig:t5rate3}
\end{figure}

The rate of events (T5 events)  normalized to the average number of active hexagons is expected to be stable in time  above the energy threshold of $3\times10^{18}$\,eV, which can be seen  in Figure \ref{fig:t5rate3}.

\begin{figure}[t]
\centering
\includegraphics[width=0.9\columnwidth]{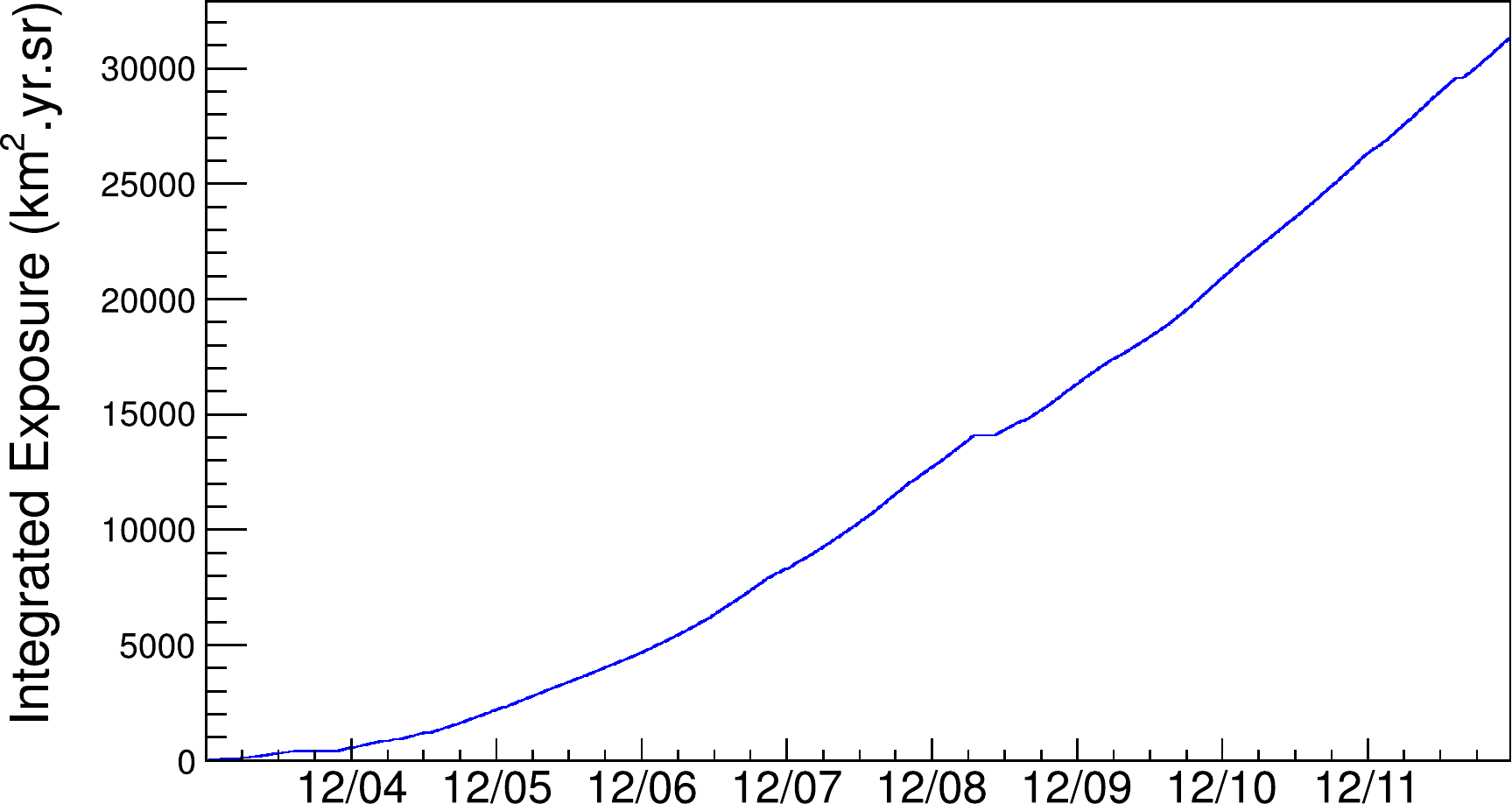}
\caption{Evolution of the exposure between 1 January 2004 and 31 December 2012.  The lack of exposure growth during an 
interval in 2009 corresponds to the data gap seen in Figure~\ref{fig:t5rate3}.}
\label{fig:Exposure}
\end{figure}

Finally the integrated exposure between 1 January 2004 and 31 December 2012 is  shown in  Figure~\ref{fig:Exposure}. Since completion of the array in 2008, the increase of the exposure has been  about 5500~km$^2$sr per year.

\subsection{Fluorescence detector performance}

The data taking of the FD can only take place under specific environmental conditions and is organized in night shifts.
As described in Section~\ref{sec:FDperformance}, the telescopes are not operated when the weather conditions are unfavorable (high wind speed, rain, snow, etc.) and when the observed sky brightness (caused mainly by scattered moonlight) is too high. As a consequence, the shifters have to continuously monitor (see Section~\ref{fd}) the atmospheric and environmental conditions and judge the operation mode on the basis of the available information.

The performance of the fluorescence and hybrid data taking is then influenced by
many effects. These can be external, e.g.,\ lightning or storms, or internal to
the data taking itself, e.g.,\ DAQ failures. For the determination of the
\emph{on-time} of the Observatory in the hybrid detection mode
it is, therefore, crucial to take into account all of these occurrences and derive
a solid description of the data taking time sequence.

Data losses and inefficiencies can occur on different levels, from the smallest
unit of the FD, i.e.,\ one single photomultiplier (pixel) readout channel, up to
the highest level, i.e.,\ the combined SD/FD data taking of the Observatory.

\begin{figure}[t]
\centering
\centerline{\includegraphics[width=0.48\textwidth]{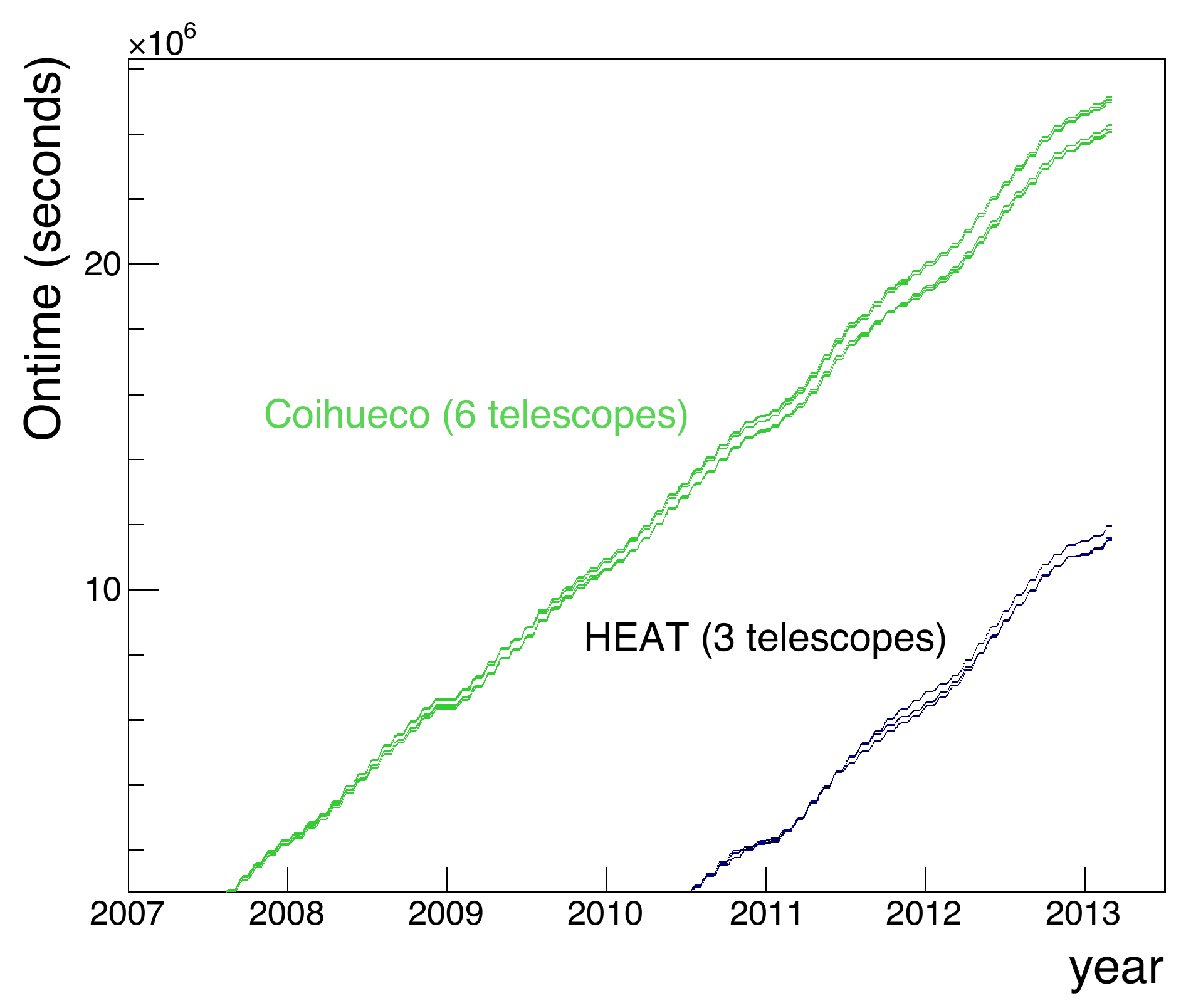}}
%\hskip -0.5cm
%\hspace{-0.5cm}
\centerline{\includegraphics[width=0.48\textwidth]{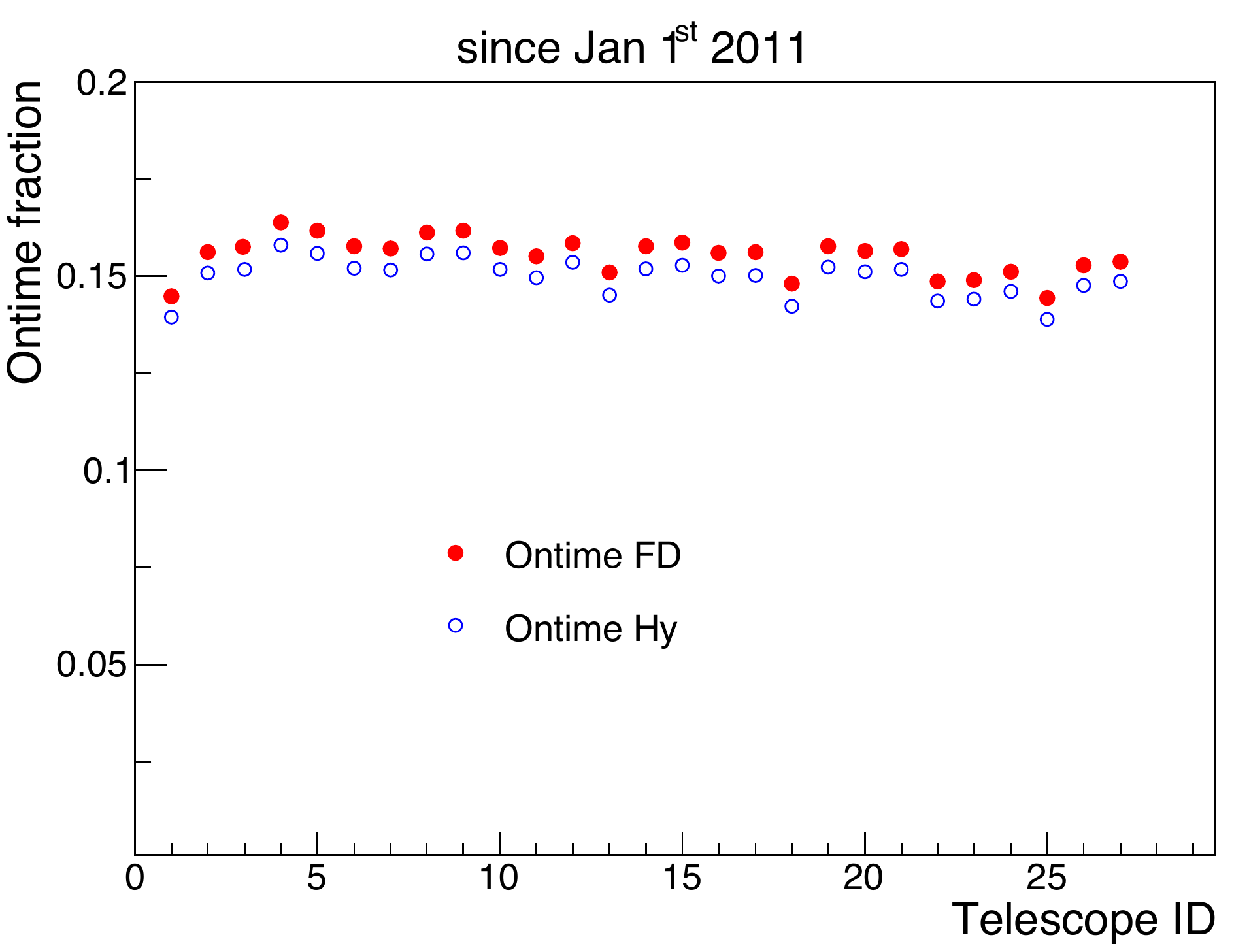}}
\caption{Top: accumulated on-time
since 1 Jul 2007 for the 6 telescopes at Coihueco and for the 3 HEAT telescopes. 
Bottom: FD and hybrid on-time of individual telescopes since 1 Jan 2011.
(1--6), (7--12), (13--18), (19--24), (25--27) for the sites of Los Leones, Los
Morados, Loma Amarilla, Coihueco and HEAT, respectively.}
\label{fig:on-time}
\end{figure}

The active time of the FD data acquisition is calculated using
a minimum bias data stream with a less restrictive trigger condition.  Since
July 2007, the relevant information concerning the status of the FD detector
has been read out from the Observatory monitoring system (see
Section~\ref{fd}).  An on-time dedicated database has been set up
by storing the average variances and the on-time fraction of individual
telescopes in time bins of 10 minutes. The information on the veto due to the
operation of the lidar or to an anomalous trigger rate on FD together with the
status of the CDAS  needed to form a hybrid event are also recorded.  The
method to calculate the on-time of the hybrid detector is described in detail
in Ref.~\cite{Abreu:2010aa}.

The accumulated on-time
is shown in Figure~\ref{fig:on-time}, top, for the six telescopes at Coihueco
and for the three HEAT telescopes.
The average FD on-time (full circles) of individual telescopes since 1 January
2011 is shown in Figure~\ref{fig:on-time}, bottom.  Requiring that the CDAS 
is active defines the hybrid on-time (empty circles).

\begin{figure}[t]
\centering
\centerline{\includegraphics[width=0.49\textwidth]{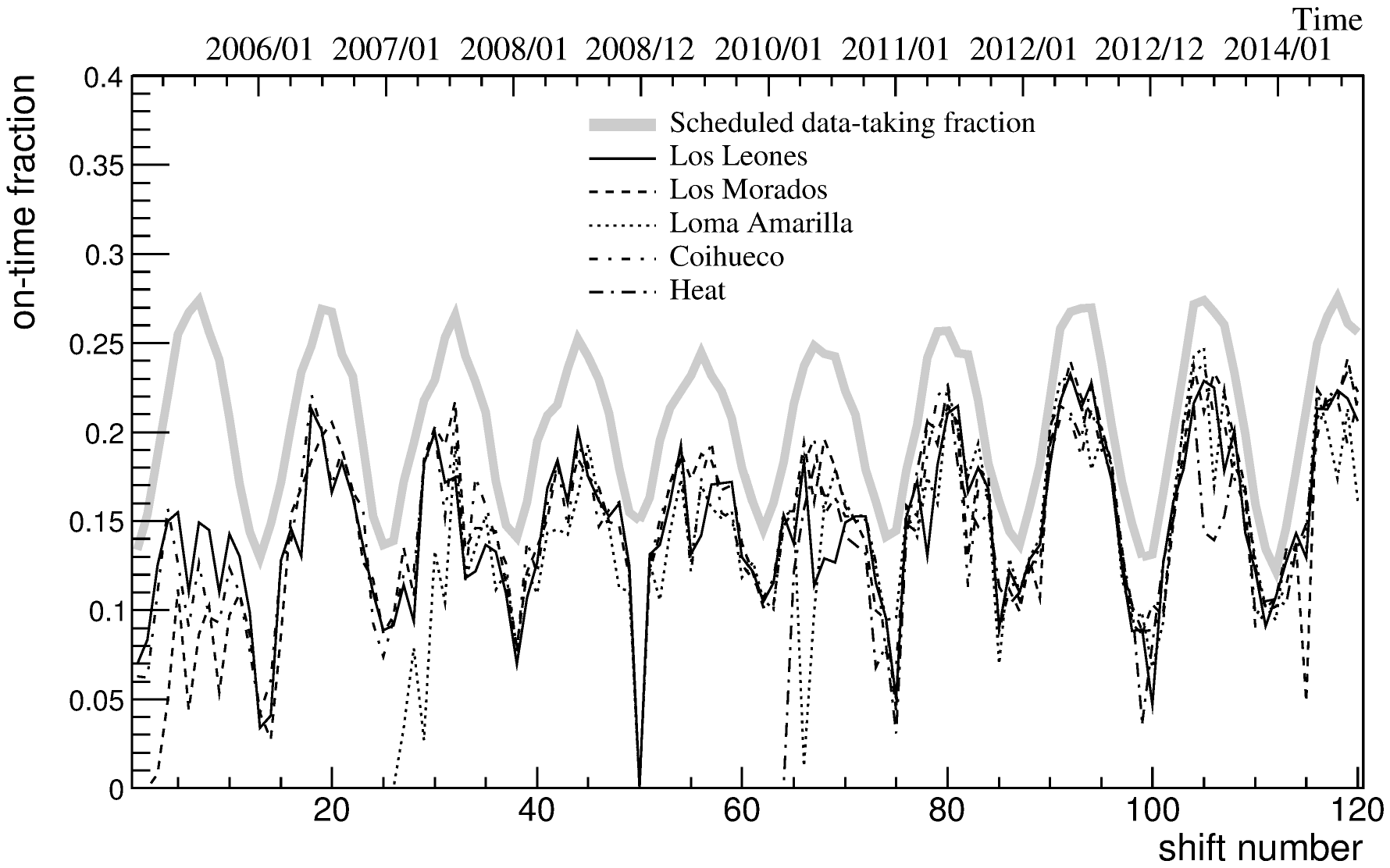}}
\centerline{\includegraphics[width=0.49\textwidth]{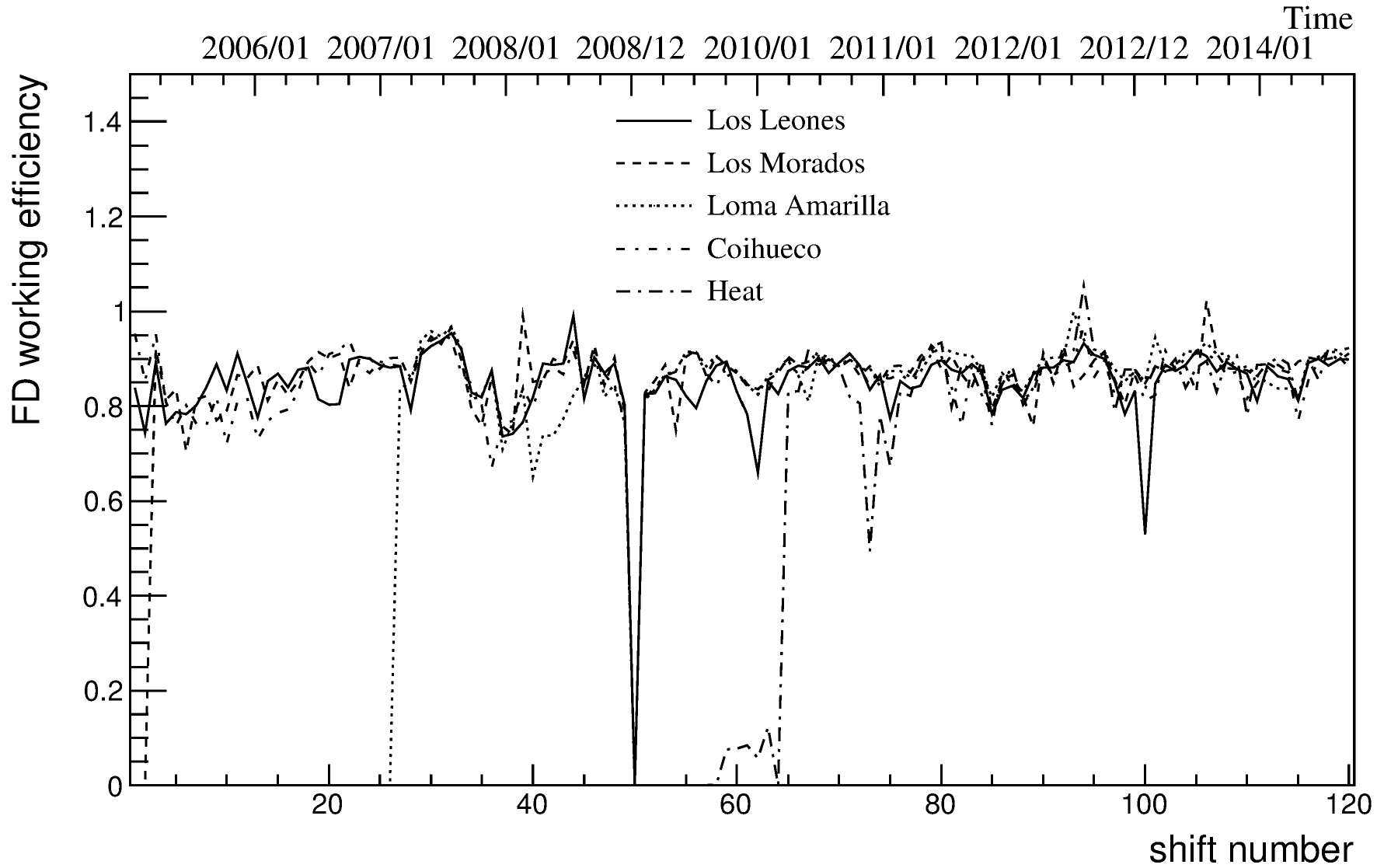}}
\caption{Top: time evolution of the average hybrid on-time fraction over nine
years of operation of the Observatory.
The thick gray line defines the scheduled
data taking time fraction defined as the time periods with
moon fraction lower than 70\,\% and with the moon being below the horizon for
more than 3 hours (see also Section~\ref{sec:FDperformance} for details).
Bottom: readiness of the FD
detector (see text for details).}
\label{fig:metrics}
\end{figure}

The time evolution of the full hybrid duty cycle over 9 years of operation is
shown in Figure~\ref{fig:metrics}, top, for all FD sites.  Time bins are taken as
the time intervals elapsed between two subsequent FD data taking shifts.
The performance of the hybrid detector is compared to the nominal DAQ time
 (see Section~\ref{sec:FDperformance}) in the top panel of
Figure~\ref{fig:metrics}.  In the bottom panel, the FD on-time is normalized to
the time with high voltage ON, leading to an average FD detector readiness of
about 85\,\% for all telescopes.  The remaining inefficiency can be ascribed to
different factors such as bad weather conditions (high wind load and/or rain)
or high variances due to bright stars/planets crossing the field of view of the
FD.

It should be noted that the FD site of Los Morados became operational in
May 2005, Loma Amarilla starting from March 2007 and HEAT since September 2009.
After the initial phase due to the start up of the running operations, the mean
on-time is about 15\,\% for all of the FD sites.  Additionally, a seasonal
modulation is visible, since higher on-time fractions are observed in the
austral winter during which the nights are longer.

\subsection{Time stability of the hybrid detector response}

%\red{This issue is currently being studied in detail using different approaches.
%The following text and figures will be updated accordingly. They are only a base for further development.}

\begin{figure}[t]
\centering
\includegraphics[width=0.49\textwidth]{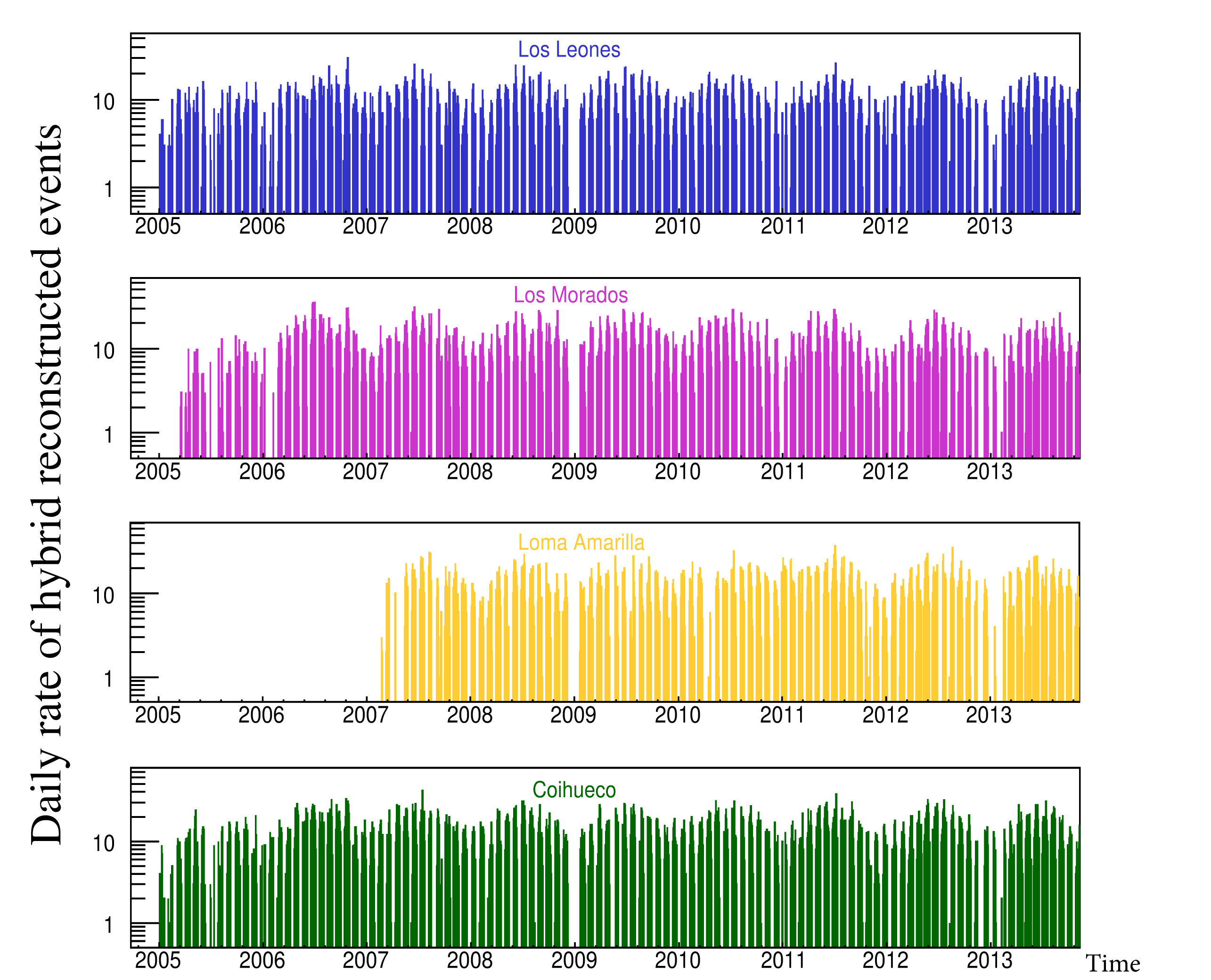}
\caption{Daily rate of hybrid reconstructed events as a function of year, starting in 2005, for (from top to bottom) Los Leones, Los Morados, Loma Amarilla and Coihueco, respectively.}
\label{fig:rate}
\end{figure}

The performance of the hybrid detector is demonstrated as a function of time using a sample
of events fulfilling basic reconstruction requirements, such as a reliable geometrical reconstruction
and accurate longitudinal profile and energy measurement.
The daily rate of  well-reconstructed hybrid events observed by individual FD sites is shown in Figure~\ref{fig:rate} as a function of time, starting in 2005.

%The collection of data was smooth and stable for all FD sites, except for the data shift coinciding with December 2008.
%%\red{(- add a comment -)}.

\begin{figure}[t]
\centering
\centerline{\includegraphics[width=0.46\textwidth]{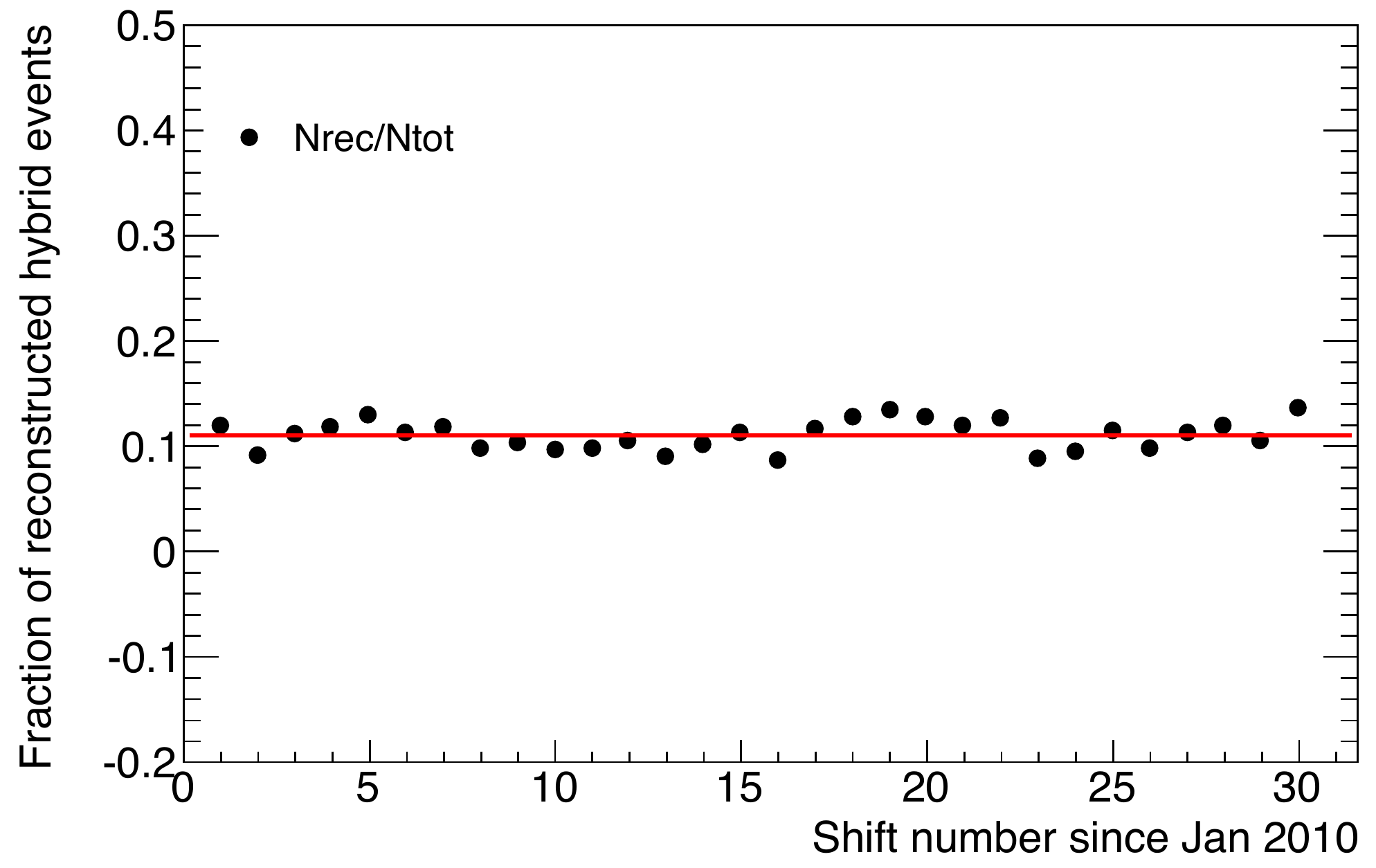}}
\centerline{\includegraphics[width=0.46\textwidth]{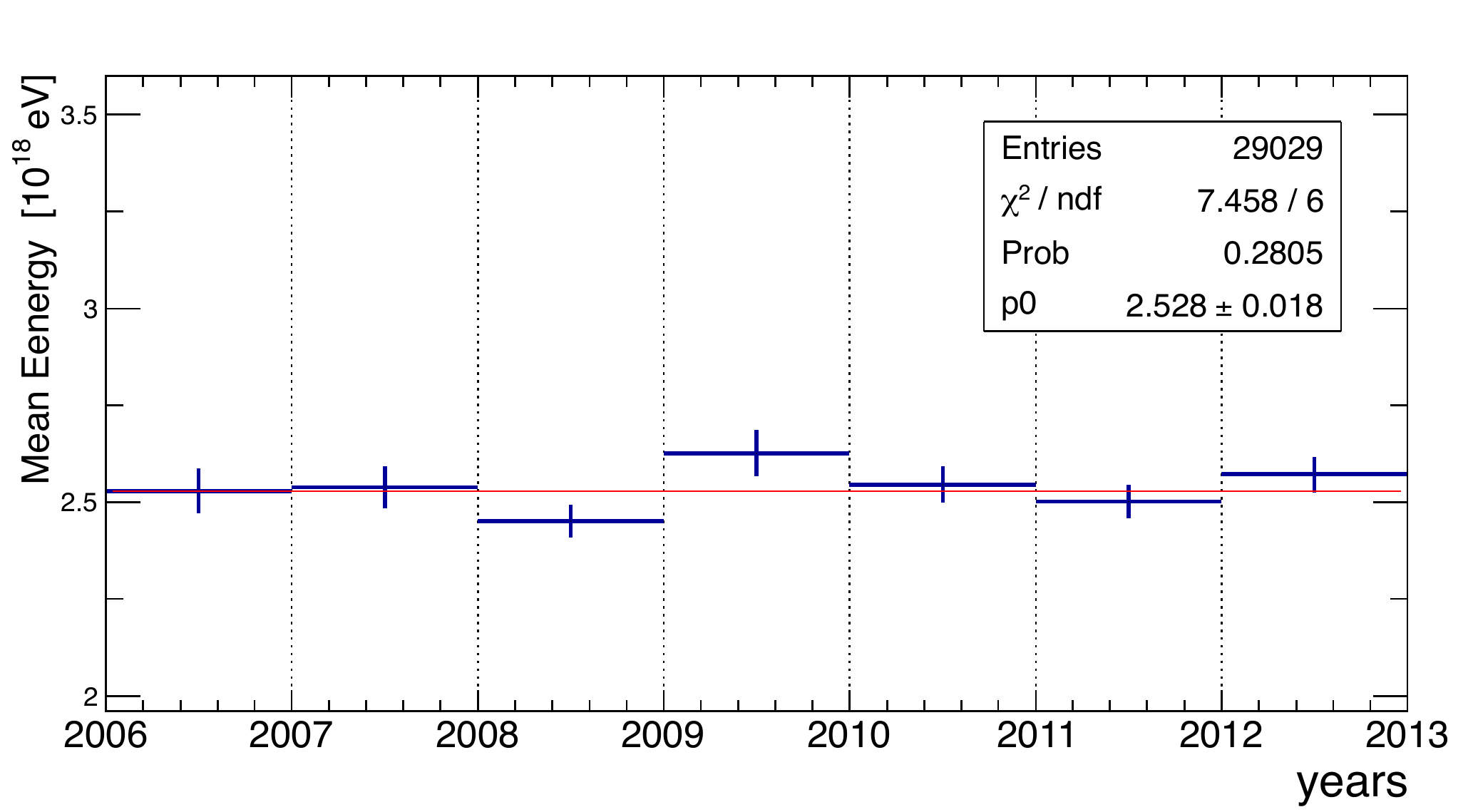}}
\caption{Top: fraction of all events that are well reconstructed hybrids since 2010.
Bottom: mean energy for reconstructed hybrid events.}
\label{fig:effontime}
\end{figure}

An important benchmark for the time stability of the hybrid detector response is the study of the effective on-time,
defined as the fraction of all events that are well reconstructed hybrids.
Its time evolution, shown in Figure~\ref{fig:effontime} (top), exhibits quite a stable behavior over time.
%it is a further confirmation of the quality of the hybrid detector performance over time.
Moreover the mean energy of the hybrid events above $10^{18}$\,eV, with distance
to the shower maximum  between 7 and 25\,km (corresponding to the 90\,\% of the
entire hybrid data sample), is shown as a function of time in Figure~\ref{fig:effontime} (bottom).
All these features demonstrate the quality of the collected hybrid data and directly assess their long term stability.

%The mean energy and $X_\text{max}$ for the selected events are shown in figure~\ref{fig:obstime}
%as a function of time.

%\begin{figure}[t]
%\centering
%\includegraphics[width=0.48\textwidth]{FD_Perf_ene-time}
%\\
%\includegraphics[width=0.48\textwidth]{FD_Perf_xmax-time}
%\caption{Mean energy (top) and $X_\text{max}$ (bottom) for reconstructed hybrid events.}
%\label{fig:obstime}
%\end{figure}

\section{Enhancements to the Auger Observatory}
\label{sec:Enhancements}

With the simultaneous and successful operation of the SD and FD, the Pierre Auger Collaboration
has demonstrated the power of hybrid measurements. Since completion of the baseline
construction, new enhancements were proposed to
further extend the science reach of the Observatory.

\subsection{High Elevation Auger Telescopes (HEAT)}
\label{sec:HEAT}

%\red{(Radomir S.)}

\begin{figure}[t]
\centering
\includegraphics[width=0.48\textwidth]{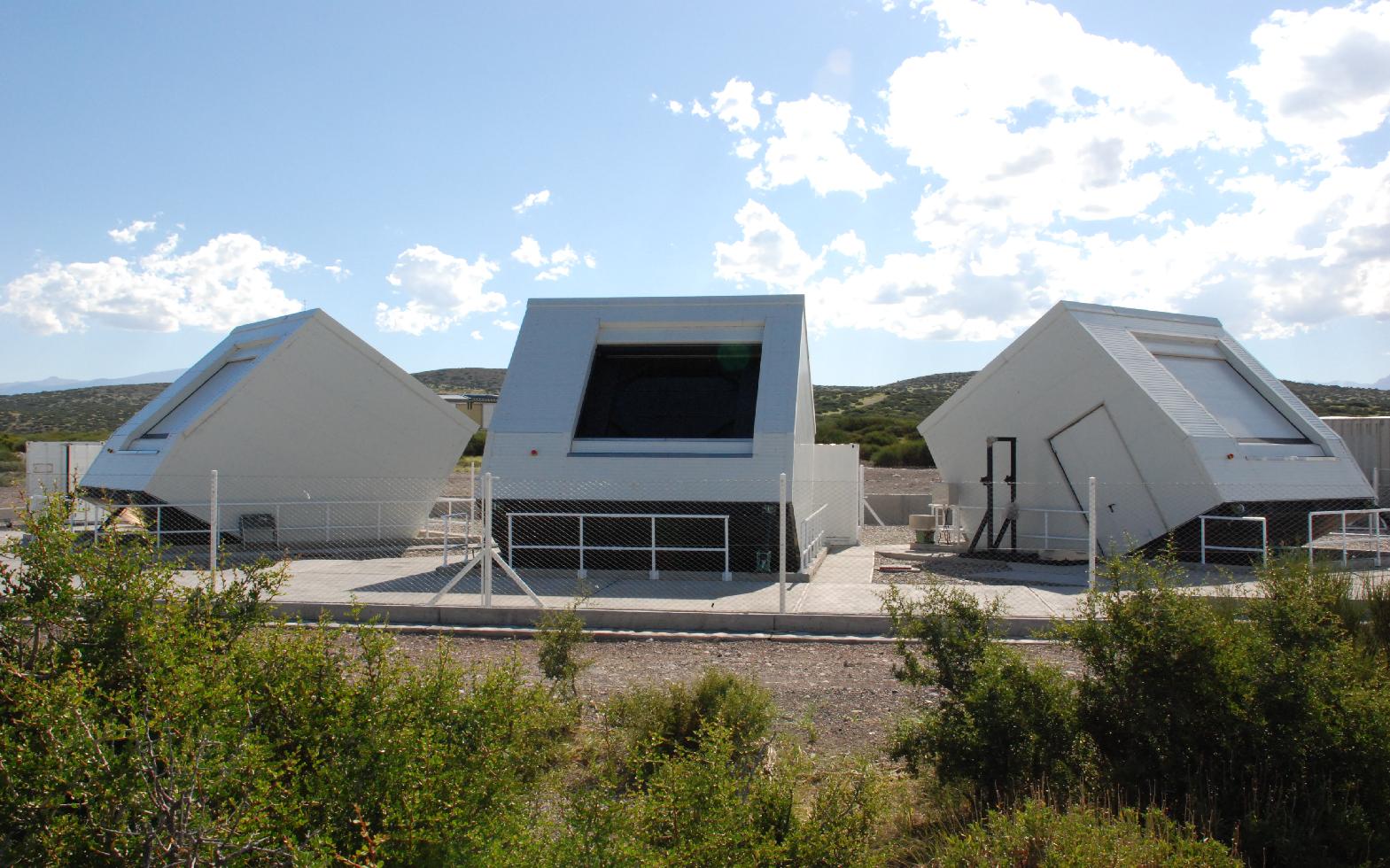}
\caption{Photo of the three HEAT telescopes in tilted mode. The container
for DAQ, slow control, and calibration hardware is behind the enclosure
of the second telescope.}
\label{fig:Heat-tilted}
\end{figure}

\begin{figure}[t]
  \subfigure[Horizontal (downward) mode for service and cross-calibration.]{
    \includegraphics[width=0.48\textwidth]{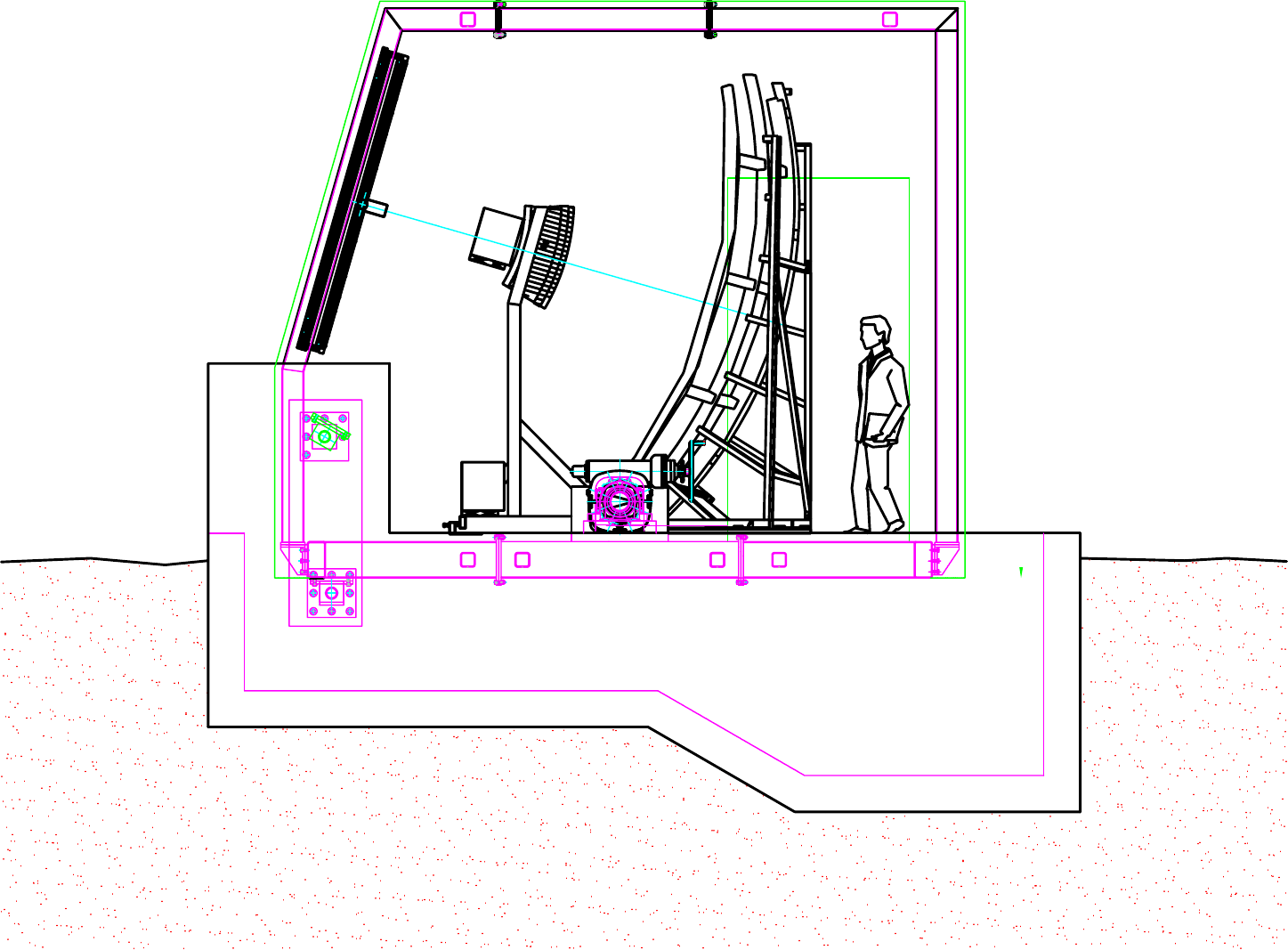}
  }
  \subfigure[Data taking (upward) mode in tilted orientation.]{
    \includegraphics[width=0.48\textwidth]{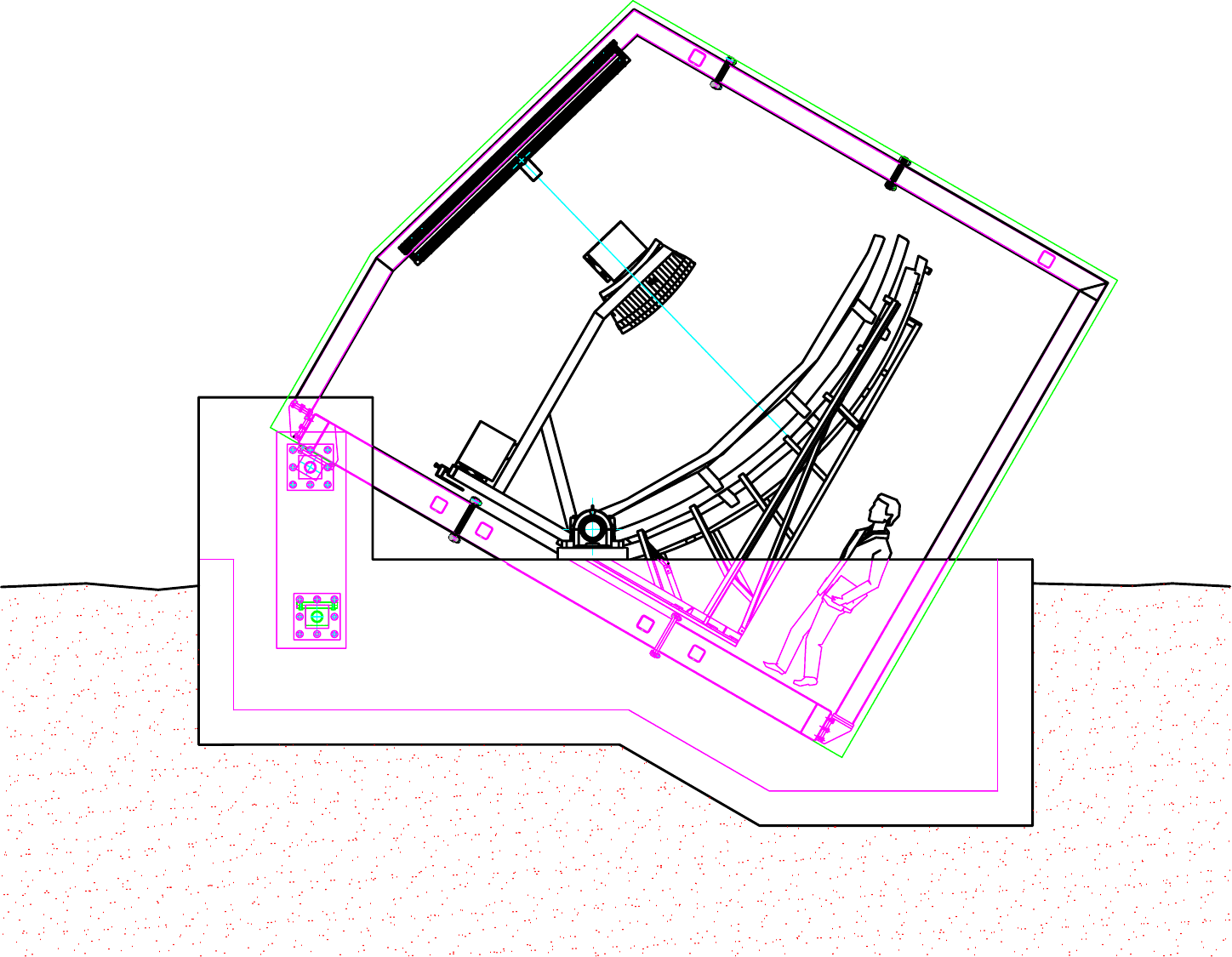}
  }
\caption{Schematic view of the cross-section of one of
the HEAT telescopes.}
\label{fig:Heat-modes}
\end{figure}

Three additional fluorescence telescopes with an elevated field of view were
built about 180\,m in front of the FD site at
Coihueco~\cite{Mathes-ICRC:2011}.
These telescopes are very similar to the original fluorescence
telescopes but can be tilted by $29^\circ$ upward with
an electrically driven hydraulic system. These three telescopes
work independently of other FD sites and form the ``fifth site"
of the Observatory. 
The HEAT telescopes were designed to cover the elevation range
from $30^\circ$ to $58^\circ$, which lies above the field of view of the other FD telescopes. 
The HEAT telescopes allow a determination
of the cosmic ray spectrum and $X_\text{max}$ distributions in the energy range
from below the second knee up to the ankle. The HEAT telescopes are depicted in Figures~\ref{fig:Heat-tilted} and \ref{fig:Heat-modes}.

The main objective of this extension was to lower the energy
threshold of hybrid data to enable an unbiased detection of nearby
low-energy showers. In combination with the SD,
information from an infilled array of water Cherenkov detectors on a 750\,m grid
(see Section~\ref{sec:amiga})
close to the HEAT site, the energy range of high quality hybrid air
shower measurements has been extended down to $10^{17}$\,eV.
In addition, close inclined shower events are detectable
without any mass dependent bias or cuts up to the highest energies.

The layout of HEAT consists of three telescope enclosures
and one container for DAQ, slow control, and calibration hardware (see Figure~\ref{fig:Heat-tilted}).
The telescope enclosure consists of three main building blocks.
First is the concrete foundation that supports the tilting
mechanism and provides stability for the whole building. Second is
the strong steel base plate, filled with concrete, on which all
the sensitive optical elements are mounted. Finally a relatively
lightweight steel container encloses the optical components and
electronics. The base plate is connected to the foundation, while the
steel enclosure is itself fixed to the plate. Similarly to the
baseline telescopes, a shutter system is mounted on the steel
enclosure, but of a different design.

The HEAT telescopes can be tilted using the hydraulic mechanism.
The telescopes are parked in the horizontal position between
the FD data taking periods to be accessible for maintenance, see Figure~\ref{fig:Heat-modes}.
The same position is used for the absolute calibration of the HEAT
telescopes and also for the cross-calibration with telescopes at Coihueco.
All three HEAT telescopes are usually moved in the upward position
before the first DAQ night and stay there during the whole data
taking period.

To ensure sufficient mechanical stability during high winds and snow
loads, all telescope components are connected to a heavy and stiff ground
plate with adjustment mounting bolts. The mechanical stability is
monitored by two types of sensors. The first type is the inclination
sensor that is used to measure the inclination variations at different
points inside the HEAT shelters. The second type is needed to measure the distance
variations between the optical components of the telescope. The distance
sensors are not only used to measure the long term variations, but also the higher
frequency variations that can take place in the telescope when subjected
to strong winds or other similar conditions. The maximal allowed deformations
and any movements after tilting lead to an angular
offset less than $0.1^\circ$.

The response of the HEAT cameras was tested at multiple elevations using
the relative calibration method (see Section~\ref{FdCalibration}).
The effect on the signal of tilting HEAT is at the percent level or
below, which matches the overall magnitude expected due to the direction of
the Earth's magnetic field as seen by the PMTs.
Also, the absolute calibration may be determined in the horizontal mode.

The Schmidt optics of the HEAT telescopes, camera body, PMTs, light
collectors, etc., are the same as in the other sites. All three spherical
mirrors are built up from hexagonal glass mirrors with vacuum-deposited
reflective coatings.

A feature that sets HEAT apart from the classic Auger telescopes is its
new electronics kit that can sample up to 40\,MHz instead of 
10\,MHz. % \cite{NewFDE}.
In practice, a sampling rate of 20\,MHz (corresponding
to a 50\,ns FADC bin size) was chosen. The higher rate improves the measurement
for close showers that have a correspondingly larger angular velocity --
precisely the showers we are interested in observing with HEAT. From this it 
follows that the first level trigger  interval was reduced to 50\,ns,
whereas the second level trigger  continues to operate every 100\,ns.
The length (in time) of the FADC traces remains the same, so the number
of bins doubles.

The trigger rate of the HEAT telescopes is high, particularly because of the
Cherenkov light from low energy showers. Therefore the T4 trigger 
has been implemented to reduce the readout of the SD array for these
low energy showers.

\subsection{Auger Muon and Infilled Ground Array (AMIGA)}
\label{sec:amiga}

%\red{(Federico Sanchez)}

\begin{figure}[t]
\centering
\includegraphics[width=0.48\textwidth]{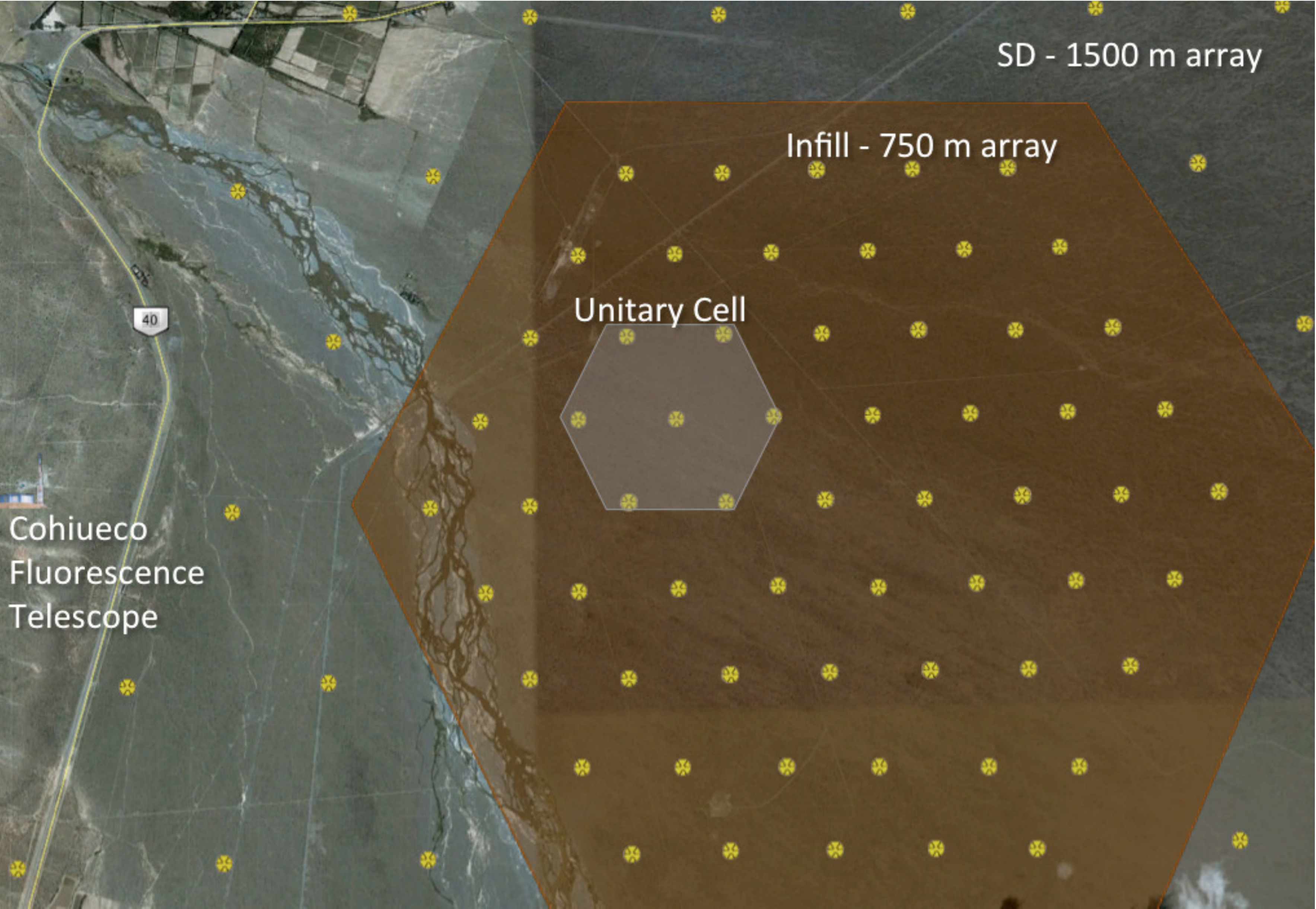}
\caption{%
          AMIGA layout: an infill of surface stations with an inter-detector spacing of 750\,m plus
          plastic scintillators of 30\,m$^2$ buried under $\approx 540\,\mathrm{g/cm^2}$ of vertical mass to measure the muon component of the showers.
          The small shaded area indicates the prototype hexagon ({\it Unitary Cell}) of the muon detector which has been fully operational since March 2015.
          Two positions of UC are equipped with extra {\it twin} scintillators to allow the detector accuracy to be assessed while in the hexagon's centre
          also $20\mathrm{\,m^2}$ were installed at $\approx 310\,\mathrm{g/cm^2}$ to experimentally analyze the shielding power of the local soil.
        }%
\label{fig:AMIGAScheme}
\end{figure}

A dedicated detector to directly measure
the muon content of air showers is being built.
The AMIGA enhancement
\cite{Suarez-ICRC:2013,Sanchez-ICRC:2011,Platino:2011zz}
is a joint system of water Cherenkov 
and buried scintillator detectors that spans an area of
23.5\,km$^2$ in a denser array with 750\,m spacing nested within the 1500\,m array (see Figure~\ref{fig:AMIGAScheme}).
The area is centered 6\,km away from the Coihueco fluorescence site.
The 750\,m array is fully efficient from $3\times10^{17}$\,eV onwards for air showers with zenith
angle $\leq 55^\circ$ \cite{Maris-ICRC:2011}.
Although the infilled area is
much smaller than the regular SD, the flux of cosmic rays increases steeply with decreasing energy
such that  this area is sufficient to observe a significant number of events and
to study the region between the second knee \cite{Nagano:1991jz} and the ankle of the cosmic ray spectrum.

The SD 750\,m array was completed in September 2011 while the
first prototype hexagon of buried scintillators, the {\it Unitary Cell}, has been fully operational since March 2015.
This engineering array consists of seven water Cherenkov detectors paired with $30\mathrm{\,m^2}$ scintillators segmented in two modules
of $10\mathrm{\,m^2}$ plus two of $5\mathrm{\,m^2}$ in each position. 
In addition, two positions of the hexagon were equipped with {\it twin} detectors (extra $30\mathrm{\,m^2}$ scintillators)
to allow the accuracy of the muon counting technique to be experimentally assessed \cite{Maldera-ICRC:2013} and one position
has $20\mathrm{\,m^2}$ of extra scintillators buried at a shallower depth to analyze the shielding features.
In total, $290\mathrm{\,m^2}$ fully equipped plastic scintillators are operative in the {\it Unitary Cell}. 
The proven tools and methods used for the analysis of the 1500\,m SD array data have been extended to reconstruct
the lower energy events.
The angular resolution for E$\geq4\times10^{17}$\,eV is better than $1^\circ$ and the energy reconstruction is
based on the lateral density of shower particles at the optimal distance of 450\,m from the core \cite{Ravignani-ICRC:2013}.
%To date, an exposure of
%$(57\pm3)\mathrm{\,km^2\,sr\,yr}$ for showers below a zenith angle of $55^\circ$ was achieved.
%As future development, there are also plans
%to include some detectors on a spacing of 433\,m to extend the observation range at least down to $10^{17}$\,eV.
%Such addition is expected to provide conclusive evidence about the presence of a second knee in the energy spectrum at $3\times 10^{17}$\,eV.

\begin{figure}[t]
\centering
\includegraphics[width=0.48\textwidth]{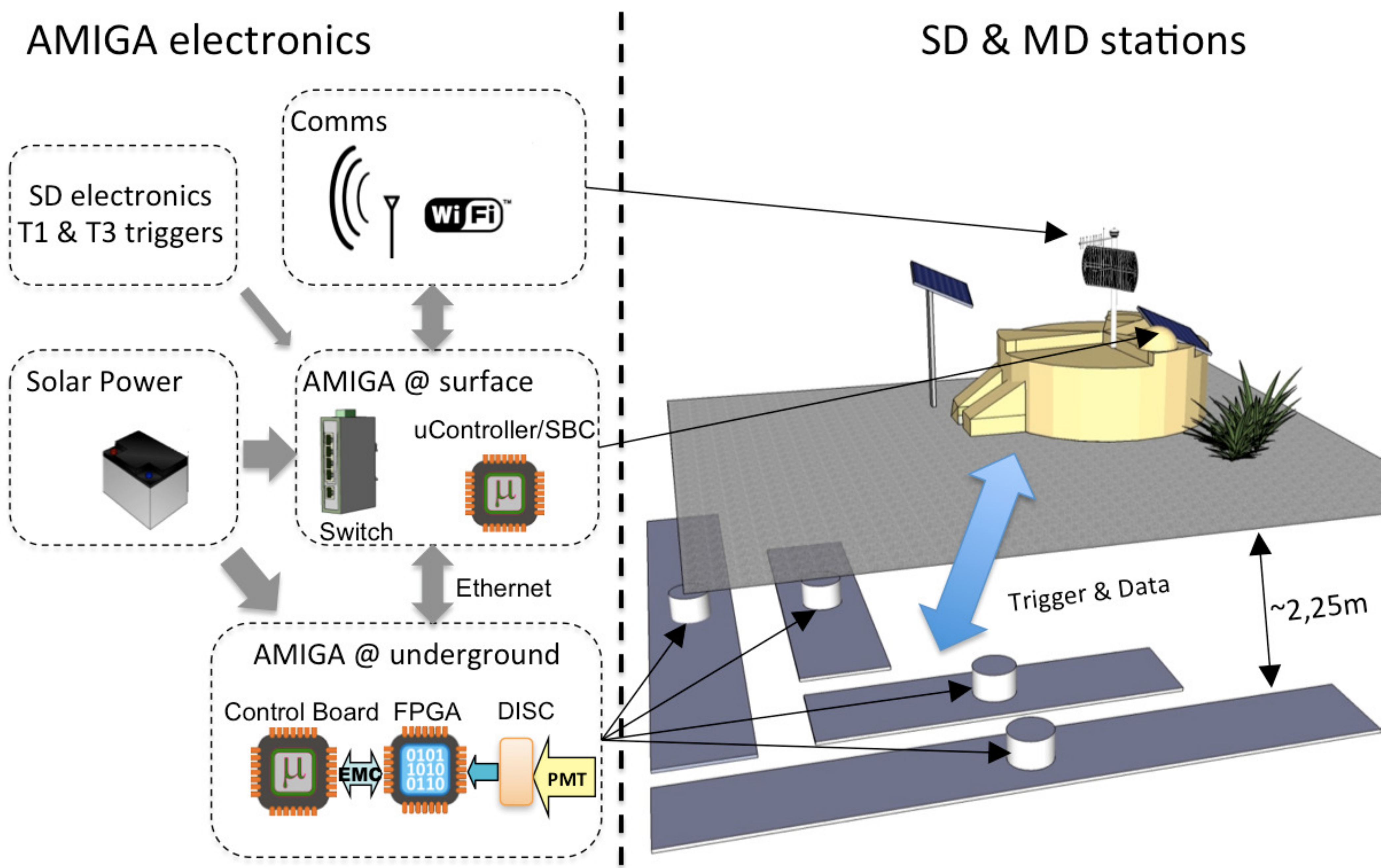}
%\subfigure[]{\label{fig:AMIGA_Layout_1}
%\includegraphics[width=0.48\textwidth]{AMIGA_Station}
%}
%\subfigure[]{\label{fig:AMIGA_Layout_2}
%\includegraphics[width=0.48\textwidth]{AMIGA_AccessPipe}
%}
\caption{
AMIGA station: SD+MD paired detectors. 
During the MD prototype phase the $30\mathrm{\,m^2}$ buried scintillators are segmented into 4 modules,
$2\times 10\mathrm{\,m^2} + 2\times 5\mathrm{\,m^2}$. %For the production phase, only 3 modules of $10\mathrm{\,m^2}$ will be used.
To avoid shadowing effects by the water Cherenkov detector, there are 5\,m of sideways clearance.
The buried front end electronics is serviceable by means of an access pipe which is filled with local soil bags.
Data are sent by a dedicated WiFi antenna.
}
\label{fig:AMIGA_Layout}
\end{figure}

\begin{figure}[t]
\centering
\includegraphics[width=0.48\textwidth]{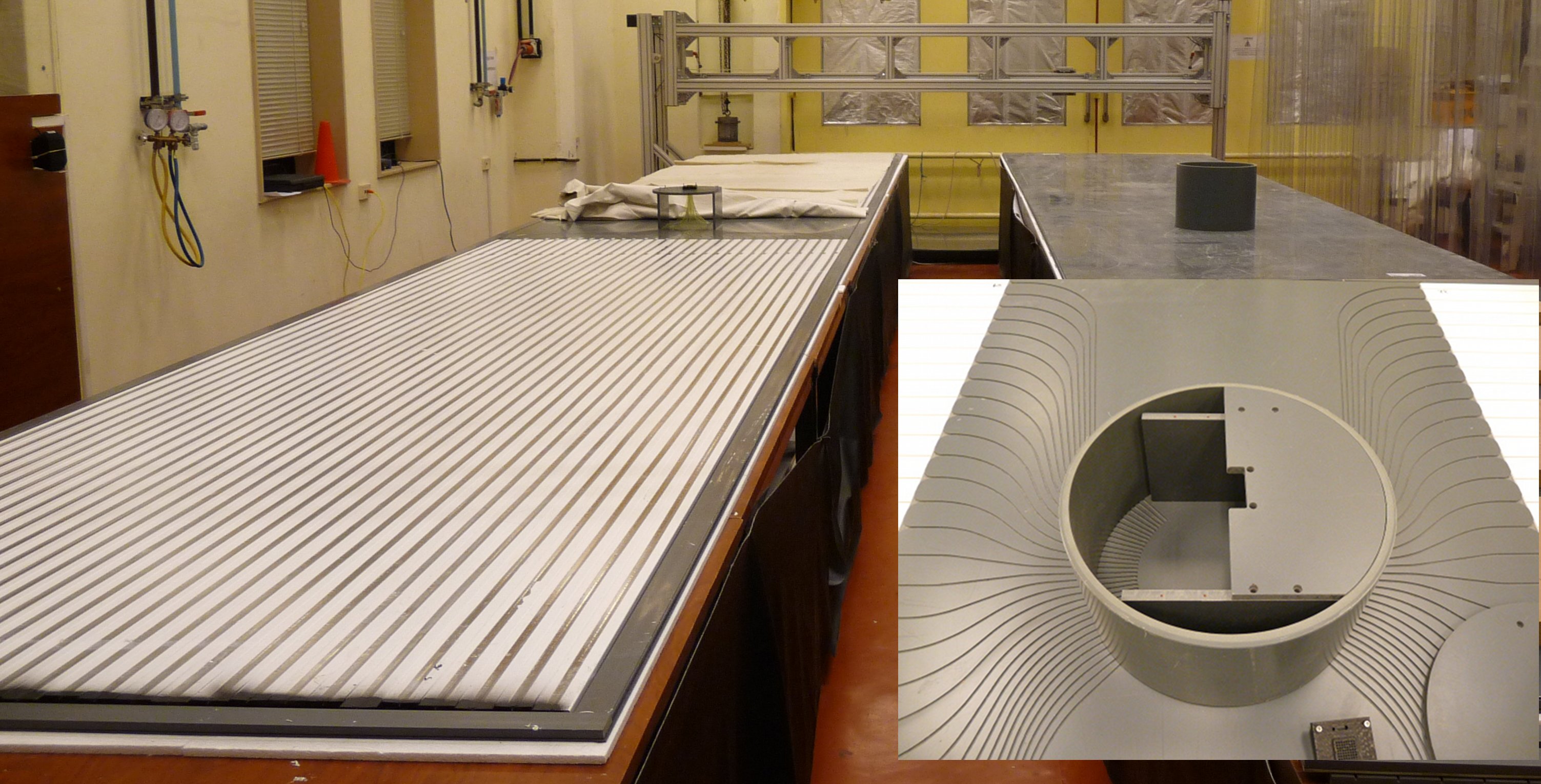}
%\subfigure[]{\label{fig:AMIGA_Assemblage}
%\includegraphics[width=0.48\textwidth]{AMIGA_ModuleAssemblage}
%}
%\subfigure[]{\label{fig:AMIGA_FEB}
%\includegraphics[width=0.37\textwidth]{AMIGA_FEB}
%}
\caption{
AMIGA scintillator detector, illustrating the 
assembly of a 10\,m$^2$ module. Strips are grouped in two sets of 32 strips on each side of the electronics dome located at
the center of the detector. The inset shows a detailed view of the manifold fiber routing and optical connector.
The multi-anode PMT and front end electronics board are hosted in the central dome. 
Once deployed, access to the buried setup
is provided by an inspection pipe.
}

\label{fig:AMIGA_10m2Module}
\end{figure}

The buried scintillators are the core of  
the detection system for the muonic component of air showers (the muon detector, MD). 
To effectively shield the electromagnetic
component, the MD is placed under $\approx 540\,\mathrm{g/cm^2}$ of vertical mass corresponding to
a depth of 2.3\,m in the local soil ($\approx 20$ radiation lengths) 
while the shallower extra scintillators are at $\approx 310\,\mathrm{g/cm^2}$ (1.3\,m).
These shieldings impose
a cutoff for vertical muons of around 1\,GeV and 0.6\,GeV respectively. 
The layout of SD+MD paired stations is shown in Figure~\ref{fig:AMIGA_Layout}.
The scintillator surface of each MD station is highly segmented. It consists of
%three 10\,m$^2$ 
modules made of 64 strips each.
Strips are 4.1\,cm wide$\times$1.0\,cm thick and 400\,cm or 200\,cm long for the $10\mathrm{\,m^2}$ and $5\mathrm{\,m^2}$
modules, respectively. They consist of extruded Dow Styron 663W polystyrene
doped by weight with 1\% PPO (2,5-diphenyloxazole) and 0.03\% POPOP (1,4-bis(5-phenyloxazole-2-yl)benzene).
They are completely wrapped with a thin white reflective layer of titanium dioxide (TiO$_2$)
except for a central groove into which a wavelength shifting (WLS) optical fiber is installed.
The light output uniformity is $\pm$5\%. Because the scintillators have an
attenuation length of ${\sim}(55\pm5)$\,mm, light is
transported to a photomultiplier tube using the WLS fiber.
The manifold of fibers of each module
ends in an optical connector matched to a 64 multi-anode PMT
from the Hamamatsu H8804 series. Scintillator strips are grouped in two sets of 32 strips on each side of the PMT
and front end electronics board (see Figure~\ref{fig:AMIGA_10m2Module}).

%Electronics description
The bandwidth of the front end electronics is set to 180\,MHz to determine the pulse width.
Signal sampling is performed by a Field Programmable Gate Array (FPGA) from the ALTERA Cyclone III series at 320\,MHz.
MD scintillator modules receive the trigger signal from their associated SD station.
The lowest level trigger  (T1) of the surface detectors is used.
Once a T1 condition is fulfilled on the surface, its MD companion freezes a 6.4\,$\upmu$s data sample into a local buffer
-- 1.6\,$\upmu$s
before and 4.8\,$\upmu$s after the T1.
Data are then moved to an external RAM capable of storing 1024 triggers \cite{Wainberg:2014}.

Incoming analog signals from each pixel of the PMT are digitized with a discriminator that provides the input to 
the FPGA.
Samples can be either a logical ``1'' or ``0'' depending on whether the incoming signal was above or below a given (programmable) 
discrimination
threshold.
This method of {\it one-bit} resolution is very robust for counting muons in a highly segmented
detector. %(a segmentation of 256 strips corresponds to 8-bits) 
This avoids missing muons due to simultaneous particle arrivals \cite{Supanitsky:2008dx}.
It relies neither on deconvolving the number of muons from an integrated signal,
nor on the PMT gain or its fluctuations, nor on the point of impact of the muon and the corresponding light attenuation along the fiber.
It also does not require a thick scintillator to control Poissonian fluctuations in the number of single photoelectron pulses per impinging muon
\cite{Wundheiler-ICRC:2011}.
The MD station power is supplied by an additional solar panel and battery box (see Figure~\ref{fig:AMIGA_Layout}).

%Comms
Whereas the data of the 750\,m array are transmitted over the same SD radio as for the regular array, a dedicated telecommunication system based on
WiFi 802.11g standard is used for MD data transmission during the prototype phase. The system is provided by an extra antenna located on the SD mast as
indicated in figure~\ref{fig:AMIGA_Layout}.
MD data are sent to the CDAS only at T3 level.
As WiFi based telecommunication has proven to satisfy the network throughput and data transfer requirements for SD T2s and SD+MD T3s data,
it is foreseen that this system will be used for the whole  AMIGA detector.

\section{Other capabilities of the Observatory}

\subsection{Space weather}
%\red{(Xavier)}

The rate of background low energy particles detected by the water Cherenkov detectors
of the Observatory is recorded every second and transmitted
to CDAS. This rate, of around 2000 
particles/s per detector, is used to monitor the stability of operations. 
The particles themselves are residual components of air
showers initiated by primary cosmic rays with a mean energy of about
90 GeV. We have observed that this rate correlates strongly with
neutron monitors measuring Forbush decreases \cite{Dasso:2012vk}.
Measuring  the flux of secondaries with great accuracy allows 
the Observatory to contribute to the ``Space Weather'' 
program \cite{Abreu:2011zza}. These data 
are available on the  Public Event 
Display of the Observatory web site \cite{publicEventDisplay}.

\subsection{ELVES}
\label{sec:Elves}
%\red{(Mussa)}

ELVES (an acronym for Emissions of Light and Very low frequency
perturbations due to Electromagnetic pulse Sources)
 are transient luminous events produced by heating,
ionization, and subsequent optical emission due to intense
electromagnetic pulses radiated by both positive
and negative lightning dischar\-ges.
These intense flashes of light appear in the night sky as rapidly
expanding quasi-circular fronts;  generated at 80-95\,km altitudes, they are
visible at distances of several hundred kilometers. The original pulse
lasts less than 20 $\mu\!$s, but the propagating light front is visible for
a few ms \cite{Inan:1991}.
The first clear observation of ELVES was made using a
high-speed photometer \cite{Boeck:1992}; more recently
such phenomena were studied using both  ground based photometers
\cite{Fukunishi:1996,Inan:1997,Newsome:2010} and satellite missions
\cite{Chen:2008}.
After the first serendipitous observation of an ELVES events during a FD shift
\cite{Mussa:2012dq} further studies, done on a pre-scaled sample of data
taken in the period 2008-2011, have shown that the FD is ideally suited
for detailed studies
of ELVES.  A new, modified third level trigger algorithm was
implemented in
March 2013 \cite{Tonachini-ICRC:2013} to increase the detection and
recording efficiency of these events. Since then, a large fraction of these events is
 regularly recorded by two or three FD eyes allowing a stereo
reconstruction of the light emission.
Also, since January 2014, when an ELVES trigger is received, the FD camera readout is extended to 300 $\mu\!$s to allow observation of the light emitted vertically above the causative lightning.

Auger detected 305 events in nine months of running in 2013, and 581 events (including 127 stereo and 20 triplets)
in all of 2014. The distance of the causative lightning ranged from 300--1000 km,
as determined by comparison with WWLLN data~\cite{Abarca:2010}. More than 40\% of these events
are correlated to lightning detected by WWLLN.

\begin{figure}[t]
\centering
\includegraphics[width=0.5\textwidth]{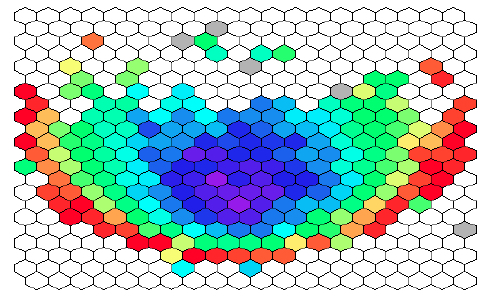}
\caption{A typical ELVES candidate event as seen in the FD.}
\label{f:elves}
\end{figure}

Figure~\ref{f:elves} shows the light propagation pattern in an ELVES candidate event:
the earliest triggered pixels are in blue, while the latest are in red.

\section{Outreach}

%\red{(Paul M)}

The scale and scope of the physics explored at the Pierre Auger Observatory offer significant 
opportunities for outreach both to the local community and beyond to the collaborating countries.  
Education, outreach and public relations have been an integral part of the Auger organization 
from the beginning when these activities were included as a level two management task group.  The goals 
of the Outreach and Education Task are to encourage and support a wide range of efforts that link Auger 
scientists and the science of particle astrophysics, particle physics and related technologies to the public 
and especially to schools.  Outreach focused on the communities surrounding the Observatory has fostered a 
remarkable amount of goodwill, which has contributed significantly to the success of the project.

The Auger Collaboration initiated outreach first locally as a way to become better integrated into the community 
during the construction phase of the Observatory. Later outreach activities spread to the participating institutions 
but on a worldwide scale, including the Internet.  

The heart of local outreach activities is the Auger Visitor Center (VC), located in the central office and data 
acquisition building on the Observatory campus.  A staff member dedicated to outreach gives 
presentations and tours to visitors that are mostly from the area but often from all over Argentina and even 
from 25 other countries worldwide.  Many of the visitors are in the area because of the proximity of the 
Las Le\~{n}as ski area and other area tourist attractions.  Over 90,000 people have attended the lectures 
in the Visitor Center since it opened in 2001.  The impact of these visits can be seen from the still increasing 
interest and the comments in the guest book.  The VC, which seats up to 50 people, is outfitted with multimedia 
equipment and contains a number of displays illustrating features of the Observatory.  These displays include 
a full size SD station,
a quarter sized model of an FD mirror, a spark chamber, a Geiger counter, a number of posters that explain the 
science and detectors of the Observatory and a library of books in several languages.

Many schools outside of Malarg\"{u}e, some over 100 km away, have trouble bringing their students to the 
Visitor Center.  A rural schools education program, funded by donations from the collaboration, sponsors a 
dedicated team of Observatory staff members and collaborators, who give presentations on Auger and science 
generally. The visiting team not only goes to these schools to give information on Auger but also helps to 
enhance the learning environment by providing learning materials or helping with infrastructure improvements 
such as connection to the Internet.  The effect of these visits on the students is enormous and creates an 
atmosphere of good will towards the Observatory.

Every two years a science fair is organized by the Observatory.  The fair targets both elementary schools and 
high schools, and is still growing.  The latest fair hosted 36 entries with schools from all over Mendoza 
Province participating.  The exhibits and presentations of the participants were judged by international members 
of the Auger Collaboration.  The interaction of the participants with the Auger scientists reinforces the connection 
between Auger and surrounding communities.

Each of the surface detector stations placed in the field was given a name.  During deployment, the stations were 
often named after the daughter or son of the nearby \emph{puestero} (farmer).  One of the results has been that 
there has been a negligible amount of vandalism even though the detectors are spread over 3000\,km$^2$, 
with a number only a few meters from a road.

Every November a collaboration meeting is scheduled to overlap with the annual Malarg\"{u}e Days 
celebration and a large group of collaborators march in the celebration parade behind a colorful Auger banner.  
These close contacts together with the local outreach activities have instilled a sense of ownership of the Observatory
by the community.

Our dedication to education is also clear from the construction of the James Cronin School, 
a secondary school in Malarg\"{u}e inaugurated in 2006.  The school, built from donated funds, was named for 
one of the founders of the Pierre Auger Observatory for his contribution to the local community.  Members of the collaboration 
have also been instrumental in helping to bring a new planetarium, one of the most modern in Argentina, to Malarg\"{u}e.

For outreach on a larger scale the Collaboration provides cosmic ray event data on the public Auger Web site 
(\url{www.auger.org}) 
%(http://www.auger.org) 
in nearly real time along with information, photos, videos and teaching materials.  
This material not only explains the mission of the Observatory, but also contains educational material on 
several aspects of the measurement of cosmic rays and the history of these measurements.  Furthermore, it provides 
manuals on how to work with and analyze the Auger public data set.  The online event display, coupled to the 
public data set, is a useful tool to provide insight to students in what is measured and how it is interpreted.  
These materials are mostly aimed at students at the high school level or above.  An online analysis interface 
called VISPA \cite{vispa} has been set up to allow students to work with and 
analyze these data. 

Outreach has been an important part of the activities of the Auger Project.  Our close relationship with the people of 
Malarg\"{u}e and the other local communities as a result of our outreach activities has not only made our work comfortable and rewarding but has, indeed, contributed to the success of the Observatory.  Among the collaborating institutions many innovative outreach ideas have sprung from our research.  Because we can easily show their continuous presence around us, cosmic rays are an effective way to excite young people about the wonders and science of the cosmos.

\section{Further developments}

Even as the Pierre Auger Observatory was under construction, new ideas for methods of air shower detection were being developed.  These ideas became more focused as the underlying phy\-sics of cosmic rays was revealed by analysis of the expanding data set.

Research and development is currently underway on two detection techniques which could complement the array of water Cherenkov and air fluorescence detectors.  The first is radio detection, 
a technique that was first pursued many years ago but is now benefitting 
from recent advances in electronics.  The second depends on the possibility that microwave radiation, arising from molecular bremsstrahlung of electromagnetic shower particles on air, is isotropic and strong enough to be effectively recorded.  If successful, one or both of these techniques could substantially extend the power of cosmic ray air shower detectors.

\subsection{Radio research program \label{sec:developments_radio}}

%\red{(J\"{o}rg H\"{o}randel)}

High-energy cosmic ray air showers generate radio emission via two processes:
one is a geomagnetic, current-induced emission mechanism~\cite{Kahn-Lerche}; 
the other is a charge-excess mechanism~\cite{Askaryan}.
The observation of air showers with radio detection techniques can be done at all times (day and night). 
Moreover, radio signals are sensitive to the development of the electromagnetic component of particle showers in the 
atmosphere of the Earth and, in particular, to the depth of the shower maximum or mass of the incoming 
cosmic ray~\cite{LOFAR:2013a}. 
Radio detection of air showers started in the 1960s, and the achievements in those days have been presented 
in reviews by Allan~\cite{Allan:1971} and Fegan~\cite{Fegan:2011fb}. 
More recent developments are based on initial studies performed by the LOPES~\cite{Falcke:2005tc} and the 
CODALEMA \cite{Ardouin:2006nb} collaborations and the LOFAR radio telescope~\cite{Schellart:2013bba}. 
In the last 10 years the radio detection technique in the MHz region has been revived and the present radio detector arrays 
for cosmic ray research are equipped with low noise and high rate analog-to-digital converters. 
Simultaneously, the number of stations within these arrays has grown from less than ten to more than one thousand. 
The questions to be addressed in the VHF band (30 -- 300\,MHz) are: can we use radio signals to determine the primary energy, the arrival direction, and the mass of cosmic rays with accuracies which are equal to or better than those obtained by other techniques? If yes, can we build a large surface detector array based on the radio detection technique
for an affordable price?

The Pierre Auger Collaboration has started a research program to answer both questions through a stepwise approach. 
Since 2009 the activities are coordinated within the Auger Engineering Radio Array (AERA), which is based on work within the Collaboration using various prototypes at the site of the Pierre Auger 
Observatory~\cite{Revenu-ICRC:2011, Kelley-ICRC:2011, Acounis:2012dg}.
As a first step, the emission mechanisms need to be understood. 
%Recently, AERA measured the relative contribution of the two main emission processes in air showers~\cite{Aab:2014esa},
%and found that the charge-excess process has a strength of $(14 \pm 2)$\% relative to the geomagnetic emission process.
The contributions of the main emission mechanisms have been measured recently~\cite{Aab:2014esa, Schellart:2014oaa}.
The fraction of the charge-excess emission relative to the geomagnetic emission varies between $\sim 5$\% for 
very inclined showers at 50~m from the shower axis to more than 20\% for vertical showers at 250~m from the shower axis.
As a second step the data obtained with radio detection stations deployed at the Observatory will be used to check 
their sensitivity with respect to the determination of the air shower parameters. 
We take advantage of the existing infrastructure of the Observatory: its surface detector, its fluorescence detector
and its low energy enhancements HEAT and AMIGA \cite{Mathes-ICRC:2011,Sanchez-ICRC:2011}. To help answer the questions, hardware and software are being developed to study the required specifications and performance of solitary radio stations as a blueprint for a large array. 
Within the same research program, a rigorous effort has been and is being made to understand the emission processes 
using our current knowledge of the development of air showers. 
Simultaneously, experiment and theory are being connected through software tools where end-to-end simulations and data analysis can be performed within the same software package \cite{Abreu:2011fb}. 
%All these activities fit into the Auger Engineering Radio Array (AERA) project, which started in 2009 and which is based on work within the Collaboration using various prototypes at the site of the Pierre Auger Observatory \cite{Revenu-ICRC:2011,Kelley-ICRC:2011}.

The scientific goals of the AERA project are as follows: 1) calibration of the radio emission from the air showers, including subdominant emission mechanisms; 2) demonstration at a significant scale of the physics capabilities of the radio technique, e.g., energy, angular, and  mass resolutions; and 3) measurement of the cosmic ray composition from 0.3 to 5\,EeV, with the goal of elucidating the transition from galactic to extragalactic cosmic rays. 

Each radio detection station is comprised of a dual polarization antenna, 
sensing the electric field in the north/south and east/west directions, associated analog and digital readout electronics, 
an autonomous power system and a communication link to a central data acquisition system. 
The antennas are sensitive between 30 and 80 MHz, chosen as the relatively radio quiet region between the 
shortwave and FM bands. AERA deployment began in 2010 with 24 stations. These stations are equipped with logarithmic 
periodic dipole antennas and are connected via a fiber optic link to a central data acquisition site. 
Stable physics data taking started in March 2011, and the first hybrid detection of cosmic ray events by radio, 
fluorescence, and surface particle detectors was recorded in April 2011. In May 2013 an additional 100 stations were installed. They are equipped with butterfly antennas. Detailed simulations and measurements demonstrated that butterfly antennas perform better for narrow pulse detection as compared to the logarithmic periodic dipole antennas~\cite{Abreu:2012pi}. 
The additional stations are connected through a wireless communication system to a central data acquisition system. 
%It is planned to install 25 further stations of the same design in Spring 2015 to complete the setup. 
%AERA will then be comprised of about 150 radio detection stations, covering an area exceeding 10 km$^2$.
AERA successfully deployed 25 additional radio stations in March 2015. AERA now includes 153 radio detection stations,
spread over an area of 17\,km$^2$. The detector spacings range from 150\,m to 750\,m, which enables the full exploitation
of radio detection of air showers as envisioned in the AERA technical design report.

\subsection{Microwave research program}

%\red{(Pedro Facal)}

Recent results of a test beam experiment at SLAC \cite{Gorham:2007af} showed that it could be possible to use microwave radiation to detect extensive air showers. This radiation, expected to be isotropic and broad in frequency, is interpreted \cite{Gorham:2007af} as molecular bremsstrahlung (MBR) produced by the scattering of low energy electrons in the weakly ionized plasma produced by the shower with the neutral molecules of the atmosphere. The Auger collaboration is pursuing an active R\&D program to determine if a detector sensitive to MBR would be a suitable alternative for the study of ultrahigh energy cosmic rays.

This R\&D program \cite{Allison-ICRC:2011, FacalSanLuis:2013qza} consists of three different setups installed at the Observatory. The AMBER and MIDAS experiments use radiotelescope style detectors intended for the observation of the shower longitudinal development  in the same manner as an FD. In the EASIER setup on the other hand, SD tanks are instrumented with smaller radio receivers that take advantage of the enhancement of the signal when the shower is observed close to its axis.

Installation of the microwave detectors was finalized in Sep\-tember 2012. A previous result by the MIDAS detector \cite{AlvarezMuniz:2012ew}, obtained in Chicago, places tight constraints on the amount of microwave signal emitted and its scaling with the energy of the shower \cite{AlvarezMuniz:2012dx}. The ongoing work to identify showers detected at the same time in the SD and in one of the microwave detectors already yielded the fist unambiguous detection of a cosmic ray shower in the EASIER setup in June 2011 \cite{FacalSanLuis:2013qza}.

\section{Final remarks}

The Pierre Auger Observatory is the world's largest cosmic ray observatory.  
%The Observatory is highly productive, with very high efficiency data accumulation, sophisticated atmospheric monitoring for accurate interpretation of showers and advances in the ability to extract longitudinal development information from surface detector information alone.  
The Observatory is highly productive, with very high efficiency data accumulation and sophisticated atmospheric monitoring for accurate interpretation of showers.  There have been recent advances enabling the extraction of longitudinal development information from surface detector information alone.  
New analysis methods using both the fluorescence detectors and the surface array have improved the determination of primary composition.  At the same time the recently completed HEAT and AMIGA infilled enhancements now extend the reach of the detector to cover the critical galactic/extragalactic transition region.   The Observatory is also serving as a test bed for advanced detector technologies for the next generation of cosmic ray experiments.

%There is increasing evidence in the Auger data that the flux suppression beyond $10^{19.5}$ eV may not, in fact, be due to the GZK mechanism but is rather the result of changing (i.e., heavier) composition. 
Current source scenarios assume that
particle acceleration takes place at sites distributed similarly to the matter distribution
in the universe, with energy loss processes leading to the observed flux suppression (GZK
effect) and arrival direction anisotropy. However, Auger data on shower development fluctuations,
as well as other composition sensitive observables, require consideration
of a rather different interpretation: that the observed flux suppression
is indicating the upper limit of the power of the accelerator. It may be that the upper
end of the cosmic ray energy spectrum is dominated by particles from a single source or
source population, possibly within the GZK horizon, for which the upper limit of particle
acceleration almost coincides with the energy of the GZK suppression.  
To answer this question, the Pierre Auger Collaboration is planning an upgrade to the Observatory to enable a determination of the primary composition on an event by event basis at the highest energies.  
%Several upgrade strategies are being explored for revisions to the surface detector stations that will identify the shower muon content, providing an effective handle on the composition.  Once approved, the upgrade is expected to be completed by 2018 with Observatory operation through 2025.
The required electron-muon discrimination in the surface detector stations will be achieved by installing a 4\,m$^2$ scintillation detector on top of each water Cherenkov detector. Once approved, the upgrade is expected to be completed by 2018 with Observatory operation through 2025.

%The Auger results are having a major impact on the field. The Collaboration has published 44 full author list journal articles. Another 5 have been submitted for publication and 20 more are in preparation. Auger is also training a cadre of future scientists, with 194 students granted PhDs based on their work in Auger. Another 150 PhD students are in the pipeline.

\section*{Acknowledgments}

The successful installation, commissioning, and operation of the Pierre Auger Observatory would not have been possible without the strong commitment and effort from the technical and administrative staff in Malarg\"{u}e. 

\begin{sloppypar}
We are very grateful to the following agencies and organizations for financial support: 
Comisi\'{o}n Nacional de Energ\'{\i}a At\'{o}mica, Fundaci\'{o}n Antorchas, Gobierno De La Provincia de Mendoza, Municipalidad de Malarg\"{u}e, NDM Holdings and Valle Las Le\~{n}as, in gratitude for their continuing cooperation over land access, Argentina; the Australian Research Council; Conselho Nacional de Desenvolvimento Cient\'{\i}fico e Tecnol\'{o}gico (CNPq), Financiadora de Estudos e Projetos (FINEP), Funda\c{c}\~{a}o de Amparo \`{a} Pesquisa do Estado de Rio de Janeiro (FAPERJ), S\~{a}o Paulo Research Foundation (FAPESP) Grants \# 2010/07359-6, \# 1999/05404-3, Minist\'{e}rio de Ci\^{e}ncia e Tecnologia (MCT), Brazil; MSMT-CR LG13007, 7AMB14AR005, CZ.1.05/2.1.00/03.0058 and the Czech Science Foundation grant 14-17501S, Czech Republic;  Centre de Calcul IN2P3/CNRS, Centre National de la Recherche Scientifique (CNRS), Conseil R\'{e}gional Ile-de-France, D\'{e}partement Physique Nucl\'{e}aire et Corpusculaire (PNC-IN2P3/CNRS), D\'{e}partement Sciences de l'Univers (SDU-INSU/CNRS), Institut Lagrange de Paris, ILP LABEX ANR-10-LABX-63, within the Investissements d'Avenir Programme  ANR-11-IDEX-0004-02, France; Bundesministerium f\"{u}r Bildung und Forschung (BMBF), Deutsche Forschungsgemeinschaft (DFG), Finanzministerium Baden-W\"{u}rttemberg, Helmholtz Alliance for Astroparticle Physics (HAP), Helmholtz-Gemeinschaft Deutscher Forschungszentren (HGF),  Ministerium f\"{u}r Wissenschaft und Forschung, Nordrhein Westfalen, Ministerium f\"{u}r Wissenschaft, Forschung und Kunst, Baden-W\"{u}rttemberg, Germany; Istituto Nazionale di Fisica Nucleare (INFN), Ministero dell'Istruzione, dell'Universit\`{a} e della Ricerca (MIUR), Gran Sasso Center for Astroparticle Physics (CFA), CETEMPS Center of Excellence, Italy; Consejo Nacional de Ciencia y Tecnolog\'{\i}a (CONACYT), Mexico; Ministerie van Onderwijs, Cultuur en Wetenschap, Nederlandse Organisatie voor Wetenschappelijk Onderzoek (NWO), Stichting voor Fundamenteel Onderzoek der Materie (FOM), Netherlands; National Centre for Research and Development, Grant Nos.ERA-NET-ASPERA/01/11 and ERA-NET-ASPERA/02/11, National Science Centre, Grant Nos. 2013/08/M/ST9/00322, 2013/08/M/ST9/00728 and HARMONIA 5 - 2013/10/M/ST9/00062, Poland; Portuguese national funds and FEDER funds within COMPETE - Programa Operacional Factores de Competitividade through Funda\c{c}\~{a}o para a Ci\^{e}ncia e a Tecnologia, Portugal; Romanian Authority for Scientific Research ANCS, CNDI-UEFISCDI partnership projects nr.20/2012 and nr.194/2012, project nr.1/ASPERA2/2012 ERA-NET, PN-II-RU-PD-2011-3-0145-17, and PN-II-RU-PD-2011-3-0062, the Minister of National  Education, Programme for research - Space Technology and Advanced Research - STAR, project number 83/2013, Romania; Slovenian Research Agency, Slovenia; Comunidad de Madrid, FEDER funds, Ministerio de Educaci\'{o}n y Ciencia, Xunta de Galicia, European Community 7th Framework Program, Grant No. FP7-PEOPLE-2012-IEF-328826, Spain; Science and Technology Facilities Council, United Kingdom; Department of Energy, Contract No. DE-AC02-07CH11359, DE-FR02-04ER41300, DE-FG02-99ER41107 and DE-SC0011689, National Science Foundation, Grant No. 0450696, The Grainger Foundation, USA; NAFOSTED, Vietnam; Marie Curie-IRSES/EPLANET, European Particle Physics Latin American Network, European Union 7th Framework Program, Grant No. PIRSES-2009-GA-246806; and UNESCO.
\end{sloppypar}

\bibliographystyle{elsarticle-num}
\bibliography{AugerSouth_bibliography}

\begin{thebibliography}{100}
\expandafter\ifx\csname url\endcsname\relax
  \def\url#1{\texttt{#1}}\fi
\expandafter\ifx\csname urlprefix\endcsname\relax\def\urlprefix{URL }\fi
\expandafter\ifx\csname href\endcsname\relax
  \def\href#1#2{#2} \def\path#1{#1}\fi

\bibitem{Rossi:1934}
B.~Rossi, {Misure sulla distribuzione angolare di intensita della radiazione
  penetrante all'Asmara}, {Supplemento a la Ricerca Scientifica} 1 (1934) 579.

\bibitem{Schmeiser:1938}
K.~Schmeiser, W.~Bothe, {Die harten Ultrastrahlschauer}, Ann.\ Phys. 424 (1938)
  161--177.
\newblock \href {http://dx.doi.org/10.1002/andp.19384240119}
  {\path{doi:10.1002/andp.19384240119}}.

\bibitem{Kolhorster:1938}
W.~Kolh{\"o}rster, I.~Matthes, E.~Weber, {Gekoppelte H{\"o}henstrahlen},
  Naturwissenschaften 26 (1938) 576.

\bibitem{Auger:1939a}
P.~Auger, R.~Maze, A.~F. Robley, {Extension et pouvoir p\'en\'etrant des
  grandes gerbes de rayons cosmiques}, Comptes Rendus 208 (1939) 1641.

\bibitem{Auger:1939sh}
P.~Auger, P.~Ehrenfest, R.~Maze, J.~Daudin, A.~F. Robley, {Extensive cosmic ray
  showers}, Rev.\ Mod.\ Phys. 11 (1939) 288--291.
\newblock \href {http://dx.doi.org/10.1103/RevModPhys.11.288}
  {\path{doi:10.1103/RevModPhys.11.288}}.

\bibitem{Linsley:1963km}
J.~Linsley, {Evidence for a primary cosmic-ray particle with energy
  $10^{20}$\,eV}, Phys.\ Rev.\ Lett. 10 (1963) 146--148.
\newblock \href {http://dx.doi.org/10.1103/PhysRevLett.10.146}
  {\path{doi:10.1103/PhysRevLett.10.146}}.

\bibitem{Afanasiev:1993}
B.~Afanasiev, et~al., {Recent Results From Yakutsk Experiment}, in: M.~Nagano
  (Ed.), {Proceedings of the Tokyo Workshop on Techniques for the Study of the
  Extremely High Energy Cosmic Rays}, {Institute for Cosmic Ray Research,
  University of Tokyo, Tokyo, Japan}, 1993, p.~35.

\bibitem{Lawrence:1991cc}
M.~Lawrence, R.~Reid, A.~Watson, {The cosmic ray energy spectrum above
  4$\times$10$^{17}$~eV as measured by the Haverah Park array}, J.\ Phys.\ G 17
  (1991) 733--757.
\newblock \href {http://dx.doi.org/10.1088/0954-3899/17/5/019}
  {\path{doi:10.1088/0954-3899/17/5/019}}.

\bibitem{Takeda:2002at}
M.~Takeda, N.~Sakaki, K.~Honda, M.~Chikawa, M.~Fukushima, et~al., {Energy
  determination in the Akeno Giant Air Shower Array experiment}, Astropart.\
  Phys. 19 (2003) 447--462.
\newblock \href {http://arxiv.org/abs/astro-ph/0209422}
  {\path{arXiv:astro-ph/0209422}}, \href
  {http://dx.doi.org/10.1016/S0927-6505(02)00243-8}
  {\path{doi:10.1016/S0927-6505(02)00243-8}}.

\bibitem{Bird:1994uy}
D.~Bird, S.~Corbato, H.~Dai, J.~Elbert, K.~Green, et~al., {Detection of a
  cosmic ray with measured energy well beyond the expected spectral cutoff due
  to cosmic microwave radiation}, Astrophys.\ J. 441 (1995) 144--150.
\newblock \href {http://dx.doi.org/10.1086/175344} {\path{doi:10.1086/175344}}.

\bibitem{Schulz-ICRC:2013}
A.~Schulz, {Measurement of the Energy Spectrum of Cosmic Rays above 3 $\times
  10^{17}$ eV with the Pierre Auger Observatory},
  \cite{ThePierreAuger:2013eja}, p.~27.
\newblock \href {http://arxiv.org/abs/1307.5059} {\path{arXiv:1307.5059}}.

\bibitem{Abraham:2008ru}
J.~Abraham, et~al., {Observation of the suppression of the flux of cosmic rays
  above $4\times 10^{19}$eV}, Phys.\ Rev.\ Lett. 101 (2008) 061101.
\newblock \href {http://arxiv.org/abs/0806.4302} {\path{arXiv:0806.4302}},
  \href {http://dx.doi.org/10.1103/PhysRevLett.101.061101}
  {\path{doi:10.1103/PhysRevLett.101.061101}}.

\bibitem{Abreu:2011ve}
P.~Abreu, et~al., {Search for First Harmonic Modulation in the Right Ascension
  Distribution of Cosmic Rays Detected at the Pierre Auger Observatory},
  Astropart.\ Phys. 34 (2011) 627--639.
\newblock \href {http://arxiv.org/abs/1103.2721} {\path{arXiv:1103.2721}},
  \href {http://dx.doi.org/10.1016/j.astropartphys.2010.12.007}
  {\path{doi:10.1016/j.astropartphys.2010.12.007}}.

\bibitem{Sidelnik-ICRC:2013}
I.~Sidelnik, {Measurement of the first harmonic modulation in the right
  ascension distribution of cosmic rays detected at the Pierre Auger
  Observatory: towards the detection of dipolar anisotropies over a wide energy
  range},  \cite{ThePierreAuger:2013eja}, p.~56.
\newblock \href {http://arxiv.org/abs/1307.5059} {\path{arXiv:1307.5059}}.

\bibitem{Almeida-ICRC:2013}
R.~de~Almeida, {Constraints on the origin of cosmic rays from large scale
  anisotropy searches in data of the Pierre Auger Observatory},
  \cite{ThePierreAuger:2013eja}, p.~60.
\newblock \href {http://arxiv.org/abs/1307.5059} {\path{arXiv:1307.5059}}.

\bibitem{Abreu:2012ybu}
P.~Abreu, et~al., {Constraints on the origin of cosmic rays above $10^{18}$ eV
  from large scale anisotropy searches in data of the Pierre Auger
  Observatory}, Astrophys.\ J. 762 (2012) L13.
\newblock \href {http://arxiv.org/abs/1212.3083} {\path{arXiv:1212.3083}},
  \href {http://dx.doi.org/10.1088/2041-8205/762/1/L13}
  {\path{doi:10.1088/2041-8205/762/1/L13}}.

\bibitem{Abraham:2007si}
J.~Abraham, et~al., {Correlation of the highest-energy cosmic rays with the
  positions of nearby active galactic nuclei}, Astropart.\ Phys. 29 (2008)
  188--204.
\newblock \href {http://arxiv.org/abs/0712.2843} {\path{arXiv:0712.2843}},
  \href {http://dx.doi.org/10.1016/j.astropartphys.2008.06.004,
  10.1016/j.astropartphys.2008.01.002}
  {\path{doi:10.1016/j.astropartphys.2008.06.004,
  10.1016/j.astropartphys.2008.01.002}}.

\bibitem{Abraham:2007bb}
J.~Abraham, et~al., {Correlation of the highest energy cosmic rays with nearby
  extragalactic objects}, Science 318 (2007) 938--943.
\newblock \href {http://arxiv.org/abs/0711.2256} {\path{arXiv:0711.2256}},
  \href {http://dx.doi.org/10.1126/science.1151124}
  {\path{doi:10.1126/science.1151124}}.

\bibitem{Collaboration:2012wt}
P.~Abreu, et~al., {Measurement of the proton-air cross-section at $\sqrt{s}=57$
  TeV with the Pierre Auger Observatory}, Phys.\ Rev.\ Lett. 109 (2012) 062002.
\newblock \href {http://arxiv.org/abs/1208.1520} {\path{arXiv:1208.1520}},
  \href {http://dx.doi.org/10.1103/PhysRevLett.109.062002}
  {\path{doi:10.1103/PhysRevLett.109.062002}}.

\bibitem{Csorgo:2012dm}
T.~Cs{\"o}rg{\"o}, et~al., {Elastic Scattering and Total Cross-Section in $p+p$
  reactions measured by the LHC Experiment TOTEM at $\sqrt{s}=7$\,TeV}, Prog.\
  Theor.\ Phys.\ Suppl. 193 (2012) 180--183.
\newblock \href {http://arxiv.org/abs/1204.5689} {\path{arXiv:1204.5689}},
  \href {http://dx.doi.org/10.1143/PTPS.193.180}
  {\path{doi:10.1143/PTPS.193.180}}.

\bibitem{Abraham:2010yv}
J.~Abraham, et~al., {Measurement of the Depth of Maximum of Extensive Air
  Showers above $10^{18}$\,eV}, Phys.\ Rev.\ Lett. 104 (2010) 091101.
\newblock \href {http://arxiv.org/abs/1002.0699} {\path{arXiv:1002.0699}},
  \href {http://dx.doi.org/10.1103/PhysRevLett.104.091101}
  {\path{doi:10.1103/PhysRevLett.104.091101}}.

\bibitem{Abreu:2013env}
P.~Abreu, et~al., {Interpretation of the Depths of Maximum of Extensive Air
  Showers Measured by the Pierre Auger Observatory}, JCAP 1302 (2013) 026.
\newblock \href {http://arxiv.org/abs/1301.6637} {\path{arXiv:1301.6637}},
  \href {http://dx.doi.org/10.1088/1475-7516/2013/02/026}
  {\path{doi:10.1088/1475-7516/2013/02/026}}.

\bibitem{Aab:2014pza}
A.~Aab, et~al., {Muons in air showers at the Pierre Auger Observatory: Mean
  number in highly inclined events}, Phys.Rev.D\href
  {http://arxiv.org/abs/1408.1421} {\path{arXiv:1408.1421}}.

\bibitem{Aab:2014dua}
A.~Aab, et~al., {Muons in air showers at the Pierre Auger Observatory:
  Measurement of atmospheric production depth}, Phys.Rev. D90 (2014) 012012.
\newblock \href {http://arxiv.org/abs/1407.5919} {\path{arXiv:1407.5919}},
  \href {http://dx.doi.org/10.1103/PhysRevD.90.012012}
  {\path{doi:10.1103/PhysRevD.90.012012}}.

\bibitem{Settimo:2013}
M.~Settimo,
  \href{http://pos.sissa.it/archive/conferences/192/062/Photon%202013\_062.pdf}{{Search
  for ultra-High Energy Photons with the Pierre Auger Observatory}}, Proc.\
  Science Photon 2013 (2013) 062.
\newline\urlprefix\url{http://pos.sissa.it/archive/conferences/192/062/Photon%202013\_062.pdf}

\bibitem{Aab:2014bha}
A.~Aab, et~al., {A search for point sources of EeV photons}, Astrophys.J. 789
  (2014) 160.
\newblock \href {http://arxiv.org/abs/1406.2912} {\path{arXiv:1406.2912}},
  \href {http://dx.doi.org/10.1088/0004-637X/789/2/160}
  {\path{doi:10.1088/0004-637X/789/2/160}}.

\bibitem{Pieroni-ICRC:2013}
P.~Pieroni, {Ultra-high energy neutrinos at the Pierre Auger Observatory},
  \cite{ThePierreAuger:2013eja}, p.~76.
\newblock \href {http://arxiv.org/abs/1307.5059} {\path{arXiv:1307.5059}}.

\bibitem{Abreu:2012zz}
P.~Abreu, et~al., {Search for point-like sources of ultra-high energy neutrinos
  at the Pierre Auger Observatory and improved limit on the diffuse flux of tau
  neutrinos}, Astrophys.J. 755 (2012) L4.
\newblock \href {http://arxiv.org/abs/1210.3143} {\path{arXiv:1210.3143}},
  \href {http://dx.doi.org/10.1088/2041-8205/755/1/L4}
  {\path{doi:10.1088/2041-8205/755/1/L4}}.

\bibitem{Abreu:2013zbq}
P.~Abreu, et~al., {Ultrahigh Energy Neutrinos at the Pierre Auger Observatory},
  Adv.High Energy Phys. 2013 (2013) 708680.
\newblock \href {http://arxiv.org/abs/1304.1630} {\path{arXiv:1304.1630}},
  \href {http://dx.doi.org/10.1155/2013/708680}
  {\path{doi:10.1155/2013/708680}}.

\bibitem{Auger:2012yc}
P.~Abreu, et~al., {A Search for Point Sources of EeV Neutrons}, Astrophys.J.
  760 (2012) 148.
\newblock \href {http://arxiv.org/abs/1211.4901} {\path{arXiv:1211.4901}},
  \href {http://dx.doi.org/10.1088/0004-637X/760/2/148}
  {\path{doi:10.1088/0004-637X/760/2/148}}.

\bibitem{Aab:2014caa}
A.~Aab, et~al., {A Targeted Search for Point Sources of EeV Neutrons},
  Astrophys.J. 789 (2014) L34.
\newblock \href {http://arxiv.org/abs/1406.4038} {\path{arXiv:1406.4038}},
  \href {http://dx.doi.org/10.1088/2041-8205/789/2/L34}
  {\path{doi:10.1088/2041-8205/789/2/L34}}.

\bibitem{design-report}
{The Pierre Auger Collaboration, Design Report},
  \url{http://www.auger.org/technical\_info/design\_report.html}, accessed 16
  Jan 2015.

\bibitem{Abraham:2004dt}
J.~Abraham, et~al., {Properties and performance of the prototype instrument for
  the Pierre Auger Observatory}, Nucl.\ Instrum.\ Meth.\ A 523 (2004) 50--95.
\newblock \href {http://dx.doi.org/10.1016/j.nima.2003.12.012}
  {\path{doi:10.1016/j.nima.2003.12.012}}.

\bibitem{Sommers:1995dm}
P.~Sommers, {Capabilities of a giant hybrid air shower detector}, Astropart.\
  Phys. 3 (1995) 349--360.
\newblock \href {http://dx.doi.org/10.1016/0927-6505(95)00013-7}
  {\path{doi:10.1016/0927-6505(95)00013-7}}.

\bibitem{Dawson:1996ci}
B.~Dawson, H.~Dai, P.~Sommers, S.~Yoshida, {Simulations of a giant hybrid air
  shower detector}, Astropart.\ Phys. 5 (1996) 239--247.
\newblock \href {http://dx.doi.org/10.1016/0927-6505(96)00024-2}
  {\path{doi:10.1016/0927-6505(96)00024-2}}.

\bibitem{Abraham:2010zz}
J.~Abraham, et~al., {Trigger and aperture of the surface detector array of the
  Pierre Auger Observatory}, Nucl.\ Instrum.\ Meth.\ A 613 (2010) 29--39.
\newblock \href {http://arxiv.org/abs/1111.6764} {\path{arXiv:1111.6764}},
  \href {http://dx.doi.org/10.1016/j.nima.2009.11.018}
  {\path{doi:10.1016/j.nima.2009.11.018}}.

\bibitem{Abreu:2010aa}
P.~Abreu, et~al., {The exposure of the hybrid detector of the Pierre Auger
  Observatory}, Astropart.\ Phys. 34 (2011) 368--381.
\newblock \href {http://arxiv.org/abs/1010.6162} {\path{arXiv:1010.6162}},
  \href {http://dx.doi.org/10.1016/j.astropartphys.2010.10.001}
  {\path{doi:10.1016/j.astropartphys.2010.10.001}}.

\bibitem{Abraham:2010mj}
J.~Abraham, et~al., {Measurement of the energy spectrum of cosmic rays above
  $10^{18}$\,eV using the {P}ierre {A}uger {O}bservatory}, Phys.\ Lett.\ B 685
  (2010) 239--246.
\newblock \href {http://arxiv.org/abs/1002.1975} {\path{arXiv:1002.1975}},
  \href {http://dx.doi.org/10.1016/j.physletb.2010.02.013}
  {\path{doi:10.1016/j.physletb.2010.02.013}}.

\bibitem{Hersil:1961zz}
J.~Hersil, I.~Escobar, D.~Scott, G.~Clark, S.~Olbert, {Observations of
  Extensive Air Showers near the Maximum of Their Longitudinal Development},
  Phys.\ Rev.\ Lett. 6 (1961) 22--23.
\newblock \href {http://dx.doi.org/10.1103/PhysRevLett.6.22}
  {\path{doi:10.1103/PhysRevLett.6.22}}.

\bibitem{Bonifazi:2009ma}
C.~Bonifazi, {The angular resolution of the {P}ierre {A}uger {O}bservatory},
  Nucl.\ Phys.\ Proc.\ Suppl. 190 (2009) 20--25.
\newblock \href {http://arxiv.org/abs/0901.3138} {\path{arXiv:0901.3138}},
  \href {http://dx.doi.org/10.1016/j.nuclphysbps.2009.03.063}
  {\path{doi:10.1016/j.nuclphysbps.2009.03.063}}.

\bibitem{Dawson:2007di}
B.~Dawson, {Hybrid Performance of the {P}ierre {A}uger {O}bservatory}, Proc.\
  30th International Cosmic Ray Conference (ICRC) 2007, Merida, Mexico\href
  {http://arxiv.org/abs/0706.1105} {\path{arXiv:0706.1105}}.

\bibitem{Allekotte:2007sf}
I.~Allekotte, et~al., {The {S}urface {D}etector System of the {P}ierre {A}uger
  {O}bservatory}, Nucl.\ Instrum.\ Meth.\ A 586 (2008) 409--420.
\newblock \href {http://arxiv.org/abs/0712.2832} {\path{arXiv:0712.2832}},
  \href {http://dx.doi.org/10.1016/j.nima.2007.12.016}
  {\path{doi:10.1016/j.nima.2007.12.016}}.

\bibitem{Bertou:2005ze}
X.~Bertou, et~al., {Calibration of the surface array of the {P}ierre {A}uger
  {O}bservatory}, Nucl.\ Instrum.\ Meth.\ A 568 (2006) 839--846.
\newblock \href {http://dx.doi.org/10.1016/j.nima.2006.07.066}
  {\path{doi:10.1016/j.nima.2006.07.066}}.

\bibitem{Allison:2005ge}
P.~Allison, et~al., {Observing muon decays in water Cherenkov detectors at the
  Pierre Auger Observatory}, {Proc.\ 29th International Cosmic Ray Conference
  (ICRC) 2005, Pune, India} 8 (2005) 299--302.
\newblock \href {http://arxiv.org/abs/astro-ph/0509238}
  {\path{arXiv:astro-ph/0509238}}.

\bibitem{Sato-ICRC:2011}
R.~Sato, {Long Term Performance of the Surface Detectors of the PierreAuger
  Observatory},  \cite{Abreu:2011pg}, p.~1.
\newblock \href {http://arxiv.org/abs/1107.4806} {\path{arXiv:1107.4806}}.

\bibitem{Rautenberg:2011zz}
J.~Rautenberg, {Remote operation of the Pierre Auger Observatory},
  \cite{Abreu:2011pg}, p.~5.
\newblock \href {http://arxiv.org/abs/1107.4806} {\path{arXiv:1107.4806}}.

\bibitem{Abraham:2009pm}
J.~Abraham, et~al., {The Fluorescence Detector of the Pierre Auger
  Observatory}, Nucl.\ Instrum.\ Meth.\ A 620 (2010) 227--251.
\newblock \href {http://arxiv.org/abs/0907.4282} {\path{arXiv:0907.4282}},
  \href {http://dx.doi.org/10.1016/j.nima.2010.04.023}
  {\path{doi:10.1016/j.nima.2010.04.023}}.

\bibitem{deOliveira:2004dh}
M.~de~Oliveira, V.~de~Souza, H.~Reis, R.~Sato, {Manufacturing the Schmidt
  corrector lens for the Pierre Auger Observatory}, Nucl.\ Instrum.\ Meth.\ A
  522 (2004) 360--370.
\newblock \href {http://dx.doi.org/10.1016/j.nima.2003.11.409}
  {\path{doi:10.1016/j.nima.2003.11.409}}.

\bibitem{Baeuml-ICRC:2013}
J.~Baeuml, {Measurement of the Optical Properties of the Auger Fluorescence
  Telescopes},  \cite{ThePierreAuger:2013eja}, p.~15.
\newblock \href {http://arxiv.org/abs/1307.5059} {\path{arXiv:1307.5059}}.

\bibitem{Becker:2007zza}
K.~Becker, A.~Behrmann, H.~Geenen, S.~Hartmann, K.~Kampert, et~al.,
  {Qualification tests of the 11000 photomultipliers for the Pierre Auger
  Observatory fluorescence detectors}, Nucl. Instrum. Meth. A576 (2007)
  301--311.
\newblock \href {http://dx.doi.org/10.1016/j.nima.2007.03.007}
  {\path{doi:10.1016/j.nima.2007.03.007}}.

\bibitem{Arqueros:2005yn}
B.~Fick, et~al., {The Central laser facility at the Pierre Auger Observatory},
  JINST 1 (2006) P11003.
\newblock \href {http://arxiv.org/abs/astro-ph/0507334}
  {\path{arXiv:astro-ph/0507334}}, \href
  {http://dx.doi.org/10.1088/1748-0221/1/11/P11003}
  {\path{doi:10.1088/1748-0221/1/11/P11003}}.

\bibitem{Brack:2013bta}
J.~Brack, R.~Cope, A.~Dorofeev, B.~Gookin, J.~Harton, et~al., {Absolute
  Calibration of a Large-diameter Light Source}, JINST 8 (2013) P05014.
\newblock \href {http://arxiv.org/abs/1305.1329} {\path{arXiv:1305.1329}},
  \href {http://dx.doi.org/10.1088/1748-0221/8/05/P05014}
  {\path{doi:10.1088/1748-0221/8/05/P05014}}.

\bibitem{BenZvi:2006xb}
S.~BenZvi, R.~Cester, M.~Chiosso, B.~Connolly, A.~Filipcic, et~al., {The Lidar
  System of the Pierre Auger Observatory}, Nucl.\ Instrum.\ Meth.\ A 574 (2007)
  171--184.
\newblock \href {http://arxiv.org/abs/astro-ph/0609063}
  {\path{arXiv:astro-ph/0609063}}, \href
  {http://dx.doi.org/10.1016/j.nima.2007.01.094}
  {\path{doi:10.1016/j.nima.2007.01.094}}.

\bibitem{Argiro:2007qg}
S.~Argiro, S.~Barroso, J.~Gonzalez, L.~Nellen, T.~C. Paul, et~al., {The Offline
  Software Framework of the Pierre Auger Observatory}, Nucl.\ Instrum.\ Meth.\
  A 580 (2007) 1485--1496.
\newblock \href {http://arxiv.org/abs/0707.1652} {\path{arXiv:0707.1652}},
  \href {http://dx.doi.org/10.1016/j.nima.2007.07.010}
  {\path{doi:10.1016/j.nima.2007.07.010}}.

\bibitem{xml}
E.~Harold, W.~Means, XML in a Nutshell, O'Reilly Media, 2004, iSBN
  0-596-00764-7.

\bibitem{python}
{P}ython, \url{https://www.python.org/}, accessed 15 Jan 2015.

\bibitem{root}
{ROOT} - data analysis framework, \url{https://root.cern.ch/drupal/}, accessed
  15 Jan 2015.

\bibitem{Sciutto:1999jh}
S.~Sciutto, {{AIRES}: A System for air shower simulations. User's guide and
  reference manual. Version 2.2.0}, downloads and manuals to the software can
  be found at http://www2.fisica.unlp.edu.ar/auger/aires/eg\_Aires.html (1999).
\newblock \href {http://arxiv.org/abs/astro-ph/9911331}
  {\path{arXiv:astro-ph/9911331}}.

\bibitem{Heck:1998vt}
D.~Heck, G.~Schatz, T.~Thouw, J.~Knapp, J.~Capdevielle, {{CORSIKA}: A {M}onte
  {C}arlo code to simulate extensive air showers}, Forschungszentrum Karlsruhe
  - Wissenschaftliche BerichteDownloads and manuals to the software can be
  found at {\url{https://web.ikp.kit.edu/corsika/}}.

\bibitem{Bergmann:2006yz}
T.~Bergmann, R.~Engel, D.~Heck, N.~Kalmykov, S.~Ostapchenko, et~al.,
  {One-dimensional Hybrid Approach to Extensive Air Shower Simulation},
  Astropart.\ Phys. 26 (2007) 420--432, for further information contact the
  authors of CORSIKA, see url at \cite{Heck:1998vt}.
\newblock \href {http://arxiv.org/abs/astro-ph/0606564}
  {\path{arXiv:astro-ph/0606564}}, \href
  {http://dx.doi.org/10.1016/j.astropartphys.2006.08.005}
  {\path{doi:10.1016/j.astropartphys.2006.08.005}}.

\bibitem{Drescher:2002cr}
H.-J. Drescher, G.~R. Farrar, {Air shower simulations in a hybrid approach
  using cascade equations}, Phys.\ Rev.\ D 67 (2003) 116001, downloads and
  manuals to the software can be found at
  {\url{http://th.physik.uni-frankfurt.de/~drescher/SENECA/}}.
\newblock \href {http://arxiv.org/abs/astro-ph/0212018}
  {\path{arXiv:astro-ph/0212018}}, \href
  {http://dx.doi.org/10.1103/PhysRevD.67.116001}
  {\path{doi:10.1103/PhysRevD.67.116001}}.

\bibitem{xerces}
{A}pache - {X}erces, \url{http://xerces.apache.org/}, accessed 15 Jan 2015.

\bibitem{xml-schema}
{XML} - {SCHEMA}, \url{http://www.w3.org/standards/xml/schema}, accessed 15 Jan
  2015.

\bibitem{cppunit}
{C}pp{U}unit - {C}++ port of {JU}nit,
  \url{http://sourceforge.net/projects/cppunit/}, accessed 16 Jan 2015.

\bibitem{buildbot}
Buildbot - the continuous integration framework, \url{http://buildbot.net/},
  accessed 16 Jan 2015.

\bibitem{cmake}
{CM}ake, \url{http://www.cmake.org}, accessed 16 Jan 2015.

\bibitem{ape}
{A}uger {P}ackage {E}nvironment,
  \url{https://svn.auger.unam.mx/trac/projects/ape/}, accessed 16 Jan 2015.

\bibitem{boehm}
B.~Boehm, Software Engineering Economics, Prentice-Hall Advances in Computing
  Science \& Technology Series, 1981, iSBN 0-13-822122-7.

\bibitem{Abreu:2011fb}
P.~Abreu, et~al., {Advanced functionality for radio analysis in the Offline
  software framework of the Pierre Auger Observatory}, Nucl.\ Instrum.\ Meth.\
  A 635 (2011) 92--102.
\newblock \href {http://arxiv.org/abs/1101.4473} {\path{arXiv:1101.4473}},
  \href {http://dx.doi.org/10.1016/j.nima.2011.01.049}
  {\path{doi:10.1016/j.nima.2011.01.049}}.

\bibitem{Kostunin:2013iaa}
D.~Kostunin, N.~Budnev, O.~Gress, A.~Haungs, R.~Hiller, et~al., {Tunka-Rex:
  Status and results of the first measurements}, Nucl. Instrum. Meth. A742
  (2014) 89--94.
\newblock \href {http://arxiv.org/abs/1310.8477} {\path{arXiv:1310.8477}},
  \href {http://dx.doi.org/10.1016/j.nima.2013.10.070}
  {\path{doi:10.1016/j.nima.2013.10.070}}.

\bibitem{Abeysekara:2013tka}
A.~Abeysekara, et~al., {The HAWC Gamma-Ray Observatory: Design, Calibration,
  and Operation}\href {http://arxiv.org/abs/1310.0074}
  {\path{arXiv:1310.0074}}.

\bibitem{Ebisuzaki:2014wka}
T.~Ebisuzaki, G.~Medina-Tanco, A.~Santangelo, {The JEM-EUSO mission}, Adv.\
  Space Res. 53 (2014) 1499--1505.
\newblock \href {http://dx.doi.org/10.1016/j.asr.2013.11.056}
  {\path{doi:10.1016/j.asr.2013.11.056}}.

\bibitem{Ardouin:2006gj}
D.~Ardouin, A.~Belletoile, D.~Charrier, R.~Dallier, L.~Denis, et~al.,
  {CODALEMA: A cosmic ray air shower radio detection experiment}, Int.\ J.\
  Mod.\ Phys. A21S1 (2006) 192--196.
\newblock \href {http://dx.doi.org/10.1142/S0217751X0603360X}
  {\path{doi:10.1142/S0217751X0603360X}}.

\bibitem{Wyszynski:2012fa}
O.~Wyszynski, A.~Laszlo, A.~Marcinek, T.~Paul, R.~Sipos, et~al., {Legacy code:
  Lessons from NA61/SHINE offline software upgrade adventure}, J.\ Phys.\
  Conf.\ Ser. 396 (2012) 052076.
\newblock \href {http://dx.doi.org/10.1088/1742-6596/396/5/052076}
  {\path{doi:10.1088/1742-6596/396/5/052076}}.

\bibitem{Sipos:2012hs}
R.~Sipos, A.~Laszlo, A.~Marcinek, T.~Paul, M.~Szuba, et~al., {The offline
  software framework of the NA61/SHINE experiment}, J.\ Phys.\ Conf.\ Ser. 396
  (2012) 022045.
\newblock \href {http://dx.doi.org/10.1088/1742-6596/396/2/022045}
  {\path{doi:10.1088/1742-6596/396/2/022045}}.

\bibitem{Abraham:2010pf}
J.~Abraham, et~al., {A Study of the Effect of Molecular and Aerosol Conditions
  in the Atmosphere on Air Fluorescence Measurements at the Pierre Auger
  Observatory}, Astropart.\ Phys. 33 (2010) 108--129.
\newblock \href {http://arxiv.org/abs/1002.0366} {\path{arXiv:1002.0366}},
  \href {http://dx.doi.org/10.1016/j.astropartphys.2009.12.005}
  {\path{doi:10.1016/j.astropartphys.2009.12.005}}.

\bibitem{Abreu:2012zg}
P.~Abreu, et~al., {Description of Atmospheric Conditions at the Pierre Auger
  Observatory using the Global Data Assimilation System (GDAS)}, Astropart.\
  Phys. 35 (2012) 591--607.
\newblock \href {http://arxiv.org/abs/1201.2276} {\path{arXiv:1201.2276}},
  \href {http://dx.doi.org/10.1016/j.astropartphys.2011.12.002}
  {\path{doi:10.1016/j.astropartphys.2011.12.002}}.

\bibitem{Keilhauer:2012yp}
B.~Keilhauer, M.~Will, {Description of Atmospheric Conditions at the Pierre
  Auger Observatory Using Meteorological Measurements and Models}, Eur.\ Phys.\
  J.\ Plus 127 (2012) 96.
\newblock \href {http://arxiv.org/abs/1208.5417} {\path{arXiv:1208.5417}},
  \href {http://dx.doi.org/10.1140/epjp/i2012-12096-8}
  {\path{doi:10.1140/epjp/i2012-12096-8}}.

\bibitem{Abreu:2012oza}
P.~Abreu, et~al., {The Rapid Atmospheric Monitoring System of the Pierre Auger
  Observatory}, JINST 7 (2012) P09001.
\newblock \href {http://arxiv.org/abs/1208.1675} {\path{arXiv:1208.1675}},
  \href {http://dx.doi.org/10.1088/1748-0221/7/09/P09001}
  {\path{doi:10.1088/1748-0221/7/09/P09001}}.

\bibitem{Keilhauer:2004jr}
B.~Keilhauer, J.~Blumer, R.~Engel, H.~Klages, M.~Risse, {Impact of varying
  atmospheric profiles on extensive air shower observation: - Atmospheric
  density and primary mass reconstruction}, Astropart.\ Phys. 22 (2004)
  249--261.
\newblock \href {http://arxiv.org/abs/astro-ph/0405048}
  {\path{arXiv:astro-ph/0405048}}, \href
  {http://dx.doi.org/10.1016/j.astropartphys.2004.08.004}
  {\path{doi:10.1016/j.astropartphys.2004.08.004}}.

\bibitem{Arqueros:2008cx}
F.~Arqueros, J.~R. Hoerandel, B.~Keilhauer, {Air Fluorescence Relevant for
  Cosmic-Ray Detection - Summary of the 5th Fluorescence Workshop, El Escorial
  2007}, Nucl.\ Instrum.\ Meth.\ A 597 (2008) 1--22.
\newblock \href {http://arxiv.org/abs/0807.3760} {\path{arXiv:0807.3760}},
  \href {http://dx.doi.org/10.1016/j.nima.2008.08.056}
  {\path{doi:10.1016/j.nima.2008.08.056}}.

\bibitem{Keilhauer:2008sy}
B.~Keilhauer, J.~Bluemer, R.~Engel, H.~Klages, {Altitude dependence of
  fluorescence light emission by extensive air showers}, Nucl.\ Instrum.\
  Meth.\ A 597 (2008) 99--104.
\newblock \href {http://arxiv.org/abs/0801.4200} {\path{arXiv:0801.4200}},
  \href {http://dx.doi.org/10.1016/j.nima.2008.08.060}
  {\path{doi:10.1016/j.nima.2008.08.060}}.

\bibitem{Monasor:2010fn}
M.~Monasor, J.~Vazquez, D.~Garcia-Pinto, F.~Arqueros, {The impact of the
  air-fluorescence yield on the reconstructed shower parameters of ultra-high
  energy cosmic rays}, Astropart.\ Phys. 34 (2011) 467--475.
\newblock \href {http://arxiv.org/abs/1010.3793} {\path{arXiv:1010.3793}},
  \href {http://dx.doi.org/10.1016/j.astropartphys.2010.10.009}
  {\path{doi:10.1016/j.astropartphys.2010.10.009}}.

\bibitem{Keilhauer:2012hu}
B.~Keilhauer, M.~Bohacova, M.~Fraga, J.~Matthews, N.~Sakaki, et~al., {Nitrogen
  fluorescence in air for observing extensive air showers}, EPJ Web Conf. 53
  (2013) 01010.
\newblock \href {http://arxiv.org/abs/1210.1319} {\path{arXiv:1210.1319}},
  \href {http://dx.doi.org/10.1051/epjconf/20135301010}
  {\path{doi:10.1051/epjconf/20135301010}}.

\bibitem{Abraham:2009bc}
J.~Abraham, et~al., {Atmospheric effects on extensive air showers observed with
  the Surface Detector of the Pierre Auger Observatory}, Astropart.\ Phys. 32
  (2009) 89--99.
\newblock \href {http://arxiv.org/abs/0906.5497} {\path{arXiv:0906.5497}},
  \href {http://dx.doi.org/10.1016/j.astropartphys.2009.06.004}
  {\path{doi:10.1016/j.astropartphys.2009.06.004}}.

\bibitem{Keilhauer:2009ax}
B.~Keilhauer, M.~Unger, {Fluorescence emission induced by extensive air showers
  in dependence on atmospheric conditions}, Proc.\ 31st International Cosmic
  Ray Conference (ICRC) 2009, Lodz, Poland\href
  {http://arxiv.org/abs/0906.5487} {\path{arXiv:0906.5487}}.

\bibitem{corsika_usersguide}
Extensive air shower simulation with corsika: A user's guide (version 74xxx
  from september 3, 2013),
  \url{http://web.ikp.kit.edu/corsika/usersguide/usersguide.pdf}, accessed 16
  Jan 2015.

\bibitem{Abreu:2013qtw}
P.~Abreu, et~al., {Techniques for Measuring Aerosol Attenuation using the
  Central Laser Facility at the Pierre Auger Observatory}, JINST 8 (2013)
  P04009.
\newblock \href {http://arxiv.org/abs/1303.5576} {\path{arXiv:1303.5576}},
  \href {http://dx.doi.org/10.1088/1748-0221/8/04/P04009}
  {\path{doi:10.1088/1748-0221/8/04/P04009}}.

\bibitem{Abreu:2011pg}
P.~Abreu, et~al., {The Pierre Auger Observatory IV: Operation and Monitoring},
  Proc.\ 32nd International Cosmic Ray Conference (ICRC) 2011, Beijing,
  China\href {http://arxiv.org/abs/1107.4806} {\path{arXiv:1107.4806}}.

\bibitem{Henyey-Greenstein}
L.~Henyey, J.~Greenstein, {Diffuse radiation in the Galaxy}, Astrophys.\ J. 93
  (1941) 70.
\newblock \href {http://dx.doi.org/10.1086/144246} {\path{doi:10.1086/144246}}.

\bibitem{BenZvi:2007px}
S.~BenZvi, B.~Connolly, J.~Matthews, M.~Prouza, E.~Visbal, et~al., {Measurement
  of the Aerosol Phase Function at the Pierre Auger Observatory},
  Astropart.Phys. 28 (2007) 312--320.
\newblock \href {http://arxiv.org/abs/0704.0303} {\path{arXiv:0704.0303}},
  \href {http://dx.doi.org/10.1016/j.astropartphys.2007.06.005}
  {\path{doi:10.1016/j.astropartphys.2007.06.005}}.

\bibitem{Chirinos-ICRC:2013}
J.~Chirinos, {Cloud Monitoring at the Pierre Auger Observatory},
  \cite{ThePierreAuger:2013eja}, p. 108.
\newblock \href {http://arxiv.org/abs/1307.5059} {\path{arXiv:1307.5059}}.

\bibitem{Winnick:2010}
M.~Winnick, {Cloud cameras at the Pierre Auger Observatory}, Ph.D.\ thesis,
  University of Adelaide, online available at
  \url{http://digital.library.adelaide.edu.au/dspace/handle/2440/67198} (2010).

\bibitem{Abreu:2013qfa}
P.~Abreu, et~al., {Identifying Clouds over the Pierre Auger Observatory using
  Infrared Satellite Data}, Astropart.\ Phys. 50-52 (2013) 92--101.
\newblock \href {http://arxiv.org/abs/1310.1641} {\path{arXiv:1310.1641}},
  \href {http://dx.doi.org/10.1016/j.astropartphys.2013.09.004}
  {\path{doi:10.1016/j.astropartphys.2013.09.004}}.

\bibitem{Rizi:2012dn}
V.~Rizi, A.~Tonachini, M.~Iarlori, G.~Visconti, {Atmospheric monitoring with
  LIDARs at the Pierre Auger Observatory}, Eur.\ Phys.\ J.\ Plus 127 (2012) 92.
\newblock \href {http://dx.doi.org/10.1140/epjp/i2012-12092-0}
  {\path{doi:10.1140/epjp/i2012-12092-0}}.

\bibitem{Jelinek:2006gf}
M.~Jelinek, M.~Prouza, P.~Kubanek, R.~Hudec, M.~Nekola, et~al., {The bright
  optical flash from GRB 060117}, Astron.\ Astrophys. 454 (2006) L119--L122.
\newblock \href {http://arxiv.org/abs/astro-ph/0606004}
  {\path{arXiv:astro-ph/0606004}}, \href
  {http://dx.doi.org/10.1051/0004-6361:20065092}
  {\path{doi:10.1051/0004-6361:20065092}}.

\bibitem{BenZvi:2007uj}
S.~BenZvi, et~al., {New method for atmospheric calibration at the Pierre Auger
  Observatory using FRAM, a robotic astronomical telescope}, Proc.\ 30th
  International Cosmic Ray Conference (ICRC) 2007, Merida, Mexico\href
  {http://arxiv.org/abs/0706.1710} {\path{arXiv:0706.1710}}.

\bibitem{Bueno-Wiencke:2012}
A.~Bueno, L.~Wiencke (Eds.), Focus Point on Interdisciplinary Science with
  Cosmic Rays, Vol. 127~8, Eur.\ Phys.\ J.\ Plus, 2012.

\bibitem{Mussa:2012dq}
R.~Mussa, G.~Ciaccio, {Observation of ELVES at the Pierre Auger Observatory},
  Eur.\ Phys.\ J.\ Plus 127 (2012) 94.
\newblock \href {http://dx.doi.org/10.1140/epjp/i2012-12094-x}
  {\path{doi:10.1140/epjp/i2012-12094-x}}.

\bibitem{Micheletti:2012fs}
M.~Micheletti, L.~Murruni, M.~Debray, M.~Rosenbusch, M.~Graf, et~al.,
  {Elemental analysis of aerosols collected at the Pierre Auger Cosmic Ray
  Observatory with PIXE technique complemented with SEM/EDX}, Nucl.\ Instrum.\
  Meth.\ B 288 (2012) 10--17.
\newblock \href {http://dx.doi.org/10.1016/j.nimb.2012.07.022}
  {\path{doi:10.1016/j.nimb.2012.07.022}}.

\bibitem{Micheletti-ICRC:2013}
M.~Micheletti, {Aerosol characterization at the Pierre Auger Observatory},
  \cite{ThePierreAuger:2013eja}, p. 104.
\newblock \href {http://arxiv.org/abs/1307.5059} {\path{arXiv:1307.5059}}.

\bibitem{Porter:1970et}
L.~Porter, J.~Earnshaw, E.~Tielsch-Cassel, J.~Ahlstrom, K.~Greisen, {A
  space-time detector for cosmic ray showers}, Nucl.\ Instrum.\ Meth. 87 (1970)
  87--92.
\newblock \href {http://dx.doi.org/10.1016/0029-554X(70)90886-4}
  {\path{doi:10.1016/0029-554X(70)90886-4}}.

\bibitem{Giller:2004}
M.~Giller, G.~Wieczorek, A.~Kacperczyk, H.~Stojek, W.~Tkaczyk,
  \href{{http://stacks.iop.org/0954-3899/30/i=2/a=009}}{{Energy spectra of
  electrons in the extensive air showers of ultra-high energy}}, J.\ Phys.\ G
  30~(2) ({2004}) 97.
\newline\urlprefix\url{{http://stacks.iop.org/0954-3899/30/i=2/a=009}}

\bibitem{Nerling:2005fj}
F.~Nerling, J.~Bluemer, R.~Engel, M.~Risse, {Universality of electron
  distributions in high-energy air showers: Description of Cherenkov light
  production}, Astropart.\ Phys. 24 (2006) 421--437.
\newblock \href {http://arxiv.org/abs/astro-ph/0506729}
  {\path{arXiv:astro-ph/0506729}}, \href
  {http://dx.doi.org/10.1016/j.astropartphys.2005.09.002}
  {\path{doi:10.1016/j.astropartphys.2005.09.002}}.

\bibitem{Roberts:2005xv}
M.~Roberts, \href{{http://stacks.iop.org/0954-3899/31/i=11/a=012}}{{The role of
  atmospheric multiple scattering in the transmission of fluorescence light
  from extensive air showers}}, J.\ Phys.\ G 31~(11) ({2005}) 1291.
\newline\urlprefix\url{{http://stacks.iop.org/0954-3899/31/i=11/a=012}}

\bibitem{Pekala:2009fe}
J.~Pekala, P.~Homola, B.~Wilczynska, H.~Wilczynski, {Atmospheric multiple
  scattering of fluorescence and Cherenkov light emitted by extensive air
  showers}, Nucl. Instrum. Meth. A605 (2009) 388--398.
\newblock \href {http://arxiv.org/abs/0904.3230} {\path{arXiv:0904.3230}},
  \href {http://dx.doi.org/10.1016/j.nima.2009.03.244}
  {\path{doi:10.1016/j.nima.2009.03.244}}.

\bibitem{Giller:2012tt}
M.~Giller, A.~Smialkowski, {An analytical approach to the multiply scattered
  light in the optical images of the extensive air showers of ultra-high
  energies}, Astropart.\ Phys. 36 (2012) 166--182.
\newblock \href {http://arxiv.org/abs/1201.4052} {\path{arXiv:1201.4052}},
  \href {http://dx.doi.org/10.1016/j.astropartphys.2012.05.021}
  {\path{doi:10.1016/j.astropartphys.2012.05.021}}.

\bibitem{Rosado:2014bya}
J.~Rosado, F.~Blanco, F.~Arqueros, {On the absolute value of the
  air-fluorescence yield}, Astropart.\ Phys. 55 (2014) 51--62.
\newblock \href {http://arxiv.org/abs/1401.4310} {\path{arXiv:1401.4310}},
  \href {http://dx.doi.org/10.1016/j.astropartphys.2014.02.003}
  {\path{doi:10.1016/j.astropartphys.2014.02.003}}.

\bibitem{Ave:2008zza}
M.~Ave, et~al., {Spectrally resolved pressure dependence measurements of air
  fluorescence emission with AIRFLY}, Nucl.\ Instrum.\ Meth.\ A 597 (2008)
  41--45.
\newblock \href {http://dx.doi.org/10.1016/j.nima.2008.08.052}
  {\path{doi:10.1016/j.nima.2008.08.052}}.

\bibitem{Ave:2012ifa}
M.~Ave, et~al., {Precise measurement of the absolute fluorescence yield of the
  337 nm band in atmospheric gases}, Astropart.\ Phys. 42 (2013) 90--102.
\newblock \href {http://arxiv.org/abs/1210.6734} {\path{arXiv:1210.6734}},
  \href {http://dx.doi.org/10.1016/j.astropartphys.2012.12.006}
  {\path{doi:10.1016/j.astropartphys.2012.12.006}}.

\bibitem{Unger:2008uq}
M.~Unger, B.~Dawson, R.~Engel, F.~Schussler, R.~Ulrich, {Reconstruction of
  Longitudinal Profiles of Ultra-High Energy Cosmic Ray Showers from
  Fluorescence and Cherenkov Light Measurements}, Nucl.\ Instrum.\ Meth.\ A 588
  (2008) 433--441.
\newblock \href {http://arxiv.org/abs/0801.4309} {\path{arXiv:0801.4309}},
  \href {http://dx.doi.org/10.1016/j.nima.2008.01.100}
  {\path{doi:10.1016/j.nima.2008.01.100}}.

\bibitem{Gora:2005sq}
D.~Gora, R.~Engel, D.~Heck, P.~Homola, H.~Klages, et~al., {Universal lateral
  distribution of energy deposit in air showers and its application to shower
  reconstruction}, Astropart.\ Phys. 24 (2006) 484--494.
\newblock \href {http://arxiv.org/abs/astro-ph/0505371}
  {\path{arXiv:astro-ph/0505371}}, \href
  {http://dx.doi.org/10.1016/j.astropartphys.2005.09.007}
  {\path{doi:10.1016/j.astropartphys.2005.09.007}}.

\bibitem{Dawson-Giller:2007}
B.~Dawson, M.~Giller, G.~Wieczorek,
  \href{http://indico.nucleares.unam.mx/contributionDisplay.py?contribId=651&confId=4}{{Influence
  of the scattered Cherenkov light on the width of shower images as measured in
  the EAS fluorescence experiments}}, Proceedings of the 30th International
  Cosmic Ray Conference, Mexico-City, Mexico 4 (2007) 401.
\newline\urlprefix\url{http://indico.nucleares.unam.mx/contributionDisplay.py?contribId=651&confId=4}

\bibitem{Gaisser-Hillas:1977}
T.~Gaisser, A.~Hillas, Reliability of the method of constant intensity cuts for
  reconstructing the average development of vertical showers, Proceedings of
  the 15th International Cosmic Ray Conference, Plovdiv, Bulgaria 8 (1977) 353.

\bibitem{Tueros-ICRC:2013}
M.~Tueros, Estimate of the non-calorimetric energy of showers observed with the
  fluorescence and surface detectors of the {P}ierre {A}uger {O}bservatory,
  \cite{ThePierreAuger:2013eja}, p.~11.
\newblock \href {http://arxiv.org/abs/1307.5059} {\path{arXiv:1307.5059}}.

\bibitem{Veberic-ICRC:2013}
D.~Veberi\v{c}, {Estimation of the Total Signals in Saturated Stations of the
  Pierre Auger Observatory},  \cite{ThePierreAuger:2013eja}, p.~23.
\newblock \href {http://arxiv.org/abs/1307.5059} {\path{arXiv:1307.5059}}.

\bibitem{Kamata:1958}
K.~Kamata, J.~Nishimura, {The Lateral and the Angular Structure Functions of
  Electron Showers}, Proc.\ Theor.\ Phys.\ Supplement 6 (1958) 93--155.
\newblock \href {http://dx.doi.org/10.1143/PTPS.6.93}
  {\path{doi:10.1143/PTPS.6.93}}.

\bibitem{Greisen:1956}
K.~Greisen, \href{https://archive.org/details/progessincosmic031401mbp}{{The
  extensive air showers}}, Progress in Cosmic Ray Physics 3 (1956) 1.
\newline\urlprefix\url{https://archive.org/details/progessincosmic031401mbp}

\bibitem{Newton:2007}
D.~Newton, J.~Knapp, A.~A. Watson, The optimum distance at which to determine
  the size of a giant air shower, Astropart.\ Phys. 26 (2007) 414--419.
\newblock \href {http://arxiv.org/abs/astro-ph/0608118}
  {\path{arXiv:astro-ph/0608118}}, \href
  {http://dx.doi.org/10.1016/j.astropartphys.2006.08.003}
  {\path{doi:10.1016/j.astropartphys.2006.08.003}}.

\bibitem{Ave:2007wf}
M.~Ave, {Reconstruction accuracy of the surface detector array of the Pierre
  Auger Observatory}, Proc.\ 30th International Cosmic Ray Conference (ICRC)
  2007, Merida, Mexico\href {http://arxiv.org/abs/0709.2125}
  {\path{arXiv:0709.2125}}.

\bibitem{Bonifazi:2007ck}
C.~Bonifazi, A.~Letessier-Selvon, E.~Santos, {A model for the time uncertainty
  measurements in the {A}uger surface detector array}, Astropart.\ Phys. 28
  (2008) 523--528.
\newblock \href {http://arxiv.org/abs/0705.1856} {\path{arXiv:0705.1856}},
  \href {http://dx.doi.org/10.1016/j.astropartphys.2007.09.007}
  {\path{doi:10.1016/j.astropartphys.2007.09.007}}.

\bibitem{Pesce-ICRC:2011}
R.~Pesce, {Energy calibration of data recorded with the surface detectors of
  the Pierre Auger Observatory: an update},  \cite{Abreu:2011pj}, p.~13.
\newblock \href {http://arxiv.org/abs/1107.4809} {\path{arXiv:1107.4809}}.

\bibitem{Dembinski-ICRC:2011}
H.~Dembinski, {The Cosmic Ray Spectrum above $4\times10^{18}$\,eV as measured
  with inclined showers recorded at the Pierre Auger Observatory},
  \cite{Abreu:2011pj}, p.~5.
\newblock \href {http://arxiv.org/abs/1107.4809} {\path{arXiv:1107.4809}}.

\bibitem{Verzi-ICRC:2013}
V.~Verzi, {The Energy Scale of the {P}ierre {A}uger {O}bservatory},
  \cite{ThePierreAuger:2013eja}, p.~7.
\newblock \href {http://arxiv.org/abs/1307.5059} {\path{arXiv:1307.5059}}.

\bibitem{Tokuno:2012mi}
H.~Tokuno, Y.~Tameda, M.~Takeda, K.~Kadota, D.~Ikeda, et~al., {New air
  fluorescence detectors employed in the Telescope Array experiment},
  Nucl.Instrum.Meth. A676 (2012) 54--65.
\newblock \href {http://arxiv.org/abs/1201.0002} {\path{arXiv:1201.0002}},
  \href {http://dx.doi.org/10.1016/j.nima.2012.02.044}
  {\path{doi:10.1016/j.nima.2012.02.044}}.

\bibitem{Array:2013dra}
T.~Abu-Zayyad, et~al., {Pierre Auger Observatory and Telescope Array: Joint
  Contributions to the 33rd International Cosmic Ray Conference (ICRC
  2013)}\href {http://arxiv.org/abs/1310.0647} {\path{arXiv:1310.0647}}.

\bibitem{Valino-ICRC:2013}
I.~Vali\~{n}o, {A measurement of the muon number in showers using inclined
  events recorded at the Pierre Auger Observatory},
  \cite{ThePierreAuger:2013eja}, p.~44.
\newblock \href {http://arxiv.org/abs/1307.5059} {\path{arXiv:1307.5059}}.

\bibitem{Aab:2014gua}
A.~Aab, et~al., {Reconstruction of inclined air showers detected with the
  Pierre Auger Observatory}, JCAP 1408~(08) (2014) 019.
\newblock \href {http://arxiv.org/abs/1407.3214} {\path{arXiv:1407.3214}},
  \href {http://dx.doi.org/10.1088/1475-7516/2014/08/019}
  {\path{doi:10.1088/1475-7516/2014/08/019}}.

\bibitem{Maris-ICRC:2011}
I.~Mari\c{s}, {The AMIGA infill detector of the Pierre Auger Observatory:
  performance and first data},  \cite{Abreu:2011pj}, p.~9.
\newblock \href {http://arxiv.org/abs/1107.4809} {\path{arXiv:1107.4809}}.

\bibitem{Ravignani-ICRC:2013}
D.~Ravignani, {Measurement of the energy spectrum of cosmic rays above
  $3\times10^{17}$\,eV using the AMIGA 750\,m surface detector array of the
  Pierre Auger Observatory},  \cite{ThePierreAuger:2013eja}, p.~4.
\newblock \href {http://arxiv.org/abs/1307.5059} {\path{arXiv:1307.5059}}.

\bibitem{Mathes-ICRC:2011}
H.-J. Mathes, {The HEAT Telescopes of the Pierre Auger Observatory. Status and
  First Data},  \cite{Abreu:2011ppj}, p.~1.
\newblock \href {http://arxiv.org/abs/1107.4807} {\path{arXiv:1107.4807}}.

\bibitem{Suarez-ICRC:2013}
F.~Suarez, {The AMIGA muon detectors of the Pierre Auger Observatory: overview
  and status},  \cite{ThePierreAuger:2013eja}, p.~4.
\newblock \href {http://arxiv.org/abs/1307.5059} {\path{arXiv:1307.5059}}.

\bibitem{Sanchez-ICRC:2011}
F.~Sanchez, {The AMIGA detector of the Pierre Auger Observatory: an overview},
  \cite{Abreu:2011ppj}, p.~5.
\newblock \href {http://arxiv.org/abs/1107.4807} {\path{arXiv:1107.4807}}.

\bibitem{Platino:2011zz}
M.~Platino, M.~Hampel, A.~Almela, A.~Krieger, D.~Gorbena, et~al., {AMIGA at the
  Auger Observatory: The scintillator module testing system}, JINST 6 (2011)
  P06006.
\newblock \href {http://dx.doi.org/10.1088/1748-0221/6/06/P06006}
  {\path{doi:10.1088/1748-0221/6/06/P06006}}.

\bibitem{Nagano:1991jz}
M.~Nagano, M.~Teshima, Y.~Matsubara, H.~Dai, T.~Hara, et~al., {Energy spectrum
  of primary cosmic rays above 10$^{17}$~eV determined from the extensive air
  shower experiment at Akeno}, J.\ Phys.\ G 18 (1992) 423--442.
\newblock \href {http://dx.doi.org/10.1088/0954-3899/18/2/022}
  {\path{doi:10.1088/0954-3899/18/2/022}}.

\bibitem{Maldera-ICRC:2013}
S.~Maldera, {Measuring the accuracy of the AMIGA muon counters at the Pierre
  Auger Observatory},  \cite{ThePierreAuger:2013eja}, p.~4.
\newblock \href {http://arxiv.org/abs/1307.5059} {\path{arXiv:1307.5059}}.

\bibitem{Wainberg:2014}
O.~Wainberg, et~al., {Digital Electronics for the Pierre Auger Observatory
  AMIGA Muon Counters}, JINST 9 (2014) T04003.
\newblock \href {http://dx.doi.org/10.1088/1748-0221/9/04/T04003}
  {\path{doi:10.1088/1748-0221/9/04/T04003}}.

\bibitem{Supanitsky:2008dx}
A.~Supanitsky, A.~Etchegoyen, G.~Medina-Tanco, I.~Allekotte, M.~G. Berisso,
  et~al., {Underground Muon Counters as a Tool for Composition Analyses},
  Astropart.\ Phys. 29 (2008) 461--470.
\newblock \href {http://arxiv.org/abs/0804.1068} {\path{arXiv:0804.1068}},
  \href {http://dx.doi.org/10.1016/j.astropartphys.2008.05.003}
  {\path{doi:10.1016/j.astropartphys.2008.05.003}}.

\bibitem{Wundheiler-ICRC:2011}
B.~Wundheiler, {The AMIGA muon counters of the Pierre Auger Observatory:
  performance and first data},  \cite{Abreu:2011ppj}, p.~9.
\newblock \href {http://arxiv.org/abs/1107.4807} {\path{arXiv:1107.4807}}.

\bibitem{Dasso:2012vk}
S.~Dasso, H.~Asorey, {The scaler mode in the Pierre Auger Observatory to study
  heliospheric modulation of cosmic rays}, Adv.\ Space Res. 49 (2012)
  1563--1569.
\newblock \href {http://arxiv.org/abs/1204.6196} {\path{arXiv:1204.6196}},
  \href {http://dx.doi.org/10.1016/j.asr.2011.12.028}
  {\path{doi:10.1016/j.asr.2011.12.028}}.

\bibitem{Abreu:2011zza}
P.~Abreu, et~al., {The Pierre Auger Observatory scaler mode for the study of
  solar activity modulation of galactic cosmic rays}, JINST 6 (2011) P01003.
\newblock \href {http://dx.doi.org/10.1088/1748-0221/6/01/P01003}
  {\path{doi:10.1088/1748-0221/6/01/P01003}}.

\bibitem{publicEventDisplay}
Public event explorer, \url{http://auger.colostate.edu/ED/}, accessed 16 Jan
  2015.

\bibitem{Inan:1991}
U.~Inan, T.~Bell, J.~Rodriguez, {Heating and ionization of the lower ionosphere
  by lightning}, Geophys.\ Res.\ Lett. 18 (1991) 705.
\newblock \href {http://dx.doi.org/10.1029/91GL00364}
  {\path{doi:10.1029/91GL00364}}.

\bibitem{Boeck:1992}
W.~Boeck, O.~Vaughan, R.~Blakeslee, B.~Vonnegut, M.~Brook, {Lightning induced
  brightening in the airglow layer}, Geophys.\ Res.\ Lett. 19 (1992) 99.

\bibitem{Fukunishi:1996}
H.~Fukunishi, Y.~Takahashi, M.~Kubota, K.~Sakanoi, U.~Inan, W.~Lyons, {Elves:
  Lightning-Induced Transient Luminous Events in the Lower Ionosphere},
  Geophys.\ Res.\ Lett. 23 (1996) 2157.

\bibitem{Inan:1997}
U.~S. Inan, C.~Barrington-Leigh, S.~Hansen, V.~Glukhov, T.~Bell, R.~Rairden,
  {Rapid lateral expansion of optical luminosity in lightning-induced
  ionospheric flashes referred to as `elves'}, Geophys.\ Res.\ Lett. 24 (1997)
  583.

\bibitem{Newsome:2010}
R.~Newsome, U.~S. Inan, {Free-running ground-based photometric array imaging of
  transient luminous events}, J.\ Geophys.\ Res. 115 (2010) A00E41.
\newblock \href {http://dx.doi.org/10.1029/2009JA014834}
  {\path{doi:10.1029/2009JA014834}}.

\bibitem{Chen:2008}
A.~Chen, et~al., {Global distributions and occurrence rates of transient
  luminous events}, J.\ Geophys.\ Res. 113 (2008) A08306.
\newblock \href {http://dx.doi.org/10.1029/2008JA013101}
  {\path{doi:10.1029/2008JA013101}}.

\bibitem{Tonachini-ICRC:2013}
A.~Tonachini, {Observation of Elves at the Pierre Auger Observatory},
  \cite{ThePierreAuger:2013eja}, p. 112.
\newblock \href {http://arxiv.org/abs/1307.5059} {\path{arXiv:1307.5059}}.

\bibitem{Abarca:2010}
S.~F. Abarca, K.~L. Corbosiero, T.~J. {Galarneau Jr.}, {An evaluation of the
  Worldwide Lightning Location Network (WWLLN) using the National Lightning
  Detection Network (NLDN) as ground truth}, Journal of Geophysical Research
  115 (2010) D18206.
\newblock \href {http://dx.doi.org/10.1029/2009JD013411}
  {\path{doi:10.1029/2009JD013411}}.

\bibitem{vispa}
{V}isual {P}hysics {A}nalysis - {VISPA},
  \url{http://vispa.physik.rwth-aachen.de/}, accessed 16 Jan 2015.

\bibitem{Kahn-Lerche}
F.~D. Kahn, I.~Lerche, Radiation from cosmic ray air showers, Proceedings of
  the Royal Society of London A: Mathematical, Physical and Engineering
  Sciences 289~(1417) (1966) 206--213.
\newblock \href {http://dx.doi.org/10.1098/rspa.1966.0007}
  {\path{doi:10.1098/rspa.1966.0007}}.

\bibitem{Askaryan}
G.~Askaryan, {Excess Negative Charge of an Electron-Photon Shower And Its
  Coherent Radio Emission}, {Soviet Physics JETP} 14 (1962) 441--443.

\bibitem{LOFAR:2013a}
S.~Buitink, et~al., {Shower X\_max determination based on LOFAR radio
  measurements}, Proc.\ 33rd International Cosmic Ray Conference (ICRC), Rio de
  Janeiro, Brazil.

\bibitem{Allan:1971}
H.~Allan, {Radio emission from extensive air showers}, Progress in Elementary
  Particles and Cosmic Ray Physics 10 (1971) 171.

\bibitem{Fegan:2011fb}
D.~J. Fegan, {Detection of elusive Radio and Optical emission from Cosmic-ray
  showers in the 1960s}, Nucl.\ Instrum.\ Meth.\ A 662 (2012) S2--S11.
\newblock \href {http://arxiv.org/abs/1104.2403} {\path{arXiv:1104.2403}},
  \href {http://dx.doi.org/10.1016/j.nima.2010.10.129}
  {\path{doi:10.1016/j.nima.2010.10.129}}.

\bibitem{Falcke:2005tc}
H.~Falcke, et~al., {Detection and imaging of atmospheric radio flashes from
  cosmic ray air showers}, Nature 435 (2005) 313--316.
\newblock \href {http://arxiv.org/abs/astro-ph/0505383}
  {\path{arXiv:astro-ph/0505383}}, \href
  {http://dx.doi.org/10.1038/nature03614} {\path{doi:10.1038/nature03614}}.

\bibitem{Ardouin:2006nb}
D.~Ardouin, A.~Belletoile, D.~Charrier, R.~Dallier, L.~Denis, et~al.,
  {Radioelectric Field Features of Extensive Air Showers Observed with
  CODALEMA}, Astropart.\ Phys. 26 (2006) 341--350.
\newblock \href {http://arxiv.org/abs/astro-ph/0608550}
  {\path{arXiv:astro-ph/0608550}}, \href
  {http://dx.doi.org/10.1016/j.astropartphys.2006.07.002}
  {\path{doi:10.1016/j.astropartphys.2006.07.002}}.

\bibitem{Schellart:2013bba}
P.~Schellart, A.~Nelles, S.~Buitink, A.~Corstanje, J.~Enriquez, et~al.,
  {Detecting cosmic rays with the LOFAR radio telescope}, Astron.\ Astrophys.
  560 (2013) A98.
\newblock \href {http://arxiv.org/abs/1311.1399} {\path{arXiv:1311.1399}},
  \href {http://dx.doi.org/10.1051/0004-6361/201322683}
  {\path{doi:10.1051/0004-6361/201322683}}.

\bibitem{Revenu-ICRC:2011}
B.~Revenu, {Autonomous detection and analysis of radio emission from air
  showers at the Pierre Auger Observatory},  \cite{Abreu:2011ppj}, p.~17.
\newblock \href {http://arxiv.org/abs/1107.4807} {\path{arXiv:1107.4807}}.

\bibitem{Kelley-ICRC:2011}
J.~Kelley, {AERA: the Auger Engineering Radio Array},  \cite{Abreu:2011ppj},
  p.~13.
\newblock \href {http://arxiv.org/abs/1107.4807} {\path{arXiv:1107.4807}}.

\bibitem{Acounis:2012dg}
P.~Abreu, et~al., {Results of a self-triggered prototype system for
  radio-detection of extensive air showers at the Pierre Auger Observatory},
  JINST 7 (2012) P11023.
\newblock \href {http://arxiv.org/abs/1211.0572} {\path{arXiv:1211.0572}},
  \href {http://dx.doi.org/10.1088/1748-0221/7/11/P11023}
  {\path{doi:10.1088/1748-0221/7/11/P11023}}.

\bibitem{Aab:2014esa}
A.~Aab, et~al., {Probing the radio emission from air showers with polarization
  measurements}, Phys.\ Rev.\ D 89 (2014) 052002.
\newblock \href {http://arxiv.org/abs/1402.3677} {\path{arXiv:1402.3677}},
  \href {http://dx.doi.org/10.1103/PhysRevD.89.052002}
  {\path{doi:10.1103/PhysRevD.89.052002}}.

\bibitem{Schellart:2014oaa}
P.~Schellart, S.~Buitink, A.~Corstanje, J.~Enriquez, H.~Falcke, et~al.,
  {Polarized radio emission from extensive air showers measured with LOFAR},
  JCAP 1410~(10) (2014) 014.
\newblock \href {http://arxiv.org/abs/1406.1355} {\path{arXiv:1406.1355}},
  \href {http://dx.doi.org/10.1088/1475-7516/2014/10/014}
  {\path{doi:10.1088/1475-7516/2014/10/014}}.

\bibitem{Abreu:2012pi}
P.~Abreu, et~al., {Antennas for the Detection of Radio Emission Pulses from
  Cosmic-Rays}, JINST 7 (2012) P10011.
\newblock \href {http://arxiv.org/abs/1209.3840} {\path{arXiv:1209.3840}},
  \href {http://dx.doi.org/10.1088/1748-0221/7/10/P10011}
  {\path{doi:10.1088/1748-0221/7/10/P10011}}.

\bibitem{Gorham:2007af}
P.~Gorham, N.~Lehtinen, G.~Varner, J.~Beatty, A.~Connolly, et~al.,
  {Observations of Microwave Continuum Emission from Air Shower Plasmas},
  Phys.\ Rev.\ D 78 (2008) 032007.
\newblock \href {http://arxiv.org/abs/0705.2589} {\path{arXiv:0705.2589}},
  \href {http://dx.doi.org/10.1103/PhysRevD.78.032007}
  {\path{doi:10.1103/PhysRevD.78.032007}}.

\bibitem{Allison-ICRC:2011}
P.~Allison, {Microwave detection of cosmic ray showers at the Pierre Auger
  Observatory},  \cite{Abreu:2011ppj}, p.~25.
\newblock \href {http://arxiv.org/abs/1107.4807} {\path{arXiv:1107.4807}}.

\bibitem{FacalSanLuis:2013qza}
P.~Facal San~Luis, {Status of the program for microwave detection of cosmic
  rays at the Pierre Auger observatory}, EPJ Web Conf. 53 (2013) 08009.
\newblock \href {http://dx.doi.org/10.1051/epjconf/20135308009}
  {\path{doi:10.1051/epjconf/20135308009}}.

\bibitem{AlvarezMuniz:2012ew}
J.~Alvarez-Mu\~{n}iz, E.~Amaral~Soares, A.~Berlin, M.~Bogdan,
  M.~Boh\'{a}\v{c}ov\'{a}, et~al., {The MIDAS telescope for microwave detection
  of ultra-high energy cosmic rays}, Nucl.\ Instrum.\ Meth.\ A 719 (2013)
  70--80.
\newblock \href {http://arxiv.org/abs/1208.2734} {\path{arXiv:1208.2734}},
  \href {http://dx.doi.org/10.1016/j.nima.2013.03.030}
  {\path{doi:10.1016/j.nima.2013.03.030}}.

\bibitem{AlvarezMuniz:2012dx}
J.~Alvarez-Mu\~{n}iz, A.~Berlin, M.~Bogdan, M.~Boh\'{a}\v{c}ov\'{a},
  C.~Bonifazi, et~al., {A Search for Microwave Emission From Ultra-High Energy
  Cosmic Rays}, Phys.\ Rev.\ D 86 (2012) 051104.
\newblock \href {http://arxiv.org/abs/1205.5785} {\path{arXiv:1205.5785}},
  \href {http://dx.doi.org/10.1103/PhysRevD.86.051104}
  {\path{doi:10.1103/PhysRevD.86.051104}}.

\bibitem{ThePierreAuger:2013eja}
A.~Aab, et~al., {The Pierre Auger Observatory: Contributions to the 33rd
  International Cosmic Ray Conference (ICRC 2013)}, Proc.\ 33rd International
  Cosmic Ray Conference (ICRC) 2013, Rio de Janeiro, Brazil\href
  {http://arxiv.org/abs/1307.5059} {\path{arXiv:1307.5059}}.

\bibitem{Abreu:2011pj}
P.~Abreu, et~al., {The Pierre Auger Observatory I: The Cosmic Ray Energy
  Spectrum and Related Measurements}, Proc.\ 32nd International Cosmic Ray
  Conference (ICRC) 2011, Beijing, China\href {http://arxiv.org/abs/1107.4809}
  {\path{arXiv:1107.4809}}.

\bibitem{Abreu:2011ppj}
P.~Abreu, et~al., {The Pierre Auger Observatory V: Enhancements}, Proc.\ 32nd
  International Cosmic Ray Conference (ICRC) 2011, Beijing, China\href
  {http://arxiv.org/abs/1107.4807} {\path{arXiv:1107.4807}}.

\end{thebibliography}

\end{document}